\documentclass{article}

\usepackage[margin=0.5in]{geometry}
\usepackage{amsmath}
\usepackage{color}
\usepackage{graphicx}
\usepackage{amssymb}

\newcommand{\bt}[1]{{\mathbf #1}}
\newcommand{\pd}[2]{\frac{\partial #1}{\partial #2}}
\newcommand{\dd}[2]{\frac{d #1}{d #2}}
\newcommand{\ptwod}[2]{\frac{\partial^2 #1}{\partial #2^2}}
\newcommand{\dtwod}[2]{\frac{d^2 #1}{d #2^2}}
\newcommand{\xv}{\mathbf{x}}
\newcommand{\w}{\omega}
\newcommand{\kv}{\mathbf{k}}
\newcommand{\pv}{\mathbf{p}}
\newcommand{\qv}{\mathbf{q}}
\newcommand{\rv}{\mathbf{r}}
\newcommand{\vv}[1]{{\bf #1}}

\begin{document}

\title{Rossby and Drift Wave Turbulence and Zonal Flows: the Charney-Hasegawa-Mima model and its extensions}
\maketitle

\begin{center}
\Large{
\author{Colm Connaughton\footnote{Mathematics Institute, University of Warwick, Gibbet Hill Road, Coventry CV4 7AL, United Kingdom\\
Centre for Complexity Science, University of Warwick, Gibbet Hill Road, Coventry CV4 7AL, United Kingdom\\
Okinawa Institute of Science and Technology Graduate University, 1919-1 Tancha Onna-son, Okinawa 904-0495, Japan\\
Kavli Institute for Theoretical Physics, Kohn Hall, University of California Santa Barbara, CA 93106, United States}, Sergey Nazarenko\footnote{Mathematics Institute, University of Warwick, Gibbet Hill Road, Coventry CV4 7AL, United Kingdom\\ Laboratoire SPHYNX, Service de Physique de l'Etat Condense, DSM, IRAMIS, CEA, Saclay, CNRS URA 2464, 91191, Gif-sur-Yvette, France} and Brenda Quinn\footnote{School of Mechanical Engineering, Tel-Aviv University, Israel } }
}
\end{center}
\vspace{5mm}

\begin{abstract}
A detailed  study of  the Charney-Hasegawa-Mima model and its extensions  is presented.  These simplest nonlinear partial differential equations suggested for both Rossby waves in the atmosphere and also drift waves in a magnetically-confined plasma  exhibits some remarkable and nontrivial properties, which in their qualitative form survive in more realistic and complicated models, and as such form a conceptual basis for understanding the turbulence and zonal flow dynamics in real plasma and geophysical systems.
Two idealised  scenarios of generation of zonal flows by small-scale turbulence are explored: a modulational instability and turbulent cascades.
A detailed study of the generation of zonal flows by the modulational instability reveals that the dynamics of this zonal flow generation mechanism differ widely depending on the initial degree of nonlinearity.  The jets in the strongly nonlinear case further roll up into K\'arm\'an-like vortex streets and saturate, while for the weaker nonlinearities, the growth of the unstable mode reverses and the system oscillates between a dominant jet, which is slightly inclined to the zonal direction, and a dominant primary wave.    
%Some of these characteristics are captured by truncated models. 
A numerical proof is provided for an the extra invariant in Rossby and drift wave turbulence -- {\em zonostrophy}.  While the theoretical derivations of this invariant stem from the wave kinetic equation which assumes weak wave amplitudes, it is shown to be relatively well-conserved for higher nonlinearities also.  Together with the energy and enstrophy, these 
three invariants cascade into anisotropic sectors in the $k$-space as predicted by the Fj\o rtoft argument. The cascades are characterised by the { zonostrophy} pushing the energy to the zonal scales.
A small scale instability forcing applied to the model has demonstrated the well-known drift wave -- zonal flow feedback loop.  The drift wave turbulence is generated from this primary instability.  The zonal flows are then excited by either one of the generation mechanisms, extracting energy from the drift waves as they grow.  Eventually the turbulence is completely suppressed and the zonal flows saturate.  The turbulence spectrum is shown to diffuse in a manner which has been mathematically predicted.  
The insights gained from this simple model could provide a basis for equivalent studies in more sophisticated plasma and geophysical fluid dynamics models in an effort to fully understand the zonal flow generation, the turbulent transport suppression and the zonal flow saturation processes in both the plasma and geophysical contexts as well as other wave and turbulence systems where order evolves from chaos.
\end{abstract}

\newpage
\tableofcontents

%%%%%%%%%%%%%%%%%%%%%%%%%%%%%%%%%%%%%%%%%%%%%%%%%%%%%%%%%%%%%%%%%%%%%%%%%%%%%%%%%%%%%%%%%%%%%%%%%%%%
\section{Introduction}
\label{sec-intro}

%ADDITIONAL REFERENCES: \cite{parker_zonal_2013} 
\subsection{Wave turbulence and zonal flows}
\label{intro_general}

The term {\em wave turbulence} is used to mean the spectral transport of energy or other conserved quantities resulting from mode coupling in an ensemble of nonlinear dispersive waves. We recognize at the outset that in some sections of the literature, the term turbulence is more narrowly interpreted to mean vortex-vortex interactions in the Navier-Stokes equations and its immediate derivatives. From this point of view, the term wave turbulence is oxymoronic since waves and vortices are conceptually separate flow structures (even if is not so obvious how to separate them in practice) and by definition only the latter can exhibit turbulence. Since this review is intended to be of interest to readers from diverse backgrounds, we ask the reader to set aside such questions of terminology and focus on the fact that spectral transport of conserved quantities is physically important in
a variety of different systems. Whether this spectral transport occurs due to vortex-vortex interactions, to wave-wave interactions or to a mixture of the two is largely immaterial in the sense that many of the concepts developed to understand vortex turbulence such as Richardson's notion of an energy cascade, Kolmogorov's 1941 theory ~\cite{Frisch1995} and Kraichnan's extension \cite{Kraichnan1967} to describe the inverse energy cascade in two dimensions are equally applicable to other forms of spectral transport. In particular, a lot is known about cascades mediated by wave-wave interactions, particularly in the limit of weakly nonlinear wave interactions where considerable analytic progress has been made \cite{Zakharov1992,Nazarenko2011}. Another important concept from the hydrodynamic turbulence literature which is very generally applicable and is of crucial importance in much of what follows is the notion of the {\em locality} of a cascade. A cascade is said to be local if the properties of the cascade in the inertial range are insensitive to the values of the forcing and 
dissipation scales provided that they are widely enough separated in scale. Physically it means that the flux at a particular scale is mediated primarily by interactions with comparable scales. It would probably be clearer to refer to this property as {\em scale-locality} but we shall follow the convention in the literature and simply refer to {\em locality}. The locality of the energy cascade is an essential ingredient of Kolmogorov's theory of hydrodynamic turbulence and is frequently discussed although in  hydrodynamic turbulence; it is rare to find a situation in which locality fails. In wave turbulence cascades, on the other hand, it is quite common to find physically interesting systems in which the assumption of a local cascade fails.

A zonal flow is a coherent, banded structure with a characteristic jet-like velocity profile which is commonly found in two-dimensional and quasi two-dimensional fluid flows subject to large scale anisotropy. We have in mind two particular examples. The first is geophysical fluid dynamics (GFD) where quasi-two-dimsionality and anisotropy is the result of the large aspect ratio and imposed rotation of planetary flows. The second is the dynamics of strongly magnetised plasmas where quasi-two-dimensionality and anisotropy are the result of a strong, externally-imposed magnetic field. Some well-known examples of zonal flows in geophysical fluid dynamics (GFD) include the rings or equatorial jets around Saturn and the belts and zones of the Jovian atmosphere~\cite{Simon1999,Sanchez2000,Galperin2004,galperin_cassini_2014}.  The latitudinally-aligned bands observed on the giant planets are zonal flows, visible due to the ammonia cloud formations.  Far from being solely extra-terrestrial phenomena however, zonal flows also exist near the 
tropopause in the Earth's atmosphere such as the polar and subtropical jet streams~\cite{Lewis1988} which eastward-bound aircraft take advantage of  and in the Earth's oceans~\cite{Maximenko2008} such as the Antarctic circumpolar current. Zonal flows are also striking features of plasma flows in strongly magnetised plasmas, particularly in magnetic confinement fusion devices such as tokomaks. In this latter context, they are observed as radially localised, toroidally symmetric, strongly sheared flows in the poloidal direction. The tendency to form zonal flows is one of the key common features which leads to important overlaps between research in GFD and fusion plasma physics.

A second key common feature linking the two problems is ability of large-scale gradients to support waves. In the GFD context, these waves are called Rossby waves and are supported by the latitudinal gradient in the coriolis force as one moves from equator to pole. They are named after Carl Gustaf Rossby, who in 1939 \cite{Rossby1939,Rossby1940} used observations of large recurring eddies in the atmosphere over the USA and analysis of data for the Earth's northern hemisphere to determine the velocity of these large-scale flow patterns in the zonal direction. It was found that they propagate westward in the atmosphere with very low frequency.  They are characterised by a very long wavelength of the order $10^5$~m, an amplitude of $10^1$m and modenumber of order 5 in the Earth's atmosphere and their velocity is proportional to the gradient of the Coriolos force.  However they are more difficult to observe in the oceans since they exist in the thermocline where they are characterised by a very long wavelength, of the order $ 5 \times 10^5$~m, and a relatively low amplitude $\sim 5\times 10^1$~m.  These waves and their stability properties are largely responsible for the unpredictable mid-latitude weather systems experienced on earth due to the creation of cyclones and anticyclones. In plasma physics, there are many types of waves. The waves most closely analogous to Rossby waves are called drift waves and they are supported on large-scale density gradients. They are small, low frequency oscillations which result from the balancing of the parallel and drift dynamics in a neutral magnetically-confined plasma~\cite{Rudakov1961}.  
 The electrons with higher energy and velocity contribute most to the dynamics parallel to the magnetic field.  In the perpendicular to the magnetic field plane, as they orbit, the ions and electrons are accelerated then decelerated in the direction of the electric field so that there is a slight velocity imbalance between opposite sides of the orbit path, resulting in the particle drifting perpendicular to the electric field -- this is so-called ${\bf E} \times {\bf B}$ drift.  Further correction to this drift is due to the ion inertia: it leads to so-called polarisation drift. The drift motion combined with the plasma inhomogeneity brings about existence of a wave motion called the drift wave.

In both GFD and plasma physics, the underlying equations of motion contain nonlinearities of hydrodynamic type. Therefore if waves reach sufficiently high amplitudes, for example due to an instability, waves can interact nonlinearly. This leads to wave turbulence and spectral transport of energy via cascades. Drift wave turbulence arises when the drift waves are unstable to steep gradients of the mean plasma profile (eg. drift-dissipative instability or ion-temperature gradient instability), whereas in the GFD context, waves can be unstable due to baroclinic instability. In both the GFD and plasma physics contexts, large-scale, coherent zonal flows are seen to coexist with small-scale, incoherent wave-turbulence fluctuations. It is generally accepted that large-scale zonal flows in tokamaks are actually generated by drift wave turbulence~\cite{Diamond2005} in the sense that they grow by nonlinear energy transfer from wave turbulence in the core of the plasma. They can suck the energy from the drift wave packets as they grow and reduce the level of drift wave turbulence in the process. Zonal flows are therefore the result of self-organisation of a turbulent state to a coherent flow. A simliar mechanism may have been responsible for the generation of the zonal flows seen in planetary oceans and atmospheres. Understanding the generation and maintenance of zonal flows and the interplay between zonal flows and wave turbulence is therefore key to understanding the dynamics of both the atmospheres and of magnetically-confined plasmas, in particular the spectral energy transfer in these systems. This transfer is now believed to be primarily nonlocal in nature.

The main reason why there is currently a large amount of practical scientific interest in the topic of turbulence-zonal flow interactions is the fact 
that the interplay between incoherent turbulence and coherent zonal flows alluded to above links the spectral and spatial transport properties of the 
flow. Small-scale turbulence is well known to enhance spatial transport, a process often referred to as eddy-diffusion. Zonal flows, on the other hand, 
act as spatial transport barriers since it is difficult to diffuse in the direction perpendicular to a zonal flow. Therefore, if energy is transferred 
spectrally from small-scale turbulence to large-scale zonal flows, spatial transport is expected to be reduced through reduced eddy diffusion and 
enhanced transport barriers and vice versa. Questions related to the parametrisation and control of spatial transport lie at the heart of several key 
contemporary problems in both GFD and plasma physics which we now summarise briefly for the purposes of motivation. 

In fusion plasmas, the primary of these contemporary problems is the problem of plasma confinement. 
Achieving plasma ignition requires a sufficiently high plasma density, temperature and confinement time being achieved simultaneously before the plasma loses its energy.  The minimum value of the product of these three quantities provides a revised version of the Lawson criterion~\cite{Lawson1957}, which must be met for sustained confinement~\cite{Dendy1993}. 
Although construction of ITER, the next step in fusion research, is already under way in Cadarache, France, there are still some underlying fundamental
issues which need to be resolved for successful ignition to occur.  One of those issues is related to spatial transport.  Early experiments established 
that the transport of plasma from the hot core to the edge region in a tokamak is much greater than can be explained by classical collisional or 
neoclassical trapped particle theory~\cite{Dendy1993}. This so-called  anomalous transport initially baffled engineers and scientists and greatly 
reduced the confinement time. The origin of anomalous transport was traced to small scale turbulence, primarily drift wave turbulence resulting from 
small scale instabilities. Fresh impetus was given to fusion research however when it was discovered experimentally~\cite{Wagner1982} in the Axially 
Symmetric Divertor EXperiment (ASDEX) divertor tokamak near Munich in Germany, that this anomalous transport was not always observed in the discharges.  
Two separate regimes were defined as L-type and H-type referring to low and high values respectively of the plasma $\beta_p$ which measures the ratio of 
the plasma pressure to the magnetic pressure.  Under certain conditions, the plasma discharge would undergo an LH transition, with enhanced plasma 
confinement times and reduced anomalous transport.  This is generally believed to be due to the generation of zonal flows~\cite{Diamond2005}.  These 
zonal flows provide transport barriers and their existence is in fact crucial in regulating the turbulence from the small scale instabilities from which 
they stem~\cite{Manfredi2001}, further strengthening the necessity to fully understand drift wave -- zonal flow turbulence.  If the ITER experiment is 
successful, fusion energy could be commercially available towards the middle of this century.

In the context of GFD, the understanding of the interaction between Rossby wave turbulence and zonal flows is deemed crucial to fully 
comprehend the bigger picture of the overall atmosphere-ocean dynamics on Earth and on planets such as Jupiter and Saturn~\cite{McIntyre2008}. A 
clear and concise understanding of the atmosphere-ocean dynamics is highly desirable, particularly for climate applications. Furthermore, it is 
believed that the cyclones and anticyclones inherent of  the weather in midlatitudes are due to the baroclinic instability of the atmospheric jet 
stream indicating that a greater understanding of such features 
could further improve the  parametrisation used in weather forecast models. A feedback mechanism similar to the one described above for the interplay between zonal flows and wave turbulence in tokamaks also exists for Rossby waves in the atmosphere. It is sometimes referred to as the {\em barotropic governor} in that the barotropic flow controls the level of turbulent behaviour~\cite{James1987}.  Barriers to mixing and transport are evident in the atmosphere at the edge of the Arctic and Antarctic winter polar vortices.  The concentration of certain gases, which are otherwise evenly distributed over the poles, can have large gradients at the edge of the polar vortex, which creates a barrier to mixing towards the equator~\cite{Hartmann1989,Terry2000}.  Likewise, the equatorial jet in the stratosphere acts as a barrier, reducing polewards transport of volcanic aerosols.  By comparison, in the troposphere where the equatorial jet does not reach, the volcanic aerosol is more evenly distributed rather than more concentrated in equatorial latitudes~\cite{Terry2000,Trepte1992}.

Theoretically, many important aspects of the interactions between zonal flows and wave turbulence can be understood in the context of a simple model system common to both fields which we refer to as the Charney--Hasegawa--Mima (CHM) model and its extensions. This model, which we introduce in the next section, is a single two-dimensional scalar partial differential equation (PDE), which expresses the local conservation of a quantity known as potential vorticity. In GFD, it is sometimes referred to as the (equivalent) barotropic potential vorticity (PV) equation. In the light of the discussion above, the CHM model suffers from several important limitations which severely restrict its usefulness as a quantitative model. Firstly, it does not possess any intrisic instabilities which can lead to the generation of Rossby/drift wave turbulence in the first instance. Secondly, as a model of plasma turbulence, at least if taken literally, it does not produce any plasma transport. One the other hand, the CHM model makes up for these deficiencies by the fact that it is, to a large extent, analytically tractable. Many questions about the interplay between zonal flows and wave turbulence can be answered in detail, thereby providing a conceptual framework for understanding such interations in more realistic models where quantitative accuracy is often achieved at the cost of reduced ability for analytic understanding and increased reliance on numerical simulations. In it is the CHM model in the framework of which the drift wave -- zonal flow feedback effect, which is presently believed to be the mechanism of the LH transitions, was first discovered in~\cite{Nazarenko051990,Nazarenko081990,Nazarenko1991}. This review is devoted to the analysis of zonal flows and Rossby/drift wave turbulence in the CHM model and its immediate extensions, an analysis which turns out to be quite non-trivial, despite the seemingly elementary nature of the models. Such analyses should be viewed as complementary to more detailed analyses and numerical simulations of more realistic models.

\subsection{Summary of review contents}
\label{intro_summary}

This review deals with the spectral aspects of the interactions between turbulence and zonal flows in the CHM model and closely related Extended Hasegawa--Mima (EHM), Hasegawa--Wakatani and Extended Hasegawa--Wakatani (HW/EHW) models. Since we are focusing on spectral transport we mostly consider homogeneous systems and say relatively little about effects arising as a result of spatial inhomogeneity which is certainly important in many applications. The subject is approached primarily from the perspective of wave turbulence theory with a particular emphasis on nonlocal cascades which turn out to play a prominent role in this system. We draw together a number of different theoretical concepts from the wave turbulence literature into a unified and coherent narrative in order to explain their relevance to the problem of zonal flow generation and maintenance. We particularly emphasise the ideas of modulational instability and inverse energy cascade~\cite{SmolyakovMalk2000} mediated by resonant wave interactions. These are two of the main candidate mechanisms for zonal flow generation, applicable to both Rossby wave and drift wave turbulence. These mechanisms need not be mutually exclusive.  Before summarising the structure of this manuscript let us briefly mention some of the topics which are not discussed. Firstly, we do not attempt to give a comprehensive description of wave turbulence theory in general. Such an account can be found in the reviews by Newell et al.~\cite{newell_wave_2001,Newell2011} and the recent book by Nazarenko ~\cite{Nazarenko2011}. We do not discuss the statistical mechanics approach to two-dimensional fluid dynamics which can be applied to the structure of large-scale zonal flows. This has been covered in considerable detail in a recent review by Bouchet and Venaille ~\cite{bouchet_statistical_2012}. While we adopt notation which is more common in the GFD literature, we retain a connection to both GFD and plasma physics applications throughout the manuscript since the focus is on common features of the CHM model which are of interest to both disciplines. We do not attempt to delve deeply into more detailed models. For more detailed discussion of zonal flow -- wave turbulence interactions in GFD see the reviews by McIntyre ~\cite{McIntyre2008} and Dritschel and McIntyre ~\cite{Dritschel2008} the references therein. For a detailed discussion of zonal flows and wave turbulence in fusion plasmas and more realistic models of their interaction see the review by Diamond et al. ~\cite{Diamond2005}.

The layout of the manuscript is as follows. In Sec.~\ref{sec-CHMAndFriends} with introduce the CHM model and summarise its properties. We establish here various 
notational conventions which we will use throughout. In this section we also introduce the Extended Hasegawa--Mima, Hasegawa--Wakatani and the Extended  Hasegawa--Wakatani models which 
are closely related models from the plasma physics literature which improve slightly on the CHM model by incorporating more a realistic response of the 
electron density to fluctuations in the plasma density. In Sec.~\ref{sec-WT} we provide a pedagogical introduction to wave turbulence theory as 
applied to the specific case of the CHM equation. This section provides a detailed derivation of the collision operator which appears on the right hand 
side of the wave kinetic equation and describes the spectral transport of energy and potential enstrophy due to resonant wave-wave interactions in the 
limit of weakly nonlinear wave turbulence. This section also discusses wave cascades, the Kolmogorov-Zakharov spectra and the important issue of 
nonlocality. To the extent that Sec.~\ref{sec-WT} discusses spectral transport in a spatially continuous framework, Sec.~\ref{sec-lowDimensionalModels} 
provides a complementary summary of finite dimensional models of spectral energy transport. This section includes a 
pedagogical review of some of the classical results about the dynamics of resonant triads as well as more recent results 
on the interplay between discreteness and resonances in finite wave systems. Sec.~\ref{sec-MI} and 
Sec.~\ref{sec-numerics} treat the subject of modulational instability in some depth, the former dealing with linear stability theory 
and the latter presenting what is known, mostly from numerical simulations, about the nonlinear development of the 
instability. These sections emphasise the ability of the modulational instability to directly couple small scale-wave to 
large-scale zonal flows, an archetypal example of spectrally nonlocal interaction between the two. Cascades are discussed 
in section~\ref{triple_cascade}, particularly the inverse cascade which transports the energy from the small-scale 
turbulence to large-scale zonal flows in a step-by-step process, similar to the inverse cascade in 2D NS turbulence 
\cite{Fjortoft1953,Kraichnan1967}. This section also provides detailed discussion of the slightly mysterious third 
invariant (in addition to the energy and potential enstrophy) of kinetic Rossby wave turbulence, sometimes referred to as 
the zonostrophy. The pivotal role played by the conservation of zonostrophy in organising the zonation process in the 
cascade scenario is demonstrated. It is shown that equivalent invariants exist 
in discrete turbulence systems where it is only one of many additional quadratic invariants~\cite{Harper2012}. Finally, Sec.~\ref{WT_ZF_loop} discusses the theory of the 
wave turbulence -- zonal flow feedback loop in the context of the forced CHM, EHM, HW and EHW models. In this section it is shown 
how nonlocal wave turbulence theory can provide a consistent, weekly nonlinear scenario for the generation and 
maintenance of zonal flows from a small-scale instability. Here, the weakly nonlinear theory is extended to the strongly nonlinear case, and numerical simulations validating the theoretical predictions are presented. This section is supported by an appendix which provides 
further detail on how nonlocal wave turbulence theory can be used to calculate spectral transport properties of 
Rossby/drift wave turbulence. We finish with a short summary and conclusions.

%%%%%%%%%%%%%%%%%%%%%%%%%%%%%%%%%%%%%%%%%%%%%%%%%%%%%%%%%%%%%%%%%%%%%%%%%%%%%%%%%%%%%%%%%%%%%%%%%%%%
\section{The Charney-Hasegawa-Mima equation and related models}
\label{sec-CHMAndFriends}
\subsection{The Charney--Hasegawa--Mima equation}
\label{sec-CHM}
As mentioned above, geophysical quasi-geostrophic (QG) turbulence and plasma drift turbulence share several important structural features and are thus frequently discussed together~\cite{HasegawaMac1979,Horton1994,Diamond2005}, in particular when discussing zonal flow formation. Both systems are quasi-two-dimensional and anisotropic, exhibit direct and inverse cascades, contain a mixture of wave and vortex dynamics and tend to spontaneously form large scale zonal flows. As a result of these common features, some basic linear and nonlinear properties of these two systems can be described by the same PDE, known as the Charney equation \cite{Charney1948} in the geophysical context and known as the Hasegawa-Mima equation \cite{HasegawaMima1978} in the plasma context.  The equation is increasingly referred to as the Charney-Hasegawa-Mima (CHM) equation. We shall adopt this term while noting that the same equation is also referred to as the barotropic potential vorticity equation or the equivalent barotropic 
equation in the geophysical fluid dynamics literature. The equivalence of the Charney and Hasegawa-Mima equations was first pointed out by Hasegawa and Maclennan~\cite{HasegawaMac1979,Horton1994}. 
%The origin and derivation of the equation is outlined  in Appendix~\ref{apxCHM}. 
It is usually written as as evolution equation for the stream-function, $\psi(\xv,t)$,
\begin{equation}
\label{eq-CHMx}
\frac{\partial}{\partial t}(\nabla^2\psi-F\psi) + \beta \frac{\partial \psi}{\partial x} +\frac{\partial \psi}{\partial x} \frac{\partial \nabla^2\psi}{\partial y} - \frac{\partial \psi}{\partial y}\frac{\partial \nabla^2\psi}{\partial x}  = 0\,.
\end{equation}
It is easy to show that this equivalent to an advection equation for a quantity, $q$, called the potential vorticity:
\begin{equation}
\frac{D\,q}{D\,t}=0,
\end{equation}
where $\frac{D}{D\,t} = \pd{}{t} + \mathbf{v}\cdot \nabla$ is the advective derivative associated with the velocity field $\mathbf{v} = (v_x, v_y) = \left(-\pd{\psi}{y}, \pd{\psi}{x}\right)$ and the potential vorticity, $q$, is given by
\begin{equation}
\label{eq-PV}
q = \nabla^2 \psi + \beta\,y - F\,\psi.
\end{equation}
This form makes clear the similarity with the two-dimensional Euler equation. Like the two-dimensional Euler equation, the CHM equation conserves two quadratic invariants, the energy and the potential enstrophy:
\begin{align}
\label{energyE}
E =& \frac{1}{2}\int[(\nabla\psi)^2+F\psi^2]\, d\xv\hspace{1.0cm}&&\mbox{(energy),}\\
\label{enstrophyQ}Q =& \frac{1}{2}\int[\nabla^2\psi - F \psi ]^2\, d\xv\hspace{1.0cm}&&\mbox{(potential enstrophy).}
\end{align}

The difference between the CHM equation and the two-dimensional Euler equation is the presence of the linear term, $\beta \frac{\partial \psi}{\partial x}$. As a consequence of this linear term, the physics of the CHM equation differs from the physics of the two-dimensional Euler equation in two fundamental respects. The first is that the linear term allows the system to support wave motions which are absent in the Euler equation. The second is that it is possible for the linear term to be large compared with the nonlinear ones. The CHM equation, unlike the two-dimensional Euler equation, therefore possesses a weakly nonlinear limit. In the weakly nonlinear limit the system is dominated by waves with the nonlinearity acting as a weak perturbation although as we shall see the cumulative effect of the nonlinearity over time can be large. To quantify the strength of the nonlinearity, introduce a characteristic length scale, $L$, a characteristic wave amplitude, $A$, and a characteristic timescale, $(\beta\, L)^{
-1}$. These are used to define dimensionless variables, $\tilde{\xv}$, $\tilde{t}$ 
and $\tilde{\psi}$ according to $\xv = L \tilde{\xv}$, $t = \tilde{t}/(\beta\,L)$, $\psi = A\,\tilde{\psi}$ and a dimensionless $F$ parameter, $F = L^{-2} \tilde{F}$. Immediately dropping the tildes, the nondimensional version of Eq.~(\ref{eq-CHMx}) is:
\begin{equation}
\label{eq-CHMx2}
\frac{\partial}{\partial t}(\nabla^2\psi-F\psi) + \frac{\partial \psi}{\partial x} +M\left[\frac{\partial \psi}{\partial x} \frac{\partial \nabla^2\psi}{\partial y} - \frac{\partial \psi}{\partial y}\frac{\partial \nabla^2\psi}{\partial x}\right]  = 0,
\end{equation}
where, following the notation adopted by Gill \cite{Gill1974}, we denote the Rossby number by 
\begin{equation}
\label{eq-RossbyNumber}
M = \frac{A}{\beta\,L^3}.
\end{equation}
Eq.~(\ref{eq-CHMx2}) has {\em exact} monochromatic wave solutions,
\begin{equation}
\label{eq-RossbyWaveSolution}
\psi(\xv,t) = A\, \sin\left(\kv.\xv - \w_\kv t\right),
\end{equation}
having dispersion relation
\begin{equation}
\label{eq-RossbyWavedispersion}
\w_\kv=-\frac{k_x}{k^2 + F}.
\end{equation}
These are the Rossby or drift waves referred to above.
The time and physical scales associated with each of the wave types differ by many orders of magnitude and the linear wave frequency of the Rossby wave is much smaller than the Coriolis frequency just as the drift wave frequency is much smaller than the ion cyclotron frequency. The analogy between the drift and the Rossby wave is given in table~\ref{Rossby_drift_analogy} with some typical orders of magnitudes of the variables~\cite{Horton1994}.

\begin{table*}
\begin{center}
\caption{Analogy between the drift wave and the Rossby wave}
\label{Rossby_drift_analogy}
\begin{tabular}{@{\extracolsep{\fill}}l l l}
\hline
Drift wave    &   Rossby wave   \\ \hline \hline
Electrostatic potential $\phi$  	 &   Variable fluid depth $\eta$ & \\
Background density $n_0$	 &   Average depth $\bar{H}$  & \\
${\bt E}\times{\bt B}$ drift   		 &   Geostrophic flow  & \\
Wavelength $\approx 1 \times 10^{-3}$m	& Wavelength $\approx 2 \times 10^{6}$m\\
Period $\approx 1 \times 10^{-3}$s & Period $\approx 5$ days\\
Ion cyclotron frequency $\omega_{ci} \approx 10^8s^{-1}$	 &   Coriolis parameter $f \approx 10^{-4} s^{-1} $& \\
$\beta=\frac{\partial}{\partial y}\ln n_0$&  $\beta=\frac{\partial f}{\partial y}$ & \\
Larmor radius $\rho_i=\frac{1}{\omega_{ci}}\sqrt{\frac{T_e}{m_i}} \approx 10^{-3}$m & Rossby radius $R = \frac{\sqrt{g\bar{H}}}{f} \approx 2\times10^6$m &\\
Drift velocity & Rossby velocity &\\
Dispersion relation $\omega_{\bt k} = -\frac{\beta k_x \rho^2}{1+\rho_i^2k^2}$ & Dispersion relation $\omega_{\bt k} = -\frac{\beta k_x}{k^2+F}$
 \end{tabular}
\end{center}
\end{table*}

Due to the nonlinearity of Eq.~(\ref{eq-CHMx}) the superposition principle does not hold. Superpositions of Rossby waves are generally not exact solutions and usually interact nonlinearly to generate extra modes and exchange energy with them. This review is about turbulence in the CHM equation, meaning the statistics of energy transfer between different scales of motion. The mechanism of energy transfer is qualitatively different in the weakly nonlinear regime as compared to the strongly nonlinear regime. In both cases, however, exchange of energy between scales of motion is most conveniently studied in Fourier space. In this review we adopt the following normalization convention for the Fourier transform pair in two dimensions:
\begin{equation}
\label{eq-FourierTransform}
\hat{\psi}_{\kv}(t) = \frac{1}{2\pi}\int \psi(\xv,t)\ \mathrm{e}^{-\mathrm{i}\,\kv \cdot \xv} \, d \xv
\hspace{1.0cm}
\psi(\xv,t) =\frac{1}{2\pi} \int \hat{\psi}_\kv(t)\ \mathrm{e}^{\mathrm{i}\, \kv \cdot \xv} \, d \kv.
\end{equation}
Since the $\psi(\xv,t)$ is a real field, its Fourier transform has the symmetry $\hat{\psi}_\kv^* = \hat{\psi}_{-\kv}$. Taking the Fourier transform of Eq.~(\ref{eq-CHMx}) gives
\begin{equation}
\label{eq-CHMk} 
\pd{\hat{\psi}_\kv}{t} = - i\, \omega_\kv\, \hat{\psi}_\kv  
 + \int d\kv_1 d\kv_2\,  T^\kv_{\kv_1\,\kv_2}\, \hat{\psi}_{\kv_1}\, \hat{\psi}_{\kv_2}\, \delta^\kv_{\kv_1\,\kv_2},
\end{equation}
where $\w_\kv$ is given by Eq.~(\ref{eq-RossbyWavedispersion}) and the nonlinear interaction coefficient, $T^\kv_{\kv_1\,\kv_2}$, is expressed as
\begin{equation}
\label{eq-Tpqr}
T^\kv_{\kv_1\,\kv_2} = -\frac{1}{4\pi}\,\frac{\left(\kv_1 \times \kv_2\right)_z (k_1^2-k_2^2)}{k^2 + F}.
\end{equation}
We use compactified notation for the Dirac delta function, $\delta^\pv_{\qv\,\rv} = \delta(\pv-\qv-\rv)$, in order to keep subsequent formulae manageable. Note that the nonlinear interaction coefficient, $T^\kv_{\kv_1\,\kv_2}$, has the symmetries:
\begin{equation}
\label{eq-Tsymmetries}
T^\kv_{\kv_2\,\kv_1} = T^\kv_{\kv_1\,\kv_2}\hspace{1.0cm}T^\kv_{-\kv_1\,\kv_2}=T^\kv_{\kv_1\,-\kv_2}=-T^\kv_{\kv_1\,\kv_2}
\end{equation}
The densities of the energy and enstrophy in Fourier space are simply
\begin{equation}
\label{eq-EandHk}
E_\kv = (k^2 + F)\, \left|\psi_\kv\right|^2 \hspace{.5cm} \hbox{and}  \hspace{.5cm} 
H_\kv = (k^2 + F)^2\, \left|\psi_\kv\right|^2.
\end{equation}
When discussing wave turbulence, it is convenient to work in so-called wave-action variables, $a_\kv$, which are defined such that the spectral wave-action density, $n_\kv$, is related to the spectral energy density, $E_\kv$, by the relation $E_\kv=\left|\w_\kv\right|\,n_\kv$. Such variables arise very naturally in the Hamiltonian formulation of the CHM equation \cite{Piterbarg1988}. Although we do not make much use of the Hamiltonian formalism in this review, we shall find the concept of wave-action useful in later sections. We therefore adopt the wave-action variables which are defined
\begin{equation}
\label{eq-aVariables}
a_\kv = \frac{k^2 + F}{\sqrt{\left|k_x\right|}}\,\hat{\psi}_\kv.
\end{equation}
In terms of the $a_\kv$, Eq.~(\ref{eq-CHMk}) becomes
\begin{equation}
\label{eq-CHMk2} 
\pd{a_\kv}{t} = - i\, \omega_\kv\, a_\kv  
 + \int d\kv_1 d\kv_2\,  W^\kv_{\kv_1\,\kv_2}\, a_{\kv_1}\, a_{\kv_2}\, \delta^\kv_{\kv_1\,\kv_2},
\end{equation}
where the nonlinear interaction coefficient, $T^\kv_{\kv_1\,\kv_2}$ in Eq.~(\ref{eq-CHMk}), is replaced by
\begin{equation}
\label{eq-Wpqr}
W^\pv_{\qv\,\rv} = -\frac{1}{2} \sqrt{\frac{\left|q_x \right| \left|r_x \right|}{\left|p_x \right|}}\,\frac{(\qv\times\rv)_z\,(q^2-r^2)}{(q^2+F)\,(r^2+F)}.
\end{equation}
$W^\pv_{\qv\,\rv}$ inherits the symmetry properties, (\ref{eq-Tsymmetries}), of the original interaction coefficient:
\begin{equation}
\label{eq-Wsymmetries}
W^\pv_{\rv\,\qv} = W^\pv_{\qv\,\rv}\hspace{1.0cm}W^\pv_{-\qv\,\rv}=W^\pv_{\qv\,-\rv}=-W^\pv_{\qv\,\rv}.
\end{equation}
One final choice of variables which we will find useful are the so-called interaction variables, 
\begin{equation}
\label{eq-interactionRepresentation}
b_\kv = a_\kv\,\mathrm{e}^{i\,\w_\kv\,t},
\end{equation}
in which Eq.~(\ref{eq-CHMk}) takes the form
\begin{equation}
\label{eq-CHMk3}
\pd{b_\kv}{t} = \int d\kv_1 d\kv_2\,  W^\kv_{\kv_1\,\kv_2}\, b_{\kv_1}\, b_{\kv_2}\, \delta^\kv_{\kv_1\,\kv_2}\,\mathrm{e}^{i\,\Omega^\kv_{\kv_1\,\kv_2} t} ,
\end{equation}
where $\Omega^\pv_{\qv\,\rv}$ is shorthand notation for $\w_\pv-\w_\qv-\w_\rv$.

\subsection{Extended Hasegawa-Mima model}
\label{intro_EHM}

%Beyond CHM, in both plasma and geophysical fluid dynamics (GFD) applications, There exists a hierarchy of models with increasing degrees of realism, achieved at a cost of increasing complexity.     
In plasmas, a slightly more complex model is the Extended Hasegawa-Mima (EHM) model that improves the description of the electron response~\cite{Dorland1990} followed by the Hasegawa--Wakatani (HW) model which incorporates an instability forcing mechanism~\cite{Hasegawa1983}.  In GFD the next level of description  is the two-layer model which includes baroclinic effects and the respective instability~\cite{McWilliams2006} but is not discussed here.  
%The three plasma models will be introduced in the next section and an extensive qualitative and quantitative study of the CHM model is presented. The two-layer GFD model will not be covered by the present review.
%Here, we will review the studies of MI in the basic nonlinear models mentioned above. We will also briefly review studies of the systems forced by a
%primary instability, such as the forced-dissipated CHM and the HW models. In these systems, MI appears at the first evolutionary stage, followed by formation of ZF which feed back on the scales of the the primary forcing and suppress the latter. This is the most basic mechanism containing all the essential features of the LH transitions.  

In the plasma context, modes with $k_x=0$ must be special because for these
modes the relation between the plasma potential and the density fluctuations (so called Boltzmann response) fails. In fact, for such modes the density and the potential fields decouple. Such an effect is taken into account in 
 the extended Hasegawa-Mima (EHM) equation, where the coupling between the flux surface averaged potentials and density fluctuations is removed.  As will be shown later, this enhances the growth of ZFs in comparison the the CHM model. 

To keep a possibility to switch between the standard CHM and the EHM models, let us introduce a switch parameter $s$ so that $s=1$ would correspond to EHM and $s=0$ would correspond to CHM.
%Introducing the spatial Fourier transform of the streamfunction, $\hat{\psi}_{\bf{k}} = \int \psi({\bt x}) \mathrm{e}^{-\mathrm{i} (\bf{k} \cdot \bf{x})} \, d \bf{x}$, 
% where $L$ is the characteristic size of the domain being considered.  	\frac{1}{L^2}
In Fourier space, the model equation  is still Eq.~(\ref{eq-CHMk})
%\begin{equation}
%\label{eq-EHMk} 
%\partial_t \psi_{\bt{k}} + i \omega_{\bt{k}} \psi_{\bt{k}} - \frac{1}{2}\sum\limits_{\bt{k_1},\bt{k_2}} T (\bt{k}, \bt{k_1}, \bt{k_2}) \psi_{\bt{k_1}} \psi_{\bt{k_2}} \delta_{\bt{k}, \bt{k_1} +\bt{k_2}} = 0,
%\end{equation}
where as before $\omega_{\bt{k}} $
%\begin{equation}
%\label{eq-driftdispersion}
%\omega_\bt{k} = \frac{-\beta k_x}{1+\rho^2 k^2 }\,,
% +i \lambda_{\bt{k}}\,,
%\end{equation}
is the  dispersion relation given by (\ref{eq-RossbyWavedispersion}), but 
% in which we added an imaginary part for a possibility to model systems with instability type forcing ($\lambda_{\bt{k}} >0$) and dissipation ($\lambda_{\bt{k}} <0$). 
the interaction coefficient  is now defined as
\begin{equation}
\label{EHMinteractioncoeff}
T (\bt{k}, \bt{k_1},\bt{k_2}) = \frac{\left(\bt{k}_1 \times \bt{k}_2\right)_z \left[\rho^2(k_2^2-k_1^2)  + \delta_{s, 1} (\delta_{\bt{k}_{2_x}, 0} - \delta_{\bt{k}_{1_x}, 0})\right]}{(1 +
  \rho^2 k^2 - \delta_{\bt{k}_x, 0} \delta_{s, 1})}\,.
\end{equation}
%In plasma physics, $\psi_\bt{k}$ is the surface averaged electrostatic potential, $\rho$ is the ion Larmor radius at the electron temperature and $\beta$ is a diamagnetic velocity proportional to the plasma density gradient. Correspondingly in GFD, these variables are respectively, the stream function, Rossby deformation radius and a constant proportional to the latitudinal gradient of the vertical rotation frequency.  In plasmas (GFD), the $y$-axis is in the radial (south--north) direction along the plasma density gradient and the $x$-axis is the poloidal (west--east) direction. 
Here  $\delta_{s, 1}$ and $\delta_{\bt{k}_x, 0}$  are the Kronecker symbol switches: they are equal to one if their two respective arguments coincide and equal to zero otherwise.  We see that the extended part of the equation (terms involving $\delta$) act only on modes with a $k_x=0$ component, enhancing coupling to the zonal modes.
%Truncate Eq.~(\ref{eq-CHMk}) to the two triads $(\bt{p},\bt{q},\bt{p}_+)$ and $(\bt{p},-\bt{q},\bt{p}_-)$.  
We also see that the difference between CHM and EHM disappears in the $\rho \to \infty$ limit.

\subsection{Forced-dissipated CHM and EHM models} 

While the simple one-field CHM and EHM models exhibit some very interesting properties of Rossby and drift wave turbulence, their major shortcoming is the inability to spontaneously generate waves. This shortcoming is amended in slightly more complex two-field models, namely the Hasegawa-Wakatani plasma   model, explained in the next section and the two-layer QG model in GFD (not discussed here) which contain forcing by primary instability mechanisms, the drift-dissipative and the baroclinic instabilities respectively.

However, one could try to model such instabilities by simply adding to the one-field CHM or EHM models, extra linear terms which would result in the same $\bf k$-space distribution of the growth and dissipation rates as those predicted by the more complicated two-field models. 

This amounts to simply modifying the expression for the dispersion relation in the respective CHM and EHM models, namely taking Eq.~(\ref{eq-CHMx2}) with
$\omega_\bt{k}$
 given by
\begin{equation}
\label{eq-driftdispersion}
\omega_\bt{k} = \frac{-\beta k_x}{1+\rho^2 k^2 }
 +i \gamma_{\bt{k}}\,.
\end{equation}
Here, to the usual dispersion relation, Eq.~\eqref{eq-RossbyWavedispersion} of the unforced model, we added an imaginary part for the possibility to model systems with an instability type forcing ($\gamma_{\bt{k}} >0$) and dissipation ($\gamma_{\bt{k}} <0$). Such instability forcing is similar in some respects to the Stabilized Negative Viscosity concept of Sukoriansky et al. \cite{sukoriansky_large_1996} and has also been investigated recently in the context of wave turbulence in the nonlinear Schrodinger equation \cite{vladimirova_phase_2012}.

\subsection{Hasegawa-Wakatani and extended Hasegawa-Wakatani models}
\label{intro_HW}

 %\textcolor{red}{!!! THIS SECTION HAS YET TO BE RE-WORDED FROM BORIS' BOOK !!!!}
%The major shortcoming of the CHM and the EHM models  is their inability to spontaneously generate  waves, making prescribed initial conditions or an external forcing necessary.  In our Eq. (\ref{eq-driftdispersion}) such an internal forcing is introduced via $\lambda_{\bt{k}}$ with it ${\bt{k}}$-dependence mimicking the growth rate function of a plasma  (eg. drift dissipative)   instability or a geophysical (eg. baroclinic) instability.

The Hasegawa-Wakatani (HW) model  is more realistic and physical than the CHM and, in particular, can spontaneously generate  waves because at the level of the linear dynamics it contains a primary instability. The HW model relaxes the constraint of the adiabatic relationship between the density and potential and instead assumes that the density response is coupled to the potential via electron dynamics in the direction parallel to the magnetic field.  The  HW model is therefore a set of coupled equations for the evolution of the density and potential.  %In addition, these equations describe the fluctuations around a mean value, as opposed to the flux-surface-averaged value, resulting in a much more physical and complex model than the CHM model.  %Furthermore, the extended HW model, as mentioned here, refers not the addition of a curvature term as in the literature~\cite{} but rather to the the case where the zonal components of density and potential have been removed from the parallel current describing 
the electron dynamics.

\begin{eqnarray}
\partial_t \hat{\psi}_\bt{k} = - \frac{\alpha z}{k^2} (\hat{\psi}_\bt{k} -
\hat{n}_\bt{k})  +
 \frac 1{2} \sum_{\bt{k}_1, \bt{k}_2} T(\bt{k}, \bt{k_1},\bt{k_2}) \hat{\psi}_\bt{k_1}
\hat{\psi}_\bt{k_2}
\delta( {\bt{k} - \bt{k}_1 - \bt{k}_2})
%\label{NonLocalHWChap-HW-phi}
\nonumber
\\
\partial_t \hat{n}_\bt{k} = - i \kappa k_y \hat{\psi}_\bt{k} + z \alpha 
 (\hat{\psi}_\bt{k} - \hat{n}_\bt{k}) \!  -\!\!
\sum_{\bt{k}_1, \bt{k}_2} \!\! R(\bt{k}_1, \bt{k}_2) 
  \hat{n}_\bt{k_1}   \hat{\psi}_\bt{k_2} 
\delta( {\bt{k} - \bt{k}_1 - \bt{k}_2})
\label{NonLocalHWChap-HW-N}
\end{eqnarray}
where $\alpha$ is a coupling parameter,  $\kappa$ is the mean density gradient (similar to $\beta$ in CHM/EHM,
$z = 1- \delta_{s,1}\delta_{k_y,0}$
and s is a switching parameter: s=0 represents the HW case and s=1 represents the EHW case. 
The interaction coefficients
% in Eqs \eqref{NonLocalHWChap-HW-phi} and \eqref{NonLocalHWChap-HW-N}
in Eqs \eqref{NonLocalHWChap-HW-N} are given by
%\begin{equation}
%\omega_k = \frac{\beta k_y}{F + k^2}\label{Omega-def-EHW}
%\end{equation}
%\begin{equation}
%R (\bt{k_1},\bt{k_2}) = \left(\bt{k}_1 \times \bt{k}_2\right)_z
%\end{equation}
%and
%\begin{equation}
%\label{eq-HWinteractioncoeff}
%T (\bt{k}, \bt{k_1},\bt{k_2}) = \frac{\left(\bt{k}_1 \times \bt{k}_2\right)_z (k_2^2-k_1^2)  }{
%   k^2 }\,.
%\end{equation}
\begin{eqnarray*}
R (\bt{k_1},\bt{k_2}) &=& \left(\bt{k}_1 \times \bt{k}_2\right)_z\,,\\
\label{eq-HWinteractioncoeff}
T (\bt{k}, \bt{k_1},\bt{k_2}) & =& \frac{\left(\bt{k}_1 \times \bt{k}_2\right)_z (k_2^2-k_1^2)  }{
   k^2 }\,.
\end{eqnarray*}

In the limit $\alpha \gg 1$ the HW/EHW system tends to the familiar CHM/EHM system, whereas in the limit $\alpha \ll 1$
it becomes the 2D Euler equation for the streamfunction $\psi$  and the passive advection
equation for $n$.
Note that for the HW/EHW the $x$-and $y$-axes have been exchanged with respect to CHM/EHM notations. This is because HW/EHW are purely plasma models and we would like to use the conventions of the plasma literature. (Since CHM is a common model for GFD and plasmas, and since it was first introduced in GFD, the geophysical conventions have been used for that model).

%%%%%%%%%%%%%%%%%%%%%%%%%%%%%%%%%%%%%%%%%%%%%%%%%%%%%%%%%%%%%%%%%%%%%%%%%%%%%%%%%%%%%%%%%%%%%%%%%%%%
\section{Wave turbulence theory for the CHM equation}
\label{sec-WT}

\subsection{Overview of the application of the wave turbulence framework to the CHM equation}
\label{subsec-WTIntro}

When a geophysical flow or plasma becomes turbulent, the nonlinear interaction between different modes in the system leads to the excitation of a very large number of degrees of freedom. The high spatio-temporal complexity of turbulent motion means that only a statistical description makes sense. That is to say, we are interested in correlation functions of the field, $a_\kv$, appearing in Eq.~(\ref{eq-CHMk2}) (or $b_\kv$ in Eq.~(\ref{eq-CHMk3})) . Wave turbulence theory is a framework for studying the statistical dynamics of ensembles of interacting dispersive waves, which includes Rossby/drift waves. Statistical descriptions of turbulent systems are typically stymied by the so-called closure problem: equations for correlation functions of any given order depend on correlation functions of higher orders. An infinite hierarchy of equations must therefore be solved in order to find even the second order correlation function.  The theory of wave turbulence has the advantage that it has a weakly nonlinear limit: the theory of {\em weak} wave 
turbulence. In this limit, analytic progress is possible because writing higher order correlation functions as products of second order correlation functions (Gaussian closure) can be shown to be asympotically self-consistent in the sense of large time and weak nonlinearity \cite{newell_closure_1968,benney_random_1969,newell_wave_2001}. Thus, although the equations of weak wave turbulence are often mistakenly considered to be approximations they are actually asymptotically exact in the weakly nonlinear limit. Of course, one can question whether a particular wave system is actually in the weakly nonlinear limit or not but one cannot question that such a limiting case exists. For detailed reading on the theory of weak wave turbulence the reader is referred to \cite{Zakharov1992,
Nazarenko2011,newell_wave_2001,Newell2011}. In this section we shall summarise the application of wave turbulence theory to the CHM equation.

The most basic statistical object of interest is the spectrum, $n_\kv$, which is the second order correlation function (the meaning of the average will discussed below):
\begin{equation}
\label{eq-spectrum}
n_\kv\,\delta^\kv_{\kv^\prime} = \langle a_\kv a_{\kv^\prime}^*\rangle = \langle b_\kv b_{\kv^\prime}^*\rangle,
\end{equation}
where the Dirac delta function, $\delta^\kv_{\kv^\prime} = \delta(\kv-\kv^\prime)$, arises as a consequence of assuming that the turbulence is statistically homogeneous in space. Statistical homogeneity is an assumption which is probably never completely true in practice although it is theoretically very convenient. Even from a theoretical perspective, recent work \cite{newell_spontaneous_2012} has shown that statistical homogeneity can be spontaneously broken in some wave turbulence systems leading to the formation of coherent structures. Nothwithstanding these caveats, we shall assume statistical homeogeneity for now.  

The main output of the theory of wave turbulence as applied to Rossby/drift waves is the Rossby wave kinetic equation, Eq.~(\ref{eq-kineticEqn}) below. This equation describes how the wave spectrum, Eq.~(\ref{eq-spectrum}), evolves in time as a result of the spectral redistribution of energy by nonlinear interactions between waves. The wave turbulence literature has a reputation for being technically difficult and indeed many of the key papers in the field contain a lot of algebra, making them seem confusing to the outsider. In reality, the ideas underlying the theory are rather straight-forward and based on standard methods from perturbation theory. Much of the algebraic complexity is simply an unavoidable consequence of the fact that one needs to go to second order in perturbation theory in order to get a non-trivial result. In the next section, we will provide a brief summary of the origin of the wave kinetic equation. It is based on the standard perturbative expansion of the solution of Eq.~(\ref{eq-CHMk3}) in powers of the nonlinearity, $\epsilon$:
\begin{equation}
\label{eq-pertExpansionb}
b_\kv(t) = b^{(0)}_\kv (t) + \epsilon\,b^{(1)}_\kv (t) + \epsilon^2\,b^{(2)}_\kv (t) + \ldots
\end{equation}
The first few terms in the expansion are straightforwardly found to be
\begin{eqnarray}
\label{eq-order0} b^{(0)}_\kv(t) &=& B_\kv\\
\label{eq-order1} b^{(1)}_\kv(t) &=& \int d\kv_1 d\kv_2 W^\kv_{\kv_1\,\kv_2}\, B_{\kv_1}\, B_{\kv_2}\, \delta^\kv_{\kv_1\,\kv_2}\,\Delta(\Omega^\kv_{\kv_1\,\kv_2}, t)\\
\label{eq-order2} b^{(2)}_\kv(t) &=& -2 \int d\kv_1 d\kv_2 d\kv_3 d\kv_4  W^\kv_{\kv_1\,\kv_2}\, W^{\kv_1}_{\kv_3\,\kv_4} B_{\kv_2}\, B_{\kv_3}\,\, B_{\kv_4}\ \delta^\kv_{\kv_1\,\kv_2}\,\delta ^{\kv_1}_{\kv_3\,\kv_4}\, E(\Omega_\kv^{\kv_2\,\kv_3\,\kv_4}, \Omega_\kv^{\kv_1\,\kv_2}, t)
\end{eqnarray}
where the $B_\kv$ are constants and all the time-dependence is encoded in the integral functions
\begin{eqnarray}
\label{eq-DeltaIntegral} \Delta(x,t) &=& \int_0^td\tau\,\mathrm{e}^{i\,x\,\tau} =  \frac{\mathrm{e}^{i\,x\,t}-1}{i\,x}\\
\label{eq-EIntegral} E(x,y,t) &=& \int_0^td\tau \Delta(x-y,\tau)\, \mathrm{e}^{i\,y\,\tau}.
\end{eqnarray}
To second order in perturbation theory, the wave spectrum is therefore given by
\begin{eqnarray}
\label{eq-perturbativenk} n_\pv(t)\,\delta^\pv_{\pv^\prime} &=& \langle B_\pv\,B_{\pv^\prime}^*\rangle\\
\nonumber &+&\epsilon\,\int d\kv_1 d\kv_2 W^{\pv^\prime}_{\kv_1\,\kv_2}\, \langle B_\pv B^*_{\kv_1}\, B^*_{\kv_2}\rangle\, \delta^{\pv^\prime}_{\kv_1\,\kv_2}\,\Delta(\Omega_{\pv^\prime}^{\kv_1\,\kv_2}, t)\\
\nonumber &+&\epsilon \int d\kv_1 d\kv_2 W^\pv_{\kv_1\,\kv_2}\, \langle B_{\pv^\prime}^* B_{\kv_1}\, B_{\kv_2}\rangle\, \delta^\pv_{\kv_1\,\kv_2}\,\Delta(\Omega^\pv_{\kv_1\,\kv_2}, t)\\
\nonumber &-& 4\, \epsilon^2 \int d\kv_1 d\kv_2 d\kv_3 d\kv_4  W^\pv_{\kv_1\,\kv_2}\, W^{\kv_1}_{\kv_3\,\kv_4} \mathrm{Re}\left[\langle B_{\pv^\prime}^*\,B_{\kv_2}\, B_{\kv_3}\,\, B_{\kv_4}\rangle \right]\ \delta^\pv_{\kv_1\,\kv_2}\,\delta ^{\kv_1}_{\kv_3\,\kv_4}\, E(\Omega_\pv^{\kv_2\,\kv_3\,\kv_4}, \Omega_\pv^{\kv_1\,\kv_2}, t)\\
\nonumber &+&  \epsilon^2 \int d\kv_1 d\kv_2 d\kv_3 d\kv_4  W^{\pv}_{\kv_1\,\kv_2}\, W^{\pv^\prime}_{\kv_3\,\kv_4} \langle B_{\kv_1}\,B_{\kv_2}\, B^*_{\kv_3}\,\, B^*_{\kv_4}\rangle\ \delta^\pv_{\kv_1\,\kv_2}\,\delta ^{\pv^\prime}_{\kv_3\,\kv_4}\, \Delta(\Omega^\pv_{\kv_1\,\kv_2}, t)\,\Delta(\Omega_{\pv^\prime}^{\kv_3\,\kv_4}, t).
\end{eqnarray}
Before proceeding, however, there are two key concepts which a reader interested in understanding the derivation of the kinetic equation should pay attention to. The first of these is the use of Wick's rule to re-express higher order correlation functions of $a_\kv$ in terms of the second order correlation function or wave spectrum, $n_\kv$, appearing in Eq.~(\ref{eq-spectrum}). The second is the use of the method of multiple scales to deal with the non-uniformity of the perturbation theory which results from resonant interactions between waves.  We first discuss each of these in turn before turning to the discussion of the kinetic equation proper.

\subsubsection{The meaning of averaging in wave turbulence and quasi-Gaussianity}
\label{subsec-averaging}
The averaging procedure used to calculate the spectrum in Eq.~(\ref{eq-spectrum}) and the higher order correlation functions of $b_\kv$ in Eq.~(\ref{eq-perturbativenk}) is usually understood to be an ensemble average with respect to independent realisations of the initial conditions, $B_\kv$. In practice, particularly in the case of steady wave turbulence, it is common to replace the ensemble average with a time average under the assumption of ergodicity. The question then arises of what statistics to assume for the $B_\kv$? If the $B_\kv$ were assumed to be Gaussian, the higher order correlation functions appearing in Eq.~(\ref{eq-perturbativenk}) could be expressed in terms of products of the second order correlation function, $n_\kv(0)$, and we would obtain a closed equation for $n_\kv(t)$. Taking into account that $\langle B_\kv\rangle=0$ and using the fact that $b_\kv^* = b_{-\kv}$ and taking $s_i=\pm 1$, one can write a general third or fourth order correlation function as:
\begin{eqnarray}
\label{eq-M3} \langle B_{s_1\kv_1}B_{s_2\kv_2}B_{s_3\kv_3} \rangle &=& Q^{(3)}_{s_1\kv_1\,s_2\kv_2\,s_3\kv_3}\\
\langle B_{s_1\kv_1}B_{s_2\kv_2}B_{s_3\kv_3}  B_{s_4\kv_4}\rangle &=& 
n_{\kv_1}\,n_{\kv_2}\,\delta^{s_1\kv_1}_{s_3\kv_3} \delta^{s_2\kv_2}_{s_4\kv_4}
+ n_{\kv_1}\,n_{\kv_3}\,\delta^{s_1\kv_1}_{s_2\kv_2} \delta^{s_3\kv_3}_{s_4\kv_4}
+n_{\kv_1}\,n_{\kv_2}\,\delta^{s_1\kv_1}_{s_4\kv_4} \delta^{s_2\kv_2}_{s_3\kv_3}
+  \label{eq-M4} Q^{(4)}_{s_1\kv_1\,s_2\kv_2\,s_3\kv_3\,\,s_4\kv_4}
\end{eqnarray}
where $Q^{(3)}$ and $Q^{(4)}$ are the appropriate third and fourth order cumulants of the field $b_\kv$. Recall that for any random variable, the cumulant of a given order is {\em defined} as the difference between a moment of that order and the value it would have if the random variable were Gaussian. In the original papers on wave turbulence by Hasselmann, Zakharov and others, it was argued that the wave field should be close to Gaussian (sometimes referred to as ity) and the cumulants can thus be neglected on the basis that the phases of individual modes can be modeled as independent random variables. This is  really an ansatz rather than a controlled approximation. It is the analogue for waves of the assumption of molecular chaos which Boltzmann appealed to in the derivation of the Boltzmann equation in the kinetic theory of particles. The neglect of the cumulants was shown to be theoretically justifiable in an asymptotic sense by Benney and Newell provided that the cumulants are smooth in the {\bf k}-space (which means that in the physical space they rapidly decay at the infinite point separations) \cite{newell_closure_1968,benney_random_1969,newell_wave_2001}. They showed that even if the cumulants are nonzero initially, they decay for large times due to the fact that they never involve resonant interactions between waves. Resonances between waves and their importance will be discussed in  section \ref{subsec-RossbyWaveKineticEquation} below. In fact one can weaken the assumption of Gaussianity by writing the $B_\kv = J_\kv\,\mathrm{e}^{i\,\theta_\kv}$ and assuming that both the amplitudes, $J_\kv$, and the phases, $\theta_\kv$ are independent random variables for each mode, $\kv$ \cite{lvov_noisy_2004,choi_probability_2004}. The random phases and amplitudes approach is related to, but not equivalent to, the quasi-Gaussian approximation. In particular, one does not have to assume that the amplitudes, $J_\kv$, have a Rayleigh distribution as would be the case if the field $b_\kv$ were Gaussian \cite{choi_anomalous_2005}. Both approaches give the same kinetic equation so we shall not go into further discussion of the distinctions here. The interested reader is referred to the detailed discussion in \cite{Nazarenko2011}. From this point on, we will simply neglect the cumulants appearing in Eqs.~(\ref{eq-M3}) and (\ref{eq-M4}). We emphasise however that, unlike in the case of Navier-Stokes turbulence or the classical kinetic theory of gases, this neglect is theoretically justifiable for large times and weak nonlinearity and the interested reader is referred to the above references for details.

\subsubsection{Non-uniformity of regular perturbation theory for a nonlinear oscillator}
\label{subsec-anharmonicOscillator} 
While analytic or numerical solutions of nonlinear wave equations like Eq.~(\ref{eq-CHMx}) may indicate that the solution remains bounded in time, the regular perturbative expansion of the solution in powers of the nonlinearity parameter, $\epsilon$, often contains terms which are proportional to $t$. These terms which are unbounded in time are known as {\em secular} terms for historical reasons. If secular terms are present, expansions like Eq.~(\ref{eq-pertExpansionb}) become inconsistent for times of the order of $\epsilon^{-1}$. Secular terms arise due to nonlinear resonances between the components of the solution at different orders in perturbation theory. There are several standard modifications of regular perturbation theory which allow resonances to be accounted for consistently. One such approach is known as multiple scale analysis. An understanding of the concept of resonance for nonlinear waves and the use of multiple scale analysis to account for their effects is essential to understanding wave turbulence in the CHM equation or any other system of interacting dispersive waves. The main ideas can get lost in the algebraic complexity of the expansion Eq.~(\ref{eq-pertExpansionb}). For this reason, we first briefly illustrate the main concepts in a much simpler example which nevertheless exhibits all of the essential features of the full problem. Consider the anharmonic (Duffing) oscillator:
\begin{equation}
\label{eq-duffing}
\dtwod{x}{t} + \w^2 x = \epsilon x^3, \hspace{1.0cm}x(0) = a, \hspace{1.0cm} \dd{x}{t}(0) = 0.
\end{equation}
As with, Eq.~(\ref{eq-pertExpansionb}), let us assume that the solution can be written as an expansion in powers of $\epsilon$ when $\epsilon$ is small:
\begin{equation}
\label{eq-perturbativex}
x(t) = x_0(t) +  \epsilon x_1(t) + \epsilon^2 x_2(t) + \ldots.
\end{equation}
Substututing this into Eq.~(\ref{eq-duffing}) yields a hierarchy of perturbative equations, the first three of which are:
\begin{eqnarray}
\label{eq-x0}\left(\dtwod{ }{t} + \w^2\right) x_0 , &=& 0, \hspace{1.0cm}x_0(0) - a =0 ,  \hspace{1.0cm} \dd{x_0}{t}(0) = 0, \\
\label{eq-x1}\left(\dtwod{ }{t} + \w^2\right) x_1 , &=& x_0^3, \hspace{1.0cm}x_1(0)=0 , \hspace{1.0cm} \dd{x_1}{t}(0) = 0, \\
\label{eq-x2}\left(\dtwod{ }{t} + \w^2\right) x_2  &=& 3 x_1\,x_0^2,\hspace{1.0cm}x_2(0)=0 , \hspace{1.0cm} \dd{x_2}{t}(0) = 0.
\end{eqnarray}
These equations can be solved sequentially. The leading order solution is
\begin{equation}
x_0(t) = A_0\,\mathrm{e}^{i\,\w\,t} +  A_0^*\, \mathrm{e}^{-i\,\w\,t},
\end{equation}
where $A_0$ is a constant which can be determined from the initial conditions. Note, however, what happens when this leading order solution is substituted into Eq.~(\ref{eq-x1}). We obtain
\begin{equation}
\label{eq-x1b}
\left(\dtwod{ }{t} + \w^2\right) x_1 = A^3\,\mathrm{e}^{3i\,\w\,t} + \left|A\right|^2A\,\mathrm{e}^{i\,\w\,t} + \mathrm{cc},
\end{equation}
where $\mathrm{cc}$ denotes the complex conjugate of the preceding terms. The solution of the homogeneous equation is
\begin{equation}
x_1(t) = A_1\,\mathrm{e}^{i\,\w\,t} +  A_1^*\, \mathrm{e}^{-i\,\w\,t},
\end{equation}
where $A_1$ is a constant. Observe that the forcing term on the right hand side of Eq.~(\ref{eq-x1b}) which came from the lower order solution, $x_0(t)$,  contains a term with the same fundamental frequency, $\w$, as the homogeneous solution. This is an elementary example of nonlinear resonance. This phenomenon is fundamental to the theory of wave turbulence. The effect of resonance can be seen clearly when we find the particular solution of Eq.~(\ref{eq-x1b}). Using, for example, the method of variation of constants,  one can find the particular solutions associated with the two forcing terms on the right hand side. The first (nonresonant) term yields the particular solution
\begin{equation}
x_1^{P1}(t) = -\frac{A^3}{4\,\w^2}\,\mathrm{e}^{3i\,\w\,t} + \mathrm{cc},
\end{equation}
which is bounded for all $t$. The second (resonant) term yields the particular solution containing a secular term growing proportional to $t$,
\begin{equation}
x_1^{P2}(t) = \frac{3\left|A\right|^2 A}{4\,\w^2}\,\mathrm{e}^{i\,\w\,t} +  \frac{3\left|A\right|^2 A}{2\,i\,\w}\,t\,\mathrm{e}^{i\,\w\,t} + \mathrm{cc},
\end{equation}
which is is unbounded as $t$ grows. Using the initial conditions to fix the constants, $A_0$ and $A_1$ we obtain
\begin{equation}
\label{eq-duffingPertSolution}
x(t) = a\,\cos(\w\,t) + \epsilon\left[\frac{a^3}{32\,\w^2}\,\cos(\w\,t) - \frac{a^3}{32\,\w^2}\,\cos(3\w\,t)  +\frac{3a^3}{8\w}\,t\,\sin(\w\,t) \right] + O(\epsilon^2).
\end{equation}
Assuming that $a$ and $\w$ are of order 1, we see that the second term becomes as large as the first term when $t \sim \epsilon^{-1}$. Such an expansion is said to be non-uniform in $t$. At times of order $\epsilon^{-1}$, the perturbation expansion breaks down. As we shall see, exactly the same effect occurs in the expansion Eq.~(\ref{eq-perturbativenk}). This is a problem for wave turbulence since we are interested in the long-time statistical dynamics of the wave field.

One approach to taking into account the effect of resonances and obtain a perturbative expansion which is uniformly valid for times greater than $\epsilon^{-1}$ is the method of multiscale analysis. We shall apply this method below to the wave turbulence problem so let us first illustrate how it works in the case of the anharmonic oscillator. For an excellent detailed introduction of the this method and its applications see \cite[chap. 11]{bender_advanced_1999}.

The idea is to assume that $x(t)$, and hence the $x_i(t)$ appearing in the perturbation expansion (\ref{eq-perturbativex}), are function of multiple timescales. That is to say, rather than thinking of $x_i(t)$, we think of $x_i(T_0, T_1, T_2, \ldots)$ where the variables $T_n = \epsilon^n\,t$ will be treated as {\em independent}. The physical motivation for this is the recognition that when $\epsilon$ is very small, the timescale for nonlinear effects to become significant is much longer than the timescale for the linear motion of the oscillator. With this assumption, we should write
\begin{equation}
\dtwod{ }{t} = \ptwod{}{T_0} + 2\epsilon\frac{\partial^2}{\partial T_1\, \partial T_0} + \epsilon^2\,\left(2\, \frac{\partial^2}{\partial T_2\, \partial T_0} + \ptwod{}{T_1}\right) +\ldots 
\end{equation}
Eqs.~(\ref{eq-x0})-(\ref{eq-x2}) are modified to
\begin{eqnarray}
\label{eq-x0c}\left(\ptwod{ }{T_0} + \w^2\right) x_0 &=& 0, \\
\label{eq-x1c}\left(\ptwod{ }{T_1} + \w^2\right) x_1 &=& x_0^3 - 2\,\frac{\partial^2\,x_0}{\partial T_1\, \partial T_0}, \\
\label{eq-x2c}\left(\ptwod{ }{T_2} + \w^2\right) x_2 &=& 3 x_1\,x_0^2 - 2\,\frac{\partial^2\,x_1}{\partial T_1\, \partial T_0} - 2\, \frac{\partial^2\,x_0}{\partial T_2\, \partial T_0} - \ptwod{x_0}{T_1}.
\end{eqnarray}
The initial conditions remain the same but must be interpreted carefully since the $x_i$ are now  functions of many variables. We will restrict subsequent discussion to capturing only the dependence on $T_0$ and $T_1$ since this is sufficient to illustrate the points we wish to make but, in principle, the method can be continued to higher orders. The zeroth order solution is
\begin{equation}
\label{eq-x0mult}
x_0(T_0, T_1) = A_0(T_1)\, \mathrm{e}^{i\,\w\,T_0} +  A_0^*(T_1)\, \mathrm{e}^{-i\,\w\,T_0},
\end{equation}
where $A_0(T_1)$ is an arbitrary {\em function} of $T_1$. The idea of multiple scale analysis is that we choose this dependence of the zeroth order solution, $x_0$, on the slow timescale, $T_1$, in order to remove the secular terms which appear at next order. Upon substitution of Eq.~(\ref{eq-x0mult}) into Eq.~(\ref{eq-x1c}), we see that the appropriate condition to impose on $A_0(T_1)$ is
\begin{equation}
\label{eq-consistency}
\pd{A_0}{T_1} = -\frac{3\,i}{2\w} \,\left| A_0\right|^2 \,A_0.
\end{equation}
If $A_0(T_1)$ satisfies this consistency condition, then no secular terms appear in the expression for $x_1$:
\begin{displaymath}
x_1(T_0, T_1) = A_1\,\mathrm{e}^{i\,\w\,T_0} -  \frac{3\,A_0(T_1)^3}{8\,\w^2}\,\mathrm{e}^{3 i\,\w\,T_0} + \mathrm{cc},
\end{displaymath}
where $A_1$ is a constant to this order in the multiple scale expansion.
It may not always be easy to solve the consistency conditions which arise in multiscale analysis. In the case of Eq.~(\ref{eq-consistency}), once we recognise that
\begin{equation}
\dd{}{T_1}\, \left| A_0\right|^2 = 0,
\end{equation}
we can write down the solution immediately:
\begin{equation}
A_0(T_1) =  A_0(0)\, \mathrm{e}^{\frac{3\left| A_0(0)\right|^2 }{2\,\w}\,i\,T_1}.
\end{equation}
Using the initial conditions to  fix $A_0(0)$, we find that the leading order term in the modified perturbation theory is
\begin{equation}
\label{eq-duffingPertSolution2}
x_0(t) = a\,\cos\left[\left(\w - \frac{3\,a^2}{8\,\w}\,\epsilon\right)\,t\right],
\end{equation}
which is bounded for all time with the effect of the nonlinearity being simply to shift the frequency of oscillation. This expression is valid up to times of order $\epsilon^{-2}\,t$ which is the timescale at which our neglect of the dependence of the solution on the timescale $T_2$ above becomes inconsistent. Note that the leading order term actually contains terms of all orders in $\epsilon$. By solving the consistency condition (\ref{eq-consistency}), we have, in effect, summed up a subset of the terms occuring at {\em all} orders in the regular perturbation theory. The connections between methods of asymptotic analysis like the method multiple timescales and the summation of perturbation series are both extensive and deep and have been developed extensively by Goldenfeld and co-workers. See for example \cite{chen_renormalization_1996} and the references therein.

A word of caution is due about not taking the above analogy between the Duffing oscillator and the weak wave turbulence too far. One can see in the equation (\ref{eq-duffingPertSolution2}) that the main effect in the correct solution for the nonlinear oscillator is a frequency shift. Basically, a similar effect is expected in the leading order of the nonlinear interaction for some four-wave systems, eg. the Nonlinear Schrodinger model. However, this effect does not lead to a redistribution of energy among the resonant modes: the latter appears in the next order of expansion of the weak wave turbulence and it cannot be captured by any finite dimensional analog model.

\subsection{The Rossby/drift wave kinetic equation}
\label{subsec-RossbyWaveKineticEquation}

Let us now return to the perturbation expansion (\ref{eq-perturbativenk}) for the wave spectrum, $n_\kv$. The reader should bear in mind the illustrative example of the previous section in order to see the key features of the derivation. Firstly, we use Eqs.~(\ref{eq-M3}) and (\ref{eq-M4}) to write the third and fourth order correlation functions appearing in Eq.~(\ref{eq-perturbativenk}) in terms of products of $n_\kv$. We neglect the cumulants for the reasons discussed in Sec.\ref{subsec-averaging}. All order $\epsilon$ terms are zero since $\langle B_\kv\rangle=0$ and the first non-trivial terms appear at order $\epsilon^2$. We obtain six terms in total which we will not write out here. By integrating out two delta functions from each of these terms, using the symmetries (\ref{eq-Wsymmetries}) of $W^\pv_{\qv\,\rv}$ (including noticing that $W^\pv_{\qv\,-\qv}=0$) and relabelling integration variables, these terms can be brought to the form
\begin{eqnarray}
\nonumber n_\pv(t)\,\delta^\pv_{\pv^\prime} &=& n^{(0)}_\pv\,\delta^\pv_{\pv^\prime} + 2\,\epsilon^2 \int d\kv_1 d\kv_2\  W^\pv_{\kv_1\,\kv_2} W^\pv_{\kv_1\,\kv_2}\, \delta^\pv_{\kv_1\,\kv_2}\, n_{\kv_1}\,n_{\kv_2}\ \Delta(\Omega^\pv_{\kv_1\,\kv_2},t)\,\Delta(\Omega_{\pv}^{\kv_1\,\kv_2},t)\  \delta^\pv_{\pv^\prime} \\
\label{eq-perturbativenk2} & & - 8\,\epsilon^2\int d\kv_1 d\kv_2 \  W^\pv_{\kv_1\,\kv_2} W^{\kv_1}_{\pv\,\kv_2}\, \delta^\pv_{\kv_1\,\kv_2}\, n_{\pv}\,n_{\kv_2}\ \mathrm{Re}\left[E(0,\Omega^\pv_{\kv_1\,\kv_2};t)\right]\   \delta^\pv_{\pv^\prime}.
\end{eqnarray}
This exercise will take the determined reader a page or two of algebra. Note that the zeroth order term, $ n^{(0)}_\pv$ is independent of time. As $t$ gets large, the quantities $\Delta$ and $E$ appearing in this equation behave (under the integral sign) as follows \cite{newell_closure_1968}:
\begin{eqnarray}
\label{eq-DeltaAsymp} \Delta(x;t)\,\Delta(-x;t) &\sim& 2\,\pi\,t\delta(x) + 2\,\mathcal{P}\left(\frac{1}{x}\right)\pd{ }{x}\\
\label{eq-EAsymp}E(0,x;t) &\sim& \left[\pi\,\delta(x) + i\,\mathcal{P}\left(\frac{1}{x}\right) \right]\,t - i\,\left[\pi \delta(x) + i \, \mathcal{P}\left(\frac{1}{x}\right)\pd{ }{x}\right],
\end{eqnarray}
where $\mathcal{P}$ denotes the Cauchy Principle Value of the integral. The important point to notice is that they contain components which grow proportional to $t$. Therefore, as $t$ gets large we find that to order $\epsilon^2$:
\begin{equation}
\label{eq-perturbativenk3}
n_\pv(t) = n^{(0)}_\pv + (\epsilon^2\,t)\,S\left[n^{0}_\kv\right] + \epsilon^2\,\mbox{[terms bounded in $t$]},
\end{equation}
where
\begin{equation}
\label{eq-S0}
S\left[n^{0}_\kv\right] = 4\pi\int d\kv_1 d\kv_2\  W^\pv_{\kv_1\,\kv_2} W^\pv_{\kv_1\,\kv_2}\, \delta^\pv_{\kv_1\,\kv_2}\, n_{\kv_1}\,n_{\kv_2}\ \delta(\Omega^\pv_{\kv_1\,\kv_2}) - 8\pi \int d\kv_1 d\kv_2\  W^\pv_{\kv_1\,\kv_2} W^{\kv_1}_{\pv\,\kv_2}\, \delta^\pv_{\kv_1\,\kv_2}\, n_{\pv}\,n_{\kv_2}\ \delta(\Omega^\pv_{\kv_1\,\kv_2}).
\end{equation}
These terms are the analogue for the statistical initial value problem for the wave spectrum, $n_\kv$, of the secular terms appearing in the perturbative solution, Eq.~(\ref{eq-duffingPertSolution}), of the initial value problem for the anharmonic oscillator discussed above. They render the perturbation theory inconsistent for times of order $\epsilon^{-2}$. Notice that, as in the case of the anharmonic oscillator, these secular terms arise as a result of resonances: the integrand in Eq.~(\ref{eq-S0}) is supported on triads of wave numbers, $(\kv, \kv_1, \kv_2)$, which satisfy the resonance conditions
\begin{equation}
\label{eq-resonances}
\kv = \kv_1 + \kv_2\hspace{1.0cm}\w_{\kv} = \w_{\kv_1} + \w_{\kv_2}.
\end{equation}
To account for the presence of resonances and extend the range of validity of the perturbation theory, we can do exactly as we did in Sec.~\ref{subsec-anharmonicOscillator} and use the method of multiple scales. There is no need for a timescale, $T_1= \epsilon\,t$ since there are no terms of order $\epsilon$ in the expansion. The relevent timescale is $T_2 = \epsilon^2\,t$. Therefore we assume that the lowest order term, $n_\kv^{(0)}$ depends on $T_2$ and, as in Sec.~\ref{subsec-anharmonicOscillator}, choose the dependence $n_\kv^{(0)}(T_2)$ so that the secular terms, Eq.~(\ref{eq-S0}) are cancelled from the expansion. In principle, we should go back to Eqs.~(\ref{eq-pertExpansionb}), write out the perturbative expansion for $\pd{n_\kv}{t}$, identify the terms which lead to the secular terms in Eq.~(\ref{eq-perturbativenk3}) and set them equal to $\pd{n_\kv^{(0)}}{T_2}$. In practice, since the terms inside the integral in Eq.~(\ref{eq-S0}) are all independent of $t$ (or $T_0$), we can just write down the consistency condition directly by inspection:
\begin{equation}
\pd{n_\kv^{(0)}}{T_2} = S\left[n^{0}_\kv(T_2)\right].
\end{equation}
This is the wave kinetic equation and is the analogue of Eq.~(\ref{eq-consistency}) for the anharmonic oscillator. It tells us how nonlinear resonances between waves cause the wave spectrum, $n_\kv$ to evolve on the slow timescale, $T_2$. That is to say, it tells us how the cumulative effect of weak resonant interactions between waves leads to a spectral redistribution of energy over time.
When the resonant conditions, Eq.~(\ref{eq-resonances}) are satisfied, one can show that $W^{\qv}_{\pv\,\rv} = \mathrm{sgn}(\w_\pv\w_{\qv})\,W^{\pv}_{\qv\,\rv}$. This allows the collision integral to be written in the compact form:
\begin{equation}
\label{eq-S}
S\left[n_\pv \right] = 
4\pi\epsilon^2 \int d\kv_1d\kv_2 \left|W^\pv_{\kv_1\kv_2}\right|^2\delta^\pv_{\kv_1\kv_2}\delta(\Omega^\pv_{\kv_1\kv_2})\,\left[n_{\kv_1}n_{\kv_2} - \mathrm{sgn}(\w_\pv\w_{\kv_2})n_\pv n_{\kv_1} - \mathrm{sgn}(\w_\pv\w_{\kv_1})n_\pv n_{\kv_2} \right].
\end{equation}
We have dropped the superscript $^{(0)}$ from the wave spectrum although strictly speaking the kinetic equation as written above is valid only up to times of order $\epsilon^{-4}$ (not $\epsilon^{-3}$ as one might naively guess since all terms of order $\epsilon^3$ can also be shown to be zero).
At this point, it is appropriate to reintroduce forcing and dissipation which are usually present in practice. Thus the main conclusion of this section is that in the weakly nonlinear limit, the wave spectrum, evolved in time according to the wave kinetic equation:
\begin{equation}
\label{eq-kineticEqn}
\pd{n_\pv}{t} = S\left[n_\pv \right] + f_\pv - \gamma_\pv\,n_\pv,
\end{equation}
where $S\left[n_\pv \right]$ is given by Eq.~(\ref{eq-S}) above, $f_\pv$ models additive forcing and $\gamma_\pv$ models multiplicative forcing or dissipation (depending on the sign). A particular example of the form of $\gamma_\kv$ will be given in Sec.~\ref{WT_ZF_loop}. Several different forms of the collision integral appear in the literature which may cause confusion. In particular, Zakharov and co-workers usually prefer to write Eq.~(\ref{eq-CHMk}) in Hamiltonian variables \cite{Piterbarg1988} prior to performing the perturbation expansion. This leads to a different interaction coefficient, often written as $V^\pv_{\qv\rv}$, appearing in the theory:
\begin{equation}
\label{eq-Vpqr}
V^\pv_{\qv\rv} = \mathrm{sgn}(p_1)\,\sqrt{\beta\left| p_1 q_1 r_1\right|}\, \left( \frac{q_2}{q^2+F} + \frac{r_2}{r^2 +F}-\frac{p_2}{p^2+F} \right).
\end{equation}
This interaction coefficient has more symmetries than the original interaction coefficients, $T^\pv_{\qv\rv}$ and $W^\pv_{\qv\rv}$ which appear in Eqs.~(\ref{eq-CHMk}) and (\ref{eq-CHMk2}) respectively, which is related to the generic symmetries of the cubic hamiltonians. Under permutations of the its arguments, $V^\pv_{\qv\rv}$  has the symmetries
\begin{equation}
\label{eq-Vsymmetries1}
V^\pv_{\qv\rv} = V^\pv_{\rv\,\qv} = -V^{\rv}_{\pv\,-\qv}.
\end{equation}
Under changes of sign of the components of its arguments it has the symmetries
\begin{equation}
V^\pv_{\qv\rv} = -V^{(p_1,-p_2)}_{(q_1,-q_2)\,(r_1,-r_2)} = V^{(-p_1,p_2)}_{(q_1,q_2)\,(r_1,r_2)} = V^{(p_1,p_2)}_{(-q_1,q_2)\,(r_1,r_2)} = V^{(p_1,p_2)}_{(q_1,q_2)\,(-r_1,r_2)}.
\label{eq-Vsymmetries2}
\end{equation}

Some lengthy but elementary algebraic manipulations show that $V^\pv_{\qv\rv} = W^\pv_{\qv\rv}$ provided that $\w_\pv=\w_\qv+\w_\rv$. The two formulations are therefore equivalent {\em on the resonant manifold} (see eg. \cite{Nazarenko2011}).  It is often convenient to work with the more symmetric form, Eq.~(\ref{eq-Vpqr}), but this is only appropriate when one is interested in the weakly nonlinear limit where resonant interactions dominate. If one is discussing quasi-resonant interactions or the strongly nonlinear regime, then the true interaction coefficient, Eq.~(\ref{eq-Wpqr}), (or Eq.~(\ref{eq-Tpqr}) if using the original streamfunction variables) must be used. When reading the literature this distinction  is important to keep in mind.

Before continuing, let us rewrite the kinetic equation in a more symmetric form which commonly appears in the literature and is more convenient for some analyses. Since the function $\psi(\xv,t)$ is real, we know that $a_{-\kv} = a_\kv^*$. We therefore need only half the Fourier space to specify $n_\pv$ and its evolution. One can therefore work solely with the half-plane $p_x\geq0$. Let us divide the domain of the collision integral, Eq.~(\ref{eq-S}) into 4 regions, depending on the signs of $k_{1x}$ and $k_{2x}$. Using the  
fact that $n_\kv=n_{-\kv}$ together with the symmetries, Eq.~(\ref{eq-Vsymmetries1}), of the interaction coefficient, we can map the 3 regions containing negative values of $k_{1x}$ and/or $k_{2x}$ onto the region $k_{1x},\,k_{2x}\geq 0$ (although the region where both $k_{1x}$ and $k_{2x}$ are negative gives no contribution since the delta function vanishes identically there). The collision integral, Eq.~(\ref{eq-S}) then takes the form
\begin{equation}
\label{eq-Ssymm}
S\left[n_\pv \right] = \int_{k_{1x},\,k_{2x}\geq 0} d\kv_1 d\kv_2 \left[\mathcal{R}^\pv_{\kv_1\kv_2} - \mathcal{R}^{\kv_1}_{\kv_2\pv} - \mathcal{R}^{\kv_2}_{\pv\kv_1} \right],
\end{equation}
where
\begin{displaymath}
\mathcal{R}^\pv_{\kv_1\kv_2} = 4\pi\epsilon^2 \left|V^\pv_{\kv_1\kv_2}\right|^2\delta^\pv_{\kv_1\kv_2}\delta(\Omega^\pv_{\kv_1\kv_2})\,\left[n_{\kv_1}n_{\kv_2} - n_\pv n_{\kv_1} - n_{\kv_2}n_\pv \right].
\end{displaymath}

\subsubsection{Breakdown of weak wave turbulence}
\label{subsec-breakdown}
While the solution, $n_\kv(t)$, of the wave kinetic equation aims to describes the evolution of the wave spectrum due to the cumulative effect of nonlinear resonances between waves, it is important to remember that Eq.~(\ref{eq-kineticEqn}) has been derived under the assumption that weak nonlinearity leads to a separation in timescales between the timescale of the linear waves and the timescale over which nonlinearity acts to redistribute wave action between different wavenumbers. Without such a separation of timescales, the multiscale analysis above would not make any sense. One can estimate the linear timescale, which we shall denote by $\tau_\mathrm{L}$, from the dispersion relation. The nonlinear (or kinetic) timescale, which we denote by $\tau_\mathrm{NL}$ is obtained from the kinetic equation itself. One finds:
\begin{equation}
\label{eq-WTtimescales}
\tau_\mathrm{L} = |\w_\kv|^{-1}\hspace{1.0cm}\tau_\mathrm{NL} = \frac{n_\kv}{|\partial_tn_\kv|}.
\end{equation}
The criterion for separation of timescales is:
\begin{equation}
\label{eq-separationOfTimescales}
\frac{\tau_\mathrm{NL}}{\tau_\mathrm{L}} \ll 1.
\end{equation}
Note, however, that this relation depends on the wavenumber, $\kv$, and on the solution of the kinetic equation, $n_\kv$, through the definitions (\ref{eq-WTtimescales}). It is possible that the solution of the kinetic equation can lead to the violation of timescale separation at certain scales, even if one assumed weak nonlinearity at the outset. Indeed, in applications it frequently occurs that $\w_\kv$, $V^\kv_{\kv_1\kv_2}$ and $n_\kv$ are power law functions of $k_x$ and $k_y$. For such systems, is is almost inevitable that the criterion (\ref{eq-separationOfTimescales}) is violated either at very large or very small values of $k_x$ or/and $k_y$. This violation is sometimes referred to in the literature as ``breakdown'' of weak wave turbulence \cite{newell_wave_2001,biven_structure_2003}. In regions of wavenumber space where Eq.~(\ref{eq-separationOfTimescales}) is violated, the wave turbulence presumably becomes strong. This reasoning leads one to suspect that weak and strong nonlinearity may coexist in the same system at different scales.

\subsubsection{Structure of the resonance curves}

\begin{figure}
\begin{center}
\includegraphics[width=0.35\textwidth]{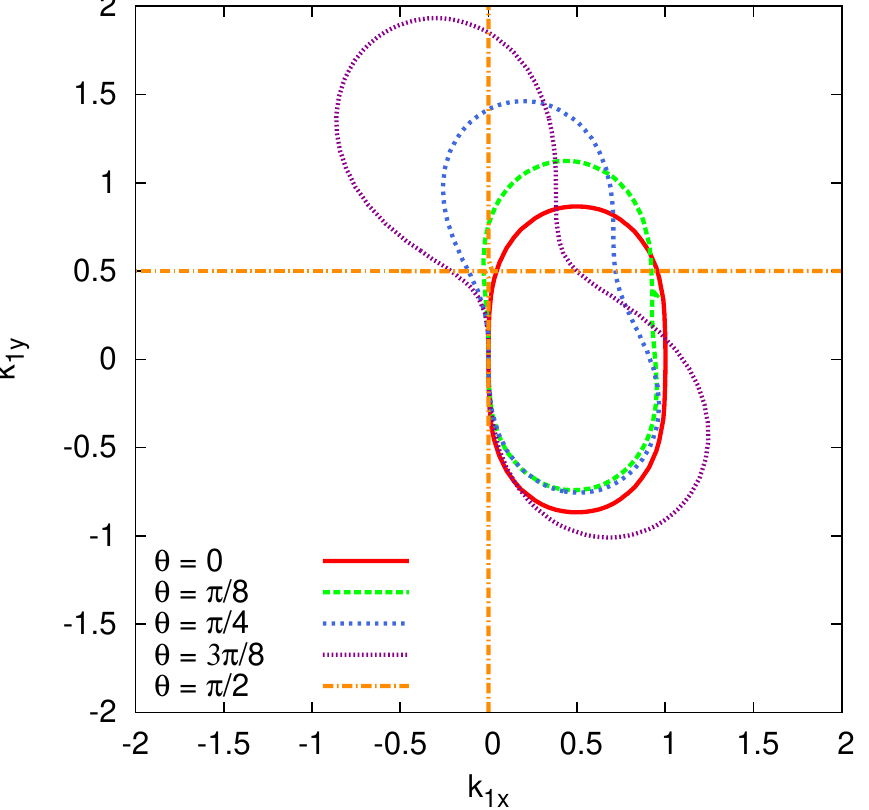}
\end{center}
\caption{Shape of the resonant manifold determined by Eq.(\ref{resonantManifolds_eqn}) with $\bf{k}=(\cos\theta,\sin\theta)$ for several values of $\theta$ for the case $F=0$.}
\label{resonantmanifolds} 
\end{figure}

The resonance conditions are central to the theory of weak wave turbulence. If one of the modes, say $\kv$, is fixed then the other two must lie on a one-dimensional curve in the wavevector space defined by Eqs.~(\ref{eq-resonances}). 
From Eq.~(\ref{eq-RossbyWavedispersion}), this curve, taken as an equation for $\kv_1=(k_{1x},k_{1y})$, is given in implicit form by the equation:
\begin{equation}
\label{resonantManifolds_eqn}
\frac{k_{1x}}{k_{1x}^2 + k_{1y}^2 + F} + \frac{k_x - k_{1x}}{(k_x-k_{1x})^2 + (k_y-k_{1y})^2 + F} -  \frac{k_x}{k_x^2 + k_y^2 + F} = 0.
\end{equation}
Because the system is anisotropic, the shape of resonant manifold depends on the direction of ${\bt k}$ as shown in figure~\ref{resonantmanifolds}. When $\kv$ is zonal, the resonant curve becomes unbounded and consists of the $k_{1y}$ axis and the line $k_{1y}=k_y$. 

\subsection{Anisotropic stationary solutions of the wave kinetic equation: the Kolmogorov-Zakharov spectra}
\label{subsec-KZSolutions}

One of the attractions of the theory of wave turbulence is that under certain conditions, exact stationary solutions of the wave kinetic equation, Eq.~(\ref{eq-kineticEqn}), can be found analytically in the physically important limiting case where the forcing and dissipation scales are infinitely separated in scale. In this limit, stationary spectra correspond to functions $n_\kv$ which make the collision integral, Eq.~(\ref{eq-S}) or (\ref{eq-Ssymm}), equal to zero. Such stationary solutions fall into two classes. The first class are equilibrium spectra. They correspond to equipartition of conserved quantities in the $\kv$-space and are usually obvious zeroes of the collision integral. The second class are stationary non-equilibrium states known as Kolmogorov-Zakharov (KZ) spectra. They correspond to constant fluxes of conserved quantities in the $\kv$-space.  The KZ spectra are the analogue for driven wave dynamics of Kolmogorov's $k^{-5/3}$ spectrum of vortex turbulence and explain the origin of the term ``wave turbulence''. They are highly non-trivial zeroes of the collision integral. In this section we will summarise what is known about the KZ solutions of the Rossby wave kinetic equation.

The theory of KZ-spectra was developed in the context of isotropic, scale invariant systems. Isotropy means that the dispersion relation, $\w_\kv$, and the wave interaction coefficient, $V^\pv_{\qv\rv}$, appearing in the collision integral are functions of $\left|\kv\right|$'s only. Scale invariance means that $\w_\kv$ and $V^\pv_{\qv\rv}$ are homogeneous functions of their arguments. That is to say, there are exponents $\alpha$ and $\beta$ such that if wave-vectors are scaled by a factor of $h$, we have
\begin{equation}
\label{eq-homogeneity}
\w_{h\kv} = h^\alpha\,\w_\kv \hspace{1.0cm} V^{h\pv}_{h\qv\,h\rv} = h^\beta\,V^\pv_{\qv\rv}.
\end{equation}
The degrees of homogeneity vary according to the application \cite{connaughton_dimensional_2003}. In the case of scale invariant systems, the stationary spectrum is a power law function of $ \left|\kv\right|$. Under these conditions, the collision integral to be simplified to involve integrals over $\left|\kv\right|$'s only and {\em} all stationary power law solutions can be found exactly using a set of ingenious transformations discovered independently by Zakharov \cite{zakharov_energy_1967} and Kraichnan \cite{Kraichnan1967}. The details of the Zakharov-Kraichnan transformations can be found in several existing reviews of the theory of wave turbulence \cite{dyachenko_optical_1992,Zakharov1992,newell_wave_2001,Nazarenko2011}. 

Rossby/drift wave turbulence is, however, neither isotropic nor scale invariant. Isotropy is broken by the second term in Eq.~(\ref{eq-CHMx2}). Scale invariance is broken by the deformation length (or ion Larmour radius) which sets a preferred scale and gives rise to the $F$ term in Eq~(\ref{eq-CHMx2}). One cannot therefore directly apply the Zakharov-Kraichnan transformations to the Rossby wave kinetic equation. A generalisation to anisotropic systems was provided by Kuznetsov \cite{kuznetsov_turbulence_1972}. Scale invariance is still required but this scale invariance is permitted to be anisotropic. Instead of assuming that the dispersion relation and interaction coefficient are homogeneous functions as in Eq.~(\ref{eq-homogeneity}), they are assumed to be bihomogeneous functions which scale differently under rescalings in the $x$ and $y$ directions. The notation is a little clumsy:
\begin{equation}
\label{eq-bihomogeneneity}
\w_{(h_xk_x, h_yk_y)} = h_x^a\,h_y^b\, \w_{(k_x,k_y)} \hspace{1.0cm} V^{(h_xp_x,h_yp_y)}_{(h_xq_x,h_yq_y)\,(h_xr_x,h_yr_y)} = h_x^u\,h_y^v\,V^{(p_x,p_y)}_{(q_x,q_y)\,(r_x,r_y)}.
\end{equation}

In the case of Rossby/drift waves, such anisotropic scale invariance is approximately recovered for zonal motions (wave-vectors $\kv$, for which $\left|k_x\right| \ll \left|k_y \right|$) in two limiting cases of very long or very short waves. We now consider these limits in some more detail.

\subsubsection{Short wave limit}
\label{subsubsec-shortWaveLimit}
In this limit we consider $F\to 0$. Equivalently we consider only waves for which $\left|\kv\right|^2 \gg F$. All wavelengths are therefore short with respect to the deformation length, $\rho=F^{-1/2}$. This is approximately the situation for mesoscale motions ($\sim$ 10's to 100's of km) in the Earth's atmosphere where the deformation length is of order 1000 km. Assuming that $k_x\ll k_y$, the dispersion relation in this limit becomes
\begin{equation}
\label{eq-omegaShortWave}
\w_\kv \sim -k_x\,k_y^{-2}.
\end{equation}
Applying similar arguments to the nonlinear interaction coefficient, Eq.~(\ref{eq-Vpqr}), we find that 
\begin{equation}
\label{eq-VpqrShortWave}
V^\pv_{\qv\rv} \sim \sqrt{\left| p_x q_x r_x\right|}\, \left( q_y^{-1} + r_y^{-1} -p_y^{-1}\right).
\end{equation}
In this limit, the pairs of exponents $(a,b)$ and $(u,v)$ of Eq.~(\ref{eq-bihomogeneneity}) take the values $(a,b) = (1,-2)$ and $(u,v) = (\frac{3}{2},-1)$.

\subsubsection{Long wave limit}
\label{subsubsec-longWaveLimit}
In this limit we consider $F\to \infty$ or waves for which  $\left|\kv\right|^2 \ll F$ so that all wavelengths are long with respect to the deformation length. This is approximately the situation for large scale motions in the Earth's oceans where the deformation length is of order 10's of km or in a large tokamak where the Larmour radius is typically small compared to the scale of the device. In this limit, assuming that $k_x\ll k_y$,
\begin{displaymath}
\w_\kv = -\frac{k_x}{F}\left[ 1 - \frac{k_x^2 + k_y^2}{F}\right] \approx -\frac{1}{F}\left(k_x - \frac{1}{F}k_xk_y^2 \right).
\end{displaymath}
In the kinetic equation, Eq.~(\ref{eq-kineticEqn}), we can ignore the leading $k_x$ term when this expression is substituted into the frequency resonance condition. This is because the delta function of the x-component of the wave-vectors (momentum conservation) ensure that these leading terms vanish. Thus the effective dispersion relation for long waves is
\begin{equation}
\label{eq-omegaLongWave}
\w_\kv \sim F^{-2}\,k_x\,k_y^2.
\end{equation}
Applying similar arguments to the nonlinear interaction coefficient, Eq.~(\ref{eq-Vpqr}), we find that in the long wave limit
\begin{equation}
\label{eq-VpqrLongWave}
V^\pv_{\qv\rv} \sim F^{-2}\,\sqrt{\left| p_x q_x r_x\right|}\, \left( q_y^3 + r_y^3 -p_y^3\right).
\end{equation}
In this limit, the pairs of exponents $(a,b)$ and $(u,v)$ of Eq.~(\ref{eq-bihomogeneneity}) take the values $(a,b) = (1,2)$ and $(u,v) = (\frac{3}{2},3)$.

We shall now summarise the application of the anisotropic Zakharov-Kraichnan transformations to these two special limits. The objective is to find the Kolmogorov-Zakharov spectra of Eq.~(\ref{eq-kineticEqn}). Readers who wish to delve into these details are advised to first familiarise themselves with how the Zakharov-Kraichnan transformations work in the isotropic case. A pedagogical discussion can be found in \cite{connaughton_numerical_2009,Nazarenko2011}.

Let us work with the collision integral in the form as written in Eq.~(\ref{eq-Ssymm}). The x-components of the wave-vectors are thus manifestly positive throughout. We shall seek spectra $n_\kv$ which make the collision integral equal to zero and which have the form 
\begin{equation}
\label{eq-anisotropicScalingSpectrum}
n_\kv = c\,k_x^x\,\left|k_y\right|^y
\end{equation}
where $c$ is a constant. Substituting this form into Eq.~(\ref{eq-Ssymm}), the integrand consists of three terms. The first can be written as
\begin{equation}
\label{eq-Rpqr}
\mathcal{R}^\pv_{\qv\rv} = c^2\left|V^\pv_{\qv\rv}\right|^2 (p_xq_xr_x)^x \left|p_yq_yr_y \right|^y\left[ p_x^{-x}\left|p_y\right|^{-y}-q_x^{-x}\left|q_y\right|^{-y}-r_x^{-x}\left|r_y\right|^{-y} \right]\,\delta(p_x-q_x-r_x)\,\delta(p_y-q_y-r_y)\,\delta(\w_\pv-\w_\qv-\w_\rv).
\end{equation}
The second and third terms, for $\mathcal{R}^\qv_{\rv\pv}$ and $\mathcal{R}^\rv_{\pv\qv}$ respectively, are written similarly. We now split the integral into three separate integrals. The Zakharov-Kraichnan transformations are changes of variables, $(\qv,\rv)\to(\qv^\prime,\rv^\prime)$, to be applied to the second and third integrals. To the second integral (involving $\mathcal{R}^\qv_{\rv\pv}$) we apply the change of variables:
\begin{eqnarray}
\label{eq-ZT1}q_x &=& \frac{p_x^2}{q_x^\prime} , \hspace{1.0cm} r_x = \frac{p_xr_x^\prime}{q_x^\prime} , \hspace{1.0cm} p_x = \frac{p_xq_x^\prime}{q_x^\prime} , \\
\nonumber q_y &=& \frac{p_y^2}{q_y^\prime} , \hspace{1.0cm} r_y = \frac{p_yr_y^\prime}{q_y^\prime} , \hspace{1.0cm} p_y = \frac{p_yq_y^\prime}{q_y^\prime}.
\end{eqnarray}
Note that the third equation here is a self-evident statement intended to make the mechanism of the transformation clear. To the third integral  (involving $\mathcal{R}^\rv_{\pv\qv}$) we apply the change of variables:
\begin{eqnarray}
\label{eq-ZT2}q_x &=& \frac{p_xq_x^\prime}{r_x^\prime}, \hspace{1.0cm}r_x = \frac{p_x^2}{r_x^\prime} , \hspace{1.0cm}p_x = \frac{p_xr_x^\prime}{r_x^\prime}, \\
q_y &=& \frac{p_yq_y^\prime}{r_y^\prime}, \hspace{1.0cm}r_y = \frac{p_y^2}{r_y^\prime} , \hspace{1.0cm}p_y = \frac{p_yr_y^\prime}{r_y^\prime}.
\end{eqnarray}
The Jacobians of these transformations  are
\begin{displaymath}
d\qv d\rv = \left|\frac{p_x}{q^\prime_x}\frac{p_y}{q^\prime_y}\right|^3 d\qv^\prime d \rv^\prime \mbox{\ \ \ and \ \ \ }d\qv d\rv = \left|\frac{p_x}{r^\prime_x}\frac{p_y}{r^\prime_y} \right|^3 d\qv^\prime d\rv^\prime 
\end{displaymath}
respectively. Using the scaling properties, Eq.~(\ref{eq-bihomogeneneity}), of $\w_\kv$ and $V^\pv_{\qv\rv}$ one finds that (we drop the primes on the transformed variables) the integrands $\mathcal{R}^\qv_{\rv\pv}$ and  $\mathcal{R}^\rv_{\pv\qv}$ are mapped onto $\mathcal{R}^\pv_{\qv\rv}$ with an additional prefactor. Specifically, Eq.~(\ref{eq-ZT1}) applied to the integral of  $\mathcal{R}^\qv_{\rv\pv}$ gives
\begin{equation}
\label{eq-Rqrp}
\mathcal{R}^\qv_{\rv\pv}\,d\qv d\rv = \left(\frac{p_x}{q^\prime_x}\right)^\alpha\left|\frac{p_y}{q^\prime_y}\right|^\beta\, \mathcal{R}^\pv_{\qv\rv}\,d\qv d\rv
\end{equation}
with $\mathcal{R}^\pv_{\qv\rv}$ given by Eq.~(\ref{eq-Rpqr}) and the exponents $\alpha$ and $\beta$ given by
\begin{equation}
\label{eq-alphaAndbeta}
\alpha = a - 2(1+u+x)\hspace{.5cm} \hbox{and} \hspace{.5cm} \beta= b - 2(1+v+y).
\end{equation}
Similarly, Eq.~(\ref{eq-ZT2}) applied to the integral of  $\mathcal{R}^\rv_{\pv\qv}$ gives
\begin{equation}
\label{eq-Rrpq}
\mathcal{R}^\rv_{\pv\qv}\,d\qv d\rv = \left(\frac{p_x}{r^\prime_x}\right)^\alpha\left|\frac{p_y}{r^\prime_y}\right|^\beta\, \mathcal{R}^\pv_{\qv\rv}\,d\qv d\rv.
\end{equation}
Combining Eqs.~(\ref{eq-Rpqr}), (\ref{eq-Rqrp}) and (\ref{eq-Rrpq}) and performing a little algebra, we conclude that, for spectra having the form (\ref{eq-anisotropicScalingSpectrum}), the collision integral can be written as
\begin{eqnarray}
\nonumber & &\int_0^\infty dq_x dr_x \int_{-\infty}^\infty dq_ydr_y c^2\left|V^\pv_{\qv\rv}\right|^2 (p_xq_xr_x)^x \left|p_yq_yr_y \right|^y\,p_x^{-\alpha}\left| p_y\right|^{-\beta}\,\delta(p_x-q_x-r_x)\,\delta(p_y-q_y-r_y)\,\delta(\w_\pv-\w_\qv-\w_\rv)\\
\label{eq-Imess}& &\times \left[ p_x^{-x}\left|p_y\right|^{-y}-q_x^{-x}\left|q_y\right|^{-y}-r_x^{-x}\left|r_y\right|^{-y} \right]\,\left[ p_x^{\alpha}\left|p_y\right|^{\beta}-q_x^{\alpha}\left|q_y\right|^{\beta}-r_x^{\alpha}\left|r_y\right|^{\beta} \right].
\end{eqnarray}
Noting that $\w_\pv = p_x^a\left|p_y\right|^b$, the exponents of the stationary spectra now become obvious since we simply need to choose the values of $x$ and $y$ so that the quantities in the square brackets correspond to the arguments of either the frequency or x-momentum delta functions. As remarked above there are no cascades associated with the y-momentum as it it not positive definite. There are four possible choices. The first two correspond to equilibrium solutions:
%\begin{equation}
\begin{alignat}{8}
\label{eq-energyEquipartition} &x=-a,&\hspace{1.0cm}&y=-b,&\hspace{1.0cm}&n_\kv=c, \,\w_\kv^{-1} &\hspace{1.0cm}&\mbox{-- equipartition of energy,}\\
\label{eq-enstrophyEquipartition} &x=-1, &\hspace{1.0cm}&y=0, &\hspace{1.0cm}&n_\kv=c\,k_x^{-1} &\hspace{1.0cm}&\mbox{-- equipartition of enstrophy.}
\end{alignat}
%\end{equation}
The second two are the constant flux KZ solutions and come from choosing either $\alpha=a$ and $\beta=b$ or $\alpha=1$ and $\beta=0$ (recall the definition (\ref{eq-alphaAndbeta}) of $\alpha$ and $\beta$). These choices give
\begin{alignat}{8}
\label{eq-energyFluxKZ} &x=-(u+1),&\hspace{1.0cm}&y=-(v+1),&\hspace{1.0cm}&n_\kv= c\,k_x^{-(u+1)}\left|k_y\right|^{-(v+1)}&\hspace{1.0cm}&\mbox{-- constant flux of energy,}\\
\label{eq-enstrophyFluxKZ} &x=\frac{a-2u-3}{2},&\hspace{1.0cm}&y=\frac{b-2v-2}{2},&\hspace{1.0cm}&n_\kv= c\,k_x^{\frac{a-2u-3}{2}}\left|k_y\right|^{\frac{b-2v-2}{2}}&\hspace{1.0cm}&\mbox{-- constant flux of enstrophy.}
\end{alignat}

With the KZ spectra in hand it is possible to make quantitative estimates of the consistency of the assumption timescale separation as discussed in Sec.\ref{subsec-breakdown}. Generalising the arguments of \cite{newell_wave_2001,biven_structure_2003} to anisotropic spectra is straightforward. For a spectrum of the form Eq.~(\ref{eq-anisotropicScalingSpectrum}), one finds from Eqs.~(\ref{eq-Ssymm}) and (\ref{eq-WTtimescales}):
\begin{equation}
\tau_\mathrm{L} \sim k_x^{-a}\left|k_y\right|^{-b},\hspace{1.0cm}\tau_\mathrm{NL}\sim k_x^{2u-a+1+x}\left|k_y\right|^{2v-b+1+y}.
\end{equation}
On the KZ spectra, Eqs.~(\ref{eq-energyFluxKZ}) and (\ref{eq-enstrophyFluxKZ}), we find that the ratio of nonlinear to linear timescales scales as:
\begin{alignat}{3}
\label{eq-energyFluxKZBreakdown} &\frac{\tau_\mathrm{L}}{\tau_\mathrm{NL}}\sim k_x^{u-2a}\,\left|k_y\right|^{v-2b}&\hspace{1.0cm}&\mbox{-- Energy cascade,}\\
\label{eq-enstrophyFluxKZBreakdown} &\frac{\tau_\mathrm{L}}{\tau_\mathrm{NL}}\sim k_x^{\frac{1}{2} (2u-3a-1)}\,\left|k_y\right|^{\frac{1}{2} (2v-3b-1)}&\hspace{1.0cm}&\mbox{-- Enstrophy cascade.}
\end{alignat}

If we now return to the specific case of Rossby/drift waves, inserting the values of $a$,$b$,$u$ and $v$ obtained in Sec.\ref{subsubsec-shortWaveLimit} above, we find that the KZ spectra in the short-wave limit are
\begin{alignat}{3}
\label{eq-energyFluxKZShortWave} &n_\kv=c\,k_x^{-\frac{5}{2}}\,\left|k_y\right|^0&\hspace{1.0cm}&\mbox{ -- Energy cascade (short wave limit),}\\
\label{eq-enstrophyFluxKZShortWave} &n_\kv=c\,k_x^{-\frac{5}{2}}\,\left|k_y\right|^{-1}&\hspace{1.0cm}&\mbox{-- Enstrophy cascade (short wave limit).}
\end{alignat}
These spectra were suggested in \cite{Monin1987}. Referring to Eqs.~(\ref{eq-energyFluxKZBreakdown}) and (\ref{eq-enstrophyFluxKZBreakdown}) above, the ratio $\tau_\mathrm{L}/\tau_\mathrm{NL}$ for these spectra scales as $k_x^{-1/2}\left|k_y\right|^3$ for the energy spectrum and as $k_x^{-1/2}\left|k_y\right|^{3/2}$ for the enstrophy spectrum. In both cases, the consistency condition $\tau_\mathrm{L}/\tau_\mathrm{NL} \ll 1$ is maintained only in the limit $k_x\to\infty$, $\left|k_y\right|\to 0$. This is at odds with the assumption made in Sec.~\ref{subsubsec-shortWaveLimit} that this spectrum describes zonal wavenumbers.  

Turning to the long wave case, the values of $a$,$b$,$u$ and $v$ obtained in Sec.\ref{subsubsec-longWaveLimit} yield the spectra
\begin{alignat}{3}
\label{eq-energyFluxKZLongWave} &n_\kv=c\,k_x^{-\frac{5}{2}}\,\left|k_y\right|^{-4}&\hspace{1.0cm}&\mbox{-- Energy cascade (long wave limit),}\\
\label{eq-enstrophyFluxKZLongWave} &n_\kv=c\,k_x^{-\frac{5}{2}}\,\left|k_y\right|^{-3}&\hspace{1.0cm}&\mbox{-- Enstrophy cascade (long wave limit).}
\end{alignat}
These spectra were suggested in \cite{Mikhailovskii1988_english,novakovskii_theory_1988}. The ratio $\tau_\mathrm{L}/\tau_\mathrm{NL}$ for these spectra scales as $k_x^{-1/2}\left|k_y\right|^{-1}$ for the energy spectrum and as $k_x^{-1/2}\left|k_y\right|^{-1/2}$ for the enstrophy spectrum. In both cases, the consistency condition $\tau_\mathrm{L}/\tau_\mathrm{NL} \ll 1$ is maintained only in the limit $k_x\to\infty$, $\left|k_y\right|\to \infty$. This is again in contradiction with the assumption of zonality made in Sec.~\ref{subsubsec-longWaveLimit}. The fact that the KZ spectra obtained in both the long and short wave limits fail to preserve the separation of timescales required for the derivation of the wave kinetic equation suggests that these spectra  do not provide the correct description of zonal scales. In the next section, we will discuss another reason why one might not expect to find these spectra in practice.

\subsection{Nonlocality of the  Kolmogorov--Zakharov spectra for Rossby/drift waves}
\label{subsec-RossbyWaveNonlocality}
In this section we discuss the important concept of scale locality in the context of turbulent cascades. A cascade is said to be local if the flux at any given inertial range scale is dominated by interactions with comparable scales. A cascade is non-local if the flux at a given inertial range scale is dominated by interactions with scales at the extremes of the inertial range, typically the forcing and/or dissipation scales. In the wave turbulence literature a distinction is made between stationary and evolutionary locality.

From a physical perspective, the easiest way to understand stationary locality conceptually is to imagine solving the kinetic equation, Eq.~(\ref{eq-kineticEqn}), with {\em finite} values for the forcing wavenumber, $\kv_f$, and dissipation wavenumber. Having thus obtained the explicit dependence of the stationary spectrum, $n(\kv,\kv_f,\kv_d)$, on $\kv_f$ and $\kv_d$, we can consider what happens when we take the limit $\ln \left(\left|\kv_f \right|/\left| \kv_d\right| \right) \to \infty$, corresponding to an infinitely wide inertial range. If $n(\kv,\kv_f,\kv_d)$ becomes independent of $\kv_f$ and $\kv_d$ in this limit then the spectrum is local and universal. If a dependence on  $\kv_f$ and/or $\kv_d$ persists in this limit (one would expect the spectrum either to diverge or to vanish) then the spectrum is nonlocal. Nonlocal cascades are explictly non-universal since a finite forcing and or/dissipation scale is required. Note that locality is a property of the solution, $n_\kv$, of the kinetic equation rather than a property of  the nonlinear interaction coefficient, Eq.~(\ref{eq-Vpqr}). It is obvious that $V^\pv_{\qv\rv}$ allows for interaction between arbitrarily widely separated scales. This does not mean, however, that the resulting turbulent cascades are scale nonlocal. Indeed, in systems such as the CHM equation where multiple cascades occur, it is quite possible for one to be local and the other nonlocal. 

We would like to know whether the KZ spectra obtained in Sec.~\ref{subsec-KZSolutions} are local or not. In practice it is not usually possible to solve for the explicit dependence of the spectrum on $\kv_f$ and $\kv_d$. In fact, the analysis of Sec.~\ref{subsec-KZSolutions} was predicated on the assumption of locality since we have assumed a universal form, Eq.~(\ref{eq-anisotropicScalingSpectrum}), for the stationary spectra in the limit of an infinite inertial range. If the cascade is local then this assumption is self-consistent and the collision integral will vanish at every value of $\kv$ when the KZ-spectrum is substituted into Eq.~(\ref{eq-Ssymm}). If not, then the collision integral will diverge when the KZ-spectrum is substituted into Eq.~(\ref{eq-Ssymm}). Such a divergence is not inconsistent with the vanishing of Eq.~(\ref{eq-Imess}) on the Kolmogorov spectrum. This is because the Zakharov-Kraichnan transformations, Eqs.~(\ref{eq-ZT1}) and (\ref{eq-ZT2}), exchange orders of integration. This can result in a cancellation of divergences and is not justified if the original integral were divergent in the first place. We can therefore check a-posteriori for the locality of the KZ-spectra obtained in Sec.~\ref{subsec-KZSolutions} by substituting Eqs.(\ref{eq-energyFluxKZ}) and (\ref{eq-enstrophyFluxKZ}) into Eq.~(\ref{eq-Ssymm}) and determining whether the collision integral is convergent or not. 

Evolutionary locality is the property that scale local perturbations to the KZ spectrum evolve in time by interaction solely with comparable scales. By contrast, evolutionary non-locality means that scale-local perturbations evolve in time by interaction with the scales at the extremities of the inertial range. This leads to singular behaviour in time if the inertial range becomes large. The technical linear stability machinery required to test for evolutionary locality is quite sophisticated and can depend on whether the perturbation is an even or an odd function of $q_y$. Details can be found in \cite{Zakharov1992,Nazarenko2011}. Evolutionary non-locality with respect to the odd perturbations is harmless: it leads to quick decay of the odd perturbations and, therefore, reinforcement of the even shape of the spectrum (see Appendix B). On the other hand, evolutionary non-locality with respect to the even perturbations 
is generally assumed to lead to the destruction of the KZ-spectrum through growth of nonlocal perturbations.

Even after passing the locality test, a KZ spectrum may still be unrealisable if the local evolution of its perturbations leads to an instability.

A detailed study of the locality and stability properties of the KZ spectra obtained above for the CHM model in the short and long wave limits was performed by Balk and Nazarenko \cite{balk_physical_1990}. This is quite a technical exercise and we shall simply quote the results here:

\small{
\begin{center}
\begin{tabular}{llllll}
\hline
Regime ($k_x\ll|k_y |$)&Cascade&KZ Spectrum&Stationary&Evolutionary&Stability\\
& & &locality&locality -- even/odd&  \\
\hline
Short waves ($\left|\kv \right|^2 \gg F$), &Energy&$k_x^{-\frac{5}{2}}$ &Yes&No/No&--\\
Short waves ($\left|\kv \right|^2 \gg F$) &Enstrophy (x-momentum)&&No&--&--\\
Long waves ($\left|\kv \right|^2 \ll F$) &Energy&$k_x^{-\frac{5}{2}}\,\left|k_y\right|^{-4}$&Yes&No/No&--\\
Long waves ($\left|\kv \right|^2 \ll F$)&Enstrophy (x-momentum)&$k_x^{-\frac{5}{2}}\,\left|k_y\right|^{-3}$&Yes&Yes/No& No\\
\hline
\end{tabular}
\end{center}
}

In the  table entries with ``--'' mean that the respective question is irrelevant, eg. if the spectrum does not satisfy stationary locality it does not make sense to ask about its evolutionary locality, and if it is nonlocal with respect to the even disturbances then it makes no sense to study its stability.

In the above discussions of the KZ spectra, we have omitted the KZ spectra with the flux of zonostrophy, an extra invariant in the Rossby and drift wave turbulence which will be discussed in section \ref{triple_cascade}. These spectra were found in~\cite{Balk_proc1990,Nazarenko011991} and soon after (in S.V. Nazarenko's PhD thesis) they were proven unrealisable in a way very similar to the enstrophy cascade spectra. Namely, in the short-wave limit the KZ spectrum with the flux of zonostrophy does not possess stationary locality, whereas in 
the long-wave limit it is stationarily local, evolutionally local with respect to the even perturbations, nonlocal with respect to the 
odd perturbations and unstable.

The lack of evolutionary locality and stability for the KZ spectra coupled with the difficulty in matching the regions of validity of these spectra, $k_x\ll\left|k_y\right|,$ with the consistency conditions for separation of timescales discussed in section \ref{subsec-breakdown} means that, despite their theoretical elegance, the KZ-spectra are probably not relevant to the turbulence of Rossby/drift waves in practice. Indeed, to the best of our knowledge, there is no experimental or numerical evidence to suggest that any of the spectra obtained in Sec.~\ref{subsec-KZSolutions} are realisable. It is important to make a distinction however between the consistency of the KZ spectra as solutions of the wave kinetic equation and the consistency of the kinetic equation itself. The problems discussed in this section with the KZ spectra do not mean that Eq.~(\ref{eq-kineticEqn}) is incorrect but rather that the true solution is something other than the KZ spectrum. Provided that this alternative solution respects the condition $\tau_\mathrm{L}/\tau_\mathrm{NL} \ll 1$, the wave kinetic theory will still apply. In Sec.~\ref{WT_ZF_loop} we shall develop a different approach to weakly nonlinear wave kinetics based on the assumption that the turbulence is strongly nonlocal.

%%%%%%%%%%%%%%%%%%%%%%%%%%%%%%%%%%%%%%%%%%%%%%%%%%%%%%%%%%%%%%%%%%%%%%%%%%%%%%%%%%%%%%%%%%%%%%%%%%%%
\section{Low dimensional models of spectral energy transfer }
\label{sec-lowDimensionalModels}
\subsection{Triad-based spectral truncations of the CHM equation}
\label{subsec-spectralTrunctions}

The discussion of wave turbulence theory in Sec.~\ref{sec-WT} was based on a continuous description of the wave field and energy transfer in the system involved an infinite number of degrees of freedom. In this section, we will take the opposite perspective and discuss models of energy transfer in Rossby/drift wave turbulence with a finite number of degrees of freedom.  We discuss two distinct but inter-related ways in which one might arrive at finite dimensional models starting from Eq.~(\ref{eq-CHMx2}). The first involves truncating a discrete spectral representation of Eq.~(\ref{eq-CHMx2}) at a finite wave-number. This generally leads to Galerkin-type models. The second involves projecting the right-hand side of Eq.~(\ref{eq-CHMx2}) onto a finite number of triads. For want of a better word, we will refer to such models as triad-based models. In some cases, these finite dimensional models are bona-fide approximations to Eq.~(\ref{eq-CHMx2}) and can be compared quantitatively with the full system. This is 
the case with Galerkin truncations which form the basis of spectral or pseudo-spectral numerical schemes.   In other cases, particularly with very low dimensional models, such models are illustrative or paradigmatic in nature and are not usually intended to be used for quantitative prediction.  Small triad-based models or predator-prey models are of this nature.

\subsubsection{Spectral truncation}

We have hitherto assumed that the system was infinite in extent. In practice one is often interested in studying systems with finite spatial extent. For theoretical and numerical studies it is common to consider a finite square domain with periodic boundary conditions on each side. In geophysical fluid dynamics it is common to solve Eq.~\ref{eq-CHMx2} on the surface of a sphere. In a finite domain, the wavenumber space is discrete rather than continuous and Eqs.~(\ref{eq-FourierTransform}) are replaced by discrete sums. For example, in the case of a bi-periodic box of size $L \times L$, discrete Fourier transforms replace their continuous versions, Eq.~(\ref{eq-FourierTransform}) and the CHM equation in Fourier space, Eq.~(\ref{eq-CHMk}), is replaced by:
\begin{equation}
\label{eq-CHMkdiscrete} 
\partial_t \hat{\psi}_\bt{k} = - i\, \omega_\bt{k}\, \hat{\psi}_\bt{k}  
 + \frac{1}{2} \sum_{\bt{k}_1, \bt{k}_2} T(\bt{k},\bt{k}_1,\bt{k}_2)\, \hat{\psi}_{\bt{k}_1}\, \hat{\psi}_{\bt{k}_2}\, \delta( {\bt{k} - \bt{k}_1 - \bt{k}_2})\,,
\end{equation}
where
$\bt{k}=(k_x,k_y)= (2\pi \bt{m}/L, 2\pi \bt{n}/L)$ is the wave vector ($\bt{m}, \bt{n} \in Z$). Modulo the multiplicative factorof $2\pi/L$, the wavevector now takes values on discrete values on the integer lattice, $\kv=(n,m) \in \mathbb{Z}^2$. In the case, of Rossby waves on a sphere, it is more appropriate to represent the wave field in terms of spherical harmonics:
\begin{equation}
\psi(\theta,\varphi,t) = \sum_{n=0}^\infty \sum_{m=-n}^n \hat{\psi}_{m\,n}(t)\,Y_{m\,n}(\theta, \varphi), 
\end{equation}
where $\theta$ and $\varphi$ are the polar angles on a sphere. The spherical case differs from the bi-periodic case in two important respects. Firstly, the wavevector, $(m,n)$, in the spherical case is not in $\mathbb{Z}^2$ but in the smaller set $n\in \mathbb{N}$ and $m \in \mathbb{Z}$ with $\left|m\right| \leq n$. Secondly, the dispersion relation is slightly different from Eq.~(\ref{eq-RossbyWavedispersion}):
\begin{equation}
\label{eq-dispersionSpherical}
\w_{m\,n} = -\frac{m}{n(n+1) + F}.
\end{equation}
For a pedagogical discussion of the spherical case, see \cite{lynch_resonant_2009}. Considering the bi-periodic case, if we introduce a spectral cut-off, $N$ (assumed even), such that $\left|n\right|\leq N/2$ and $\left|m\right|\leq N/2$ then we end up with a $(N+1)\times(N+1)$ dimensional set of ordinary differential equations for the $\hat{\psi}_\kv(t)$. This is the departure point for all spectral and pseudo-spectral numerical algorithms. The important point for the purposes of much of the discussion below, is that the wavenumber, $\kv$, is now a discrete, integer-valued vector.

\subsubsection{Triad-based truncation}

\label{sec-triadBasedTruncation}
Another approach to spectral trunction, which applies equally well regardless of whether the wavevector space is discrete or continuous, is based on the fact that, when written in Fourier space, waves interact in triads. To understand energy transfer within and between triads, one can project the equation onto some preselected set of triads of interest. Here ``project'' means that triads appearing on the right-hand side of Eq.~(\ref{eq-CHMk}) which are not members of this preselected set of triads are discarded, for example by setting the corresponding interaction coefficient, $T^\pv_{\qv\rv}$, to zero.  Unlike the Galerkin methods described above, this is usually not a controlled approximation to the original system since one does not know a-priori that the triads which are thrown away are neglible in any sense. Nevertheless we shall see below that, under some circumstances, truncations involving even very small numbers of triads can capture certain aspects of the full dynamics. If the number of triads kept 
in this procedure is progressively increased, one might expect the truncated model to be a progressively better approximation to the original system. 

The simplest triad-based truncation involves projecting onto a single triad which we shall denote $(\pv,\qv,\rv)$. Here $\pv=\qv+\rv$. We shall start from Eq.~(\ref{eq-CHMk}). The following discussion applies equally well to Eqs.~(\ref{eq-CHMk2}) or (\ref{eq-CHMk3}) or to the EHM model if we take the interaction coefficient to be given by Eq.~(\ref{EHMinteractioncoeff}). We assume that $\hat{\psi}_\kv(t)$ is supported only on the selected triad $(\pv,\qv,\rv)$ :
\begin{displaymath}
\hat{\psi}_\kv(t) = \psi_\pv(t)\,\delta^\kv_\pv + \psi_\qv(t)\,\delta^\kv_\qv + \psi_\rv(t)\,\delta^\kv_\rv + \psi_\pv(t)^*\,\delta^\kv_{-\pv} + \psi_\qv(t)^*\,\delta^\kv_{-\qv} + \psi_\rv(t)^*\,\delta^\kv_{-\rv}. 
\end{displaymath}
The inclusion of the complex conjugate terms is necessary because of the fact that $\psi(\xv,t)$ is real. This form is now substituted into Eq.~(\ref{eq-CHMk}). When all the terms proportional to $\delta^\kv_\pv$, $\delta^\kv_\qv$ and $\delta^\kv_\rv$ are gathered together we get
\begin{eqnarray}
\nonumber \dd{\psi_\pv}{t} &=& i\w_\pv\,\psi_\pv + 2 T^\pv_{\qv\,\rv}\,\psi_\qv\,\psi_\rv ,\\
\label{eq-3wave0} \dd{\psi_\qv}{t} &=& i\w_\qv\,\psi_\qv + 2 T^\qv_{-\rv\,\pv}\,\psi_\pv\,\psi_\rv^* ,\\
\nonumber \dd{\psi_\rv}{t} &=& i\w_\rv\,\psi_\rv + 2 T^\rv_{\pv\,-\qv}\,\psi_\pv\,\psi_\qv^*.
\end{eqnarray}
Some care is required with the signs of the interaction coefficients in these equations. Referring to Eq.~(\ref{eq-Tpqr}), the sign of the interaction coefficient, $T^\pv_{\qv\,\rv}$, determined by the numerator  $(\qv\times\rv)_z\,(q^2-r^2) = N^\pv_{\qv\,\rv}$. Bearing in mind that $\pv=\qv+\rv$, it is easy to show that $N^\pv_{\qv\,\rv} + N^\qv_{-\rv\,\pv} + N^\rv_{\pv\,-\qv}=0$. Therefore, two of the coefficients in Eqs.~(\ref{eq-3wave0}) must be positive and the other negative or vice versa. By symmetry, the second and third equations should share the same sign. We will therefore assume that $T^\pv_{\qv\,\rv}<0$,  $T^\qv_{-\rv\,\pv}>0$ and $T^\rv_{\pv\,-\qv}>0$ (if this is not the case, we can reverse the signs by $\psi\to -\psi$). Assuming that the interaction coefficient does not vanish, let us now make the following change of variables:
\begin{eqnarray}
\nonumber \psi_\pv &=& \frac{1}{2}\left| T^\qv_{-\rv\,\pv} \right|^{-\frac{1}{2}} \left|T^\rv_{\pv\,-\qv}\right|^{-\frac{1}{2}}\, \mathrm{e}^{i\w_\pv t}\,A_\pv , \\
\label{eq-3WavePrimitiveVariables}\psi_\qv &=& \frac{1}{2}\left| T^\pv_{\qv\,\rv} \right|^{-\frac{1}{2}} \left|T^\rv_{\pv\,-\qv}\right|^{-\frac{1}{2}}\, \mathrm{e}^{i\w_\qv t}\ A_\qv , \\
\nonumber\psi_\rv &=& \frac{1}{2}\left| T^\qv_{-\rv\,\pv} \right|^{-\frac{1}{2}} \left|T^\pv_{\qv\,\rv}\right|^{-\frac{1}{2}}\,\mathrm{e}^{i\w_\rv t}\, A_\rv.	
\end{eqnarray}
The result is the very simple dynamical system
\begin{eqnarray}
\nonumber \dd{A_\pv}{t} &=& - A_\qv\,A_\rv\,\mathrm{e}^{-i\,\Delta^\pv_{\qv\,\rv}t} , \\
\label{eq-detuned3wave} \dd{A_\qv}{t} &=& A_\pv\,A_\rv^*\,\mathrm{e}^{i\,\Delta^\pv_{\qv\,\rv}t} , \\
\nonumber \dd{A_\rv}{t} &=&  A_\pv\,A_\qv^*\,\mathrm{e}^{i\,\Delta^\pv_{\qv\,\rv}t},
\end{eqnarray}
where $\Delta^\pv_{\qv\,\rv} = \w_\pv-\w_\qv-\w_\rv$. These equations are known as the 3-wave equations or the 3-mode truncation (3MT). They describe the nonlinear dynamics of a single triad in isolation. Of particular interest is the case in which the triad is resonant so that $\Delta^\pv_{\qv\,\rv}=0$. The result is the resonant 3-wave equations:
\begin{equation}
\label{eq-3wave} \dd{A_\pv}{t} = - A_\qv\,A_\rv , \hspace{0.5cm} \dd{A_\qv}{t} = A_\pv\,A_\rv^* , \hspace{0.5cm}\dd{A_\rv}{t} =  A_\pv\,A_\qv^*.
\end{equation}
These equations have been derived in many different contexts including nonlinear optics \cite{ABDP1962}, geophysical fluid dynamics \cite{mccomas_resonant_1977}, plasma physics \cite{Sagdeev1969}, electronics \cite{manley_general_1956} and mechanics \cite{lynch_resonant_2003}. They are a paradigmatic example of resonant mode coupling in nonlinear systems with quadratic nonlinearity.

One can do the same thing with multiple triads although the process of writing down the corresponding dynamical system and identifying the number of independent interaction coefficients rapidly becomes quite cumbersome. For example, we shall make extensive use in Sec.~\ref{sec-MI} of a special triad-based truncation containing four modes which results from considering the pair of triads, $(\pv, \qv, \rv_-)$ and $(\pv,\rv_+,-\qv)$ where $\rv_\pm = \pv\pm\qv$. As in the case of the 3-mode truncation, it is convenient to introduce interaction representation variables $\Psi_{\bt{k}}(t) = \psi_{\bt{k}}(t) {\rm e}^{i \omega_{\bt{k}}\, t}$. The resulting equations are

\begin{eqnarray}
\nonumber \partial_t \Psi_{\bt{p}}&=& T(\bt{p},\bt{q},\bt{p}_-)\, \Psi_{\bt{q}} \Psi_{\bt{p_-}} {\rm e}^{i\, \Delta_-\, t} +  T(\bt{p},-\bt{q},\bt{p}_+)\, {\Psi}_{\bt{q}}^* \Psi_{\bt{p_-}} {\rm e}^{i\, \Delta_+\, t} , \\
\nonumber \partial_t \Psi_{\bt{q}} &=& T(\bt{q},\bt{p},-\bt{p}_-) \Psi_{\bt{p}} {\Psi}_{\bt{p_-}}^* {\rm e}^{-i\, \Delta_-\, t} + T(\bt{q},-\bt{p},\bt{p}_+) {\Psi}_{\bt{p}}^* \Psi_{\bt{p_+}} {\rm e}^{i\, \Delta_+\, t} , \\
\label{4MT}\partial_t \Psi_{\bt{p}_-}&=& T(\bt{p}_-,\bt{p},-\bt{q}) \Psi_{\bt{p}}\, {\Psi}_{\bt{q}}^* {\rm e}^{-i\, \Delta_-\, t} , \\
\nonumber \partial_t \Psi_{\bt{p}_+}&=& T(\bt{p}_+,\bt{p},\bt{q}) \Psi_{\bt{p}}\, \Psi_{\bt{q}} {\rm e}^{-i\, \Delta_+\, t}\,.
\end{eqnarray}
where $\Delta_{\pm} = \omega_{\bt{p}} \pm \omega_{\bt{q}} -\omega_{\bt{p_{\pm}}}$.   We refer to Eqs. (\ref{4MT}) as the four mode truncation (4MT) of the CHM/EHM models. The dynamics of some special systems constructed from two resonant triads are explored in detail in \cite{KL2007,BK2009}.

\subsection{Interplay between resonant manifolds and wave-vector quantization in finite systems}
\label{subsec-discreteness}

In an infinite system where wave modes are indexed by a continuous wavevector, we have argued in Sec.~\ref{sec-WT} that the theory of weakly nonlinear wave turbulence becomes asymptotically exact in the weakly nonlinear limit. For finite systems, the discreteness of  wavevectors leads to  some subtleties. In a finite system, there is a minimum spacing, $\Delta\,k$, between wavevectors which can be set equal to 1 provided we are not worried about how to take the continuous limit. We shall focus on the case of a bi-periodic box in which $\kv=(n,m) \in \mathbb{Z}^2$ and the dispersion relation is given by Eq.~(\ref{eq-RossbyWavedispersion}). The discussion is equally applicable to the spherical case, in which  $\kv = (n,m)$ indexes the spherical harmonics and the dispersion relation is given  by Eq.~(\ref{eq-dispersionSpherical}). When the wavevectors are integer valued, then the resonance conditions, Eq.\eqref{eq-resonances}, become a problem of Diophantine analysis (the search for integer-valued solutions to 
polynomial equations in several variables). For example, in the case of Rossby/drift waves in a periodic box, if we fix $\pv=(n,m) \in \mathbb{Z}^2$, we need to find $\qv = (u,v) \in \mathbb{Z}^2$ such that
\begin{equation}
\label{eq-discreteDispersionReln}
\frac{n}{n^2+m^2+F} = \frac{u}{u^2+v^2 +F} + \frac{n-u}{(n-u)^2 + (m-v)^2 + F}.
\end{equation}
Such Diophantine equations have far fewer solutions than their real-valued counterparts and it is generally quite difficult to find them. Thus, in finite system, exact resonances can be quite rare. Indeed, for certain dispersion relations, it can be shown that there are no resonant triads with integer-valued wavevectors at all. This occurs, for example, for capillary waves \cite{kartashova_wave_1998}. The situation for Rossby/drift waves is not so extreme. A complete enumeration of all integer-valued solutions of Eq.\eqref{eq-discreteDispersionReln} with $F=0$ was recently provided in \cite{bustamante_complete_2013}. An efficient algorithm for the spherical case Eq.~({\ref{eq-dispersionSpherical}), again with $F=0$, was provided in \cite{kartashova_laminated_2007}. For finite $F$,  solutions exist only if $F$ is a rational number. To the best of our knowledge, the problem of finding the full set of solutions in the case of finite values of $F$ remains open. The conclusion of these analyses is that although there are many integer-valued resonances for Rossby waves, such triads are nevertheless sparse in the space of real-valued resonances.

A consequence of this sparseness is that, in discrete systems, exact resonant triads often exist in isolation or in finite groups of triads. These groups are referred to as resonant clusters. Two triads belong to the same cluster if they share at least one common mode. If it were truly the case that only resonant interactions are important then each such cluster would evolve independently of all the others.
A triad, $(\pv,\qv,\rv)$, is said to be isolated if it does not share any common modes with any other resonant triads. For the dispersion relation Eq.~(\ref{eq-discreteDispersionReln}) with $F=0$, the lowest wavenumber isolated triad which does not exhibit any special symmetries is $\pv=(9,23)$, $\qv=(1,-11)$, $\rv=(8,34)$. Clusters consisting of two resonant triads may have single mode in common or may have two modes in common. These are referred to as "butterflies" and "kites" respectively owing to their appearance when drawn as graphs. The dynamics of butterflies and kites have been studied in some detail in \cite{bustamante_dynamics_2009}. Small clusters have been suggested as a possible explanation of the unusual periods of certain observed atmospheric oscillations \cite{kartashova_model_2007}. 

The question in the context of turbulence is whether there exist large clusters capable of distributing energy over a large range of scales in a discrete system. Energy transfer in such clusters is referred to as finite dimensional wave turbulence. For the dispersion relation (\ref{eq-dispersionSpherical}), for Rossby waves on a sphere with $F=0$, such a large exactly resonant cluster has been shown to exist numerically \cite{lvov_finite-dimensional_2009}. For Eq.\eqref{eq-discreteDispersionReln} with $F=0$ exactly resonanct clusters containing several hundreds of modes were found in \cite{bustamante_complete_2013}. For Eq.\eqref{eq-discreteDispersionReln} with finite $F$ the question of classification of resonant clusters is unknown. Numerical explorations in   \cite{bustamante_complete_2013} indicate, however, that for general values of $F$ large exactly resonant clusters are rare.

A consequence of the absence or rarity of large exactly resonant clusters in discrete systems is that it is often necessary to rely on approximate resonances to account for energy transfer. Approximate resonance is possible due to the phenomenon known as nonlinear resonance broadening. This is an effect whereby the frequency of a wave acquires a correction to its linear value which depends on the amplitude. We have already seen the origin of resonance broadening in our discussion of the anharmonic oscillator in Sec.~\ref{subsec-anharmonicOscillator}. Eq.~(\ref{eq-duffingPertSolution2}) shows that the effect of nonlinearity is to modify the frequency by a weakly amplitude dependent correction. The same effect happens in a statistical sense in wave turbulence and results in a frequency modification which depends on the spectrum, $n_\kv$. Explicit formulae for nonlinear frequency modification and further references can be found in \cite{newell_wave_2001} 
(see Eq. 2.25 of this reference in particular). As a result of this nonlinear frequency modification, triads which are not exactly in resonance can then interact at finite amplitude if the frequency mismatch is less than this correction.  Such triads are known as quasi-resonant triads and satisfy the broadened resonance conditions
\begin{figure}
\begin{center}
\includegraphics[width=0.9\textwidth]{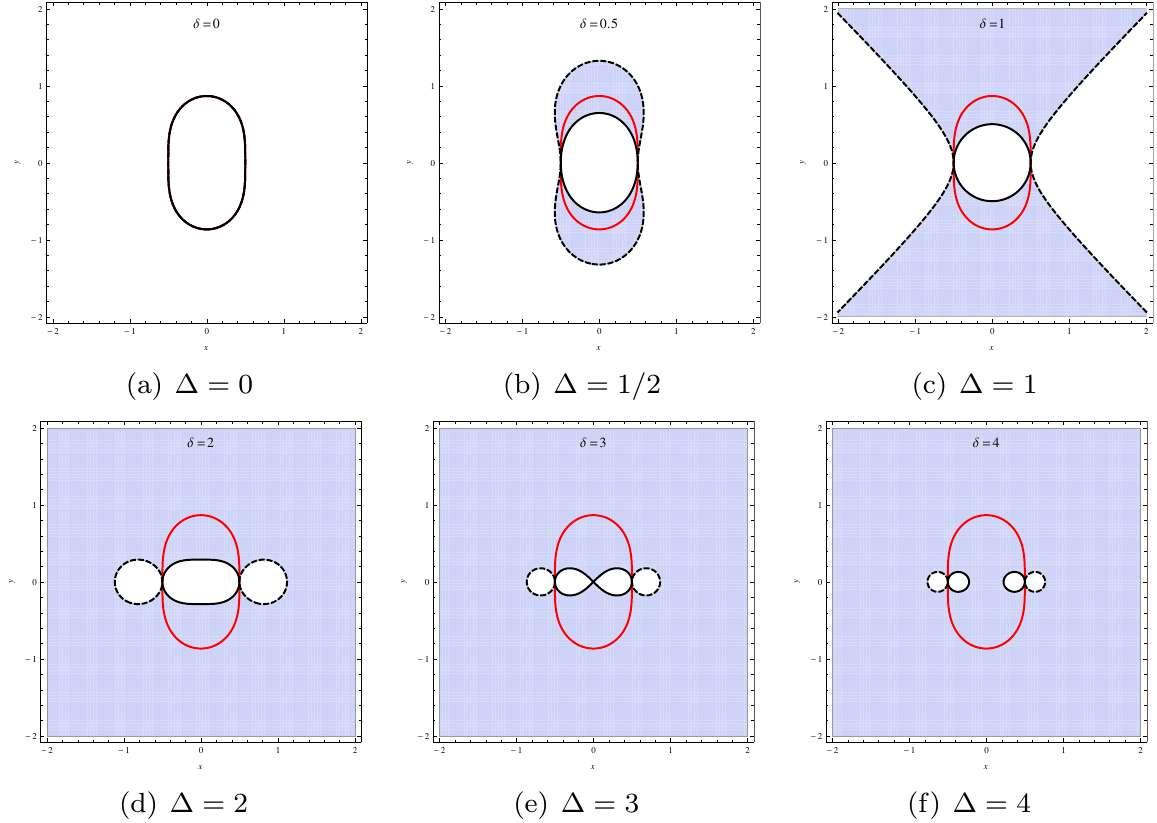}
\end{center}
\caption{\label{fig-QRSet} Shaded regions correspond to the resonant set defined by Eq.\eqref{eq-quasiresonance} with $\kv_3=(1,0)$, $F=0$ for different values of $\Delta$. The exactly resonant manifold is the red solid line. }
\end{figure}
\begin{equation}
\label{eq-quasiresonance}
\left\{
\begin{array}{l}
\kv_3 = \kv_1 + \kv_2,\\
\left|\omega(\kv_3)-\omega(\kv_1)-\omega(\kv_2)\right| \leq \Delta \, ,
\end{array}
\right. 
\end{equation}
where $\Delta$ is a characteristic value for the resonance broadening taken to be positive. Although Eq.\eqref{eq-quasiresonance} provides only a kinematic description of resonance broadening, the analogous dynamical effect is strikingly visible in linear stability analyses of weakly nonlinear waves \cite{dyachenko_decay_2003,Connaughton2010}  where it is found that the set of unstable perturbations lie in a neighbourhood around the set of exactly resonant perturbations. See also Fig.~\ref{qMaps_F_0} of Sec.~\ref{sec-MI}. The set of quasi-resonant modes is often pictured as a ``thickened'' or broadened version of the exactly resonant manifold. The structure of this set as a function of the detuning, $\Delta$, can be found by finding the boundaries given by the pair of curves
\begin{eqnarray}
\label{eq-plusBoundary}\omega(\kv_3)-\omega(\kv_1)-\omega(\kv_3-\kv_1) &=& \Delta\,,\\
\label{eq-minusBoundary}\omega(\kv_3)-\omega(\kv_1)-\omega(\kv_3-\kv_1) &=& -\Delta.
\end{eqnarray}
The details can be found in \cite{harris_percolation_2013}. The structure of the quasiresonant set for the dispersion relation (\ref{eq-RossbyWavedispersion}) is shown for a range of values of $\Delta$ in Fig.~\ref{fig-QRSet}. It is seen that the set becomes quite complicated for larger values of $\Delta$. In particular it is found that although the quasiresonant set does indeed look like a ``thickened'' version of the exact resonant curve for small values of $\Delta$, it actually becomes unbounded for a finite value of $\Delta$. For weakly nonlinear systems, the parameter $\Delta$ is expected to be small since amplitudes are small. It is shown in \cite{harris_percolation_2013}, however, that even for arbitrarily small $\Delta$, it is possible for find resonant curves for which the quasiresonant set becomes unbounded. This is related to the fact that for the dispersion relation (\ref{eq-RossbyWavedispersion}), the resonant curves themselves are unbounded for zonal modes (see Fig.~\ref{resonantmanifolds}).

The importance of quasiresonant interactions is strikingly illustrated by the case of capillary wave turbulence in a bi-periodic box.  We have already remarked that for this system there are no exact resonances.  Direct numerical simulations illustrate that the cascade of energy to small scales is entirely absent when the level of nonlinearity is sufficiently small. This phenomenon has been termed ``frozen turbulence'' \cite{pushkarev_kolmogorov_1999,pushkarev_weak_2000,connaughton_discreteness_2001}. As the nonlinearity is increased the resonance broadening eventually becomes larger than the frequency mismatches caused by the fact that the wavevectors are forced to lie on an integer-valued lattice. At this point, the cascade is able to proceed. We can infer from these results that quasiresonant interactions are essential in ``linking up'' isolated triads and clusters  to form conduits for spectral energy transfer in finite-dimensional turbulence when sufficiently large exactly resonant clusters are absent.

\subsection{Nonlinear dynamics of resonant triads}
\label{subsec-triad}

In this section we return to the resonant 3-wave equations, (\ref{eq-3wave}), and summarise some results about their solutions which will be useful later. Firstly, we note that Eqs.~(\ref{eq-3wave}) are a Hamiltonian system. They can be summarised as
\begin{equation}
\label{eq-3wave1}
i\,\dd{A_\kv}{t} = \dd{H}{A_\kv^*},\hspace{2.0cm}\mbox{where $\kv \in \left\{\pv,\qv,\rv\right\}$}
\end{equation}
and
\begin{equation}
H = i\,\left[A_\qv^*A_\rv^*A_\pv - A_\qv A_\rv A_\pv^*\right] = 2\,\mathrm{Im}\left[ A_\qv A_\rv A_\pv^*\right].
\end{equation}
The Hamiltonian, $H$, is obviously conserved. It is easy to check that, in addition, Eq.~(\ref{eq-3wave1}) conserves the quantities
\begin{equation}
\label{eq-ManleyRoweInvariants}
I_1 = \left|A_\pv \right|^2 + \left|A_\qv \right|^2 \hspace{0.5cm}\mbox{and}\hspace{0.5cm}\left|A_\pv \right|^2 + \left|A_\rv \right|^2,
\end{equation}
which are often referred to as Manley-Rowe invariants following their application in electronics \cite{manley_general_1956}. To solve Eqs.~(\ref{eq-3wave}), one introduces phase and amplitude variables as follows:
\begin{displaymath}
A_\kv(t) = C_\kv(t)\,\mathrm{e}^{i\theta_\kv(t)\,t}, \hspace{2.0cm}\mbox{where $\kv \in \left\{\pv,\qv,\rv\right\}$}.
\end{displaymath}
Upon substitution into Eqs.~(\ref{eq-3wave}) and equating real and imaginary parts, we replace the three complex equations by the following six real equations:
\begin{eqnarray}
\label{eq-amplitudes}\dot{C}_\qv = C_\pv\,C_\rv\,\cos \varphi , \hspace{1.0cm} 
\dot{C}_\rv , &=& C_\pv\,C_\qv\,\cos \varphi , \hspace{1.0cm}
\dot{C}_\pv = -C_\qv\,C_\rv\,\cos \varphi , \hspace{1.0cm} \\
\label{eq-phases}\dot{\theta}_\qv = -\frac{C_\pv\,C_\rv}{C_\qv}\,\sin\varphi , \hspace{1.0cm}
\dot{\theta}_\rv , &=& -\frac{C_\pv\,C_\qv}{C_\rv}\,\sin\varphi , \hspace{1.0cm}
\dot{\theta}_\qv = -\frac{C_\qv\,C_\rv}{C_\pv}\,\sin\varphi,
\end{eqnarray}
where $\varphi(t) = \theta_\qv(t) + \theta_\rv(t) - \theta_\pv(t)$ is called the dynamical phase \cite{BK2009}. We notice that the righthand sides of the equations for the phases depend only on the dynamical phase, $\varphi$, and on the amplitudes, $C_\kv$. The individual $\theta_\kv$'s do not appear. Therefore, if we knew the $C_\kv$'s and the dynamical phase we could reconstruct all the phases by direct integration from Eqs.~(\ref{eq-phases}). This suggests that we write an evolution equation for $\dot{\varphi}$:
\begin{equation}
\label{eq-dynamicalPhase}
\dot{\varphi} = -C_\pv C_\qv C_\rv\,\left(\frac{1}{C_\qv^2} + \frac{1}{C_\rv^2} - \frac{1}{C_\pv^2} \right)\,\sin \varphi.
\end{equation}
The system can be solved completely by solving Eqs.~(\ref{eq-amplitudes}) and (\ref{eq-dynamicalPhase}): it possesses only 4 independent degrees of freedom rather than the six which one might expect naively. Since we have already identified 3 conserved quantities, $H$, $I_1$ and $I_2$, it is clear that Eqs.~(\ref{eq-3wave}) are completely integrable. The identification of the number of conserved quantities and the number of independent degrees of freedom for more complicated resonant clusters is a nontrivial task which we shall discuss in more detail in Sec.~\ref{triple_cascade}. In terms of the amplitudes and dynamical phase, the conserved quantities are
\begin{eqnarray}
\label{eq-H} H = 2\,C_\pv\,C_\qv\,C_\rv\,\sin \varphi &\Rightarrow& \sin\varphi = \frac{H}{2\,C_\pv\,C_\qv\,C_\rv} , \\
\label{eq-Cq} I_1  = C_\pv^2 + C_\qv^2 &\Rightarrow& C_\qv = \sqrt{I_1-C_\pv^2} ,\\
\label{eq-Cr} I_2  = C_\pv^2 + C_\rv^2 &\Rightarrow& C_\rv = \sqrt{I_2-C_\pv^2}.
\end{eqnarray}
To solve the problem, we use these equations to write a single equation for $C_\pv^2$. Multiplying the first of Eqs.~(\ref{eq-amplitudes}) by $C_\pv$ one obtains
\begin{displaymath}
C_\pv\,\dot{C_\pv} = \frac{1}{2}\dd{C_\pv^2}{t} = -C_\pv\,C_\qv\,C_\rv\,\cos \varphi.
\end{displaymath}
Squaring and defining $u=C_\pv^2$ we get
\begin{displaymath}
\frac{1}{2}\,\left(\dd{u}{t}\right)^2 = 2\,u\,C_\qv^2,C_\rv^2\,\cos^2\varphi = 2\,u\,(I_1-u)\,(I_2-u)\left(1 - \frac{H^2}{4\,u\,(I_1-u)\,(I_2-u)}\right),
\end{displaymath}
where we have used Eqs.~(\ref{eq-H}), (\ref{eq-Cq}) and (\ref{eq-Cr}). Thus we have reduced the problem the the problem of a particle moving in a
potential well:
\begin{equation}
\label{eq-particleInWell}
\frac{1}{2}\,\left(\dd{u}{t}\right)^2 - V(u) = E,
\end{equation}
where 
\begin{displaymath}
V(u) = 2\,u\,(I_1-u)\,(I_2-u)\hspace{1.0cm}\mbox{and}\hspace{1.0cm}E = -\frac{H^2}{2}.
\end{displaymath}
The general solution of Eq.~(\ref{eq-particleInWell}) can be written down in terms of elliptic functions (see for example \cite{holm_stepwise_2002}). Here we will simply point out one particularly simple special case which occurs when $H=0$ and $I_1=I_2=I$. In this case, Eq.~(\ref{eq-particleInWell}) can be rearranged to give a separable equation which can be easily solved:
\begin{displaymath}
\dd{u}{t} = 2\,(I-u)\,\sqrt{u} \hspace{0.5cm}\Rightarrow\hspace{0.5cm} \frac{du}{(I-u)\,\sqrt{u}} = 2\,dt \hspace{0.5cm}\Rightarrow\hspace{0.5cm}u(t) = I\,\tanh^2\left[\sqrt{I}\,(t-t_0) \right].
\end{displaymath}
\begin{figure}
\begin{center}
\includegraphics[width=0.5\textwidth]{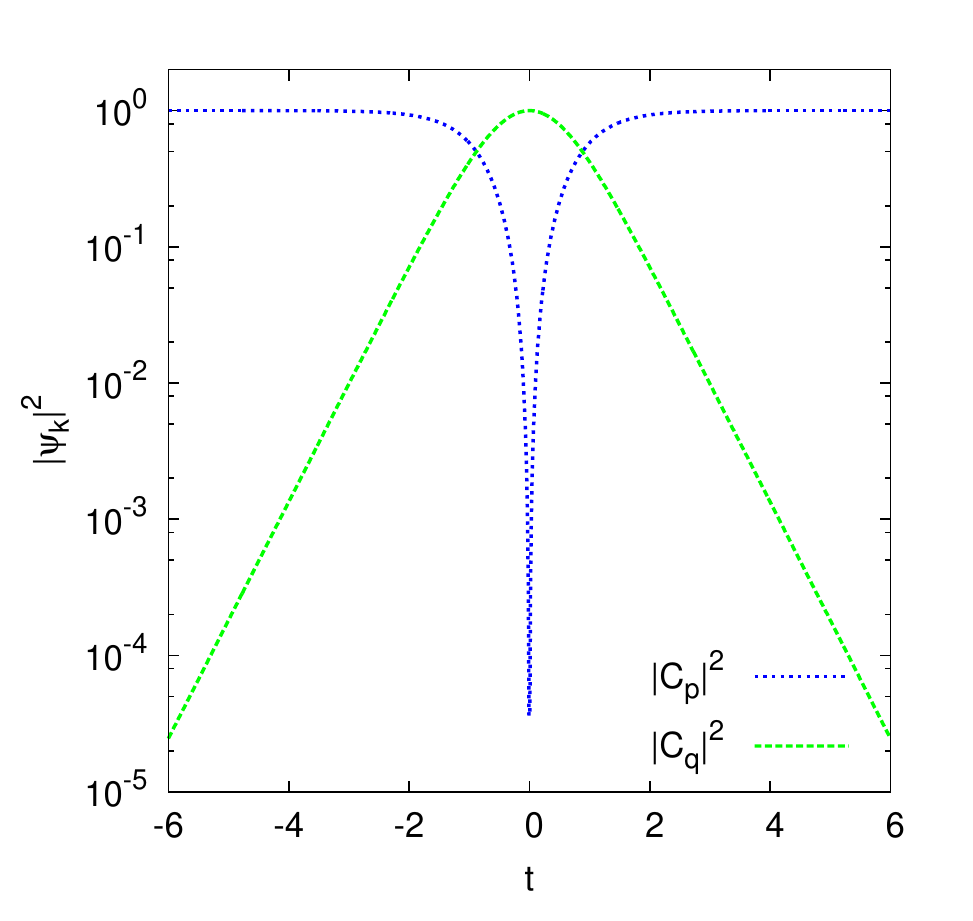}
\end{center}
\caption{\label{fig-homoclinic} Plot of the solution (\ref{eq-homoclinicOrbit}) of the resonant 3-wave equations, (\ref{eq-3wave}). }
\end{figure}
This gives the following expressions for the amplitudes (we set the reference time, $t_0=0$):
\begin{equation}
\label{eq-homoclinicOrbit}
C_\pv^2(t) = I\,\tanh^2\left[\sqrt{I}\,t \right]\hspace{1.0cm} C_\qv(t) = C_\rv(t) = I\,(1-\tanh^2\left[\sqrt{I}\,t \right]).
\end{equation}
This solution is plotted in Fig.\ref{fig-homoclinic}. It describes a situation in which all of the energy is in mode $\pv$ at $t=-\infty$ with modes $\qv$ and $\rv$ infinitessimally small. The latter two grow exponentially by absorbing energy from mode $\pv$. This represents an {\em instability} of the mode $\pv$. This instability is known as the resonant decay instability \cite{Sagdeev1969}. From the dynamical systems perspective, the resonant decay instability corresponds to a homoclinic orbit of the system (\ref{eq-3wave}): maximum energy exchange occurs at $t=0$ and the system subsequently returns to its initial configuration at time goes to $+\infty$. We shall see similar homoclinic orbits appearing in connection with the modulational instability in Sec.\ref{sec-MI}. We have already mentioned that triad-based truncations are not necessarily good approximations to the original system. If a triad or cluster is isolated and the level of nonlinearity is sufficiently low, then the system of equations (\ref{eq-3wave}) can track the solution of the full PDE quite accurately over reasonable timescales. This is illustrated in Fig.~\ref{fig-compareTriadWithCHM}. The numerical solution of the full CHM equation using a pseudospectral code is computed for an initial condition localised on the isolated exactly resonant triad  $\pv=(9,23)$, $\qv=(1,-11)$, $\rv=(8,34)$ mentioned above with a Rossby number of $M=0.1$. This solution is compared with the solution, (\ref{eq-homoclinicOrbit}) of the 3-wave equations (suitably rescaled according to Eqs.~(\ref{eq-3WavePrimitiveVariables}) to recover the primitive variables). The left panel shows an initial condition corresponding to the decay instability. It is clear that full system is well described by the solution of the triad equations for some period of time significantly longer than the timescale of the instability itself. The right panel shows the oscillatory behaviour which is typical of generic initial conditions. Again, the solution of the triad equations tracks the full system very well for many periods. 

\begin{figure}
\begin{center}
\includegraphics[width=0.45\textwidth]{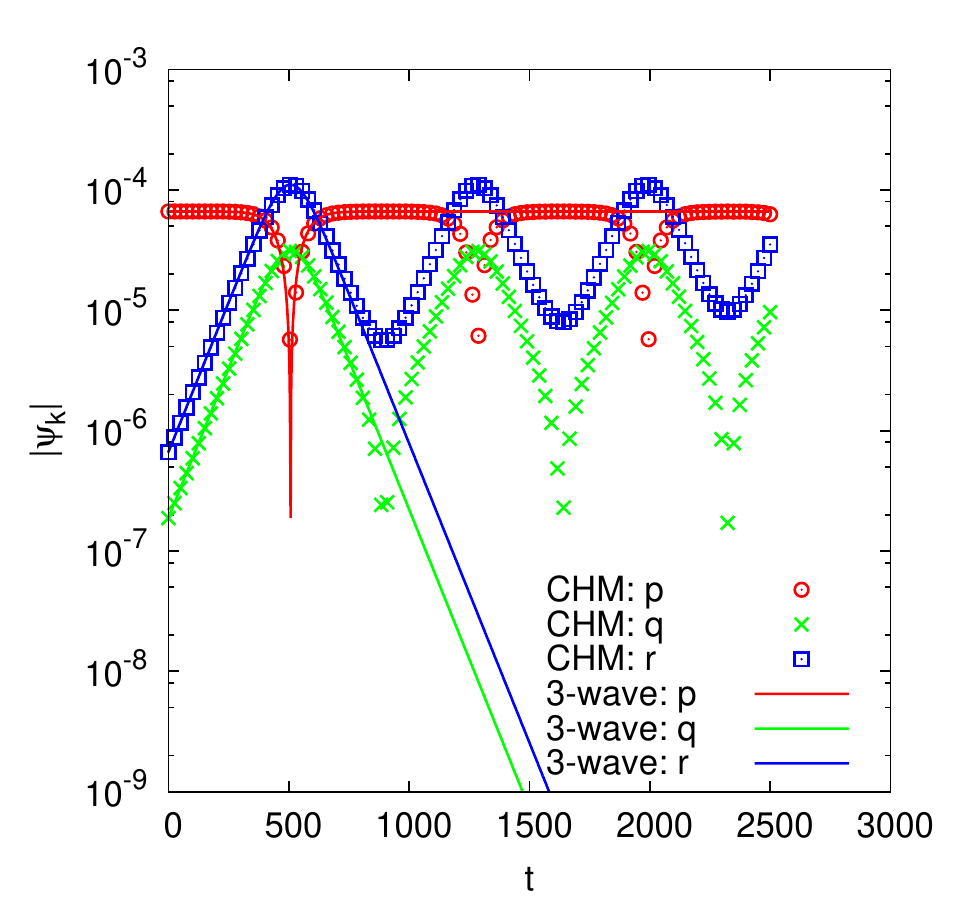}
\includegraphics[width=0.45\textwidth]{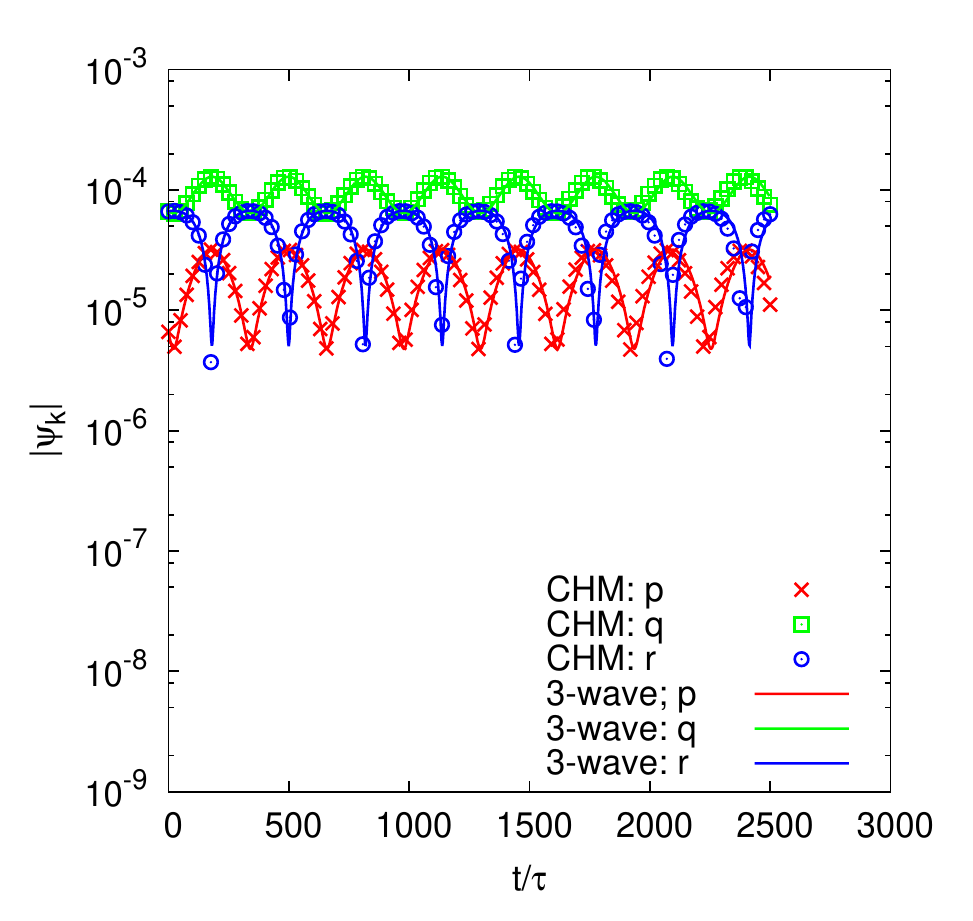}
\end{center}
\caption{\label{fig-compareTriadWithCHM} Comparision of the solution of the 3-wave equations, Eqs.~(\ref{eq-3wave}), (solid lines) with a numerical solution of the full CHM equation (symbols) for an initial condition supported on a single, isolated and exactly resonant triad. Left panel shows an initial condition corresponding to the resonant decay instability (c.f. Eqs.~\ref{eq-homoclinicOrbit}). The right panel shows the oscillatory dynamics which are typical of a generic initial condition.}
\end{figure}

To close this discussion of the nonlinear dynamics of resonant triads we note that variations on the 3-wave equations can be analysed using modifications of the methods discussed above. The case of a single triad with external forcing is analysed in \cite{harris_externally_2012}. An analogue of Eqs.~(\ref{eq-3wave}) for systems with cubic nonlinearity in which modes interact in quartets is discussed in \cite{ABDP1962}. The nonresonant case, Eqs.~(\ref{eq-3wave0}) is discussed in \cite{craik_wave_1988}.  Such resonant wave interactions have been proposed to play a role in the spatial spreading of drift wave turbulence in tokamaks \cite{gurcan_nonlinear_2006}.

\subsection{Predator-Prey models of wave-zonal flow interaction}
\label{subsec-predatorPrey}
We finish our discussion of low-dimensional models of Rossby/drift wave turbulence with a brief discussion of a different type of model known as the predator-prey model \cite{diamond_self-regulating_1994,malkov_nature_2001,Diamond2005}. A detailed discussion can be found in \cite{Diamond2005}. The predator-prey model was introduced \cite{diamond_self-regulating_1994} as a conceptual model of the self-regulating interaction between wave turbulence and zonal flows in tokamaks, a topic which we will explore in much greater detail in Sec.\ref{WT_ZF_loop}. The conceptual setting is a system in which there is a primary instability, such as the drift dissipative or ITG instability, which generates wave turbulence at small scales. There is also the potential to generate a larger scale zonal flow which can grow by extracting energy from the small scale turbulence via the eddy shearing mechanism \cite{Biglari1990,gurcan_effect_2012}. The intensity of the small scale turbulence is denoted by $N(t)$ and the intensity of the zonal flow is denoted by $U(t)$. The model equations for the dynamics are
\begin{eqnarray}
\label{eq-PPN} \dd{N}{t} &=& \gamma\,N - \alpha_1\,N^2 - \sigma\,U\,N ,\\
\label{eq-PPU} \dd{U}{t} &=& \sigma\,U\,N - \alpha2\,U - \alpha_\mathrm{NL}(U)\,U.
\end{eqnarray}
Here $\gamma$ is the growth rate of the primary instability and $\sigma$ is the strength of the shearing effect which exchanges energy between the turbulence and the zonal flows. This describes a nonlocal interaction between small scale turbulence and large scale flow. The terms with $\alpha$ are all damping terms of some kind. In the second term in Eq.~(\ref{eq-PPN}), $\alpha_1$ describes the damping of turbulence at the instability scale due to scale-local eddy-eddy interactions transfering energy to other scales. This can be thought of as the cascade term. In Eq.~(\ref{eq-PPU}), the term with $\alpha_2$ describes collisional damping of the large scale flow and in the final term $\alpha_\mathrm{NL}(U)$ is intended to model nonlinear effects which can lead to damping of the large scale flow. Some physical estimates of the parameters of the model can be found in Table 1 of \cite{diamond_self-regulating_1994}. In the simplest case, one can simply take $\alpha_\mathrm{NL}(U) = \alpha_3\,U$ where $\alpha_3$ is a constant. In this case, in addition to the trivial fixed point $(0,0)$, the physically relevant fixed points, $(N_*,U_*)$, of Eqs.~(\ref{eq-PPN}) and (\ref{eq-PPU}) are
\begin{eqnarray}
P_1 &=& \left(\frac{\gamma}{\alpha_1}, 0\right) ,\\
P_2 &=& \left( \frac{\alpha_3\,\gamma+\alpha_2\,\sigma}{\alpha_1\,\alpha_3+\sigma^2}, \frac{\gamma\,\sigma-\alpha_1\alpha_2}{\alpha_1\,\alpha_3+\sigma^2}\right) .
\end{eqnarray}
The first fixed point corresponds to a state in which there is no large scale flow ($U=0$) and the turbulence saturates due to a balance of the instability growth and the local cascade process. The second fix point, which becomes physically meaningful when the instability growth rate becomes high enough, $\gamma>(\alpha_1\,\alpha_2)/\sigma$, describes a state with a large scale flow existing in balance with small scale turbulence. The saturation level is set by a combination of the energy input from the instability, the shear effect and the damping. The fixed point $P1$ is interpreted as corresponding to the L-mode and $P2$ as corresponding to the H-mode. A stability analysis of the full parameter space is quite complicated. In \cite{diamond_self-regulating_1994}, the nonlinear damping of the zonal flow was neglected and it was shown that $P1$ is stable for $\gamma<(\alpha_1\,\alpha_2)/\sigma$ and $P2$ is stable for $\gamma>(\alpha_1\,\alpha_2)/\sigma$. When nonlinear damping is included, limit cycle solutions become possible \cite{malkov_nature_2001} in which the large scale flow exhibits periodic growth and decay. 

Predator prey models like Eqs.~(\ref{eq-PPN}) and (\ref{eq-PPU}), although comparable in complexity to the 3-mode truncation of the CHM equation, Eqs.~(\ref{eq-3wave}), described above, are therefore quite different in their structure and interpretation. The triad equations have the advantage of being directly derived from the underlying PDE but, having no free parameters, do not exhibit any bifurcations. With the exception of the homoclinic orbit, Eq.(\ref{eq-homoclinicOrbit}), the solutions of the 3-wave equations always exhibit periodic oscillations. The predator-model on the other hand, while it is difficult to obtain directly from an underlying PDE (which in many plasma regimes would not be the CHM equation anyway), contains a rich phenomenology capable of providing a paradigmatic model of how the L-H transition works. In Sec.~\ref{WT_ZF_loop}, we will describe a model based on nonlocal wave turbulence (and is thus rooted in the CHM equation) which captures certain aspects of the predator-prey phenomenology within the context of wave kinetics. 

We close this section by mentioning a final class of low dimensional models which can be thought of as intermediate between predator-prey models and large triad-based truncations. These are the so-called shell models. Shell models model the energy exchange between ranges of wave numbers which are grouped together into representative ``shells''. The interactions between the shells are obtained from phenomenological arguments constrained by symmetry considerations. Such models enjoyed a period of popularity in hydrodynamic turbulence (see the review \cite{biferale_shell_2003} which did not extend to the wave turbulence literature. One of the few exceptions is a shell model of drift wave turbulence \cite{gurcan_wave-number_2009} which has recently been used to obtain a prediction for the isotropized spectrum of drift wave turbulence which seems to be in reasonable agreement with some observations.

%%%%%%%%%%%%%%%%%%%%%%%%%%%%%%%%%%%%%%%%%%%%%%%%%%%%%%%%%%%%%%%%%%%%%%%%%%%%%%%%%%%%%%%%%%%%%%%%%%%%

\section{Modulational Instability: linear theory}
\label{sec-MI}
Since monochromatic waves of arbitrary amplitude are exact solutions of the CHM equation (\ref{eq-CHMx}), it is natural to study their stability. In this section we will consider a particular type of instability known as modulational instability (MI). Modulational instability is the name given to a process whereby a monochromatic wave, which we shall refer to as the carrier wave, can become unstable when its amplitude is modulated by a long wavelength perturbation. When modulational instability occurs, the modulation grows by extracting energy from the carrier wave. Situations in which the wavelength of the modulation is very much greater than the wavelength of the carrier wave are of particular interest because in these situations, the modulational instability provides a very effient mechanism for transferring energy from small scales to large. In the language of Sec.~\ref{subsec-RossbyWaveNonlocality}, modulational instability is an example of a scale-nonlocal process. Modulational instability is a very generic type of instability in the theory of nonlinear waves which has been widely studied at least since the seminal work on the subject by Benjamin and Feir.  We shall focus on the particular case of the CHM equation. Like in the Benjamin-Feir instability, the most unstable disturbance in CHM is actually not much longer than the carrier wave. However, we will ignore this aspect of absence of scale separation, and we will keep using the term ``modulational instability'', as it is the most commonly used term for this process in the current literature. There has been a resurgence of interest in the MI in plasma and GFD applications in recent years
~\cite{Diamond2005,Manin1994,SmolyakovShev2000,Onishchenko2004,Onishchenko2008,Champeaux2001,Connaughton2010,Gallagher2012}. This is related to the fact that MI has been recognised as a natural mechanism for initiating the formation of zonal flows.

The Rossby wave stability problem was first formulated and analysed in the geophysical context  by
Lorenz ~\cite{Lorenz1972} and Gill ~\cite{Gill1974} in the early 1970's and the main points of the theory were established. The analytical study of Gill \cite{Gill1974} in particular was very thorough and complete with virtually all important theoretical questions arising in the linear theory answered. A nonlinear analytical theory of MI for CHM model was developed in~\cite{Manin1994} using a scale separation technique.
Linear MI theory for EHM was first considered~\cite{Champeaux2001} using a scale separation technique similar to the
one of ~\cite{Manin1994}.
Recently, both linear and nonlinear MI theory within the CHM model was thoroughly revisited and a detailed numerical
study of the nonlinear MI development was carried out by~\cite{Connaughton2010}.
In a follow-up paper by~\cite{Gallagher2012} a similar study was performed for the EHM model.

Below we will overview and summarise these studies. In the formulation of the problem described below, we shall treat the linear theory of the MI as an initial value problem. That is to say that at an initial moment in time we assume that there is already a carrier wave in the system with wavevector $\pv$ and amplitude $\Psi_0$ which is subject to a small perturbation having wavevector, $\qv$. With reference to our choice of non-dimensionalization discussed in Sec.~\ref{sec-CHM}, we shall take the characteristic wave amplitude to be given by $\Psi_0$ and the characteristic length scale to be $L=p^{-1}$. In keeping with the notation of \cite{Gill1974,Connaughton2010}, the Rossby number, Eq.~(\ref{eq-RossbyNumber}), is then:
 \begin{equation}
M = \frac{\Psi_0 p^3}{\beta }\,.
\label{M}
\end{equation}
Although the nonlinear term in Eq.~(\ref{eq-CHMx}) vanishes for a monochromatic carrier wave, it is correct to think nevertheless of $M$ as a measure of nonlinearity in the MI evolution of the of the {\em perturbed} carrier wave.

The essential dynamics of the MI in the CHM model can be described in terms of the coupling of four modes having wavevectors $\pv$ (the carrier wave), $\qv$ (the perturbation) and $\qv$, and $\pv_\pm = \pv \pm \qv$ ( often referred to as``side-band'' modes). For the purposes of linear stability analysis, the dynamics of these four modes will be described using the four-mode truncation, Eqs.~(\ref{4MT}), introduced in Sec.~\ref{sec-triadBasedTruncation}. Strictly speaking, these four modes are coupled to other modes and do not form a closed system.  Indeed, even the linear problem only strictly closes with the inclusion of all the satellites $\pm \bt{q} + m \bt{p}$ where $m$ is an integer~\cite{Gill1974}.  However, in considering the linear instability it is traditional to truncate the system to the four modes only with a justification that the higher--order satellites are less excited
in the linear eigenvectors, which turns out to be a very good approximation for weak
primary waves and remains quite reasonable for strong ones.

\subsection{Linear dynamics of modulations}

We begin from the linearisation of Eq.~(\ref{eq-CHMk}) about the pure carrier wave solution.  We introduce the vector notation, 
$\bt{\Psi} = (\Psi_\bt{p},\Psi_\bt{q},\Psi_{\bt{p}_+},\Psi_{\bt{p}_-})$. A monochromatic
primary wave is given by $\bt{\Psi}_0 = (\Psi_0,0,0,0)$ where $\Psi_0$ is a complex constant representing the initial amplitude of the primary wave i.e. $\Psi_0=\Psi_{\bt{p}}|_{t=0}$ and is an exact solution of Eq.~(\ref{eq-CHMk}).  The idea is to determine how stable this solution is to small perturbations, comprised of modes $\bt{q}$, $\bt{p}_+$ and $\bt{p}_-$, by taking $\bt{\Psi}=\bt{\Psi}_0 + \epsilon \bt{\Psi}_1$ where $\bt{\Psi}_1=(0,\widetilde{\psi}_{\bt{q}},\widetilde{\psi}_\bt{p_+},\widetilde{\psi}_\bt{p_-})$.  Linearising Eqs~(\ref{4MT}) for the 4MT model at first order in $\epsilon$ gives
\begin{eqnarray}
\nonumber \partial_t \widetilde{\psi}_\bt{q} &=& T(\bt{q},\bt{p},-\bt{p}_-)\, \Psi_0\, {\widetilde{\psi}}_\bt{p_-}^* {\rm e}^{-i\, \Delta_-\, t} + T(\bt{q},-\bt{p},\bt{p}_+)\, {\Psi}_0^*\,\widetilde{\psi}_\bt{p_+} {\rm e}^{i\, \Delta_+\, t} ,\\
\nonumber \partial_t {\widetilde{\psi}}_{\bt{p}_+}^* &=& T(\bt{p}_+,\bt{p},\bt{q})\, \Psi_0\, \widetilde{\psi}_\bt{q}\, {\rm e}^{-i\, \Delta_+\, t} , \\
\label{eq-linear4MT}  \partial_t {\widetilde{\psi}}_{\bt{p}_-}^* &=& T(\bt{p}_-,\bt{p},-\bt{q})\, {\Psi}_0^*\, \widetilde{\psi}_\bt{q}\, {\rm e}^{i\, \Delta_-\, t}.
\end{eqnarray}

%\subsection{Dispersion relation for perturbations}
Now solutions are sought of the form:
\begin{equation}
\widetilde{\psi}_\bt{q}(t) = A_\bt{q} {\rm e}^{-i\, \Omega_\bt{q}\,t}\,,\;\;\;\;
\widetilde{\psi}_{\bt{p}_+}(t) = A_{\bt{p}_+} {\rm e}^{-i\, \Omega_{\bt{p}_+}\,t} \,, \;\;\;\;
\widetilde{\psi}_{\bt{p}_-}(t) =  A_{\bt{p}_-} {\rm e}^{-i\, \Omega_{\bt{p}_-}\,t},
\label{freq-match}
\end{equation}
which requires for consistency that 
\begin{equation}
\Omega_{\bt{p}_+} = \Omega_\bt{q}+\Delta_+ \quad \hbox{ and} \quad {\Omega}_{\bt{p}_-}^* = -\Omega_\bt{q}+\Delta_-.
\label{freq-match1}
\end{equation}
Writing Eqs~(\ref{eq-linear4MT}) in matrix format gives,
\begin{equation}  
\label{matrixlinear4MTpm}
\begin{pmatrix} 
i\Omega_\bt{q} & T(\bt{q},-\bt{p},\bt{p}_+) {\Psi}_0^* & T(\bt{q},\bt{p},-\bt{p}_-) \Psi_0\\ 
T(\bt{p}_+,\bt{p},\bt{q}) \Psi_0 & i(\Omega_\bt{q}+\Delta_+)   & 0\\
T(\bt{p}_-,\bt{p},-\bt{q}) {\Psi}_0^* & 0 & -i(-\Omega_\bt{q}+\Delta_+)
\end{pmatrix}
\begin{pmatrix} 
A_\bt{q} \\ 
A_{\bt{p}_+} \\
{A}_{\bt{p}_-}^* 
\end{pmatrix} = 0,
\end{equation}
and setting the determinant of the $3 \times 3$ matrix to zero, gives a cubic expression for the dispersion relation of the modulation frequency $\Omega_{\bt{q}}$,
\begin{eqnarray}
\nonumber\Omega_\bt{q}(\Omega_\bt{q}+\Delta_+)( -\Omega_\bt{q}+\Delta_-) + T(\bt{q},-\bt{p},\bt{p}_+)\, T(\bt{p}_+,\bt{p},\bt{q})\, \left|\Psi_0\right|^2 ( -\Omega_\bt{q}+\Delta_-)\\
\label{eq-MIDispersion}  - T(\bt{q},\bt{p},-\bt{p}_-)\, T(\bt{p}_-,\bt{p},-\bt{q})\, \left|\Psi_0\right|^2 ( \Omega_\bt{q}+\Delta_+)  = 0, \quad
\end{eqnarray}
which can be solved numerically for $\Omega_{\bt q}$, which is then used to find the corresponding eigenvectors
\begin{equation}
\label{4MTEigenvector}
\begin{pmatrix}
A_\bt{q}\\
A_{\bt{p}_+}\\
A_{\bt{p}_-}
\end{pmatrix} = 
\begin{pmatrix}
1\\
\frac{T(\bt{p}_+,\bt{p},\bt{q})\, \Psi_0}{-i\,(\Omega_\bt{q} + \Delta_+)}\\
\frac{T(\bt{p}_-,\bt{p},-\bt{q})\, \Psi_0}{i\,({\Omega}_\bt{q}^* - \Delta_-)}\,
\end{pmatrix}.
\end{equation}
Instability occurs when a root $\Omega_{\bt q}$ has a positive  imaginary part, which is the growth rate of the instability, $\gamma = \Im (\Omega_{\bt q}) >0$. Then $\tau=\frac{1}{\gamma}$ is the characteristic growth time.  
%Substituting Eq.~(\ref{eq-EHMinteractioncoeff}) for the interaction coefficient back in to the dispersion relation, Eq.~(\ref{eq-MIDispersion}) and performing some algebra, gives the more usual form of the dispersion for the CHM equation~\cite{Diamond2005,Lorenz1972,Gill1974,Manin1994,SmolyakovShev2000,Onishchenko2004},
%% CAN I CHANGE THIS TO NOINDENT AND NOT SMALLER??
%\small 
%\begin{eqnarray}
%\label{eq-MIdispersion}
%& & (q^2 + F)\Omega_{\bt{q}} + \beta  q_x + \\  \nonumber & & \left|\Psi_0\right|^2 \left|\bt{p} \times \bt{q} \right|^2(p^2-q^2) \left[ \frac{p^2_+-p^2}{(p^2_+ +F)(\Omega_{\bt{q}}+\omega)+\beta {p_+}_x}- \frac{p^2_- -p^2}{(p^2_-+F)(\Omega_{\bt q}+\omega)+\beta  {p_-}_x}\right] =0
%\end{eqnarray}
%\normalsize
The case of a purely meriodional primary wave ($\pv=(p,0)$) subject to a purely zonal perturbation ($\qv =(0,q)$)  is of physical interest. In this case, the dispersion relation, Eq.~(\ref{eq-MIDispersion}), simplifies somewhat (see \cite{Connaughton2010}):  
\begin{equation}
\label{eq-MIOrthogonalCase}
\Omega^{\prime\,3} + \frac{s^4 \left[{2 M^2(1-s^2)(1+f)^2(s^2+f+1)-(s^2+f) } \right]}{(1+f)^2(s^2+1+f)^2 ({s^2 + f})}  \Omega^\prime =0.
\end{equation}
where $s=q/p$, $\Omega^\prime = \Omega_\qv/p$ and $f=F/p^2$.
The roots are
\begin{eqnarray}
\Omega^\prime&=&0\\
\Omega^\prime&=&\frac{\pm i  s^2}{(1+f)(s^2+1+f)} \left[\frac{2 M^2(1-s^2)(1+f)^2(s^2+f+1)-(s^2+f) }{s^2 + f}\right]^{1/2}.
\end{eqnarray}

\subsection{Cases with purely meridional carrier wave}

The structure of the instability depends strongly on the parameter $M$.
For illustration, let us consider the case where the carrier wave is purely meridional, ${\bf p} = (p_x, 0)$. As seen in figure~\ref{qMaps_F_0}, for $M \to 0$ the instability concentrates narrowly around two oval-shaped curves. Each of these ovals appears to be the resonant curve on which the carrier wave, the modulation and one of the satellites are in exact three-wave resonance: %~\cite{Connaughton2010}.
\begin{eqnarray}
 {\bt p} = {\bt p}_- + {\bt q}, &\quad
\label{eq-resonanceConditionPlus}\omega({\bt p}) = \omega({\bt p}_-) +\omega({\bt q}) , \\
\hbox{or} \quad 
 {\bt p} = {\bt p}_+ - {\bt q}, &\quad
\label{eq-resonanceConditionMinus}\omega({\bt p}) = \omega({\bt p}_+) - \omega({\bt q}).
\end{eqnarray}

\begin{figure}
\begin{center}
\includegraphics[width=\textwidth]{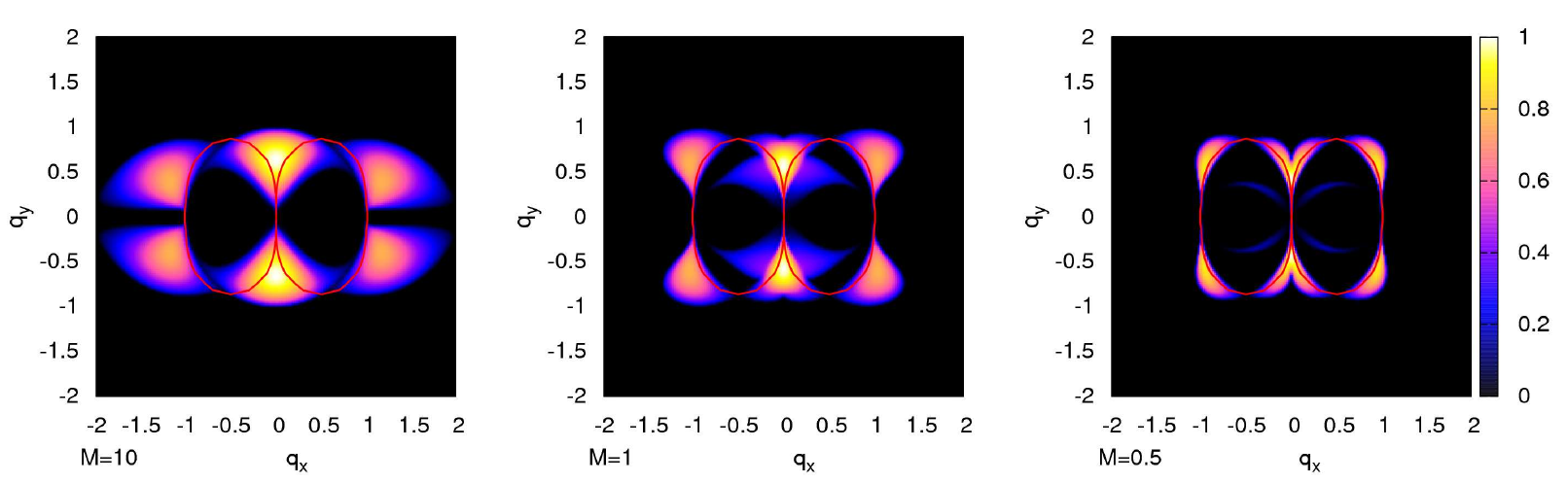}
\end{center}
\caption{\label{qMaps_F_0}  Growth rate of the modulational instability as a function of ${\bt q}$ for a fixed meridional primary wavevector, ${\bt p}=(1,0)$ and $\rho=\infty$ for various values of the nonlinearity $M$.}
\end{figure}

\begin{figure}
\begin{center}
\includegraphics[width=\textwidth]{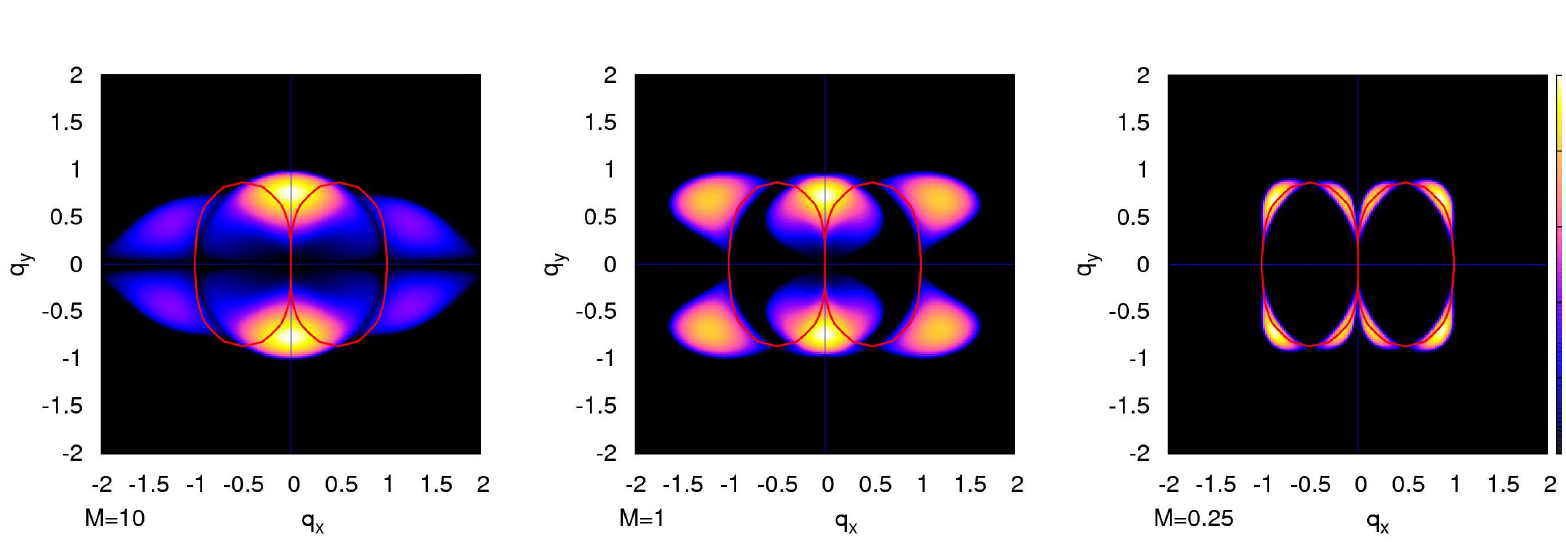}
\end{center}
\caption{\label{qMaps_F_1}  Growth rate of the modulational instability as a function of ${\bt q}$ for a fixed meridional primary wavevector, ${\bt p}=(1,0)$ and $F=1$ for various values of the nonlinearity $M$. The curves are qualitatively similar to the $F=0$ curves shown in Fig.~\ref{qMaps_F_0}.}
\end{figure}
In the strong interaction limit, $M \to \infty$, the $\beta$-term in the CHM equation is unimportant and the MI problem is reduced to the well-known instability of the Kolmogorov (sine-shaped) plane-parallel shear flows described by the Euler equation for the incompressible fluid ~\cite{Arnold1960,Meshalkin1962}. The maximum instability is obtained when the primary and secondary waves are perpendicular~\cite{Gill1974} and the traditional approach is to select a meridional flow such that  $\bt{p}$ is along the $x$-axis and the perturbation $\bt{q}$ along the $y$-axis.  In this situation, looking at figure~\ref{qMaps_F_0}, the modulation wave vector $\bt{q}$ is equally close to both branches of the resonant manifold. However, in the weak nonlinearity limit,  the maximally unstable perturbation bifurcates off the zonal axis at a critical value of $M_c \approx 0.53474$, tending to a point on the resonant manifold located at an angle of $5\pi/6$ with the $x$-axis as $M$ reduces further as shown in figure~\ref{theta_max}.
\begin{figure}
\begin{center}
\includegraphics[width=0.35\textwidth]{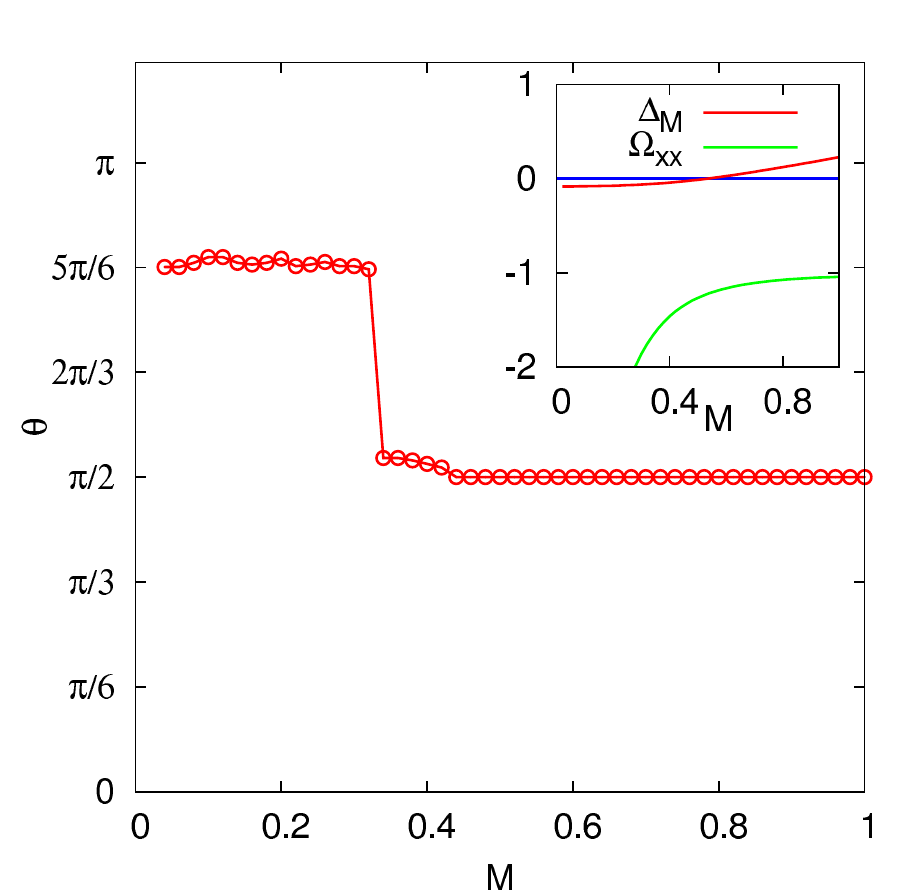}
\end{center}
\caption{ Angle, $\theta$, between the $\bf{q}$ wave-vector of the maximally unstable perturbation and the $x$-axis as a function of $M$. Inset plots $\Delta_M$ and $\Omega_{xx}$ as a function of $M$ illustrating the transition of the maximum growth rate for on-axis perturbations from a local maximum to a saddle point at $M\approx 0.53$.}
\label{theta_max}
\end{figure}

\subsection{Cases with purely zonal modulation}

%\subsection{General discussion of $M$, $\rho$ and $s$}

Fixing modulation ${\bf q}$ to be purely zonal, one can study the instability for different ${\bf p}$. This results in the most familiar statement in the MI papers on plasmas~\cite{Manin1994,SmolyakovShev2000,Onishchenko2004} that the primary wave will be unstable if it lies within the cone 
%\begin{equation}
%\label{cone}
$1/\rho^2 + p^2_x -3p^2_y > 0$.
% p_y < \frac{1}{\sqrt{3}} p_x.
%\end{equation}
When $\rho=\infty$, this cone reduces to $p_y < \frac{1}{\sqrt{3}} p_x$.  
However, this is true only in the double limit $M\gg1, p\gg q$.
Removing the scale separation condition $p\gg q$ but still keeping $M\gg1$
arrives at a more general instability region~\cite{Gill1974}
%\begin{equation}
$\cos^2 \phi < ({1+ {q^2}/{p^2}})/{4},
$
%\end{equation} 
where $\phi$ is the angle between $\bf{p}$ and $\bf{q}$.

Figure~\ref{kMaps_F_0} shows the instability growth rate 
 as a function of ${\bf p}$ for a fixed zonal modulation ${\bf q}=(0,1)$ and $F=0$ (ie. $\rho = \infty$). 
  As $M \to 0$, a region of stable wavenumbers inside the cone $p_y < \frac{1}{\sqrt{3}} p_x$ becomes larger such that unstable wavevectors require a larger $p_x$.  Furthermore, for large $M$, an instability exists for some wavenumbers outside the cone which are very close to the zonal direction.  
%A similar plot for $F=2$ in figure~\ref{kMaps_F_2} shows that the most unstable wavevectors for a given zonal perturbation are those which lie in the $x$-axis, i.e. the meridional wavevectors. 
% It is important to keep this in mind that, even for large $M$, t
such that the maximum growth 
% occurs outside of the cone, 
occurs for the primary wave orientations closer to zonal
 than to the the meridional direction, see figure~\ref{kMaps_F_0}  for $M=10$.
% This fact is easy to overlook if one considers only the limit $M \to \infty$ (as is common in the plasma literature) because in this limit, the growth rate maximum is for the meridional primary waves.
 
On the other hand, the choice of the primary wave direction is often dictated not by
 the maximum growth rate of the modulational instability, but by the structure of the
 primary instability creating the Rossby and drift waves.
% (ITG instability in plasmas and the baroclinic instability in GFD).
 
\begin{figure}
\begin{center}
\includegraphics[width=0.7\textwidth]{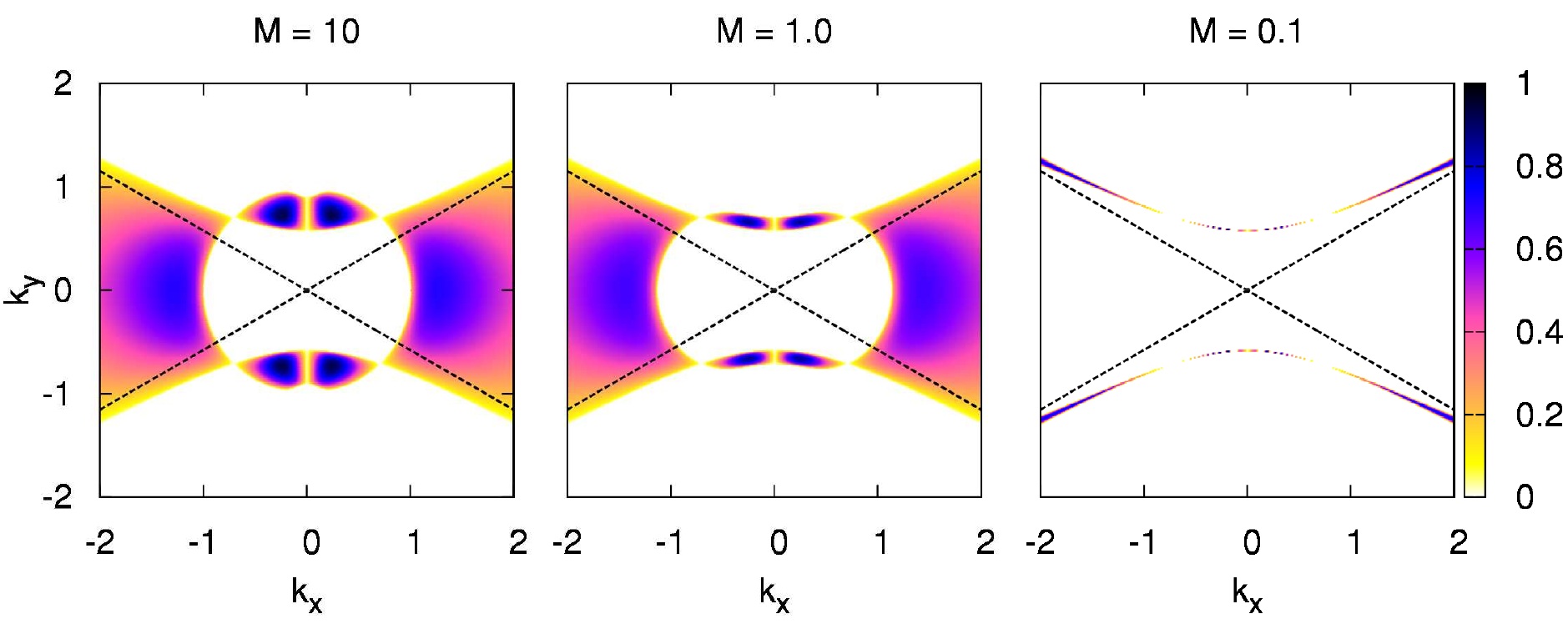}
\end{center}
\caption{\label{kMaps_F_0}
Growth rate of the modulational instability, given by  Eq.~\ref{eq-MIDispersion}
as a function of $\bf{p}$ for a fixed zonal modulation wavevector, $\bf{q}=(0,1)$ and $\rho=\infty$ for various levels of $M$. The dashed line is the cone defined by $p_y < \frac{1}{\sqrt{3}} p_x$}
\end{figure}

In the limit of weak nonlinearity $M\ll 1$, the dynamics are completely wave dominated, and the instability is again close to the resonant curves~\eqref{eq-resonanceConditionPlus} and
\eqref{eq-resonanceConditionMinus}.  The nonlinear terms allow waves to interact weakly and exchange energy. Since the nonlinearity is quadratic, wave interactions are triadic i.e. three-wave resonances
are allowed by the dispersion relation given by Eq.~(\ref{eq-MIDispersion}).
Any triad of waves having wavevectors ${\bt k}$, ${\bt k}_1$ and ${\bt k}_2$ 
interact only if they satisfy the resonance conditions, Eq.~(\ref{eq-resonances}).

In fact, figure~\ref{qMaps_F_0} shows two resonant curves corresponding to the two resonant triads, $({\bt p},{\bt q},{\bt p}_+)$ and $({\bt p}_-,{\bt q},{\bt p})$ where the sidebands ${\bt p}_{\pm} = {\bt p} \pm {\bt q}$.
An interesting feature of instability for $M \ll 1$ is evident in figure~\ref{kMaps_F_0},
that for fixed zonal ${\bf q}$ the unstable region becomes narrow and collapses onto the sides of the cone i.e. onto the lines $p_y = \pm p_x /\sqrt{3}$~\cite{Connaughton2010}. It should be noted that these resonant manifolds are relevant even for higher levels of nonlinearity because as can be seen from figure~\ref{qMaps_F_0}, the unstable modulations still concentrate close to the resonant curves.

\subsubsection{Finite $\rho$ effects  and EHM}

First of all we remind the reader that the EHM case is identical to CHM unless the radius 
$\rho$ is finite and one of the modes (most likely $\bf q$) is purely zonal.
Thus, all results described above for the cases $\rho = \infty$ or/and
when the modulation was not purely zonal apply equally to the CHM and the EHM models.

%A finite deformation radius is obtained in the QG system under a reduce-gravity approximation.  
In CHM, when $\rho$
is finite, there are two regimes~\cite{Connaughton2010}, depending on the value of $M$:
For $M>\sqrt{\frac{2}{27}}$  the finite radius reduces the growth rate of the instability but cannot suppress it, whereas for
 $M \le \sqrt{\frac{2}{27}}$  there is a range of intermediate radii which completely suppress the instability.

Modulational instability within the EHM model was considered by~\cite{Champeaux2001}. Their WKB-type method implied (explicitly) scale separation between the carrier wave and the modulation and (implicitly) strong nonlinearity $M \gg 1$ (see discussion in~\cite{Connaughton2010}).
It was shown that for EHM the instability is  stronger for purely zonal $\bf q$. 
This conclusion is generally true also for finite values of $M$~\cite{Gallagher2012}.

 Figure~\ref{EHMgrowth} shows the amplitude of the perturbation
mode $\bt{q}$ obtained by solving numerically the 4MT  system (\ref{4MT})  for the EHM and CHM equations.  Clearly, the effect of the extension to the CHM equation is that the growth rate associated with the EHM model is larger than that for the CHM model.
\begin{figure}[h]
\begin{center}
\includegraphics[width=0.45\textwidth]{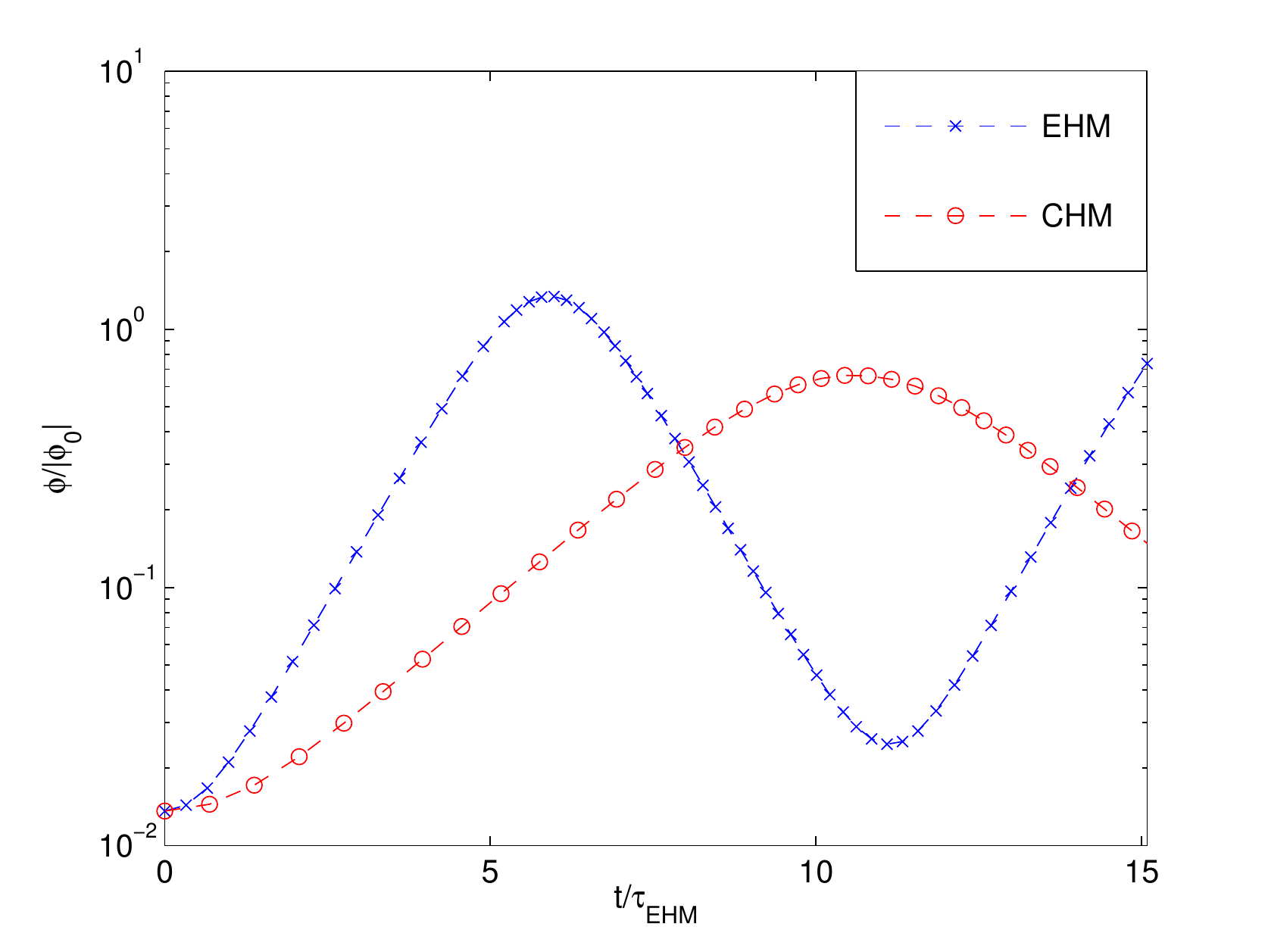}
\caption{The growth of the zonal mode for identical initial conditions in ODE simulations of the EHM and CHM
systems for the case $\bt{p}=(10,0)$, $\rho = 0.6$, $\psi_0=0.01$ and $\beta = 10 $. 
Both have been
normalised using the analytical linear growth time of the EHM system. }
\label{EHMgrowth}
\end{center}
\end{figure}

\subsection{Relationship between modulational instability and zonostrophic instability}

It has recently been shown by Srinivasan and Young \cite{srinivasan_zonostrophic_2012} that a state of spatially homogeneous turbulence in the CHM equation is unstable to  coherent zonal perturbations. Such perturbations can grow by extracting energy from the turbulent background to produce coherent jets from an initially jetless state of incoherent turbulence. Srinivasan and Young have coined the term zonostrophic instability to describe this phenomenon. The growth rate of the zonostrophic instability is a function of the spectrum of the forcing of the turbulent background; see 
Eq.(45) of \cite{srinivasan_zonostrophic_2012}. Zonostrophic instability has been proposed as another candidate mechanism for the formation of zonal jets. A related, but seemingly distinct, pattern forming instability has recently been suggested \cite{parker_zonal_2013} as another mechanism for generating zonal jets from statistically homogeneous background turbulence. While the concept of zonostrophic instability as described in  \cite{srinivasan_zonostrophic_2012} may not, at first sight, have much in common with the deterministic initial value problem outlined above to describe the modulational instability, in fact the two mechanisms are closely related. It has been shown by Parker and Krommes \cite{parker_zonal_2014} that the dispersion relation for zonostrophic instability reduces to Eq.~(\ref{eq-MIOrthogonalCase}) in the case where the background spectrum is taken to be a purely meridional monochromatic wave with wavevector $\pv$. The modulational instability is therefore a special case of the zonostrophic instability of an arbitrary background spectrum. The first study of the instability of a broadband Rossby/drift wave spectrum to large-scale zonal perturbations was done by Manin and Nazarenko~\cite{Manin1994}.

%%%%%%%%%%%%%%%%%%%%%%%%%%%%%%%%%%%%%%%%%%%%%%%%%%%%%%%%%%%%%%%%%%%%%%%%%%%%%%%%%%%%%%%%%%%%%%%%%%%%
\section{Modulational Instability: nonlinear stage}
\label{sec-numerics}
%\subsubsection{CHM }

%To test the linear predictions and to study the nonlinear evolution,
%DNS of the CHM model have been performed using a
%standard pseudospectral method with resolution up to $1024^2$, de-aliasing and
%hyperviscosity parameters $\nu_n = 4.5e^{-30}$. 

In this section we will describe the nonlinear development and saturation of the modulational instability described in the previous section at the level of linear stability theory. Although the 4MT, Eqs~(\ref{4MT}), was used as the departure point
for the linear stability analysis, it is a fully nonlinear set of equations
in its own right with its own nonlinear dynamics. Unlike the case of the 3MT described in Sec. \ref{subsec-triad}, analytic results for the  4MT are not readily available. The 4MT system can, however,  be easily solved numerically. In addition to checking the linear instability predictions
against DNS, the extent to which the nonlinear  dynamics
of the 4MT captures the behaviour of the full PDE will also be explored. In all cases, the initial condition is chosen to be along the unstable eigenvector of the 4MT.

\begin{figure}
\begin{center}
\includegraphics[width=0.8\textwidth]{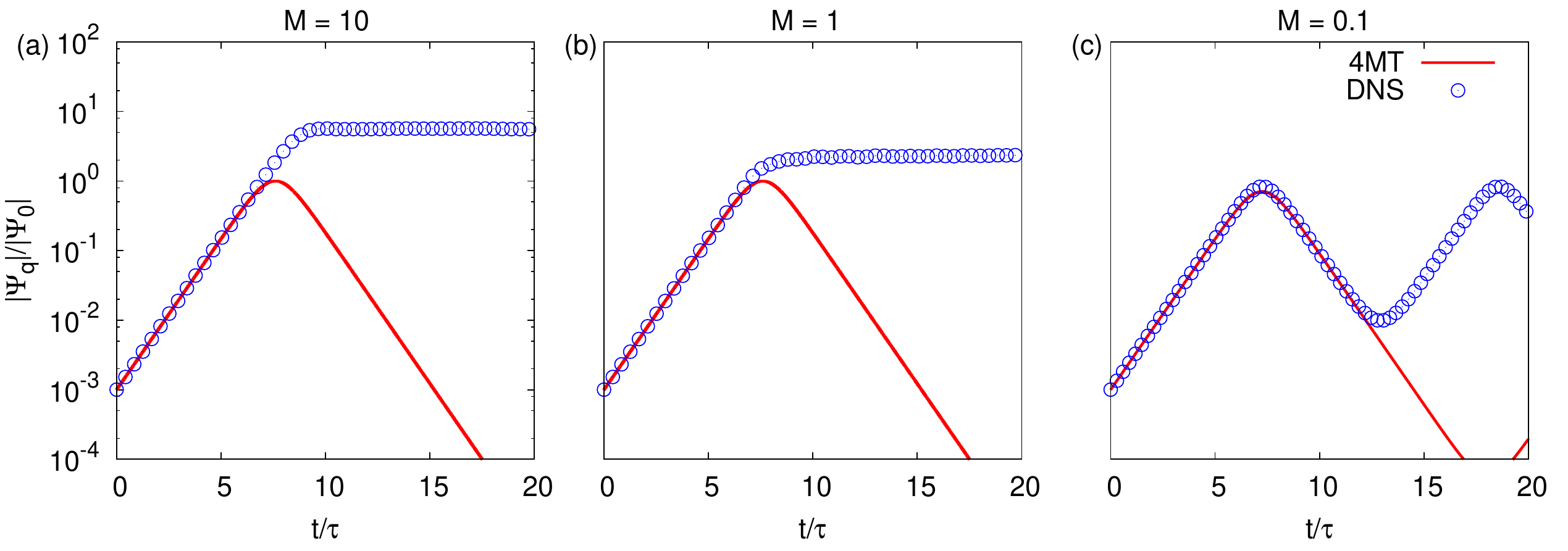}
\caption{Comparison of the growth of the zonal mode $\bf{q}$ obtained by DNS versus solving 4MT system.  In each case the primary wavenumber is ${\bf p} = (10,0) $ and the modulation wavenumber is ${\bf q} = (0,1) $.  The nonlinearity levels are (A) $M=10$, (B) $M=1.0$ and (C) $M=0.1$ and time is scaled by the corresponding $\tau$.}
\label{growth}
\end{center}
\end{figure}

Immediately from figure~\ref{growth}, it can be seen that can see that the initial stage of evolution agrees very well with the predictions of the linear theory obtained from the 4MT.  Moreover, the 4MT works
rather well beyond the linear stage, particularly in the $M = 0.1$ case, where the initial
growth reverses in agreement with the (periodic) behaviour of the four-mode system.
For $M =1$, the  growth does not reverse, but rather experiences a saturation
at the level where the four-mode system reaches maximum and reverses. The most
surprising behaviour is seen for M =10 where the linear exponential growth continues
well beyond the point of reversal of the four-wave system, even though the system is
clearly nonlinear at these times and follows a self-similar evolution (see below).

A series of snapshots of the vorticity field for the nonlinearity levels $M=0.1$ %$M=1$ 
and $M=10$ are shown in figures~\ref{snapshotsM0_1}
%, ~\ref{snapshotsM1} 
and~\ref{snapshotsM10} respectively.  The cases simulated are for ${\bf p} = (10,0) $ and ${\bf q} = (0,1) $.  The evolution of the mean zonal velocity $\overline u(y)$ averaged over $x$,
obtained from DNS is shown in figure~\ref{zonalU} for times close to the formation of the jet.
Below we will consider these figures in the context of three different stages of the nonlinear evolution.

\subsection{Initial nonlinear stage: sharpening of jets}
%Sharpening, self-similar form, final width of jet. Coarsening of jets and PV staircase...\\

The nonlinear dynamics show zonal jets self-focusing and becoming very narrow with respect to the initial modulation wavelength.
% such as those in figures~\ref{zonalU}~(a)~and~(b).  
This self-focusing was predicted theoretically~\cite{Manin1994} for large $M$ and $q \ll p$ where self-similar solutions were obtained describing a collapse of the jet width.  This feature cannot be described by the 4MT because such an harmonic jet shape involves strong contributions from higher harmonics ${\bf p} \pm n {\bf q}$. 

It was shown in \cite{Connaughton2010} that  the zonal velocity $\overline u$   in  a run with $M=10$ and $\rho = \infty$ can be fitted with a
self-similar shape $\overline u(y,t) = a(t) \, f(b(t) y)$, where $a(t) = u_0 \, e^{\gamma_{\bt q} t}$ and $b(t) = e^{1.85 t}$.  
The nonlinear growth at the self-similar stage continues with the same exponential law, ${\rm e}^{\gamma_{\bt q} t}$, as in the linear dynamics.  The self-similarity must stop when the scale separation property breaks down due to the jet narrowing, at which point
a roll-up into vortices occurs.
For smaller $M$, the extension of the growth rate beyond the linear stage is not observed and the amplitude of the zonal mode decreases after reaching a maximum in correspondence with the solution of the 4MT. The self-focusing is still observed but much reduced and the self-similar stage is not clearly seen.

\begin{figure} 
\begin{center}
\includegraphics[width=0.8\textwidth]{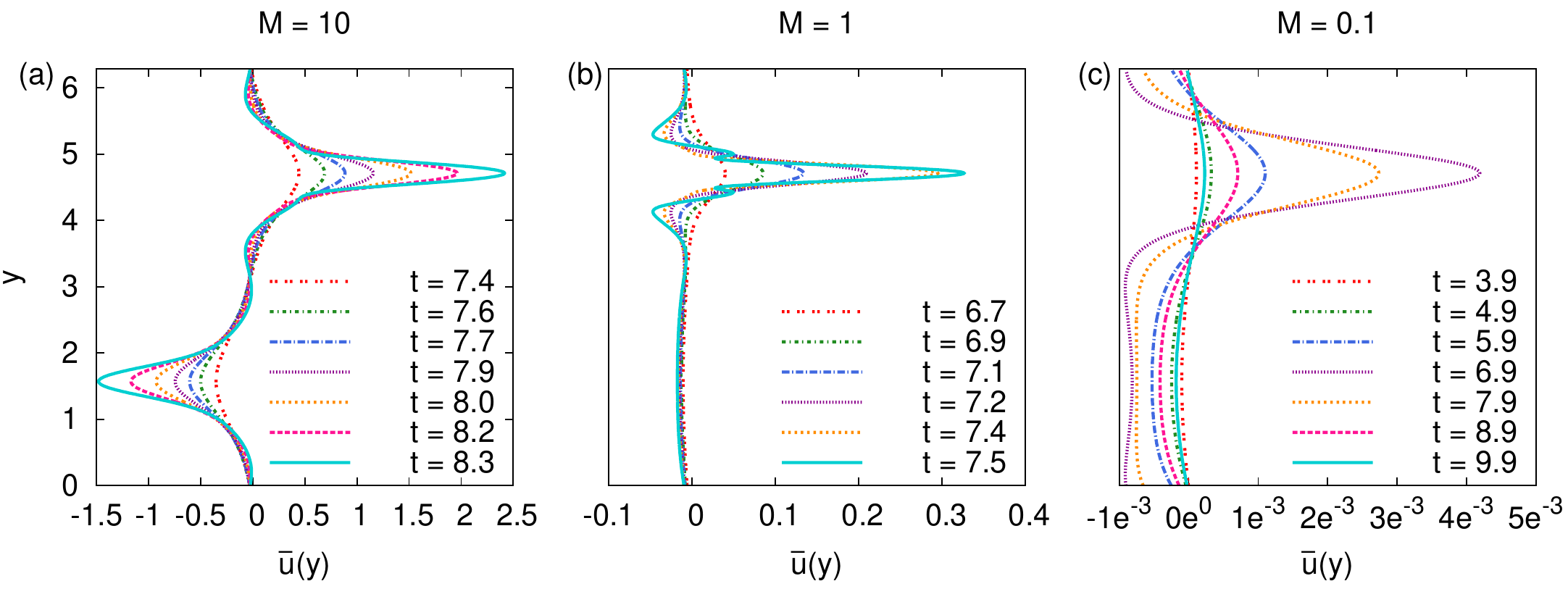}
\caption{Mean zonal velocity profiles for CHM ($s=0$); (a) $M=10$, (b) $M=1.0$ and (c) $M=0.1$.}
\label{zonalU}
\end{center}
\end{figure}

\subsection{Intermediate nonlinear stage: vortex roll-up vs oscillation}

\begin{figure}[ht]
\begin{center}
\includegraphics[width=0.8\textwidth]{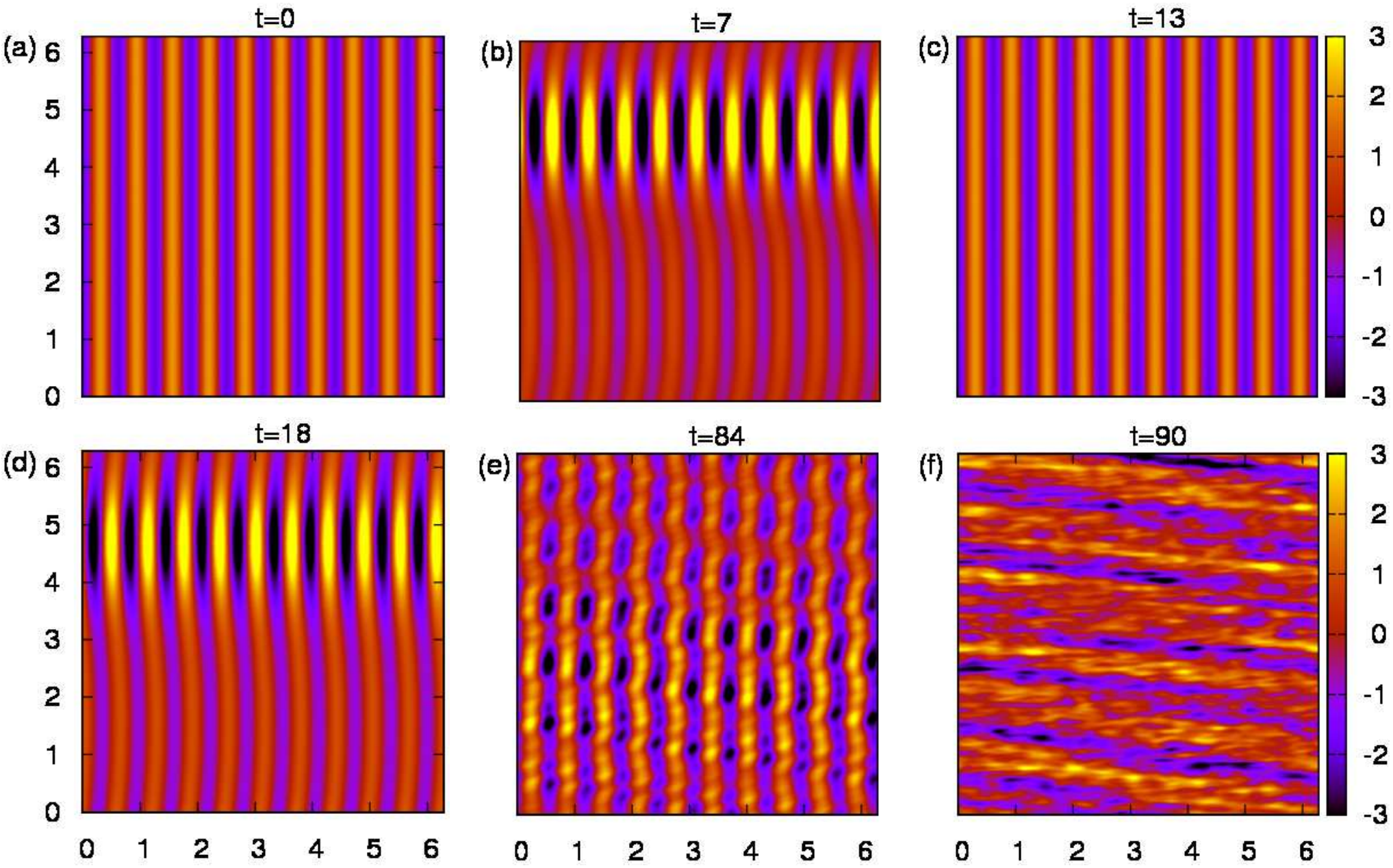}
\caption{Vorticity snapshots for a run with ${\bf p} = (10,0) $ and ${\bf q} = (0,1) $, $M=0.1$ and $\rho=\infty$. The growth stage is followed by growth reversal, saturation and transition to irregular off-zonal jet structures.
%The horizontal axis is $x$ and the vertical is $y$.
}
\label{snapshotsM0_1}
\end{center}
\end{figure}
%\begin{comment}
\begin{figure} [ht]
\begin{center}
\includegraphics[width=0.8\textwidth]{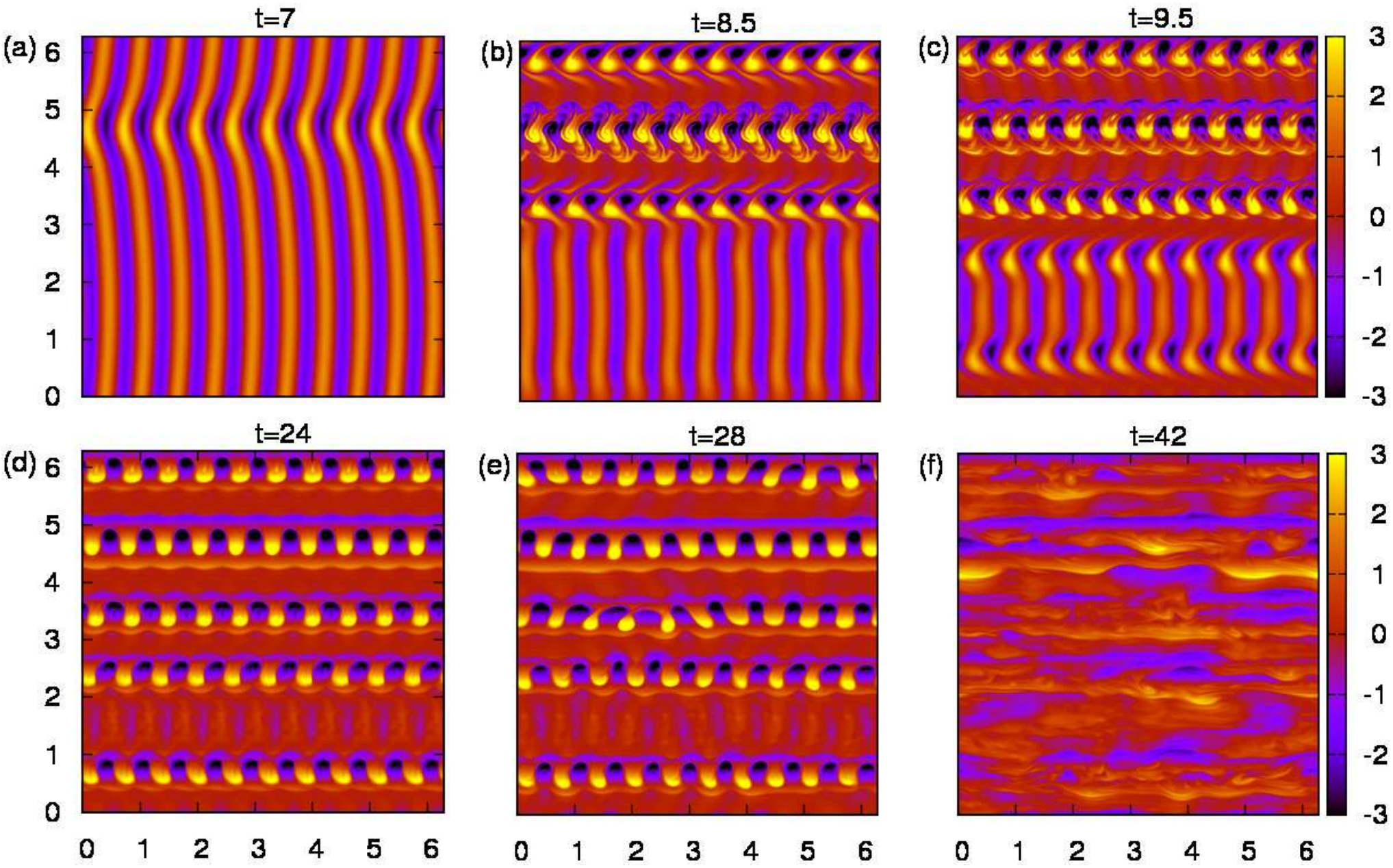}
\caption{Vorticity snapshots showing the linear growth, saturation and transition to turbulence of a zonal perturbation to a meridional primary wave having $M=1$ and $\rho=\infty$. The horizontal axis is $x$ and the vertical is $y$.}
\label{snapshotsM1}
\end{center}
\end{figure}
%\end{comment}
\begin{figure}[ht]
\begin{center}
\includegraphics[width=0.8\textwidth]{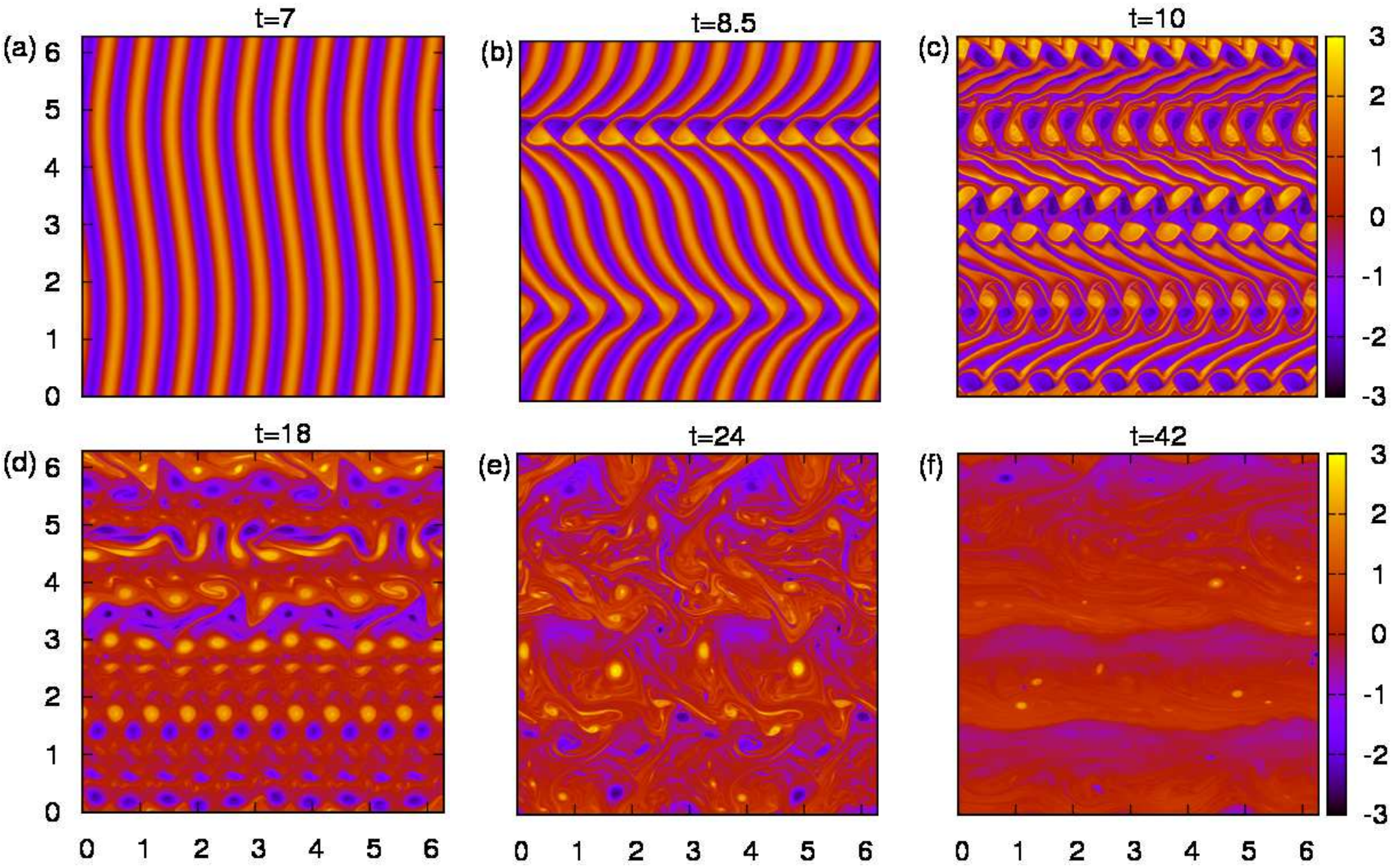}
\caption{Vorticity snapshots for a run with ${\bf p} = (10,0) $ and ${\bf q} = (0,1) $, $M=10$ and $\rho=\infty$. 
 The growth stage is followed by saturation to a vortex street state slowly deteriorating  to turbulence with a PV staircase structure.
  }
\label{snapshotsM10}
\end{center}
\end{figure}

It is evident from figures~\ref{snapshotsM0_1}
%, ~\ref{snapshotsM1} 
and~\ref{snapshotsM10} that there are two qualitatively different types of behaviour, namely vortex roll-up followed by saturation or the oscillatory wave dynamics. 
For weak waves, $M < M_*$, the  oscillatory behaviour is seen where the system returns close to the initial state.  The 4MT model captures this effect very well.   This oscillation between a Rossby wave and a zonal flow has been reported from numerical studies of a barotropic fluid~\cite{Mahanti1981}, motivated by observations of such periodic behaviour in the troposphere of the southern hemisphere~\cite{Webster1975} and similar behaviour has also been noted for the EHM model~\cite{Manfredi2001} for drift turbulence.
For strong waves, $M > M_*$, the nonlinear evolution is vortex dominated and the vorticity of the initial primary wave rolls into vortices and organises itself into K\'arm\'an-like vortex streets.  This corresponds to the jet velocity saturating. At the moment of the vortex roll-up energy spills beyond the four modes and the 4MT description fails.

 It is therefore natural to seek the critical level of nonlinearity $M_* $ which distinguishes between the two types of behaviour.  If the maximum jet strength, as predicted by the 4MT, exceeds the value of the Rayleigh-Kuo necessary instability condition $\partial_{yy} \overline u(y) - \beta >0$~\cite{Rayleigh1913,Kuo1949,McWilliams2006,Connaughton2010}, then the vortex roll-up occurs and the jet strength saturates for a long time. %(although figure~\ref{growth} shows time as far as $20$~characteristic timescales, simulations were actually run until $100$ characteristic times). 
 At this point, the behaviour of the system starts to depart from that of the 4MT.  If however, the maximum jet strength remains below the Rayleigh-Kuo threshold, then the system's growth reverses and follows the 4MT dynamics for a longer time.

This simple picture permits a qualitative physical estimate for the saturated velocity of the jet  $u_{max}  \approx 3 \frac{M \beta } {p^2}$ and respectively $M_* \sim \frac{1}{3}$, which agrees with the DNS result for the oscillatory-saturated transition at   $M_* \approx 0.25-0.35$~\cite{Connaughton2010}.
%Let us start with the latter, see Fig.~\ref{fig-growth_earlyTime}.
%\begin{comment}
%Since the $x$-periodicity is preserved, the spacing of the stable vortex street i.e. the distance between each vortex is equal to the wavelength of the original primary wave and since the vortices are approximately round, the $y$-spacing between vortices is equivalent to the $x$-spacing.  Thus, the width of the saturated jet is of the order of the wavelength of the initial primary wave and the jet saturation velocity is of the order of the velocity amplitude of the initial primary wave, $\overline u_{max}  \sim \frac{M \beta } {p^2}$.  In fact, numerical results for $M=10$ and $M=1$ from figures~\ref{zonalU}(a)~and~(b), give the estimate, $u_{max}  \approx 3 \frac{M \beta } {p^2}$.  Estimating $\partial_{yy} \overline u(y)$ as $p^2 u_{max}$ and substituting into the Rayleigh-Kuo condition above arrives at the estimate, $p^2 u_{max}  - \beta  >  0$, $\frac{3M\beta}{p^2} > \frac{\beta}{p^2} \implies M >  M_* \sim \frac{1}{3}$.
Numerical explorations in figure~\ref{Mcrit} reveal that $M_* \approx 0.25-0.35$ but this boundary is not sharp.  For $M=0.25$ the dynamics are definitely wave-dominated although some elongated, fuzzy vortices are still apparent, whereas for $M=0.35$ streets of round vortices are clearly formed with some wave-like oscillations still present.
\begin{figure}[ht]
\begin{center}
\includegraphics[width=0.35\textwidth]{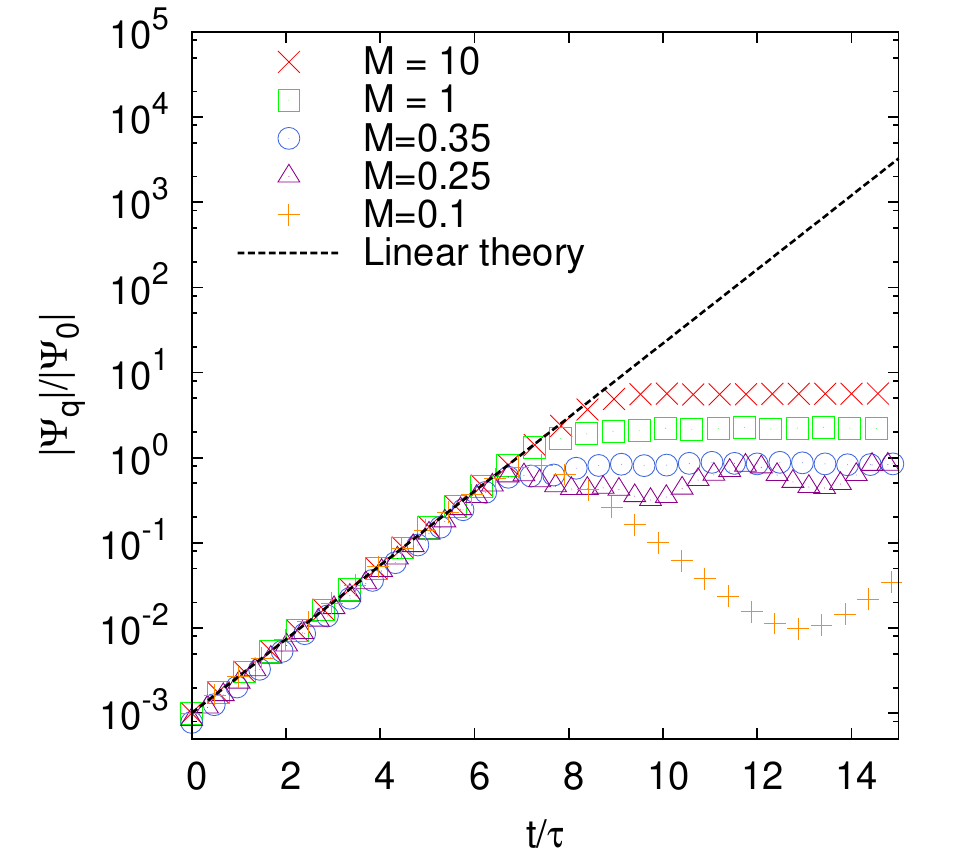}
\caption{Growth of zonal perturbations due to modulational instability of a meridional primary wave having ${\bf p}=(10,0)$ for several different values of $M$. The amplitude of the zonal mode has been scaled by $\Psi_0$ and time has been scaled by $\tau$, the inverse of the instability growth rate.}
\label{Mcrit}
\end{center}
\end{figure}
%\end{comment}

\subsection{Late nonlinear stage: turbulence}

In all cases the orderly nonlinear motion of the intermediate stages
is followed by disordered turbulent states at the late evolution stages.
However, the properties of such turbulence are different for
the cases {$M < M_*$} and {$M > M_*$}, as clearly seen in the last
frames of the figures~\ref{snapshotsM0_1}
%, ~\ref{snapshotsM1} 
and~\ref{snapshotsM10}.

%\begin{comment}
%The initial stage of the evolution agrees very well with the predicted growth from the linear stability analysis of the 4MT as shown in figure~\ref{growth}.  In fact for the $M=0.1$ case, the 4MT works very well even beyond the linear stage in that it predicts the reversal in the growth of the modulation amplitude.  The oscillating behaviour of this mode is evident in the vorticity snapshots of figures~\ref{snapshotsM0_1}(a)-(d) where the modulation becomes stronger, weakens and subsequently grows again.  
%%For $M=1$, the system's growth does not reverse, but rather experiences a saturation
%at the level where the 4MT system reaches maximum and reverses.  
%\end{comment}

For weak waves, $M < M_*$, the  quasi-oscillatory behaviour where the system periodically returns close to the initial state is not sustained and a transition to an anisotropic turbulent state occurs.
The dominant jet structures observed in such a turbulent state in figure~\ref{snapshotsM0_1}(f) are
off-zonal. This effect may be connected to the off-zonal ``striations'' reported
for ocean observations~\cite{Maximenko2008} although these ocean striations only become evident in the averaged data since they are sufficiently weak.

For strong waves, $M > M_*$,
at the final stages the vortex streets break up due
to a vortex pairing instability which is followed by a transition to turbulence.  Such turbulence is anisotropic with a pronounced zonal jet component and a well-formed potential vorticitys staircase is evident in figure~\ref{snapshotsM10}(f)~\cite{Dritschel2008}.

\subsection{Finite deformation radius}

With a finite Larmor/deformation radius $\rho$, some vortex streets are still evident.
In fact, vortex roll-up occurs quicker at the stage where the 4MT model still predicts a linear growth i.e. well before the 4MT prediction for the zonal mode reaches its maximum.
% for $M = 1$ but they quickly become unstable within approximately two timescales.  The vorticity plots in figures~\ref{snapshotsM1}~and~\ref{snapshotsM1F200} compare the turbulent behaviour for finite and infinite deformation radius.
\begin{figure}[ht]
\begin{center}
\includegraphics[width=0.8\textwidth]{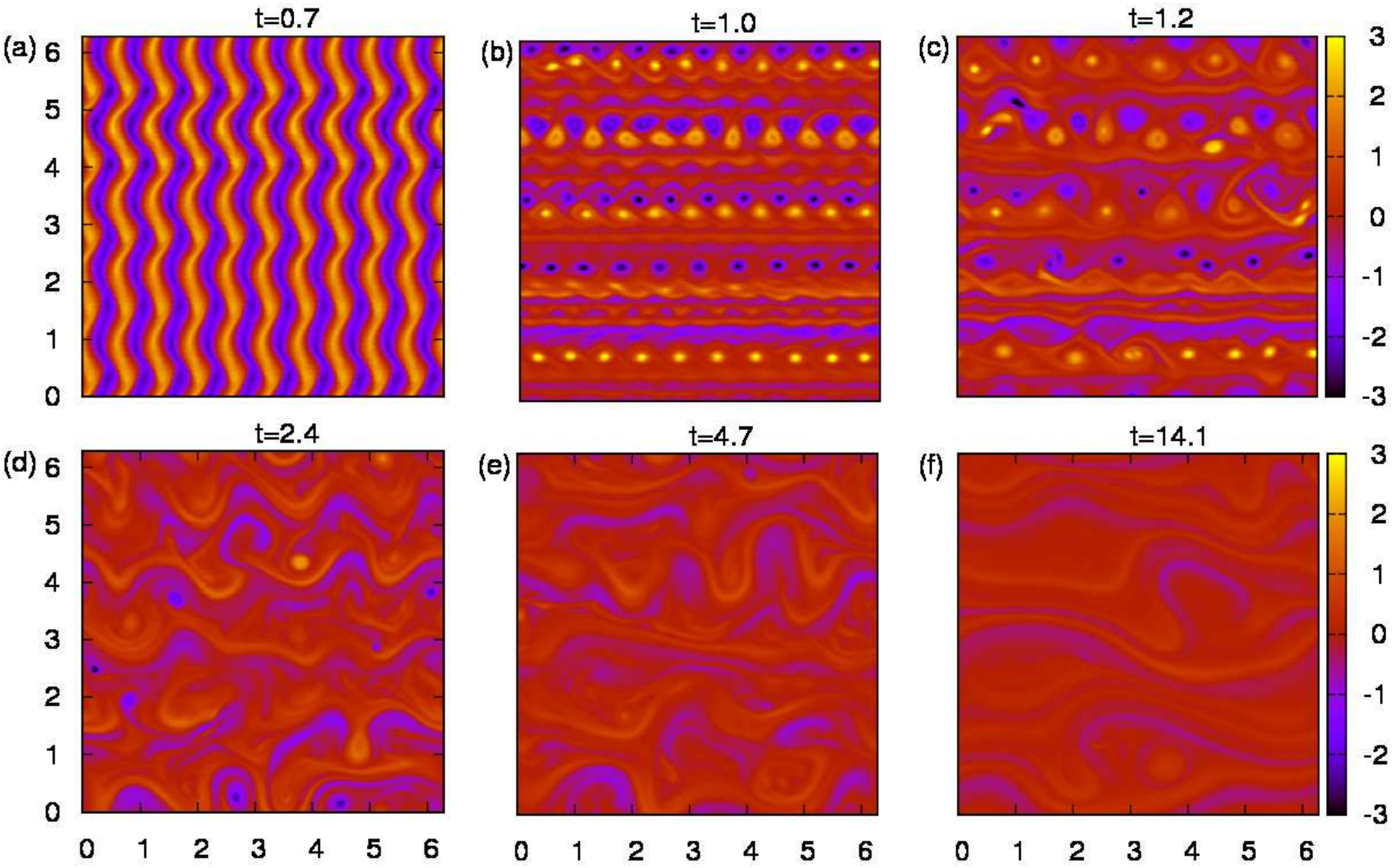}
\caption{ Vorticity snapshots showing the transition to turbulence of a zonal perturbation to a meridional primary wave having $M=1$ and $F=200$.}
\label{snapshotsM1F200}
\end{center}
\end{figure}
This is evident in figure~\ref{growthF} where the predicted linear growth rate is now observed for only one to two timescales.  
\begin{figure}[ht]
\begin{center}
\includegraphics[width=0.8\textwidth]{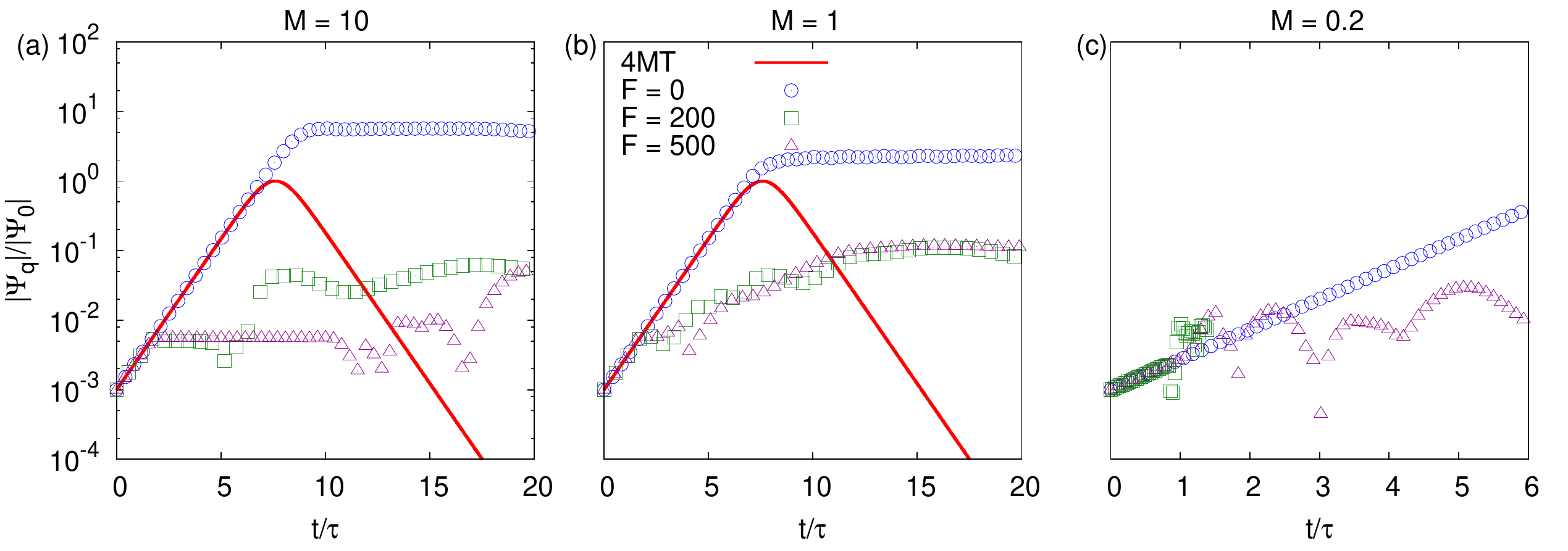}
\caption{Comparison of the growth of the zonal mode $\bf{q}$ obtained by DNS versus solving 4MT system for various $F=\frac{1}{\rho^2}$.  In each case the primary wavenumber is ${\bf p} = (10,0) $ and the modulation wavenumber is ${\bf q} = (0,1) $.  The nonlinearity levels are (a) $M=10$, (b) $M=1.0$ and (c) $M=0.2$.}
\label{growthF}
\end{center}

\end{figure}

\subsection{EHM}

Similarly, numerical results show that two distinct qualitative regimes are also observed by varying $\rho$ in the EHM model.  
%(NEED TO DEFINE $M_\rho$ HERE OR INCORPORATE INTO DEFINITION OF M)\\
\begin{figure}
\begin{center}
\includegraphics[width=0.4\textwidth]{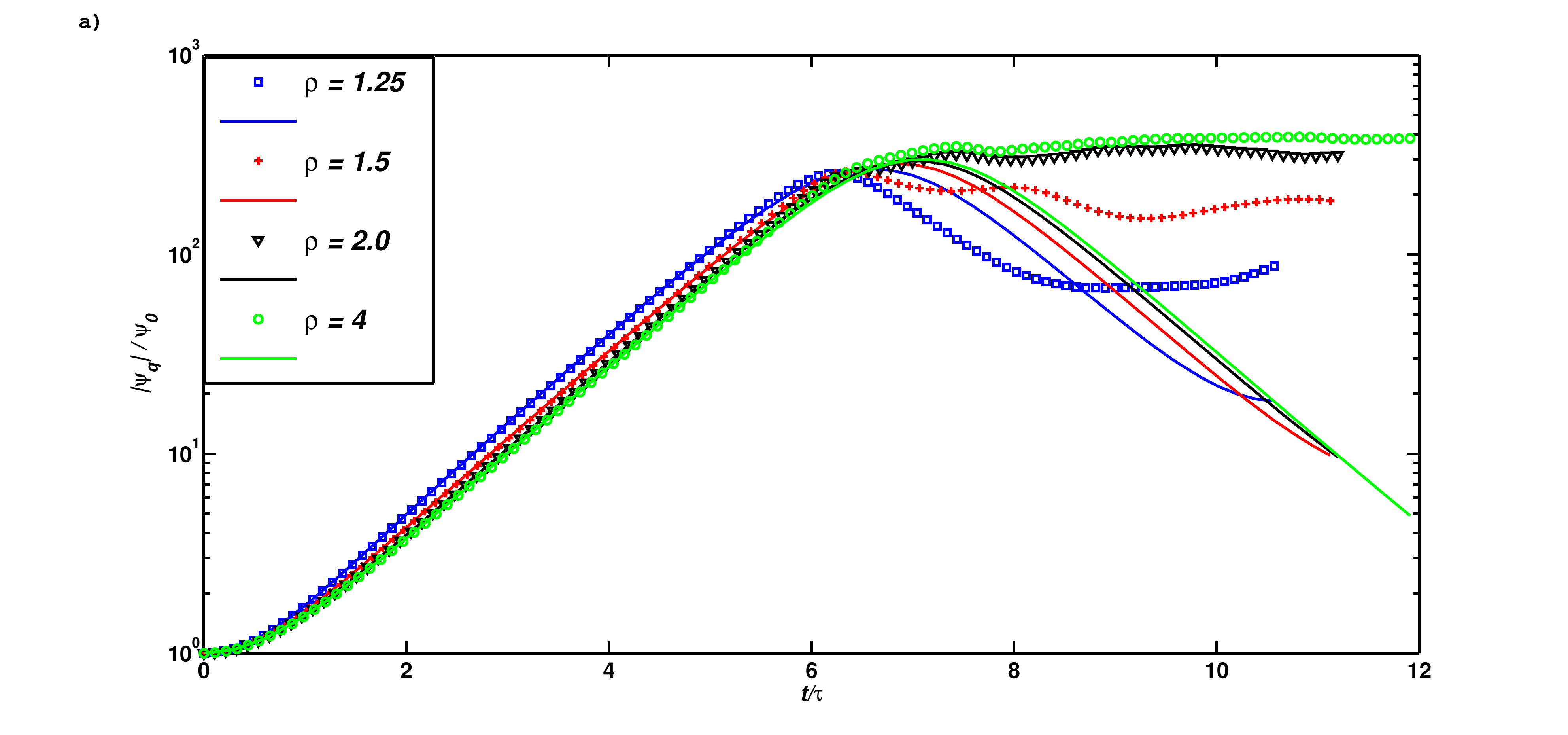}
\includegraphics[width=0.4\textwidth]{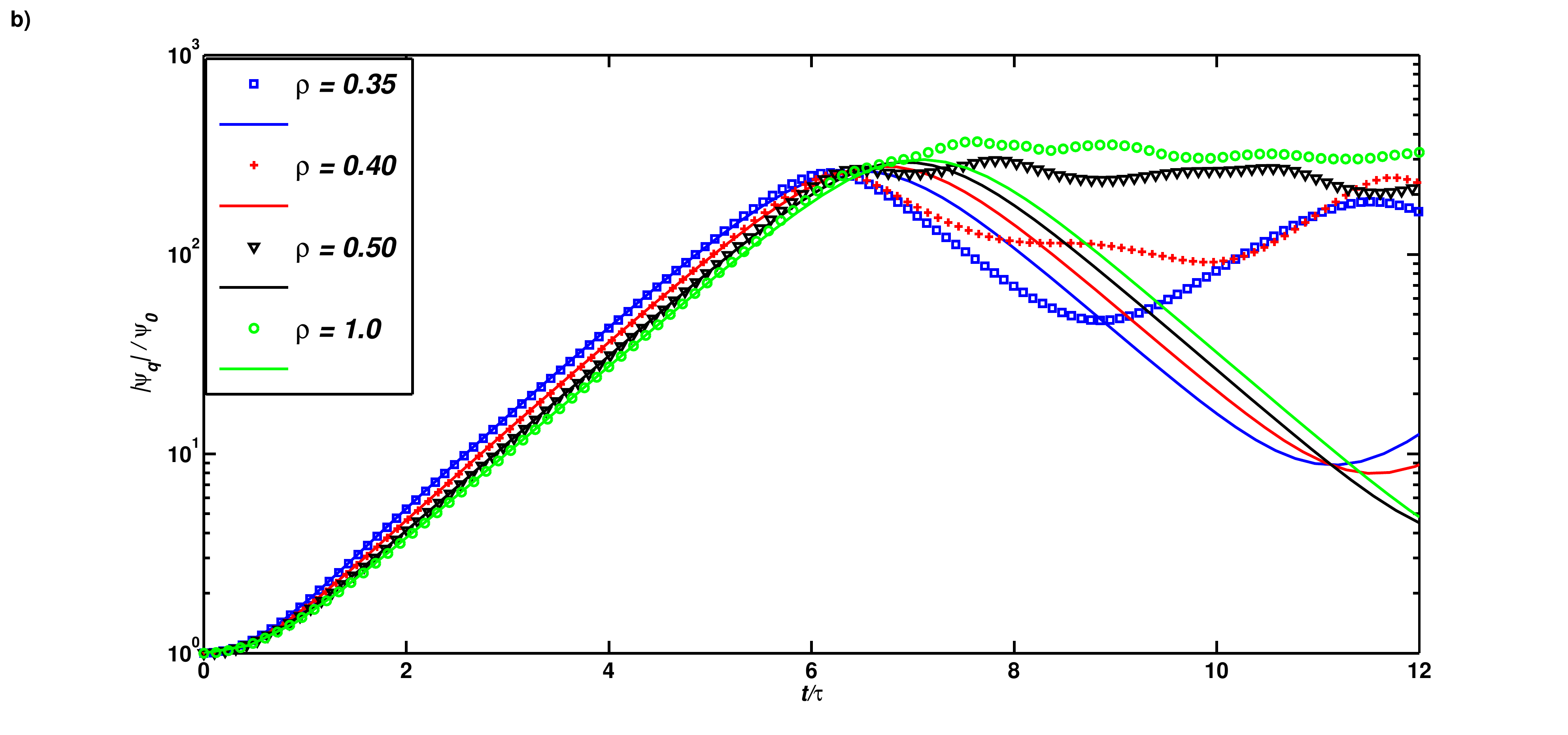}
\caption{The amplitude of the zonal mode for various $\rho$ with (a) $\psi_0 = 0.001$ ($M=0.1$) and (b) $\psi_0 = 0.01$ ($M=1$), other parameters were identical, $\beta =10$, $\bt{p}=(10,0)$, $\bt{q}=(0,1)$. Each case has been scaled by its own linear growth time. ODE predictions are shown with solid lines, full simulations with markers. }
\label{EHM_Mcrit}
\end{center}
\end{figure}
%To ensure that the constructed parameter $M_{\rho}$ allows both regimes to be explored, independent of the carrier wave amplitude, two cases for which $\psi_0$ differed by an order of magnitude are considered. The first case had a very small initial amplitude for the pump wave, $\psi_0 = 0.0001$, leading to a weak  nonlinearity and oscillatory behaviour, similar to that observed in CHM. 
Figure \ref{EHM_Mcrit}(a) shows the amplitude of the zonal mode for several
different values of $\rho$ at the same value of the nonlinearity parameter $M=0.1$.  The runs all have a linear phase in which they grow at the rate predicted by Eq.~\eqref{eq-MIDispersion} yet increasing $\rho$ gradually changes the behaviour of the system so that the zonal mode
saturates rather than oscillates,  similar to CHM. This change in behaviour occurs between $\rho = 1.5$ and $\rho =2$. Note that transition to the saturated regime occurs in EHM at smaller levels of the nonlinearity than in the CHM model. Indeed, in CHM the case $M=0.1$ exhibited an oscillitary behaviour even for $\rho =\infty$. Figure \ref{EHM_Mcrit}(b) shows the second case in which  the initial
pump wave amplitude was ten times larger, $M=1$. In this case the switch from saturating to oscillatory behaviour occurs between $\rho = 0.4$ and $\rho =0.5$. 
%Despite the two sets of simulations beginning with very different drift wave amplitudes, the change in behaviour happened for roughly the same value of $M_\rho$. 

\subsection{Instability and  zonal jets in HW/EHW model}

As we have already mentioned, the  
%\eqref{NonLocalHWChap-HW-phi} and 
\eqref{NonLocalHWChap-HW-N} has a great advantage over the CHM/EHM in that it contains a primary instability capable of spontaneously generating of waves.
On the other hand, this  property adds an extra complexity which prevents one from using the standard
MI approach based on the four-mode rtuncation. The problem is that if we start with a single monochromatic
pump wave, perturb it with a modulation and two side-band modes and linearise the system, we would end up with a system of equations with time dependent coefficients which has no simple exponential (or complex exponential) solution.

However, the 4MT approach is still useful for a qualitative and even some quantitative understanding of the nonlinear dynanics of the HW/EHW model.
Let us truncate the system~\eqref{NonLocalHWChap-HW-N}
%\eqref{NonLocalHWChap-HW-phi} and 
by allowing the wave vectors  $\bt{k}$, $\bt{k}_1$ and $\bt{k}_2$
take only values $\bt{p}$, , $\bt{q}$, $\bt{p}_{\pm}=\bt{p}\pm\bt{q}$ and their negatives (taking into account that $\psi_{-\bt{k}} = \overline \psi_{\bt{k}}$).  From Eqs.~\eqref{NonLocalHWChap-HW-N} we have the following system,
%\eqref{NonLocalHWChap-HW-phi} and 
%\begin{equation}
%\partial_t \Psi_{\bt{p}}= T(\bt{p},\bt{q},\bt{p}_-)\, \Psi_{\bt{q}} \Psi_{\bt{p_-}} {\rm e}^{i\, \Delta_-\, t} +  T(\bt{p},-\bt{q},\bt{p}_+)\, \overline{\Psi}_{\bt{q}} \Psi_{\bt{p_-}} {\rm e}^{i\, \Delta_+\, t}
%\end{equation}
\begin{eqnarray}
\nonumber 
\partial_t \hat{\psi}_\bt{p} &=& 
- \frac{\alpha z}{p^2} (\hat{\psi}_\bt{p} - \hat{n}_\bt{p})  +
 T(\bt{p},\bt{q},\bt{p}_-)\, \hat{\psi}_{\bt{q}} \hat{\psi}_{\bt{p_-}} 
+  T(\bt{p},-\bt{q},\bt{p}_+)\, \overline{\hat{\psi}}_{\bt{q}} \hat{\psi}_{\bt{p_-}} \,,\\
\nonumber \partial_t \hat{\psi}_{\bt{q}} &=& 
- \frac{\alpha z}{q^2} (\hat{\psi}_\bt{q} - \hat{n}_\bt{q})  +
T(\bt{q},\bt{p},-\bt{p}_-) \hat{\psi}_{\bt{p}} \overline{\hat{\psi}}_{\bt{p_-}} 
+ T(\bt{q},-\bt{p},\bt{p}_+) \overline{\hat{\psi}}_{\bt{p}} \hat{\psi}_{\bt{p_+}} \,,\\
\nonumber
\partial_t \hat{\psi}_{\bt{p}_-}&=& 
- \frac{\alpha z}{p_-^2} (\hat{\psi}_\bt{p_-} - \hat{n}_\bt{p_-})  +
T(\bt{p}_-,\bt{p},-\bt{q}) \hat{\psi}_{\bt{p}}\, \overline{\hat{\psi}}_{\bt{q}} \,,\\
 \label{4MT-HW-psi}
 \partial_t \hat{\psi}_{\bt{p}_+}&=& 
- \frac{\alpha z}{p_+^2} (\hat{\psi}_\bt{p_+} - \hat{n}_\bt{p_+})  +
T(\bt{p}_+,\bt{p},\bt{q}) \hat{\psi}_{\bt{p}}\, \hat{\psi}_{\bt{q}} \,.
\end{eqnarray}

\begin{eqnarray}
\nonumber
\partial_t \hat{n}_\bt{p} = - i \kappa p_y \hat{\psi}_\bt{p} + z \alpha 
 (\hat{\psi}_\bt{p} - \hat{n}_\bt{p}) \!  -\!\!
 R(\bt{p}, \bt{q}) 
(\hat{n}_\bt{p_-}  \hat{\psi}_\bt{q} - \hat{n}_\bt{q}  \hat{\psi}_\bt{p_-} 
+
\hat{n}_\bt{p_+}  \overline{\hat{\psi}}_\bt{q} -
 \hat{\psi}_\bt{p_+}  \overline{\hat{n}}_\bt{q}) \,,
\\
\nonumber
\partial_t \hat{n}_\bt{q} = - i \kappa q_y \hat{\psi}_\bt{q} + z \alpha 
 (\hat{\psi}_\bt{q} - \hat{n}_\bt{q}) \!  -\!\!
 R(\bt{p}, \bt{q}) 
(\hat{n}_\bt{p}  \overline{\hat{\psi}}_\bt{p_-} - \hat{\psi}_\bt{p}  \overline{\hat{n}}_\bt{p_-}
-\overline{\hat{n}}_\bt{p}  {\hat{\psi}}_\bt{p_+} +
\overline{\hat{\psi}}_\bt{p}  {\hat{n}}_\bt{p_+} 
)\,,
\\
\nonumber
\partial_t \hat{n}_\bt{p_-} = - i \kappa p_{-,y} \hat{\psi}_\bt{p_-} + z \alpha 
 (\hat{\psi}_\bt{p_-} - \hat{n}_\bt{p_-}) +
 R(\bt{p}, \bt{q}) 
(\hat{n}_\bt{p}  \overline{\hat{\psi}}_\bt{q} - 
\hat{\psi}_\bt{p}  \overline{\hat{n}}_\bt{q})
\,, \hspace{2.3cm}
\\
\partial_t \hat{n}_\bt{p_+} = - i \kappa p_{+,y} \hat{\psi}_\bt{p_+} + z \alpha 
 (\hat{\psi}_\bt{p_+} - \hat{n}_\bt{p_+}) -
 R(\bt{p}, \bt{q}) 
(\hat{n}_\bt{p}  {\hat{\psi}}_\bt{q} - 
\hat{\psi}_\bt{p}  {\hat{n}}_\bt{q})
\,, \hspace{2.3cm}
\label{4MT-HW-N}
\end{eqnarray}

\subsubsection{Primary  instability in the HW/EHW system}

Let us first of all consider the primary instability in the HW/EHW model which is called the drift-dissipative instability.
Substituting into Eq.\eqref{NonLocalHWChap-HW-N}  a plane wave solution, $\hat \psi_{\bf k},  \hat n_{\bf k} \sim e^{-i \omega_{\bf k} t}$ and linearising, we get
\begin{eqnarray}
\omega_{\bf k}=\frac{1}{2}[-ib\pm ib(1-4i\omega_{*}/b)^{1/2}],
\end{eqnarray}
where $b=\alpha z (1+(\rho k)^{-2})$, and 
$\omega_{*}=k_{y} \kappa/(1+\rho^2 k^{2})$.
The plus sign is an unstable mode $\gamma_{\bf k} = \Im \omega_{\bf k} >0$ with maximum growth rate at ${\bf k} \eqsim (1/\rho, 0)$.
%$\omega_{k}$ is given by Eq.~(\ref{eq-Rossbydispersion}).

%The maximum growth rate corresponds to $b\simeq 4\omega_{k}$  and
%\begin{eqnarray}\label{eq:c1hw}
%\alpha=\frac{4k^{2}k_{y}\kappa}{(1+k^{2})^{2}}
%\end{eqnarray}

\subsubsection{Secondary instability in the HW system - MI}

The growth rate of the primary instability $\gamma_{\bf k}$ has a  wide distribution around its maximum. 
%The developing MI is consequently different than in systems with very narrow $\gamma_{\bf k} $. With narrow $\gamma_{\bf k} $ one can consider there to be a dominant carrier wave that becomes unstable. With wider $\gamma_{\bf k} $ it is more 
It is natural to assume that the dominant modes will be those close to the instability
maximum and that they should have comparable strengths. In terms of the 4MT this means that the modes ${\bf p} $, ${\bf p}_+ $ and ${\bf p}_- $
have similar amplitudes and growth rates whereas the mode ${\bf q} $ is initially weak.
Since ${\bf p} $, ${\bf p}_+ $ and ${\bf p}_- $ must be close to the primary instability maximum, they must also be close to each other. Therefore, there arises a scale separation $q \ll k$. Furthermore, for nearly meridional 
${\bf p} $, ${\bf p}_+ $ and ${\bf p}_- $ with $\bf q$ close to the resonant curve, $\bf q$ is nearly zonal. 
Thus the fastest growing
$\bf q$-mode should have the smallest zonal wave number allowed by the box.

Note that all zonal $\bf q$ are neutrally stable to the primary instability mechanism, 
 $\gamma_{\bf q} = 0$, ie. that the linear terms would vanish from the mode-$\bf q$ equation in Eqs.~\eqref{4MT-HW-psi}.
Let us substitute into the right-hand side of this equation
$\hat \psi_{\bf p} = \psi_{\bf p}(0) e^{-i\omega_{\bf p} t},\, \hat \psi_{{\bf p}_-} = \psi_{\bf p}(0) e^{-i\omega_{{\bf p}_-}  t},\, \hat \psi_{{\bf p}_+}  = \psi_{\bf p}(0) e^{-i\omega_{{\bf p}_+} t}$. 
Taking into account that the real parts of the frequencies of the triads ${\bf p} $, ${\bf p}_+ $, ${\bf q} $  and ${\bf p} $, ${\bf p}_- $, ${\bf q} $
are in near resonance and that for the imaginary part we have $\gamma_{\bf p} \approx  \gamma_{\bt{p_-}} \approx  \gamma_{\bt{p_+}} $.
This gives the following equation for the mode $\bf q$,
$$
\partial_t \hat{\psi}_{\bt{q}} =
\lambda \, e^{2 \gamma_{\bf p} t},
$$
where
\begin{equation}
\label{lambda}
\lambda= T(\bt{q},\bt{p},-\bt{p}_-) \hat{\psi}_{\bt{p}}(0)  \overline{\hat{\psi}}_{\bt{p_-}}(0)  
+ T(\bt{q},-\bt{p},\bt{p}_+) \overline{\hat{\psi}}_{\bt{p}}(0)  \hat{\psi}_{\bt{p_+}}(0)
\end{equation}
which has a solution:
\begin{equation}
\hat{\psi}_{\bt{q}} =\hat{\psi}_{\bt{q}} (0) +
\frac{\lambda}{2 \gamma_{\bf p} } \, e^{2 \gamma_{\bf p} t}.
\end{equation}
We conclude that the MI  growth rate for the zonal mode in the HW/EHW system is  the double of the maximum growth rate
of the primary instability, $\gamma_{\bf p} = \gamma_{max}$. This growth occurs only after a delay needed for the second term in Eq.~\eqref{lambda} to overtake the first one. This kind of the behavior was recently observed in the direct numerical simulations of the MI in the HW model \cite{quinn_modulational_2014}.

%\afterpage{\clearpage}

%, and in EHW -- easier than in HW.

%%%%%%%%%%%%%%%%%%%%%%%%%%%%%%%%%%%%%%%%%%%%%%%%%%%%%%%%%%%%%%%%%%%%%%%%%%%%%%%%%%%%%%%%%%%%%%%%%%%%
\section{Invariants and anisotropic cascades in Rossby and drift wave turbulence}
\label{triple_cascade}
%NOTES\\
%1. Intro for Rossby

Rossby wave and drift wave turbulence can be considered as a system of weakly interacting waves, with a quadratic leading order nonlinearity.  However there are three qualitatively different regimes of weak wave turbulence (WT), namely kinetic, discrete and mesoscopic~\cite{Lvov2010,Nazarenko2011}.  which are characterized by the relationship between the nonlinear frequency broadening $\Gamma$ and the frequency spacing $\Delta_{\omega}$.  

For small wave amplitudes, the nonlinear frequency broadening is much less than the frequency spacing, $\Gamma \ll \Delta_{\omega}$. This is discrete WT with a finite number of modes where only waves that are in exact resonance can interact and exchange energy. In the case of Rossby and drift waves these interactions take place between triads of waves whose wavevectors and frequencies must satisfy the resonance conditions
\begin{eqnarray}
\nonumber
\mathbf{k}-\mathbf{k}_{1}-\mathbf{k}_{2}&=&0 , \\
\label{resonance_conditions}\omega(\mathbf{k})-\omega(\mathbf{k}_{1})-\omega(\mathbf{k}_{2})&=&0\,.
\end{eqnarray}
The wavevectors are discrete variables and, as a result, the resonance conditions (\ref{resonance_conditions}) may be hard to satisfy and any $\mathbf{k}$ may be a member of only a few resonant triads or a single triad. Resonant triads connected via common modes can be grouped together to form clusters ranging in size from two triads joined via one mode to a multiple-triad cluster involving a complicated network of interconnected triads~\cite{Kartashova2011}.

If the wave amplitudes are gradually increased, while still retaining a sufficiently low nonlinearity level, the nonlinear frequency broadening will eventually become large, $\Gamma \gg \Delta_{\omega}$.  This condition may also be realized by taking the infinite box limit while keeping the wave amplitudes constant. Then the wave system will be in the kinetic turbulence regime, where the wavevectors are continuous variables and consequently may belong to infinitely many resonant triads.  Therefore, for larger amplitudes or large boxes, the number of coupled resonant triads increases considerably eventually creating an infinite cluster containing all the possible triads in the system~\cite{Nazarenko2011}.

Both the discrete and kinetic regimes may coexist if $\Gamma \sim \Delta_{\omega}$, and the system may oscillate in time or in the $\mathbf{k}$-space, between the two regimes giving rise to the mesoscopic regime~\cite{Korotkevich2005,Lvov2010,Nazarenko2011} where it has been hypothesized that the wave energy undergoes sandpile behaviour~\cite{Nazarenko2006} summarized as follows: suppose that the WT has initially very weak or zero intensity, so that it is in the discrete regime. If wave energy is permanently supplied at small $\mathbf{k}$, it will accumulate until the resonance broadening $\Gamma$ becomes of the order of the frequency spacing $\Delta_{\omega}$. After that the turbulence cascade is released to higher $\mathbf{k}$ in the form of an `avalanche' characterised by predominately kinetic interactions. In the process of the avalanche release, the mean amplitude of the waves is reduced so that the value of broadening $\Gamma$ becomes less than the frequency spacing $\Delta_{\omega}$.  At this point 
the system returns to the energy accumulation stage in the discrete WT regime and the cycle repeats.

\subsection{Conservation Laws and Cascdes in the kinetic wave turbulence}

All kinetic wave turbulence systems conserve energy and momentum but studies of this regime~\cite{Balk_proc1990,Nazarenko011991,Balk1991} have found that drift and Rossby wave turbulence conserves not only the usual energy $E$ and enstrophy $Q$ but also a third quadratic invariant in Fourier space that is not a linear combination of the other two, now commonly referred to as {\em zonostrophy}, $Z$.  It was previously assumed that all wave systems conserve only the energy and enstrophy or that they conserve an infinite number of extra invariants.  This discovery was the first example of a wave system with a finite number of additional invariants~\cite{Nazarenko011991}.  The invariant also contains a cubic term whose form depends on the nonlinear interacaction coefficient, but for small amplitudes the cubic term can be neglected~\cite{Balk2006}. Later we will see that zonostrophy is also conserved in discrete wave turbulence and that there are actually many more quadratic invariants in that case.  

%\subsubsection{Kinetic equation}

In terms of the waveaction variable, which is defined as  
\begin{equation}
\label{waveaction_a}
a_{\bt k} = \frac{k^2+F}{\sqrt{\beta |k_x|}}\hat{\psi}_{\bt k}\,,
\end{equation}
the dynamical equation~\eqref{eq-CHMkdiscrete} can be rewritten
\begin{equation}
\label{dynamic_eqn_a}
\partial_t a_{\bt k} = - \mathrm{i}\, \omega_{\bt k}\, a_{\bt k} + \frac{1}{2}  \sum_{\bt{k}_1, \bt{k}_2} V^{\bt k}_{12}\, \delta( {{\bt k} - {\bt k}_1 - {\bt k}_2}) a_{\bt{k}_1}\, a_{\bt{k}_2}\, d{\bt k}_1d{\bt k}_2
\end{equation}
where
% \begin{equation}
% \label{Rossbydispersion}
% \omega_\bt{k} = -\frac{\beta k_x}{k^2 + F}
% \end{equation}
% is the anisotropic dispersion relation for the linear solutions,  $\bt{k}=(k_x,k_y)$, $k = \left| \bt{k} \right|$ and
% is the canonical variable~\cite{Zakharov1988} for this particular Hamiltonian system and
\begin{equation}
\label{interactioncoeff}
V^{\bt k}_{12} \equiv V^{\bt k}_{{\bt k}_1, {\bt k}_2} = \frac{\mathrm{i}}{2}\sqrt{|\beta k_x  k_{1x}k_{2x}|}\left(\frac{k_{1y}}{k_1^2+F}+\frac{k_{2y}}{k_2^2+F}- \frac{k_y}{k^2+F}\right)
\end{equation}
is the nonlinear interaction coefficient~\cite{Zakharov1988}.  
Like we said before, this symmetric form of the interaction coefficient is valid only on the resonant manifold, which will be sufficient for both  
the kinetic and the disctere wave turbulence regimes considered belos.

Let us return to the wave kinetic equation~\cite{Longuet-Higgins1967}, the main equation governing the kinetic turbulence regime, which in symmetric form is
\begin{equation}
\label{wave_kinetic_eqn_sym}
\dot{n}_{\bt k}= \int_{k_{1x}, k_{2x}>0} \left(\mathcal{R}_{12k}-\mathcal{R}_{k12}-\mathcal{R}_{2k1}   \right)\, d {\bt k}_1 d {\bt k}_2 \, ,
\end{equation}
where
\begin{equation}
\mathcal{R}_{12k}= 2\pi \left| V^{\bt k}_{12} \right|^2 \delta^k_{12} \delta(\Omega^k_{{\bt k}_1 {\bt k}_2})(n_{{\bt k}_1} n_{{\bt k}_2} - n_{\bt k} n_{{\bt k}_1} -  n_{\bt k} n_{{\bt k}_2} ).
\end{equation}
Since in this section we are considering waves in a bi-periodic box, we should think of the kinetic equation as arising in the large box limit, $L \to \infty$ and re-define the waveaction spectrum in terms of the finite-fox Fourie coefficients as 
 $\displaystyle n_{\mathbf{k}}=\lim_{L \to \infty}\left(\frac{L}{2\pi}\right)^{2}|a_{k}|^{2}$ (the latter is consistent with the infinite-box definition in terms of the Fourier transform (\ref{eq-spectrum})).

%%%%%%%%%%%%%%%%%%%%%%%%%%%%%%%%%%%%%%%%%%%%%%%%%%%%

Recall that since the CHM equation in physical space is for the streamfunction $\psi$, a real function, the wavevectors $\mathbf{k}$ and $-\mathbf{k}$ represent the same mode via the property of the Fourier transform of real functions $\hat{\psi}_{-\mathbf{k}}=\hat{\psi}_{\mathbf{k}}^{*}$.  Thus we can choose to work with only half of the Fourier space e.g. $k_{x}\geq 0$ and will further neglect $k_{x}=0$ since this corresponds to zero-frequency zonal flows which are not waves.

In terms of waveaction, the energy and momentum are defined respectively as, 
\begin{eqnarray}
\label{Energy and enstrophy}
E&=&\int_{k_{x}>0}|\omega_{\textbf{k}}|n_{\textbf{k}}\,d\textbf{k},\\
M&=&\int_{k_{x}>0}\textbf{k}\,n_{\textbf{k}}\,d\textbf{k} , \nonumber
\end{eqnarray}
such that $\omega_{\bt k}$ can be thought of as the density of the energy and ${\bt k}$ the density of the momentum.  In particular, the enstrophy corresponds to the $x$-momentum
\begin{equation}
\label{enstrophy_WT}
Q =  \int k_x\, n_{\bt k} \, d {\bt k}\,.
\end{equation}
Let us first of all recall how do the energy and the momentum conservation laws come about.
Using Eq.~(\ref{wave_kinetic_eqn_sym}), the rate of change of energy can be written as~\cite{Nazarenko2011}:
\begin{eqnarray}
\nonumber
\frac{\partial E}{\partial t} & = & \int \omega_{\bt k} \frac{\partial n_{\bt k}}{\partial t} \hspace{1mm} d {\bt k}\\
\nonumber
& = & \int \int \int \left(\omega_{\bt k}\mathcal{R}_{12k}-\omega_{\bt k}\mathcal{R}_{k12}-\omega_{\bt k}\mathcal{R}_{2k1}   \right)\, d {\bt k}_1 d {\bt k}_2 d {\bt k}\\
\label{E_dt}
& = & \int \mathcal{R}_{12k}(\omega_{\bt k}-\omega_{{\bt k}_1}-\omega_{{\bt k}_2}) d {\bt k}_1 d {\bt k}_2 d {\bt k} \, ,
\end{eqnarray}
where a change of variables ${\bt k}_1, {\bt k}_2, {\bt k} \mapsto {\bt k},{\bt k}_1, {\bt k}_2 $ and ${\bt k}_1, {\bt k}_2, {\bt k} \mapsto {\bt k}_2,{\bt k}, {\bt k}_1 $ have been taken for the second and third terms in the parentheses respectively.  Similarly for the enstrophy
\begin{equation}
\label{Q_dt}
\frac{\partial Q}{\partial t} =  \int \mathcal{R}_{12k}({\bt k}-{\bt k}_1-{\bt k}_2) d {\bt k}_1 d {\bt k}_2 d {\bt k}\,.
\end{equation}
By virtue of the resonance conditions in Eqs~\eqref{resonance_conditions}, the terms within the parentheses of Eqs~\eqref{E_dt} and~\eqref{Q_dt} equal zero, showing that energy and enstrophy remain constant in the system.  Generally, one can write for a conserved quantity $\Phi$ with density $\varphi_{\mathbf{k}}$,
\begin{equation}
\label{general_invar}
\Phi=\int_{k_{x}>0}\varphi_{\mathbf{k}}n_{\mathbf{k}}d\mathbf{k}\,,
\end{equation}
and it follows that 
\begin{eqnarray}
\frac{\partial \Phi}{\partial t}&=&\int_{k_{x}>0}\varphi_{\mathbf{k}}\dot{n}_{\mathbf{k}}d\mathbf{k}=
%\int\int\int_{k_{x},k_{1x},k_{2x}>0}(\varphi_{\mathbf{k}}\mathcal{R}_{12\mathbf{k}}-\varphi_{\mathbf{k}}\mathcal{R}_{\mathbf{k}12}-\varphi_{\mathbf{k}}\mathcal{R}_{2\mathbf{k}1})d\mathbf{k}_{12}d\mathbf{k}\\
\int\int\int_{k_{x},k_{1x},k_{2x}>0}\mathcal{R}_{12\mathbf{k}}(\varphi_{\mathbf{k}}-\varphi_{1}-\varphi_{2})d\mathbf{k}_{12}d\mathbf{k} = 0\,,
\end{eqnarray}
if the spectral density satisfies 
\begin{equation}
\label{general_resonance_condition}
\varphi_{\mathbf{k}}-\varphi_{\mathbf{k}_{1}}-\varphi_{\mathbf{k}_{2}}=0\,.
\end{equation}

Such a quantity was found to exist~\cite{Balk_proc1990,Balk1991,Nazarenko011991} in 1990.  
The original theory~\cite{Balk_proc1990,Nazarenko011991} was limited to considering either very large or very small scales (respectively longer or shorter than the Rossby deformation radius or the Larmor radius) or to the scales which are already
anisotropic and are close to zonal,
%Besides, the conservation of the extra invariant is based
%on the weakness of nonlinearity and on the randomness of phases, conditions too of the validity of the wave kinetic equation, which even if present initially, can break down later during the zonation process.
but soon after, the  invariant was generalised to the whole of the $k$-space~\cite{Balk1991}. It has the form
$$
Z=\int_{k_{x}>0}\eta_{\mathbf{k}}n_{\mathbf{k}}\, d\mathbf{k}\,,
$$
which
\begin{equation}
\label{zonostrophy_general}
\eta_{\mathbf{k}}=\arctan\frac{k_{y}+k_{x}\sqrt{3}}{\rho k^{2}}-\arctan\frac{k_{y}-k_{x}\sqrt{3}}{\rho k^{2}}\,.
\end{equation}
This extra invariant was named {\em zonostrophy} in~\cite{NazarenkoQuinn2009}  because it is intimately related to the system's zonation and it causes energy to cascade to the zonal scales, as we will see below.
The Rossby and drift wave system is a special case of wave turbulence in that the wave kinetic equation exactly conserves an extra quadratic invariant that is not a linear combination of the energy and momentum. 
% Actually, the invariant also contains a cubic term whose form depends on the nonlinear action between the waves but it was shown that for long times, the cubic term can be neglected~\cite{Balk2006}.  
% Recently the physical space representation of this invariant was determined, a step towards deducing its actual physical meaning~\cite{Balk2009} since it is always referred to in Fourier space.  
Furthermore, it was shown that the zonostrophy could be more significant than the enstrophy since it is less sensitive to dissipation than enstrophy is, thought to be due to the fact it is based on large scale modes, as is the energy while the enstrophy integral is based mainly on small scale modes~\cite{Balk2006, Balk2009}.

This was a significant achievement because the extra invariant of such a kind appears to be
unique for Rossby and drift wave systems and is not observed in any other known nonlinear wave model.
% \cite{Balk1997}. Besides, its conservation has revealed interesting geometrical properties
% of the wave dispersion relation \cite{Balk1997}.
% However, an alternative zonation argument was put forward in \cite{Balk2005}.

\subsubsection{Small-scale limit}
The general expression for zonostrophy appeared to have a form for which the Fj{\o}rtoft  cascade argument cannot be used since it is neither scale invariant nor sign-definite.
%\vspace{2mm}
%{\bf Small-scale, $\rho \to \infty$}
In the small-scale limit, $\rho k\rightarrow \infty$, after substracting the part corresponding to the energy invariant, the zonostrophy density becomes
\begin{equation}
\label{Z_small}
\eta_{\mathbf{k}}^{s} =-\lim\limits_{\rho \rightarrow \infty}\frac{5\rho^{5}}{8\sqrt{3}}\left(\varphi_{\mathbf{k}}-2\frac{\sqrt{3}\omega}{\beta \rho}\right)=\frac{k_{x}^{3}}{k^{10}}\left(k_{x}^{2}+5k_{y}^{2}\right).
\end{equation}
The integral~\eqref{general_invar} with the density~\eqref{Z_small} is an exact invariant of the kinetic equation~\eqref{wave_kinetic_eqn_sym} and thus is an approximate invariant of the small-scale CHM equation.  The zonostrophy conservation was examined numerically~\cite{NazarenkoQuinn2009} and found to be conserved within 0.1\% for a weakly nonlinear initial condition.

%\subsubsection{EZQ boundaries}

Below we present a generalisation of the famous Fj\o rtoft argument to triple cascade systems following the method introduced in~\cite{Balk_proc1990,Nazarenko011991} (see a pedagogical discussion in~\cite{Nazarenko2011}). We will start with a brief review of the original
Fj\o rtoft argument for 2D turbulence.
Fj\o rtoft~\cite{Fjortoft1953} deduced the nature of the spectral evolution of the conserved quantities, energy and enstrophy, and demonstrated that when the initial energy is concentrated at the intermediate scale then for later times, more energy is acquired at the larger scale than at the smaller scale i.e. the energy cascade is an inverse one from larger to small wavenumbers.  Conversely, the enstrophy cascade is a direct one to larger wavenumbers.  The proof is {\em ad absurdum} and it is summarised as follows.\\
Consider 2D isotropic turbulence described by the Navier-Stokes equations, which is generated at some intermediate forcing scale $k_0$ and dissipated outside the inertial ranges at both very large and very small scales for which $k_- \ll k_0$ and $k_+ \gg k_0$ respectively.  The expressions for the conserved quantities, energy and enstrophy in the absence of forcing and dissipation are equivalent to those for the CHM model, Eqs~(\ref{energyE}) and~(\ref{enstrophyQ}).  For steady-state turbulence, the dissipation rates of the energy and enstrophy must be equal the rate at which they are produced.  According to Eqs~(\ref{energyE}) and~(\ref{enstrophyQ}) with $F=0$, the ratio of the $k$-space densities is $k^2$ such that the enstrophy dissipation rate $\mu$ is related to the energy dissipation rate $\varepsilon$ as $\mu \sim k_0^2 \varepsilon$.

If it is assumed that the energy is dissipated at small scales, $k_+$ at a rate $\varepsilon$ then the enstrophy would have to be dissipated at the rate $k_+^2\varepsilon$.  However this is impossible since it is only produced at a rate of $k_0^2 \varepsilon \ll k_+^2\varepsilon$.  Since the energy must be dissipated at either $k_+$ or $k_-$ then it can be concluded that it is dissipated at $k_-$ via an inverse cascade through the scales.   Similarly, if it is assumed that the enstrophy is dissipated at $k_-$, then the energy would have to be dissipated at a rate of $\frac{1}{k_-^2}\mu$.  Again this rate is a lot greater than $\frac{1}{k_0^2}\mu$, the rate at which it is produced, concluding that the enstrophy must be dissipated at smaller scales or larger wavenumbers $k_+$, than which it is produced via a direct cascade through the inertial range.  

A similar argument can be adopted to determine the direction of the cascade directions of the three conserved quantities of the small-scale CHM model by examining in turn the ratio of the densities of each pair of the invariants~\cite{NazarenkoQuinn2009}.  It is also required that each of the three densities are positive since positive and negative amounts of the same quantity would not rule out a possibility of dissipation of that quantity with a net total of zero  with the result that the Fj\o rtoft argument could not be applied.  
%Furthermore, Eqs~\ref{energy_mom_WT},~\ref{enstrophy_WT} and~\ref{zonostrophy_WT} show that the three invariants are linearly related to the waveaction spectrum or turbulence intensity $n_k$ so that the density ratios are functions of ${\bf k}$ rather than the turbulence intensity.

Let the turbulence be produced near ${\bf k}_0 = (k_{0_x}, k_{0_y})$ and dissipated in regions which are separated in scales from the forcing scale as with the 2D Navier-Stokes turbulence.  To divide the 2D $\bf k$-space into three non-intersecting regions implies anisotropy. Each invariant will flow to that sector in the 2D $\bf k $-space where its density prevails over the others.  The boundaries of these sectors are defined where the ratio of the spectral densities of each pair of invariants remains constant to its initial value as detailed below.

\begin{itemize}
\item {\bf E-Q boundary}: The energy density is $\frac{k_x}{k^2}$ and that of enstrophy is $k_x$ giving the ratio $k^2$ as with the 2D Navier-Stokes turbulence.  Equating this to its initial value gives
\begin{equation}
\label{E_Q_boundary}
k^2 \sim k_0^2 ,
\end{equation}
which in the 2D $\bf k$-space amounts to a circle of radius $k_0$ centered at the origin.  Using the same argument as before, energy must be dissipated at the larger scales and enstrophy at the small scales.

\item {\bf E-Z boundary}: By the inequality $k^{2}\leq k_{x}^{2}+5k_{y}^{2}\leq 5k^{2}$ we can assume $k_{x}^{2}+5k_{y}^{2}$ scales as $k^{2}$ and replace the small-scale zonostrophy density~(\eqref{Z_small}) with a simpler expression as $\eta_{\textbf{k}}^s\sim \frac{k_{x}^{3}}{k^{8}}$.
Thus the ratio of the energy to zonostrophy densities is
\begin{equation}
\label{E_Z_boundary}
\frac{k^3}{k_x} \sim \frac{k_0^3}{k_{0x}}\,.
\end{equation}
Since it is already known that the energy accumulates at large scales then this boundary slices that region in two.  The zonostrophy density is dominant in the region adjacent to the $k_x$ axis and the energy density in the area adjacent to the $k_y$ axis so that the zonostrophy is effectively pushing the energy to zonal scales.

\item {\bf Q-Z boundary}: Equating the ratio of the enstrophy density $k_x$ to the zonostrophy density ${{k_x}^3}/{{k}^{8}}$ to the initial value of this ratio gives the boundary separating the enstrophy and the zonostrophy cascades,
\begin{equation}
\label{3b}
\frac{k^4}{k_x} \sim \frac{k_0^4}{k_{0x}}\,.
\end{equation}
\end{itemize}
These regions are shown in figure~\ref{triple_sketch} where it can be seen that the Q-Z boundary cuts the $k_x$ axis at 
\begin{equation}
\label{kx_star}
k_x^* \sim \frac{k_0^{\frac{4}{3}}}{k_{0x}^{\frac{1}{3}}}.
\end{equation}
thereby restricting the zonostrophy to not too large wavenumbers unless the initial wavenumber is approximately a zonal scale, $k_{0_y} \gg k_{0_x}$.  In particular, if $k_{0_y} =k_{0_x}$ then $k_x^* = 2^{\frac{1}{6}} k_0$, which means that the maximum allowed wavenumber for the zonostrophy cascade is practically the same as the initial scale.  In other words, in this case the zonostrophy can only cascade to the larger scales and not to small scales.
\begin{figure}[ht]
\includegraphics[width=0.45\textwidth,angle=-90]{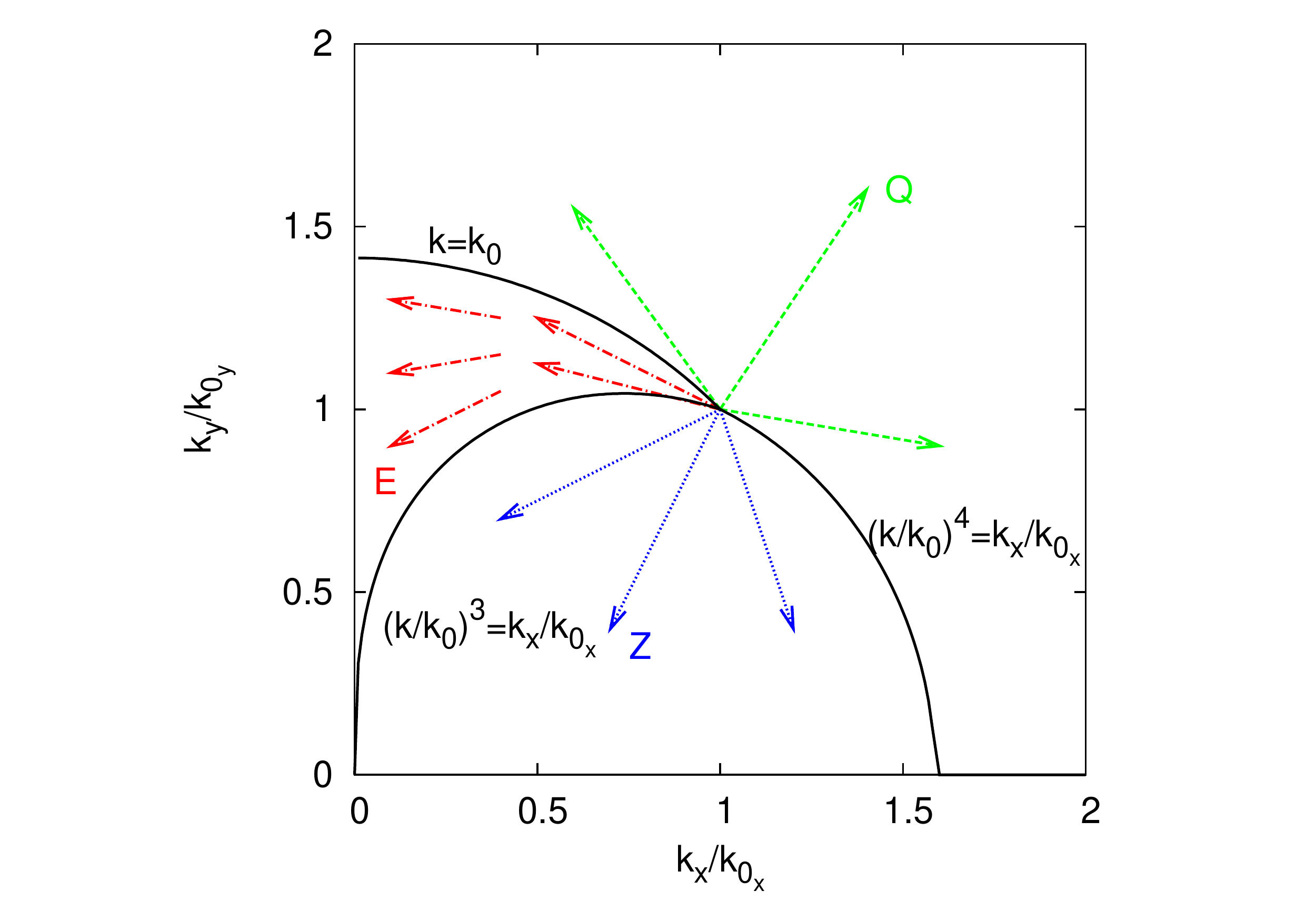}
\begin{centering}
\caption{\label{triple_sketch} Non-intersecting sectors for triple cascade as predicted by the generalised Fj\o rtoft argument.}
\end{centering}
\end{figure}

Fj{\o}rtoft's argument for the directions of the energy and enstrophy cascades in 2D turbulence is in the presence of forcing and dissipation. However, the boundaries have also been determined in terms of centroids for non-dissipative evolving turbulence~\cite{Harper2012}, which is useful for visualizing the direction of the energy and enstrophy cascades in $\mathbf{k}$-space (as will be done below).
The centroids of $E$, $Q$ and $Z$ are difined respectively as:
% For the large-scale limit, since $|k_{y}|\gg |k_{x}|$ Eq.~\eqref{Z_large} can be replaced with:
%\begin{equation}
%\label{Z_large_positive}
%\varphi_{\mathbf{k}} \sim \frac{k_{x}^{3}}{k_{y}^{2}},
%\end{equation}
%so as to make it a positive invariant.  In this limit, the three invariants are,
%\begin{equation}
%E=\int\limits_{0}^{\infty}k_{x}k_{y}^{2}n_{\mathbf{k}}d\mathbf{k},\label{E}
%\end{equation}
%\begin{equation}
%Q=\int\limits_{0}^{\infty}k_{x}n_{\mathbf{k}}d\mathbf{k},\label{Omega}
%\end{equation}
%\begin{equation}
%Z=\int\limits_{0}^{\infty}\frac{k_{x}^{3}}{k_{y}^{2}}n_{\mathbf{k}}d\mathbf{k}.\label{Z}
%\end{equation}
%They can be written in terms of centroids as follows:
\begin{equation}
{\bt k}_{E}=\frac 1 E \int\limits_{0}^{\infty}{\mathbf{k} \omega_{\mathbf{k}} n_{\mathbf{k}}d\mathbf{k}},\label{Energy centroid}
\end{equation}
\begin{equation}
{\bt k}_Q=\frac 1 Q \int\limits_{0}^{\infty}{\mathbf{k}k_{x}n_{\mathbf{k}}d\mathbf{k}},\label{Enstrophy centroid}
\end{equation}
\begin{equation}
{\bt k}_{Z}=\frac 1 Z \int\limits_{0}^{\infty}{\mathbf{k} \eta_{\mathbf{k}}n_{\mathbf{k}}d\mathbf{k}}.\label{Zonostrophy centroid}
\end{equation}
%Application of Fj{\o}rtoft for the forced/dissipated system yields the boundaries: $C_{EQ}:y=\text{const.}, C_{EZ}:y=\sqrt{x}$ and $C_{ZQ}: y=x$ (see figure \ref{fig: Centroids diagram}). Then using the Cauchy-Schwartz inequality, inequalities can be obtained in order to determine the direction of the fluxes of the three invariants. For derivation see appendix \ref{app: Centroids working}. The results are summarised as follows:
%\renewcommand{\arraystretch}{1.5}
%\begin{center}
%\begin{tabular}{|c | c|}
%\hline
%Inequality & What it shows\\
%\hline
%$\frac{k_{Q y}}{k_{Q x}}>(\frac{Z}{Q})^{1/2}$ & $Q$ is bounded below by $C_{ZQ}$\\
%\hline
%$k_{Q y}^{2}<\frac{E}{Q}$ & $k_{y}$ coordinate of $Q$ is bounded above by $C_{EQ}$\\
%\hline
%$k_{Ey}^{2}>\frac{E}{Q}$ & $k_{y}$ coordinate of $E$ is bounded below by $C_{EQ}$\\
%\hline
%$\frac{k_{Ey}^{2}}{k_{Ex}}>(\frac{E}{Z})^{1/2}$ & $E$ is bounded below by $C_{EZ}$\\
%\hline
%$\frac{k_{Zy}}{k_{Zx}}<(\frac{Q}{Z})^{1/2}$ & $Z$ is bounded above by $C_{ZQ}$\\
%\hline
%$\frac{k_{Zy}^{2}}{k_{Zx}}<(\frac{E}{Z})^{1/2}$ & $Z$ is bounded above by $C_{EZ}$\\
%\hline
%\end{tabular}
%\end{center}

%\subsubsection{Numerical Study}
\vspace{2mm}
{\bf Numerical study}

A pseudo-spectral simulation ohe CHM model was employed in~\cite{NazarenkoQuinn2009} to numerically test the theoretical predictions of both the conservation of zonostrophy and the triple cascade behaviour of the small scale turbulence ($F=0$).  The initial condition was taken to be a Gaussian distribution of the stream function defined as 
\begin{equation}
\label{triple_ic}
\hat \psi({\bt k},0)  = A {\rm e}^{\left({\frac{|{\bt k} - {\bt k}_0|^2}{k_*^2} + i \phi_{\bt k}}\right)} +\hbox{image}\,,
\end{equation}
where ${\bt k}_0$ is the wavenumber on which the Gaussian is initially centred, $A$ is the constant amplitude, $k_*$ is the variance and $\phi_\textbf{k}$ are random independent phases.  The `image' refers to the mirror-reflected spectrum with respect to the $k_x$ axis.

Zero numerical dissipation has been modelled rather than a forced-dissipated steady state turbulence because there appears to be no physically meaningful dissipation which acts selectively on nearly zonal and nearly meridional scales only.  While the triple cascade theory presented in the previous section is based on a forced-dissipative turbulence, it is assumed it is also valid for freely-decaying turbulence as is the case of the dual cascade in  2D Navier-Stokes turbulence.  As such, simulations have been carefully monitored to avoid a bottleneck accumulation of turbulence at the maximum wavenumber.

The conservation of energy, enstrophy and zonostrophy and their associated cascades are shown in figure~\ref{triple_weak_NL}.  Due to the slow weakly nonlinear evolution of this system, any quantity proportional to the turbulence intensity could appear to be conserved, so in addition to the three invariants, a non-conserved quantity, $\int |\psi_{\bt k}|^2 \, d{\bt k}$ is also plotted in figure~\ref{triple_weak_NL}(a) to demonstrate the true conservation of the zonostrophy.

\begin{figure}[ht]
\begin{centering}
\includegraphics[width=0.75\textwidth]{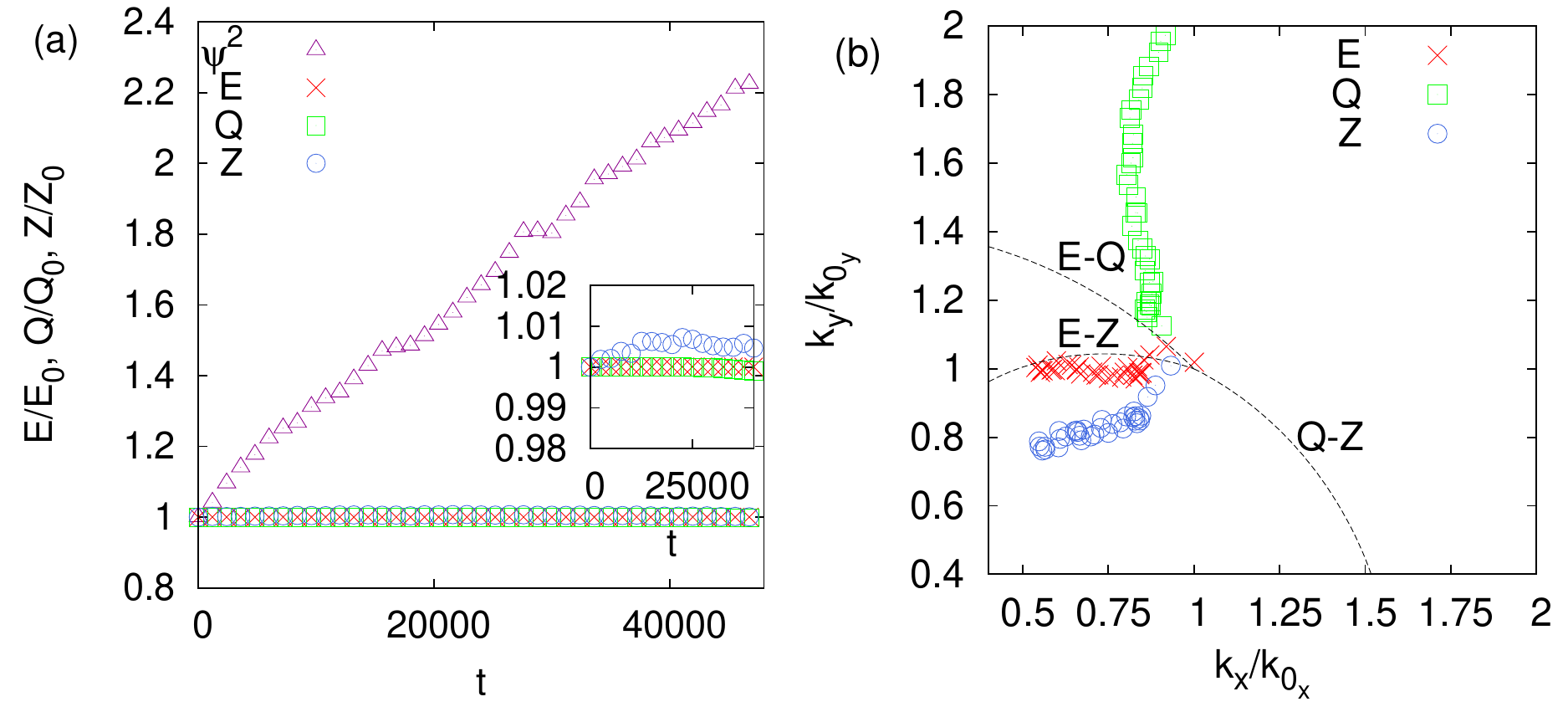}
\caption{\label{triple_weak_NL}  Weak nonlinearity. (a) Conservation of energy $E$, enstrophy $Q$ and zonostrophy $Z$. The inset shows the zoom on the values around 1.  Non-conserved quantity $\int \psi^2 \, dx \, dy$ is also shown and (b) the cascades of each invariant tracked by their centroids}
\end{centering}
\end{figure}

It is clear that all three invariants, $E$, $Q$ and $Z$ are well conserved, the energy to within $0.01\%$, the enstrophy to within $0.15\%$ and the zonostrophy is conserved to within $1\%$.  This is a numerical demonstration of the conservation of the zonostrophy invariant.  It should be remembered that $Z$ is precisely conserved by the wave kinetic equation, Eq.~(\ref{wave_kinetic_eqn_sym})
and therefore its conservation by the dynamical CHM equation, Eq.~(\ref{eq-CHMkdiscrete}), is subject to the conditions of applicability of the kinetic equation, namely weak nonlinearity and the random phases. It is not clear {\em apriori} how well these conditions are satisfied throughout the $\bf k$-space, particularly near the zonal scales.

The centroids of $E$, $Q$ and $Z$ are normalised by their initial values so that figure~\ref{triple_weak_NL}(b) shows the three centroid paths starting from the same point.  Each invariant cascades into its predicted sector.  The enstrophy and zonostrophy cascades are well within their respective sector while the energy follows the boundary of its sector with the zonostrophy sector. 
% MENTION FJORTOFT WITH STRONG INEQUAlITIES?????

\begin{figure}[ht]
\begin{centering}
\includegraphics[width=0.7\textwidth]{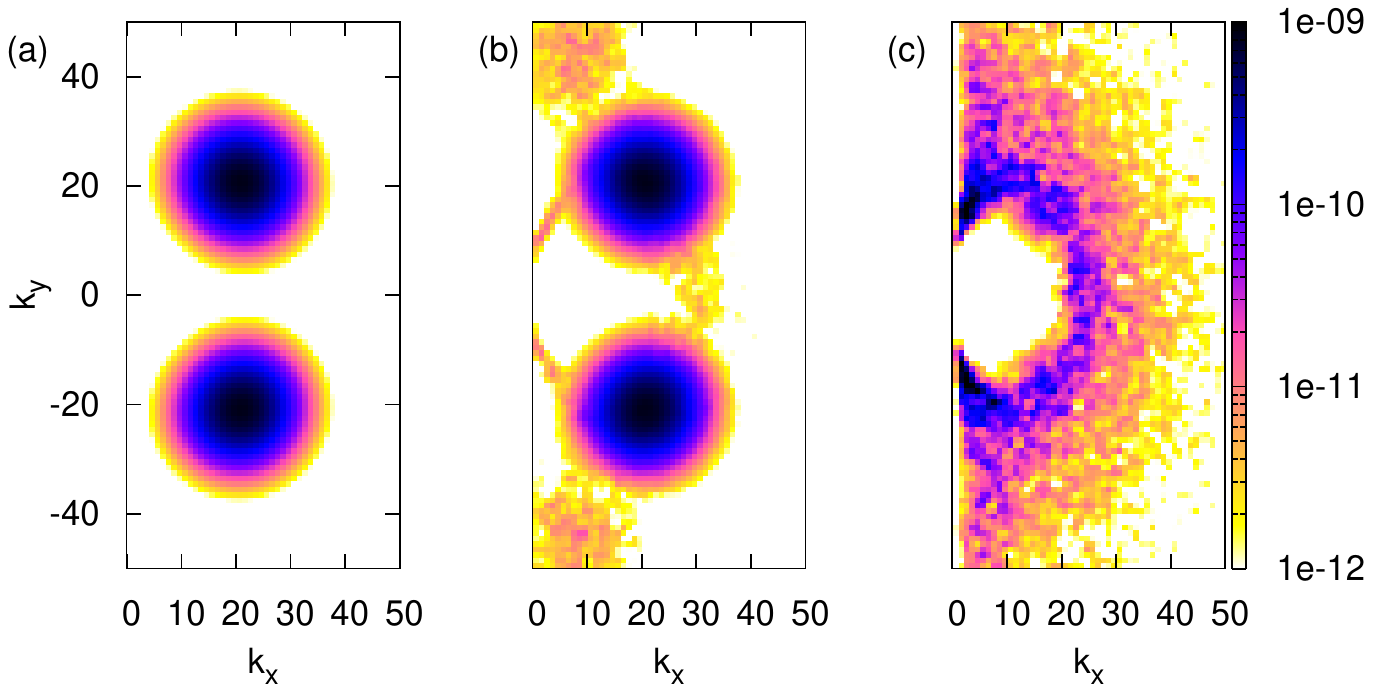}
\caption{\label{2dE_weak_NL}  Weak nonlinearity. 2D energy spectrum at (a) $t=0$  (b) $t=100$ and~(c)~$t=40000$}
\end{centering}
\end{figure}
Three successive frames of the energy spectrum in 2D $k$-space are shown in figure~\ref{2dE_weak_NL} and of vorticity distributions in $x$-space at corresponding times in figure~\ref{w_weak_NL}.  The initial spectrum, which represents the Gaussian spot as defined by Eq.~\ref{triple_ic}, centered at ${\bt k_0}$ and its mirror image in figure~\ref{2dE_weak_NL}(a), seems to grow branches in panel~(b) which stretch toward the origin forming a closed band which subsequently starts shrinking in size.
 These features are probably indicative of the structure of the anisotropic inverse energy cascade process.  In the respective vorticity $x$-plots in figure~\ref{w_weak_NL}, the initial dominant shortwave components are evident in panel~(a), propagating at $\pm 45^\circ$ corresponding to the position of the initial maxima in the spectrum. These waves evolve into a more disordered turbulent state with a predominant zonal orientation as seen in panel~(c), similar to the off-zonal flows observed for the weekly nonlinear MI study.
\begin{figure}[ht]
\begin{centering}
\includegraphics[width=0.8\textwidth]{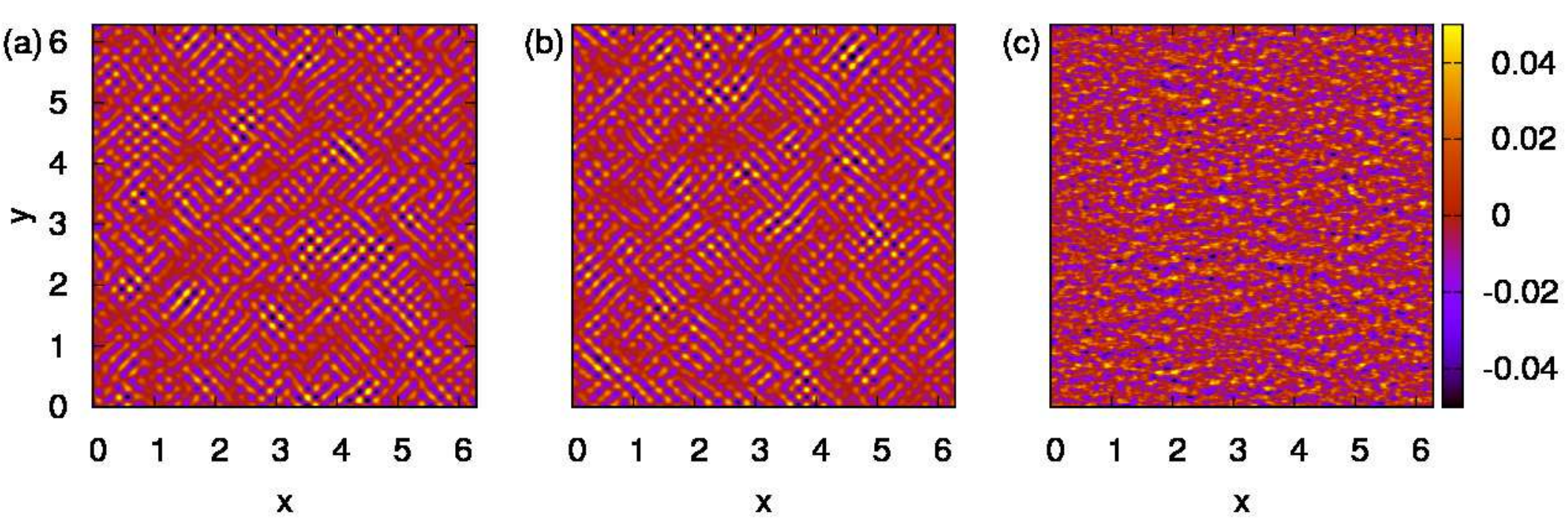}
\caption{\label{w_weak_NL} Weak nonlinearity. Vorticity distribution at (a) $t=0$  (b) $t=100$ and~(c)~$t=40000$}
\end{centering} 
\end{figure}

\subsubsection{Large-scale limit, $\rho \to 0$}
\label{sec:Large-scale limit}
%\vspace{2mm}
%{\bf Large-scale, $\rho \to 0$}

In the large-scale limit, $\rho^{2}k^{2}\ll 1$ the zonostrophy density originally found in ~\cite{Balk_proc1990,Nazarenko011991} is
\begin{equation}
\label{Z_large}
\eta_{\mathbf{k}}^L=\frac{k_x^3}{(k_{y}^{2}-3k_{x}^{2})}\,.
\end{equation}
However, it was found recently by
Saito and Ishioka  \cite{saito_angular_2013} that the exact limiting expression of the general formula for the zonostrophy, Eq. (\ref{zonostrophy_general})
\begin{equation}
\label{Z_large1}
\varphi_{\mathbf{k}}= 
- 2 \sqrt 3 \, \rho  k_x - 8 \sqrt 3 \, \rho 
\frac{k_x^3}{(k_{y}^{2}-3k_{x}^{2})} + O(\rho^2) \quad \hbox{for} \;\;\; |k_y| > \sqrt 3 k_x \, ,
\end{equation}
\begin{equation}
\label{Z_large2}
\varphi_{\mathbf{k}}= -\pi
- 2 \sqrt 3 \, \rho  k_x - 8 \sqrt 3 \, \rho 
\frac{k_x^3}{(k_{y}^{2}-3k_{x}^{2})} + O(\rho^2) \quad \hbox{for} \;\;\; |k_y| < \sqrt 3 k_x \, ,
\end{equation}
and
\begin{equation}
\label{Z_large3}
\varphi_{\mathbf{k}}= -\pi/2
- 2 \sqrt 3 \, \rho  k_x - 8 \sqrt 3 \, \rho 
\frac{k_x^3}{(k_{y}^{2}-3k_{x}^{2})} + O(\rho^2) \quad \hbox{for} \;\;\; |k_y| = \sqrt 3 k_x \, ,
\end{equation}
Performing a similar expansion in the dispersion relation (\ref{eq-RossbyWavedispersion}), we have
\begin{equation}
\label{w_large}
\omega_{\mathbf{k}}= - \rho^2 k_x + \rho^4 k_x k^2 + O(\rho^6)  \, ,
\end{equation}
which can be replaced with a simpler expression for an effective dispersion relation
which satisfies the same three-wave resonance condition:
\begin{equation}
\label{w_large1}
\omega_{\mathbf{k}}= k_x k^2 + O(\rho^2)  \, .
\end{equation}
Note that the order $\rho^1$ is missing in this expression.
Thus,
we actually have two extra invariant for the dispersion relation 
$\omega_{\mathbf{k}}= k_x k^2$, because the resonant conditions for the 
quantities~(\ref{Z_large1})-(\ref{Z_large3}) must be satisfied independently
in the orders $\rho^0$ and $\rho^1$ respectively.
Namely, the order $\rho^1$ gives the zonostrophy invariant~(\ref{Z_large}),
whereas the order $\rho^0$ gives another invariant \cite{saito_angular_2013} 
$$\Lambda = \int \zeta_{\mathbf{k}}\, n_\kv \, d\mathbf{k}$$
with spectral
density $\zeta_{\mathbf{k}}$ defined as
\begin{equation}
\label{eta_large1}
\zeta_{\mathbf{k}}= 1
 \quad \hbox{for} \;\;\; |k_y| > \sqrt 3 k_x \, ,
\end{equation}
\begin{equation}
\label{eta_large2}
\zeta_{\mathbf{k}}= 0
 \quad \hbox{for} \;\;\; |k_y| < \sqrt 3 k_x \, ,
\end{equation}
and
\begin{equation}
\label{eta_large3}
\zeta_{\mathbf{k}}= 1/2
 \quad \hbox{for} \;\;\; |k_y| = \sqrt 3 k_x \, .
\end{equation}
We will call this extra invariant {\em semi-action} because its density coincides with the one of the wave action in the sector where it is not zero. The existence of the semi-action invariant has profound consequences for large-scale Rossby/drift wave turbulence. Let us consider a non-dissipative unforced system and suppose that the initial spectrum is zero inside of the sector $|k_y| < \sqrt 3 k_x $.
Then it is clear that for any time the spectrum must remain zero inside the sector 
$|k_y| < \sqrt 3 k_x $ -- else the conservation of the semi-action would be violated
(ie. it would take a finite positive value which is different from its initial zero value).

On the other hand, if we consider a forced-dissipated system  then it is clear that the semi-action must be dissipated 
inside the  sector $|k_y| > \sqrt 3 k_x $ because (like the spectral density of the semi-action itself) its dissipation density is zero outside of this sector.

Let us now consider the effect of the zonostrophy invariant on the turbulence properties.
In the limit $|k_y| \gg k_x$ the integral (\ref{Z_large}) becomes positive and the triple cascade argument can be used~\cite{Balk_proc1990,Nazarenko011991} (For a pedagogical discussion of this limit see~\cite{Nazarenko2011}, section 8.2.3). The application of this argument is similar to what we have done above for the small-scale turbulence, and the outcome  is also similar:  that the inverse cascade is anisotropic with tendency to create large-scale zonal flows. In fact,  the limit $|k_y| \gg k_x$  can be relaxed to $|k_y| > \sqrt 3 k_x $ taking into account the property that the spectrum excited only inside such a sector will not propagate outside of it and, therefore, the zonostrophy integral will remain sign-definite (positive). See the result in figure 3 of~\cite{Nazarenko011991}.
 Actually in the case wnen the forcing is in the sector $|k_y| < \sqrt 3 k_x $ the triple cascade argument can also be used because the energy and the potential enstrophy invariants restrict the flux of zonostrophy to the sector 
$|k_y| < \sqrt 3 k_x $ where the zonostrophy is also sign-definite (negative).
See the result in figure 2 of~\cite{Nazarenko011991}.

Finally in this section we would like to mention the numerical results of 
Saito and Ishioka  \cite{saito_angular_2013} for the large-scale CHM turbulence evolving out of an isotropic initial spectrum. It appears that such a spectrum tends to accumulate in the sectors $|k_y| > \sqrt 3 k_x $ and near the lines $|k_y| = \sqrt 3 k_x $, ie. the lines on which the spectral density of the zonostrophy invariant is singular and the density of the semi-action invariant is discontinuous. This agrees with the simple analytical argument presented above that the state where the spectrum is finite only  in the sectors $|k_y| > \sqrt 3 k_x $ is stable: the energy and the other invariants cannot leave this sector and dissipate outside of it. Another qualitative explanation of the spectrum near the lines $|k_y| = \sqrt 3 k_x $ will be given in section \ref{sec:Nonlocal evolution of CHM turbulence} when we consider consequences of nonlocal interactions with zonal flows.

\subsection{Discrete WT}
In the discrete turbulence regime, resonant and quasi-resonant clusters are described by a system of dynamical equations which 
%\begin{equation}
%\label{Hamiltonian}
%i\dot a_{\mathbf{k}}=\delta \mathcal{H}/\delta a_{\mathbf{k}}^{*}=\omega_{\mathbf{k}} a_{\mathbf{k}}+\sum\limits_{1,2}(V_{12}^{\mathbf{k}}a_{1}a_{2}\delta_{12}^{\mathbf{k}}+2V_{{\mathbf{k}}2}^{1*}\delta_{{\mathbf{k}}2}^{1}a_{2}^{*} a_{1}),
%\end{equation}
%where $V_{12}^{\mathbf{k}}$ is the non-linear interaction coefficient. In 
in terms of the interaction representation variable, $b_{\mathbf{k}}$
%=a_{\mathbf{k}}e^{i\omega_{\mathbf{k}}t}$, Eq.~\eqref{dynamic_eqn_a} 
can be rewritten as:
\begin{equation}
\label{dyn_eqn_b}
\mathrm{i}\dot b_{\mathbf{k}}=\sum\limits_{1,2} (V_{12}^{\mathbf{k}}\delta_{12}^{\mathbf{k}}b_{1}b_{2}e^{-i \omega_{12}^{\mathbf{k}} t}+2V_{{\mathbf{k}}2}^{1*}\delta_{{\mathbf{k}}2}^{1}b_{2}^{*}b_{1}e^{i\omega_{\mathbf{k}2}^{1} t}),
\end{equation}
where the triad detuning parameter, $\omega_{12}^{\mathbf{k}}=\omega_{\mathbf{k}}-\omega_{\mathbf{k}_{1}}-\omega_{\mathbf{k}_{2}},$ measures the deviation of each triad from exact resonance. This set of equations can be divided into independent subsets, called {\em  clusters}. 
Each cluster evolves independently of the others and conservation properties hold independently for each cluster. This division and independence of the clusters from each other become asymptotically precise as the nonliarity tends to zero, in which case the quasi-resonances sharpen and tend to exact wave resonances.
The independence of clusters  from each other arises from the property that in finite-sized systems the resonant manifold splits into a set of isolated resonant or quasi-resonant clusters ranging in size from individual triads to much larger multiple-triad clusters.

The simplest possible cluster consisting of one quasi-resonant triad ois described by the following equations,
\begin{eqnarray}
\label{Triad}
\dot{b}_{1}&=&W^{*}b_{2}^{*}b_{3}\,e^{\mathrm{i} \omega_{12}^{3} t},\\
\dot{b}_{2}&=&W^{*}b_{1}^{*}b_{3}\,e^{\mathrm{i} \omega_{12}^{3} t},\nonumber\\
\dot{b}_{3}&=&-Wb_{1}b_{2}\,e^{-\mathrm{i} \omega_{12}^{3} t},\nonumber
\end{eqnarray}
where $W=2\mathrm{i}V_{12}^{3}$.

For a cluster consisting of $m=1, \ldots, M$ triads and $n=1, \ldots, N$ modes with amplitudes $b_{n}(t)$, it has been shown that finding quadratic invariants is equivalent to an underlying basic linear algebra problem~\cite{Harper2012}, consisting of finding the null space of a rectangular $M \times N$ cluster matrix $A$ in which the rows correspond to the triads and the columns -- to the wave modes. In each row there only three nonzero entries: one $-1$ entry at the position of the active mode (the one that decays) and two  $1$ entries at the positions of the passive mode (the ones arising from the active mode's decay).  Then the three-wave resonance conditions for $\mathbf{k}$ for the $m$th triad, defined in~\eqref{resonance_conditions} can be put into the following matrix form:
\begin{equation}
\label{resonance_matrix_form}
{\sum\limits_{n=1}^N} A_{m n} {\mathbf{k}}_{n} = {\mathbf{0}}\,,\qquad m \,\,\mathrm{fixed}, \,\,m = 1, \ldots, M
\end{equation}
where for each fixed $m$ the set $\{A_{m n}\}_{n=1}^N$ contains exactly two elements with value $1$, one element with value $-1,$ and the remaining elements are equal to zero.
If ${\mathbf \varphi} \equiv  (\varphi_1 , \varphi_2 , \ldots, \varphi_N)^{T}$ is the null space of the cluster matrix then
\begin{equation}
\sum\limits_{n=1}^{N} A_{mn}\varphi_{n}={0},
\end{equation}
for all triads in the cluster and there exists a quadratic invariant of the system, $I$ defined as
\begin{equation}
\label{I}
I= {\sum\limits_{n=1}^N} \varphi_n\,|b_n(t)|^2 \,.
\end{equation}
By taking the time derivative of Eq.~\eqref{I} and substituting for $\dot{b}_{n}$ using \eqref{dyn_eqn_b},
%with $\tilde{V}_{12}^{\mathbf{k}}=V_{12}^{\mathbf{k}}\delta_{12}^{\mathbf{k}}e^{-i\omega_{12}^{\mathbf{k}} t}$:
%\begin{eqnarray}
%\frac{dI}{dt}=\sum\limits_{n=1}^{N}\varphi_{n}\left(\dot b_{n}b_{n}^{*}+b_{n}\dot{b}_{n}^{*}\right),\\
%&=&\sum\limits_{n=1}^{N}\varphi_{n}b_{n}^{*}\sum\limits_{1,2}^{N}(-\mathrm{i})\left(\tilde{V}_{12}^{{\mathbf{k}}}b_{1}b_{2}+ 2\tilde{V}_{{\mathbf{k}}2}^{1*}b_{2}^{*}b_{1}\right)+c.c\,,\\
%&=&-\mathrm{i}\sum\limits_{1,2,3}\varphi_{3}b_{3}^{*}b_{1}b_{2}\tilde{V}_{12}^{3}+\varphi_{1}b_{1}^{*}\tilde{V}_{12}^{3*}b_{2}^{*}b_{3}+\varphi_{2}b_{2}^{*}\tilde{V}_{12}^{3*}b_{1}^{*}b_{3}\nonumber\\
%&&-\varphi_{3}b_{3}b_{1}^{*}b_{2}^{*}\tilde{V}_{12}^{3*}-\varphi_{1}b_{1}b_{2}b_{3}^{*}\tilde{V}_{12}^{3}-\varphi_{2}b_{2}b_{1}b_{3}^{*}\tilde{V}_{12}^{3}\nonumber\\
%&=&-\mathrm{i}\sum\limits_{1,2,3}\left(\varphi_{3}-\varphi_{1}-\varphi_{2}\right)\left(b_{3}^{*}b_{1}b_{2}\tilde{V}_{12}^{3}+c.c\right).\nonumber
%\end{eqnarray}
%By virtue of Eq.~\eqref{resonance_matrix_form}, 
it follows that if $\varphi_{3}-\varphi_{1}-\varphi_{2}=0$ for every term in the sum then $\dot I \equiv 0$ (see the details in ~\cite{Harper2012}).  The condition for existence of an invariant is similar therefore for discrete WT as for kinetic WT, namely the spectral density of the quadratic invariant must satisfy the  resonance conditions~\eqref{resonance_conditions} for each triad in the cluster (recall that for the resonant triads both wavenumber and the frequencies are in exact resonance, whereas for quasi-resonant triads the frequency resonance is approximate).

The number of independent quadratic invariants of a cluster in discrete WT is equal to $J \equiv N - M^* \geq N - M$ where $M^*$ is the number of linearly independent triads/rows in $A$, so $M^{*}\le M$.  The resonance conditions are much harder to satisfy in a discrete system and therefore there are less triads than in the kinetic or quasi-resonant case. Hence $N-M$ increases and there are many more invariants in discrete WT than in kinetic WT.  Typically in kinetic WT, the only invariants that remain are the ones corresponding to the physical energy with density $\varphi_{\mathbf{k}}=\omega_{\mathbf{k}},$ the momenta with density $\varphi_{\mathbf{k}}$ equal to each of the $d$ components of the wave vector $\mathbf{k}$ and, in the case of Rossby/drift waves, zonostrophy.

It is natural to ask how the energy, momentum and zonostrophy appear in discrete clusters.  Starting with the smallest, for an isolated, exactly resonant triad the number of linearly independent quadratic invariants $J = N - M = 2$ which correspond to the Manley-Rowe invariants and as a result the energy with density $\omega_{\mathbf{k}}$, the two components of momentum with density ($k_{x}, k_{y}$) and the zonostrophy with density $\varphi_{\mathbf{k}}$ will not be independent of one another.  Only two may be linearly independent.  To see this consider the Manley-Rowe invariants,
\begin{eqnarray}
I_{1}&=&|{b_{1}}|^{2}+|{b_{3}}|^{2},\nonumber\\
I_{2}&=&|{b_{2}}|^{2}+|{b_{3}}|^{2},\nonumber
\end{eqnarray}
Seeking an extra invariant
\begin{equation}
I=\varphi_{1}|{b_{1}}|^{2}+\varphi_{2}|{b_{2}}|^{2}+\varphi_{3}|{b_{3}}|^{2}.\nonumber
\end{equation}
and substituting in $\varphi_{3}=\varphi_{1}+\varphi_{2}$, we have
\begin{eqnarray}
I&=&\varphi_{1}(|b_{1}|^{2}+|b_{3}|^{2})+\varphi(|b_{2}|^{2}+|b_{3}|^{2}),\nonumber\\
&=&\varphi_{1}I_{1}+\varphi_{2}I_{2}\,,\nonumber
\end{eqnarray}
ie. any such invariant is linearly dependent on the two Manley-Rowe invariants.

Two triads connected by one mode, commonly referred to as a butterfly (see bottom right of figure~\ref{discrete_small_scale}),
%as shown in figure~\ref{butterfly},
have three invariants in total. As a result zonostrophy does not appear as an extra invariant to the energy, and the two momentum components.  For larger, three-triad clusters with four invariants, the zonostrophy does appear as an extra invariant since $k_{x}, k_{y}, \omega_{\mathbf{k}}$ and $\eta_{\mathbf{k}}$ are linearly independent of one another. For larger clusters there may be many more invariants than these four.  Unlike the triple cascade previously presented for the kinetic WT regime, it is not yet clear how the discrete regime is affected by the presence of extra invariants.  Numerical simulations of Rossby and drift WT in large discrete clusters could determine how the behaviour is different from their counterparts in kinetic WT and how the presence of numerous additional quadratic invariant affects the turbulent cascades. There is currently no understanding of additional invariants in  the mesoscopic WT regime.

\subsubsection{Small-scale Rossby and drift wave clusters}
In the small-scale limit, $\rho k\rightarrow \infty$, the wave dispersion relation, Eq.~\eqref{eq-RossbyWavedispersion} reduces to $-\frac{\beta k_x}{k^2}$.
Considering the region $1 \le k_{x}\le 100$ and $-100\le k_{y}\le 100$, a total of thirty-four clusters are found numerically, seventeen clusters plus their mirror images. This consists of twenty-four isolated triads, four butterflies, two triple-chains, two seven-triad clusters and two thirteen-triad clusters~\cite{Harper2012}; 
see figure~\ref{discrete_small_scale}.
% as shown in figure~\ref{discrete_small_scale}.
\begin{figure}[ht]
\centering
\includegraphics[width=0.8\textwidth]{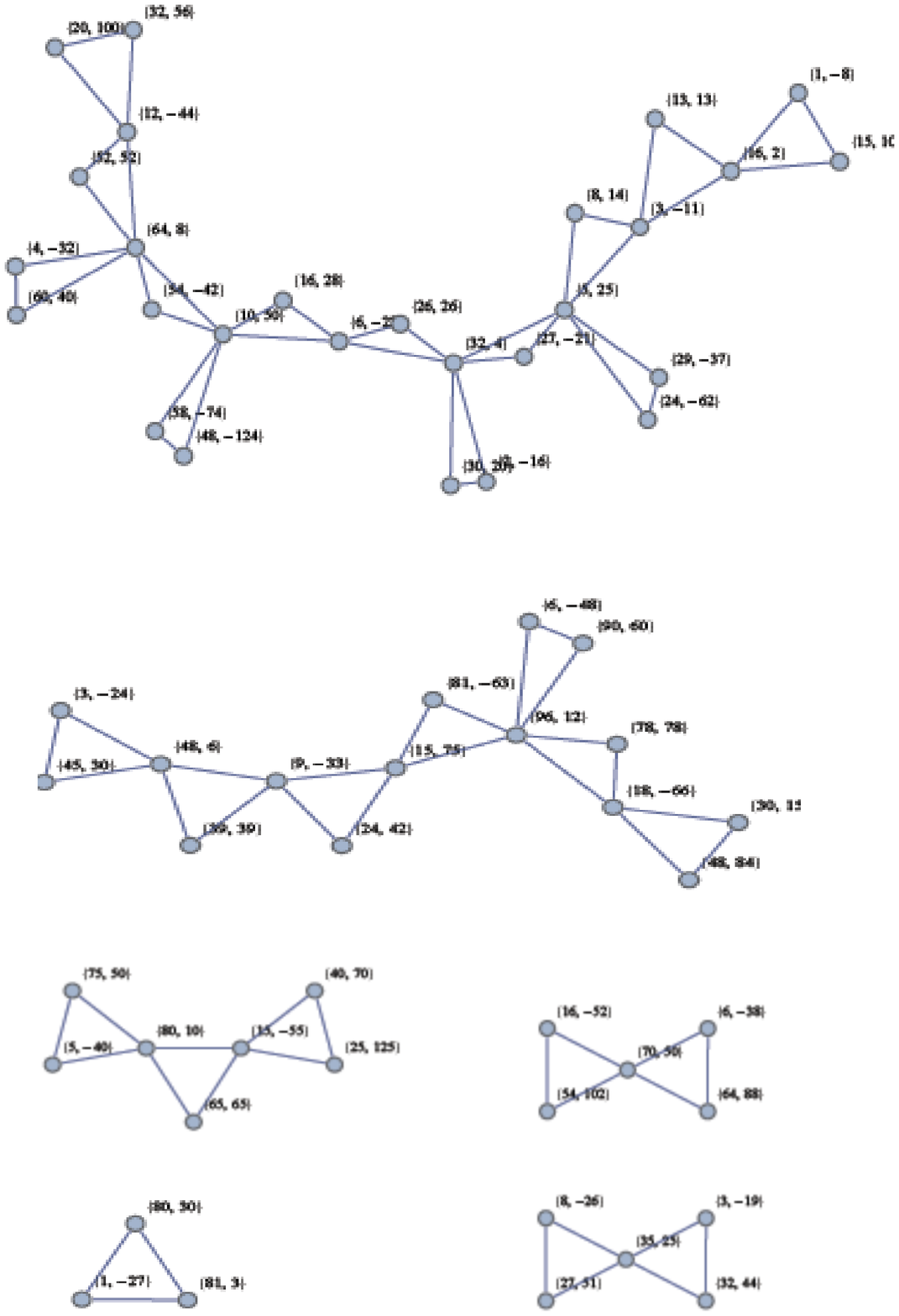}
\vspace{-1cm}
\caption{Small-scale Rossby waves in the region $1 \le k_{x}\le 100$, $-100\le k_{y}\le 100$.}
\label{discrete_small_scale}
\end{figure}

\subsubsection{Large-scale Rossby and drift wave clusters}
For large scale waves, $\rho^{2}k^{2}\ll 1$, the leading order term in the Taylor expansion of Eq.~\eqref{eq-RossbyWavedispersion} in $\rho^{2}k^{2}$ is
\begin{equation}
\omega_{\mathbf{k}}=- k_{x}\rho^{2}(1-\rho^{2}k^{2}).
\end{equation}
Since the first part in this expression is proportional to $k_{x}$; for the purpose of finding the resonances, the simpler expression satisfying the same resonant condition is taken:
\begin{equation}
\omega_{k}=k_{x}k^{2}.
\end{equation}
  In the region $1 \le k_{x}\le 20$ and $-20\le k_{y}\le 20$, a total of four clusters are found numerically. 
This consists of two isolated triads, one ten-triad cluster and one 104-triad cluster, the latter coupling 178 wave modes~\cite{Harper2012}. Note that, even in a much smaller domain for the large-scale limit, a much larger cluster exists than for the small-scale limit indicating that the resonance conditions are much easier to satisfy in the large-scale limit than in the small-scale limit.
The 104-triad cluster has in total 76 quadratic invariants, which is a huge number comparing to the invariants of non-resonant systems, and which must have a profound effect on the dynamics! Exploring such an effect could be done by a numerical experiment, which is an exciting subject for future study.
\section{Wave Turbulence -- Zonal Flow feedback loop}
\label{WT_ZF_loop}
The initial evolution of Rossby and drift wave turbulence and the subsequent generation of a zonal flow by MI or the anisotropic inverse cascade, as told so far, is only half the story however.  It is natural to ask how this large-scale zonal flow affects the turbulence at small-scales, which generates it.  It is now generally accepted in both the fields of plasma physics and GFD that in the turbulence of Rossby and drift waves the dominant interaction is with a zonal flow rather than a neighbouring-scale interaction~\cite{Nazarenko051990,Nazarenko081990,Diamond2005,Shats2005,Onishchenko2008,Nazarenko2011}, such that studies of drift wave turbulence now imply the study of the drift wave turbulence - ZF feedback mechanism \cite{Nazarenko051990,Nazarenko081990,Nazarenko1991,Biglari1990,Diamond2005} which can be summarised as follows.

Rossby waves and drift waves are produced by a primary instability, for example, the ion-temperature-gradient (ITG) instability in plasmas or the baroclinic instability in the atmospheres and oceans. These waves are predominantly meridional.  A ZF is generated via a secondary modulational instability of these waves and they grow by a direct interaction with the small scale waves.  The growing ZF extracts energy from the wave turbulence and thereby eventually suppresses it at the small scales.  

The WT-ZF feedback mechanism was first discovered theoretically~\cite{Nazarenko051990,Nazarenko081990,Nazarenko1991} within a WT approach 
based on a wave-kinetic equation and postulating nonlocality of the scale
interactions (due to divergence of the collision integral), occurring directly with large zonal scales.
The original dynamical model was CHM, Eq.~\eqref{eq-CHMx} with an instability forcing i.e. $\gamma_{\bf k} \ne 0$ in Eq.~\eqref{eq-driftdispersion}. 

A brief summary of the extended explanation of these early works which was given recently in~\cite{NonlocalDWarxiv2010} is presented here and the mechanism was confirmed numerically by DNS of the forced/dissipated CHM~\cite{ConnaughtonEPL2011} and EHM models.

On the other hand, a candidate mechanism for suppressing small-scale turbulence via shearing of small vortices by the ZF was put forward in~\cite{Biglari1990}. Their argument evoked a mechanism typical for 2D incompressible fluid and therefore it was relevant to the strongly turbulent vortex-dominated regime rather than WT. The conditions under which the WT or the vortex dominated regimes can be realised in real fusion devices, 
as well as the respective laws for saturated ZF strength were discussed 
in~\cite{ConnaughtonEPL2011}. 

\subsection{Nonlocal evolution of CHM turbulence}
\label{sec:Nonlocal evolution of CHM turbulence}

To study the nonlocal interaction between wavenumbers ${\bt k}$ and zonal flows i.e. waves with a zonal wavenumber, the points of most interest are those two where the resonant manifold crosses the $k_{1x}=0$ axis, as shown in figure~\ref{resonantmanifolds}. %(\textcolor{red}{Picture of resonant manifolds will be in MI section??})
One of those points is the origin which describes the interaction with large-scale zonal flows such that ${\bt k}_1=(0,0)$.  The other point can be determined by setting $k_{1_x}=0$ in the resonance Eq.~\eqref{resonance_conditions}, which gives the wavevector ${\bt k}_1=(0,2k_y)$.  The interactions with the wavenumbers in the vicinity of these points give the main contributions to the collision integral, Eq.~(\ref{eq-Ssymm}).

\subsubsection{Nonlocal interaction with large-scale zonal flows}

Looking first at the interactions of ${\bt k}$ with the large-scale zonal flows near the origin.  These large-scale waves travel with the $x$-component of the phase speed,
\begin{equation}
\label{phaseSpeed}
c_{\bt k} = \frac{\omega_{\bt k}}{k_x} = -\frac{\beta}{k^2+F}.
\end{equation}
In the large-scale limit, $k \ll F$, $c_{\bt k} \to -\beta/F$ and the absolute value of this is known as the drift velocity or Rossby velocity.  It is convenient therefore to work in a coordinate frame moving with this velocity, in which the Doppler-shifted frequency is given as 
\begin{equation}
\label{OmegaDoppler}
\Omega_{\bt k} = \omega_{\bt k} - c_{\bt k}k_x = \frac{\beta k_x k^2}{F(k^2 + F)}\,.
\end{equation}
In this moving reference frame, the kinetic equation becomes
\begin{eqnarray}
\nonumber
\frac{\partial n_{\bt k}}{\partial t} = 	4\pi \int\hspace{-0.5mm} &  \left| V^{\bt k}_{12} \right|^2 \delta({\bt k}-{\bt k}_1 - {\bt k}_2) \delta( \Omega_{\bt k} - \Omega_{{\bt k}_1} - \Omega_{{\bt k}_2}) \times\\
\label{kinetic_eqn_doppler}
& \left[  n_{{\bt k}_1} n_{{\bt k}_2} - n_{\bt k} n_{{\bt k}_1} \hspace{.5mm} \mathrm{sign} (\omega_{\bt k} \omega_{{\bt k}_2}) -  n_{\bt k} n_{{\bt k}_2} \hspace{.5mm} \mathrm{sign}(\omega_{\bt k} \omega_{{\bt k}_1})  \right]\, d {\bt k}_1 d {\bt k}_2.
\end{eqnarray}
A detailed analysis, as outlined in the appendix~\ref{apx_nonlocal}, reduces this kinetic equation to a diffusion equation
\begin{equation}
\label{diffusion_kyOmega}
\frac{\partial n_{\bt k}}{\partial t} =\frac{\partial \Omega_{\bt k}}{\partial k_x} \left(\frac{\partial}{\partial k_y} {D}_{\bt k} \left(\frac{\partial n_{\bt k}}{\partial k_y}\right)_\Omega\right)_\Omega
\end{equation} 
where $()_\Omega$ means $\Omega$ being held constant and the diffusion coefficient is
\begin{equation}
\label{diffTensorDtilde}
{D}_{\bt k} = 2\pi \left|\frac{\partial \Omega_{\bt k}}{\partial k_x}\right|^{-2} \int^{\infty}_{-\infty}  \left[ \left| V^{\bt k}_{{\bt k}_1\, {\bt k}-{\bt k}_1} \right|^2 k_{1y}^2 \right]_{k_{1x}=-\sigma_{\bt k} k_{1y}} n_{{\bt k}_1}d k_{1y} \,.
\end{equation}
Eq.~\ref{diffusion_kyOmega} therefore describes diffusion in the $k_y$ direction along the curves of constant $\Omega(k_x,k_y)$ which are shown in figure~\ref{levelsets_Omega}(a).

\begin{figure}
\begin{center}
(a)\includegraphics[width=0.2\textwidth]{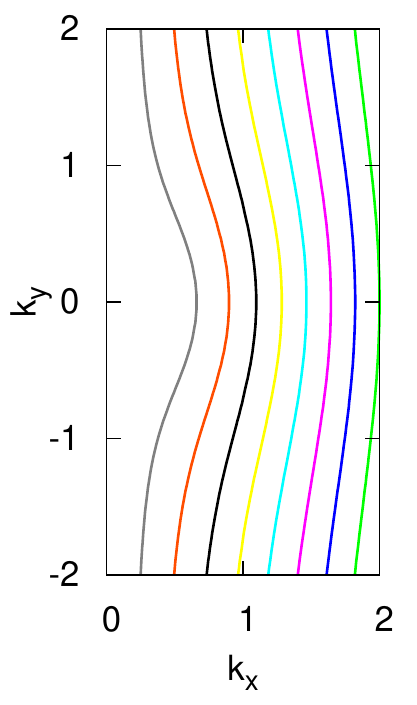}
\hspace{15mm}
(b)\includegraphics[width=0.2\textwidth]{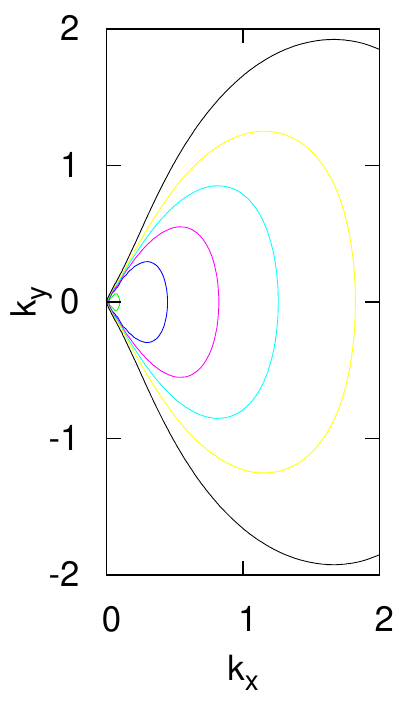}
\end{center}
\caption{The level sets of (a) $\Omega_{\bt k}$ and (b) ${Z}_{\bt k}$.}
\label{levelsets_Omega} 
\end{figure}

The mathematically predicted evolution of the turbulence spectrum, Eq.~(\ref{diffusion_kyOmega}) provides some insight into the features of the nonlocal turbulence.  The energy of the small scale waves with dispersion $\omega_{\bt k}$ is defined as,
\begin{equation}
E = \int |\omega_{\bt k}| n_{\bt k} d{\bt k}\,.
\end{equation}
As the spectrum of these small scales diffuses along the curves of constant $\Omega_{\bt k}$, it loses energy since $\omega_{\bt k}$ decreases along these curves.  However for an unforced, dissipation-less turbulence, the total energy of the system must be conserved, therefore this energy from the small scales has to go somewhere and must therefore be transferred to larger scales.  Neither is there dissipation at these large scales so the energy spectrum will accumulate there.  Furthermore, since the diffusion coefficient $D_{\bt k}$
%$\widetilde{D}_{\bt k}$ 
given by Eq.~(\ref{diffTensorDtilde}) is directly proportional to $k_y$, the rate of transfer of energy from the small wavepackets to the large scales is also increased.

For a forced and dissipative turbulence regime, the evolution of the spectrum will be determined by the forcing and dissipation terms, collectively denoted as $\gamma_{\bt k}$, added to the RHS of Eq.~(\ref{diffusion_kyOmega})  i.e.  
\begin{equation}
\label{diffusion_forced}
\frac{\partial n_{\bt k}}{\partial t} =\frac{\partial \Omega_{\bt k}}{\partial k_x} \left(\frac{\partial}{\partial k_y} {D}\left(\frac{\partial n_{\bt k}}{\partial k_y}\right)_\Omega\right)_\Omega + \gamma_{\bt k} n_{\bt k}\,.
\end{equation}
Note from Eq.~(\ref{diffTensorDtilde}) that the diffusion coefficient $D$,
%$\widetilde{D}$
depends on the spectrum of the zonal flow, $n_{{\bt k}_1}$.    If this spectrum is given when ${D}$ is a fixed function of $k_y$ and $\Omega$.  The spectrum $n_{\bt k}$ evolves along the curves of constant $\Omega_{\bt k}$ and it's magnitude is controlled by the
eigenvalue $\Gamma_\Omega$ of the eigenvalue problem on each curve $\Omega_{\bt k} =$~const. maximum value of $\Gamma_{\bt k}$ on that curve, growing exponentially if $\Gamma_\Omega > 0$ and being damped exponentially if $\Gamma_\Omega < 0$~\cite{Nazarenko051990,Nazarenko081990}.  Positive values of $\Gamma_\Omega$ occur on the curves $\Omega =$~const passing  near the maximum of the forcing $\gamma_{\bt k}$ while negative $\Gamma_\Omega$ occurs far away from this maximum where dissipation terms are dominant.

For positive $\Gamma_\Omega$, the spectrum $n_{\bt k}$ grows on the curve of constant $\Omega_{\bt k}$ but this growth cannot be sustained indefinitely as it has already been stated that diffusion of the spectrum along these curves results in energy transfer to large scales.  Since this results in a large diffision coefficient and augmented diffusion to the dissipation region, the growth of the spectrum due to positive $\Gamma_\Omega$ is gradually relaxed until a balance is achieved between the growth and the diffusion. 
This corresponds to the moment when the region of positive $\Gamma_\Omega$ shrinks to zero and the maximum value of  $\Gamma_\Omega$ lowers to zero. Consequently, the large scale zonal flow saturates at a level which can be estimated by balancing the forcing and diffusion terms of Eq.~(\ref{diffusion_forced}), which gives
\begin{equation}
\label{balanceTerms}
n_{\bt p} \sim \frac{\beta\,\gamma_{max}}{4\pi p^3 |V^{\bt k}_{{\bt k}_1\, {\bt k}-{\bt k}_1}|^2},
\end{equation}
where for large scales $p_x \sim p_y \sim p$.  Again, using the fact that $n_{\bt p}=\frac{E_{\bt p}}{\omega_{\bt p}}$ whereby in the large-scale limit, 
\begin{equation}
\label{largescaleE}
E_{\bt p} \approx \frac{F}{p^2}\psi^2
\end{equation}
and $U_{ZF}=\frac{\partial \psi}{\partial y} \sim p\psi$, another expression can be obtained for $n_{\bt p}$
\begin{equation}
\label{approx_np}
n_{\bt p} \sim \frac{F^2 U_{ZF}^2}{p^5 \beta}\,.
\end{equation}
Equating Eqs~(\ref{balanceTerms}) and~(\ref{approx_np}) and deducing from Eq.~(\ref{interactioncoeff}) that $|V^{\bt k}_{{\bt k}_1\, {\bt k}-{\bt k}_1}|^2 \sim \frac{\beta p^3}{F}$, then the large scale zonal flow is estimated to saturate at the velocity
\begin{equation}
\label{Uzf_sat}
U_{ZF} \sim \sqrt{\beta\, \gamma_{max} L /(\pi F)}\,,
\end{equation}
where $L \sim \frac{1}{p}$.

This estimate, Eq.~(\ref{Uzf_sat}) is valid for weak turbulence only, $U_{ZF} \ll \beta$, since it's derivation stems from the kinetic equation.  For strong turbulence, the $\beta$ term is negligible and Eq.~(\ref{Uzf_sat}) is invalid.  An alternative zonal flow saturation velocity for the strong turbulence regime will be determined later.

\subsubsection{Nonlocal interaction with small-scale zonal flows}
A similar analysis is  performed in appendix~\ref{apx_nonlocal} for the nonlocal interaction with small-scale zonal flows. It yields the diffusion equation
\begin{equation}
\label{eq-ssDiffusion4}
\pd{n_\vv{\kappa}}{t} =   \pd{Z_\kv}{k_y} \pd{}{k_x} \left( \left( \pd{Z_\kv}{k_y} \right)^{-1} \left( \pd{\Omega_\kv}{k_y} \right)^{-1}\, \tilde{B}(\kv)\, \left(\pd{n_\vv{\kappa}}{k_x}\right)_Z\right)_Z,
\end{equation}
where $\left(\pd{\cdot }{k_x}\vphantom{\frac{1}{2}}\right)_Z$ means differentiation with respect to $k_x$ with $Z_\kv$ being held constant. 
Here
\begin{equation}
\label{Btilde}
\tilde{B}({\bt k})  =  4\pi \left| \frac{\partial \Omega_{\bt k}}{\partial k_y} \right|^{-1}\int^{\infty}_{-\infty} dq_x \left[ \left| V^{\bt k}_{{\bt q}, {\bt k}-{\bt q}} \right|^2 n(-q_x,q_y)\, q_x^2 \right]_{q_y=2k_y} \,.
\end{equation}
Eqs~\eqref{eq-ssDiffusion4} and~\eqref{Btilde} describe diffusion along curves of constant ${Z}_{\bt k}$, defined as
\begin{equation}
{Z}_{\bt k} =  \arctan{\left(\frac{(k_y+\sqrt{3}\,k_x)\sqrt{F}}{k^2}\right)} - \arctan{\left(\frac{(k_y-\sqrt{3}\,k_x)\sqrt{F}}{k^2}\right)}\\
\label{Ztilde}-\frac{2\sqrt{3\,F}k_x}{k^2+F}\,,
\end{equation}
and which are shown in figure~\ref{levelsets_Omega}(b). Note that near ${\bf k} =0$ these curves focus onto the same lines $k_y = \pm \sqrt 3 \, k_x$. We have already poined out the signifinance of this line in 
section \ref{sec:Large-scale limit} when we discussed the role of the zonostrophy invariant in the large-scale limit of CHM turbulence. In fact it is easy to see that the expression for ${Z}_{\bt k}$ is nothing but the linear combination of the spectral densities of the zonostrophy (\ref{zonostrophy_general}) and the energy. Now we have another perspective of why the numerically observed spectrum tends to acummulate near the lines $k_y = \pm \sqrt 3 \, k_x$:
the nonlocal interaction with the zonal flows leads to the spectrum evolution along the curves, which compresses the turbulence onto these lines, thereby enhancing its value.

%\begin{figure}
%\begin{center}
%\includegraphics[width=0.2\textwidth]{images/shell}
%\caption{The level sets of $\widetilde{Z}_{\bt k}$}
%\label{levelsets_shell}
%\end{center}
%\end{figure}

%\vspace{3mm}
\subsection{Qualitative WKB description of zonal flow growth}

The foregoing rigorous mathematical theory is based on the wave kinetic equation and is therefore valid only in the weak wave turbulence regime which requires that the quasi-zonal scale turbulence be weak.  However an alternative theory which is similar to Wentzel-Kramers-Brillouin (WKB) theory and which does not require a weak turbulence assumption, was introduced in~\cite{Dyachenko1992}, extended in~\cite{Nazarenko2011} and is now presented.  The presentation of both theories provides a unified picture of both strong and weak wave turbulence regimes.
% WKB theory is a method for finding an approximate solution to linear PDE with spatially varying coefficients.

Consider a drift wave packet propagating on the background zonal flow with a velocity profile $U(y)$ which varies randomly in $y$. The correlation length $L$ corresponds to a typical wavelength profile in $U(y)$.  Again nonlocal drift turbulence is assumed, so that interactions between the drift waves themselves are negligible in comparison to their interactions with the zonal flow. 

By averaging the Fourier transformed CHM Eq.~(\ref{eq-CHMk}) over the characteristic times of the small scales, the evolution equation for the slow large scales, i.e. the zonal flow is obtained.  The evolution equation for the small scale drift and Rossby waves is obtained by considering  Eq.~(\ref{eq-CHMk})  at large ${\bt k}$, neglecting the terms describing the interactions with the other high  ${\bt k}$'s (ie. ignoring the mutual interactions within the small-scale component condidering it to be weak compared to the interactions with the large scales),  multiplying the resulting equation by $\hat{\psi}_{\bt k}$ and averaging over the fast times.  The resulting  evolution equation is ~\cite{Dyachenko1992}:
\begin{equation}
\label{n_conservation}
\left(\frac{\partial}{\partial t} + \frac{\partial \bar{\omega}_{\bt k}}{\partial {\bt k}}\frac{\partial}{\partial {\bt x}} + \frac{\partial \bar{\omega}_{\bt k}}{\partial {\bt x}}\frac{\partial}{\partial {\bt k}}\right) n_{\bt k} = 0,
\end{equation}
 $\bar{\omega}_{\bt k}$ is the total frequency of the drift wave packet due to its own linear dispersion plus that due to the motion of the large scale zonal flow~\cite{Dyachenko1992},
\begin{equation}
\label{omega_bar}
\bar{\omega}_{\bt k} = \frac{k_x(Uk^2+\beta)}{k^2+F}\,,
\end{equation}
and $n_{\bt k} \equiv n({\bt k},{\bt x},t)$
is the local density of the small-scale waveaction which appears to be the local density of the enstrophy. It is related to the local density of the small-scale energy, $e_{\bt k}$ , via the usual 2D turbulence relation: $n_{\bt k} = k^2 e_{\bt k}$. Recall that from WT we would expect 
a different relation, $n_{\bt k} =  e_{\bt k}/\omega_{\bt k}$. Naturally, the two expressions coincide when the assumptions of the scale separation and the turbulence weakness are fulfilled simultaneously. Indeed in this case, as we established above, the waves evolve along the ${\bt k}$-space curves $\Omega_{\bt k} =$~constant, which happen to be the same as the curves $k^2 \omega_{\bt k} =$~constant.

The term within the parentheses of Eq.~\eqref{n_conservation}
is a time derivative along the wave rays~\cite{Dyachenko1992,Nazarenko2001} defined as
\begin{equation}
\label{Dt_ray}
D_t = \frac{\partial}{\partial t} + \dot {\bt x}\cdot\nabla + \dot {\bt k} \cdot\nabla\, ,
\end{equation}
where
\begin{equation}
\label{rayss}
\dot {\bt x} = 
\frac{\partial \bar{\omega}_{\bt k}}
{\partial {\bt x}} \, , \quad  \dot {\bt k} = \frac{\partial \bar{\omega}_{\bt k}}{\partial {\bt x}}.
\end{equation}
 
It then follows from Eqs~(\ref{n_conservation}) and~(\ref{Dt_ray}) that
\begin{equation}
\label{ky_dot}
\dot k_x = -\frac{\partial \bar{\omega}}{\partial x} = 0, \quad \dot k_y = -\frac{\partial \bar{\omega}}{\partial y} = \frac{k_x k^2U^{\prime}}{k^2+F} \, ,
\end{equation}
where $^{\prime}\equiv\frac{\partial}{\partial y}$ and
\begin{equation}
\label{y_dot}
\dot y = \frac{\partial \bar{\omega}}{\partial k_y} = \frac{2k_x k_y(F\,U-\beta)}{(k^2+F)^2}\,.
\end{equation}

\subsubsection{Weak zonal flow}

Assuming that the zonal flow is sufficiently weak, i.e $U \ll \beta$,
the drift wavepacket  can travel through many correlation lengths in the $y$-direction.  The wavepacket will experience a random walk in $k_y$ so that the evolution of the mean waveaction $n_{\bt k}$ will be described by a diffusion equation,
\begin{equation}
\label{diffusion_rw}
n=\frac{\partial}{\partial k_y}\left(D_{rw}\frac{\partial n_k}{\partial k_y}\right)+\gamma_kn \, , 
\end{equation}
where $\gamma_k$ describes the instability in the unstable region where $\gamma_k>0$ and $\gamma_k<0$ in the dissipative region.  The diffusion coefficient $D_{rw}$ can be estimated from a standard random walk argument,
\begin{equation}
\label{D_rw_exp}
D_{rw}=\left(\frac{1}{\dot k_y}\right)^2\tau ,
\end{equation}
% diffusion coefficient has units distance^2/time.  This exp is (dt/dy)^2 t.  From Einstein PhD thesis
where $\tau$ is the correlation time estimated as 
\begin{equation}
\label{tau_rw}
\tau =\frac{L}{|\dot y |}\,.
\end{equation}  
Inserting Eq.~(\ref{y_dot}) into Eq.~(\ref{tau_rw}) and then into Eq.~(\ref{D_rw_exp}), the random walk diffusion coefficient is estimated to be
\begin{equation}
\label{D_rw}
D_{rw}=\frac{k^4U^2}{2\beta} \, ,
\end{equation}
where the estimates $k_x \sim k_y (\sim k \sqrt{2})$ and $U^\prime \sim U/L$ have been used.  Then the characteristic time of the diffusion process is
\begin{equation}
\label{tau_diff}
\tau_{dif}=\frac{k_y^2}{D}=\frac{\beta L}{k^2U^2}\,.
\end{equation} 
Assuming that the zonal flow is excited by drift wave turbulence via an inverse cascade until a saturated value is reached, the saturation value can be determined by balancing the diffusion and instability terms in Eq.~(\ref{diffusion_rw}) which gives 
\begin{equation}
\label{balance_rw}
\frac{D}{k_y^2}=\gamma_{max},
\end{equation}
where $\gamma_{max}$ is the maximum value of the growth rate of the underlying instability.  Substituting Eq.~(\ref{D_rw}) into Eq.~(\ref{balance_rw}), the estimated saturation velocity of the zonal flow is
\begin{equation}
\label{U_dif}
U_{dif}=\frac{\sqrt{\beta \gamma_{max} L }}{k}.
\end{equation}
This estimate coincides with our previously obtained Eq.~(\ref{Uzf_sat})  if $k \sim \sqrt F$.

\subsubsection{Strong zonal flow}

In the case of a strong zonal flow the $\beta$-effect is less important so that the DWP will travel through ${\rm O}(1)$ correlation lengths, getting carried from the unstable region to the dissipative region in a shorter time than it would take to cross one zonal flow oscillation of length $L$.  The characteristic time for these rapid distortions follows from Eq.~(\ref{ky_dot}),
\begin{equation}
\label{tau_rd}
\tau_{rd} \sim \frac{k_y}{\dot k_y} \sim \frac{(k^2+F) L}{k^2 U}.
\end{equation} 
This rapid distortion will occur when $\tau_{rd}$ is less than the time needed for the wavepacket to cross one ZL oscillation, $\tau_{rd} < \tau_{dif} $ or equivalently
% which is equivalent to detering the critical zonal flow velocity as
\begin{equation}
\label{U_rd}
U > \frac{\beta }{k^2+F}\,.
\end{equation} 
Using Eq.~(\ref{U_dif}) in this expression, we get the following condition for when rapid distortion is expected,
\begin{equation}
\label{condRD}
L \gamma_{max} > \frac{k^2\beta}{(k^2+F)^2}\,.
\end{equation}
%From Eq.~\ref{U_rd}, it can be said that when $k \lesssim \sqrt{F}$, rapid distortion occurs when $U \gtrsim \beta$ so the assumption of $U \ll \beta$ in obtaining the weak zonal flow saturation estimate is justified.  Considering the specific limit when $U=\frac{\beta}{F}$, there will be stagnation points because according to Eq.~\ref{ky_dot}, $\frac{\partial y}{\partial t} = 0$ at these points.
%   It is not yet clear what special role is played by these regions in forming the profile $U(y)$.
Saturation ZF velocity in the rapid distortion  regime is expected when the inverse of its characteristic time becomes similar to the instability growth rate, $\gamma_{max} \sim \frac{1}{\tau_{rd}}$ which gives
\begin{equation}
\label{Uzf_satrd}
U_{rd}=\left(1+\frac{F}{k^2} \right)L \gamma_{max}.
\end{equation}

From Eq.~(\ref{condRD}) it is clear that transition between the diffusive and rapid distortion regimes for zonal flow saturation occurs when the strength of the instability reaches $\gamma_c=\frac{k^2\beta}{(k^2+F)^2L}$; see figure~\ref{transition}.
\begin{figure}
\begin{center}
\includegraphics[width=0.4\textwidth]{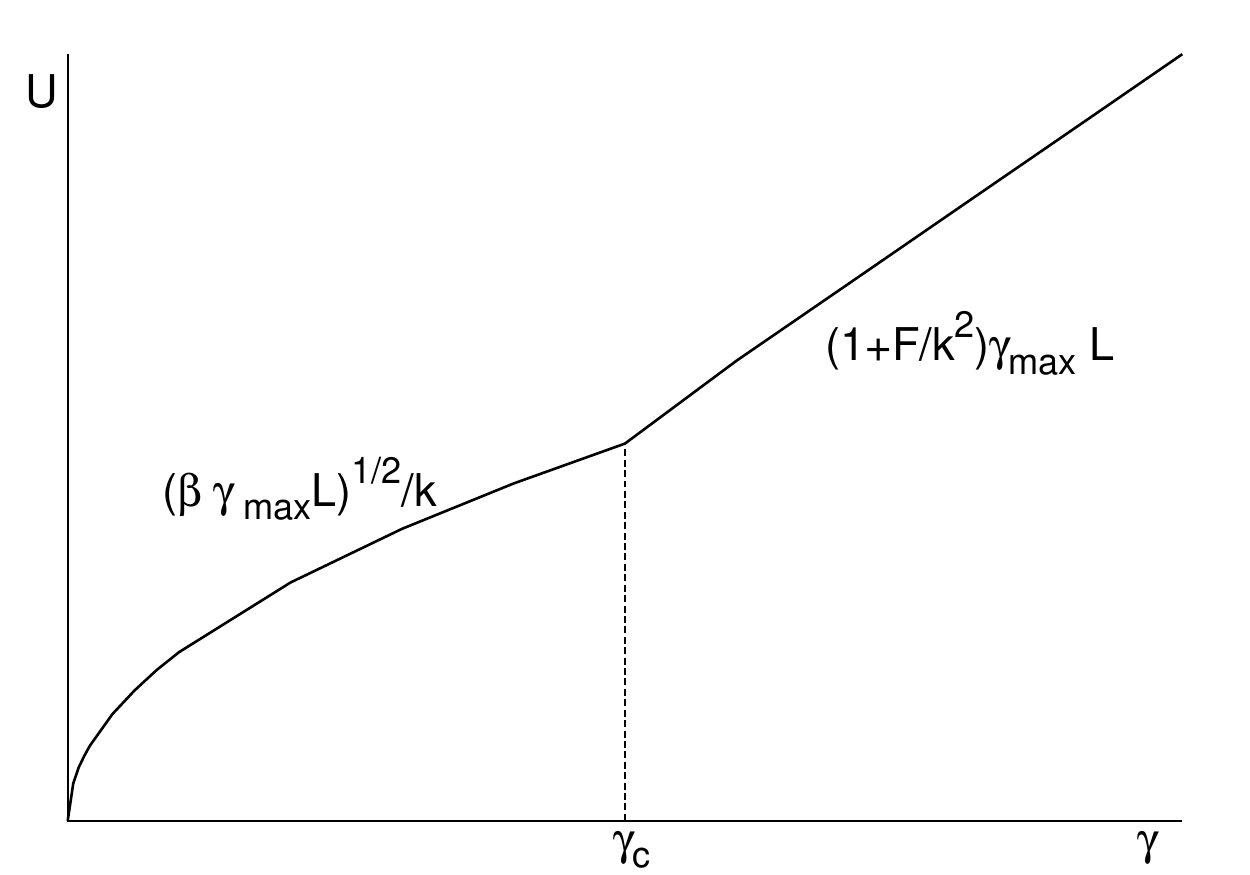}
\caption{Zonal flow velocity $U$ as a function of the instability growth rate $\gamma$}
\label{transition} 
\end{center}
\end{figure}

\subsection{Numerical validation of the WT/ZF feedback loop}

The presented above picture of the  drift/Rossby wave turbulence coupled with ZF was tested numerically in~\cite{ConnaughtonEPL2011}.  As previously mentioned, the CHM equation does not contain small-scale instabilities but  the forcing can be modeled by modifying the dispersion as in Eq.~\eqref{eq-driftdispersion}.  For the instability forcing $\gamma_{\bf k}$, the simplest form was sought in~\cite{ConnaughtonEPL2011} which retains the key features of the relevant primary instabilities mentioned above, namely that they act at small scales and primarily generate meridional waves.  Here, it is modeled as $\gamma_{\bt k} =\gamma^+_{\bt k} -\nu_m k^{2 m}$ where $\gamma^+_{\bt k}$ is the instability forcing term and $\nu_m$ a hyperviscosity coefficient.  Dissipation $\nu_m (-\Delta)^m$  is a typical hyperviscosity for turbulence simulations with $m \in \mathbb{N}$.
In various physical situations, the region of instability which leads to drift waves turns out to be approximately the same~\cite{Nazarenko081990,Horton1986,Drake1984} and also coincides with the region determined by the linear instability analysis of the two-layer QG PV model.   A dimensionless parameter is introduced
\begin{equation}
\label{chi}
\chi = \frac{\gamma_{max}  \sqrt {F} }{\beta}\, ,
\end{equation}
which measures the nonlinearity of the system (here $\gamma_{max} = \rm{max}(\gamma_{\bt k})$).  Results are presented for an idealised narrowband forcing and a more physical-type instability with a broadband  forcing.  
%
%A linear analysis of the Hasegawa-Wakatani model or a two-layer baroclinic model 
%baroclinic modes 
%

\subsubsection{Narrowband forcing}

The instability forcing is required to act at small scales and to generate the meridional drift or Rossby waves.  For simplicity, an idealized narrowband forcing at a single mode, ${\bt k}$ at the scale of the ion gyroradius or deformation radius was chosen with $\gamma_{\bt k} > 0$ and constant~\cite{ConnaughtonEPL2011}.  In order to introduce mode coupling and allow the development of turbulence with a single-mode forcing, the numerical simulations were started from an initial condition consisting of spatial white noise of very low amplitude.

Results presented in figure~\ref{narrowforcing} are for $\gamma^+=1.2$ at ${\bt k}_i=(50,0)$, $m=8$, $\nu_m=1.0 \times 10^{-29}$ and $F=2500$ and $\beta=1\times10^5$ which corresponds to 
$\chi = 3.75 \times 10^{-4}$.  The $\bf k$-space has been divided into the zonal sector for which $|k_x|<|k_y|$ and a meridional sector for which the opposite is true, minus the forcing mode.  Figure~\ref{narrowforcing} shows the evolution of the energy contained in the forcing mode, zonal and meridional sectors.  The energy of the forced mode grows exponentially while the initial energies of the zonal and meridional sectors remains constant at the level of the background noise.  After $t=10$, when the amplitude of the forcing mode reaches a high enough level to trigger nonlinear mode coupling, a sharp growth of the zonal and meridional energies takes place.  A short time after however, the energies in the forcing mode and in the meridional sector are suddenly suppressed while that of the zonal energy remains at the highest level attained.  A steady state is reached whereby the energy of the forcing mode 
is fully suppressed while that of the zonal energy  subsequently saturates at a  level which is much higher than the one of the meridional sector.  
The inset in figure~\ref{narrowforcing} shows the saturated value of the ZF as a function of $\beta$ in terms of $U_{rms} (\nabla U)_{rms} \sim U^2/L$ and the comparison with the theoretical predictions arising from Eq.~(\ref{Uzf_sat})  and Eq.~(\ref{Uzf_satrd}) for the weak and strong turbulence cases respectively (dashed lines). We see that the saturated ZF level agrees well with the theoretical predictions for both the weak and the strong regimes.  
%In this case, although it is of the correct order, the estimate is a little high but it should be noted that it is an approximate estimate not taking into account some constant factors.  

\begin{figure}[ht]
\begin{center}
\includegraphics[width=0.4\textwidth]{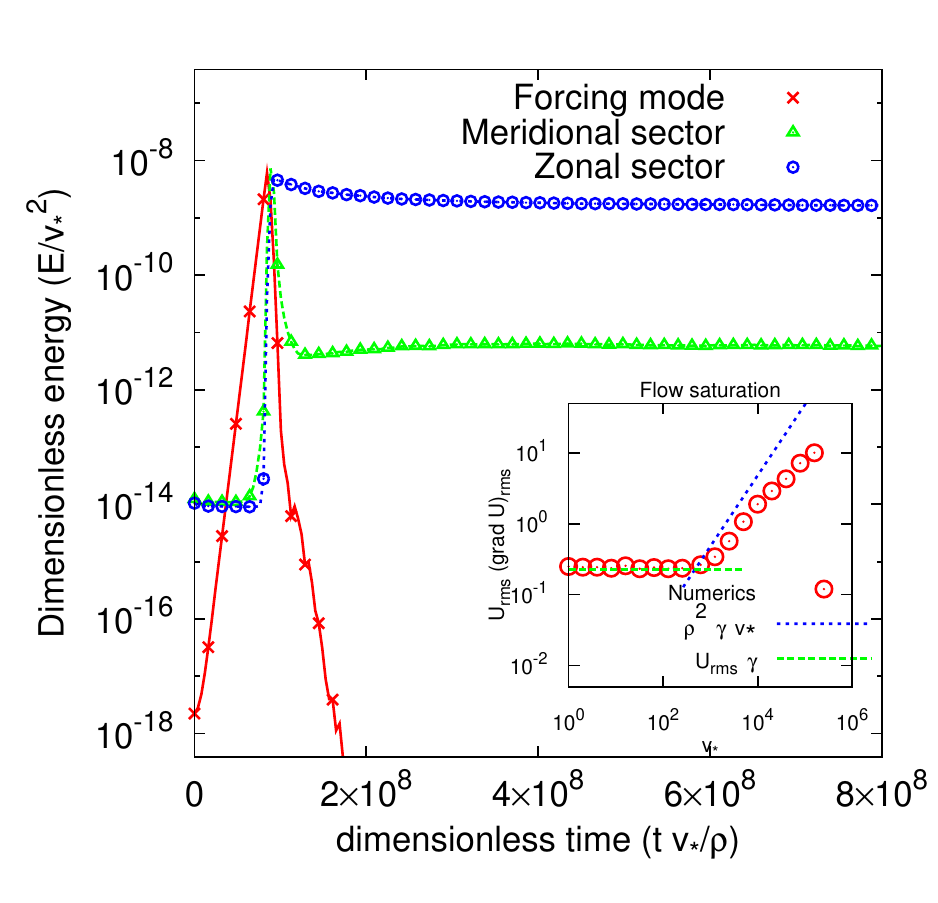}
\end{center}
\caption{ Evolution of the energy contained in the forcing mode, zonal and meridional sectors for narrowband forcing in the CHM for $\chi  = 3.75 \times 10^{-4}$.
The inset  shows the saturated value of the ZF as a function of $\beta$. The horizontal and the inclined dashed lines are the theoretical predictions arising from Eq.~(\ref{Uzf_sat})  and Eq.~(\ref{Uzf_satrd}) for the weak and strong turbulence cases respectively.}
\label{narrowforcing}
\end{figure}

The corresponding 2D energy spectrum is shown in figure~\ref{2dspectrum_idealised}.  In panel~(a), the wave amplitude of the forcing mode at $(50,0)$ has reached a high enough level to become modulationally unstable~\cite{Manin1994,Onishchenko2004,Connaughton2010}.  This triggers the nonlinear interactions, initially with modes which lie on the resonant (blue, dotted) curve which grow from the background noise.  The maximally unstable perturbations to the LHS of the resonant curve have a significant zonal component and are equivalent to those in figure~\ref{qMaps_F_1}.  The observed rapid decrease in the meridional energy corresponds to the zonal flow-induced diffusion along the (red, solid) curves of constant $\Omega$ which is evident in figure~\ref{2dspectrum_idealised}(b).  While the diffusion is not {\em exactly} along the curves, the deviation could be attributed to the fact that the scale-separation between the forcing and zonal modes is not as pronounced as is assumed in the derivation of 
Eq.~(\ref{diffusion_kyOmega}).  Diffusion along the (green, dashed) curves of constant ${Z}_{\bt k}$ shows evidence of interactions with small scale zonal flows and again highlights the importance of the zonostrophy invariant in drift and Rossby wave turbulence theory~\cite{Nazarenko1991}.

\begin{figure}[ht]
\begin{center}
\includegraphics[width=0.65\textwidth]{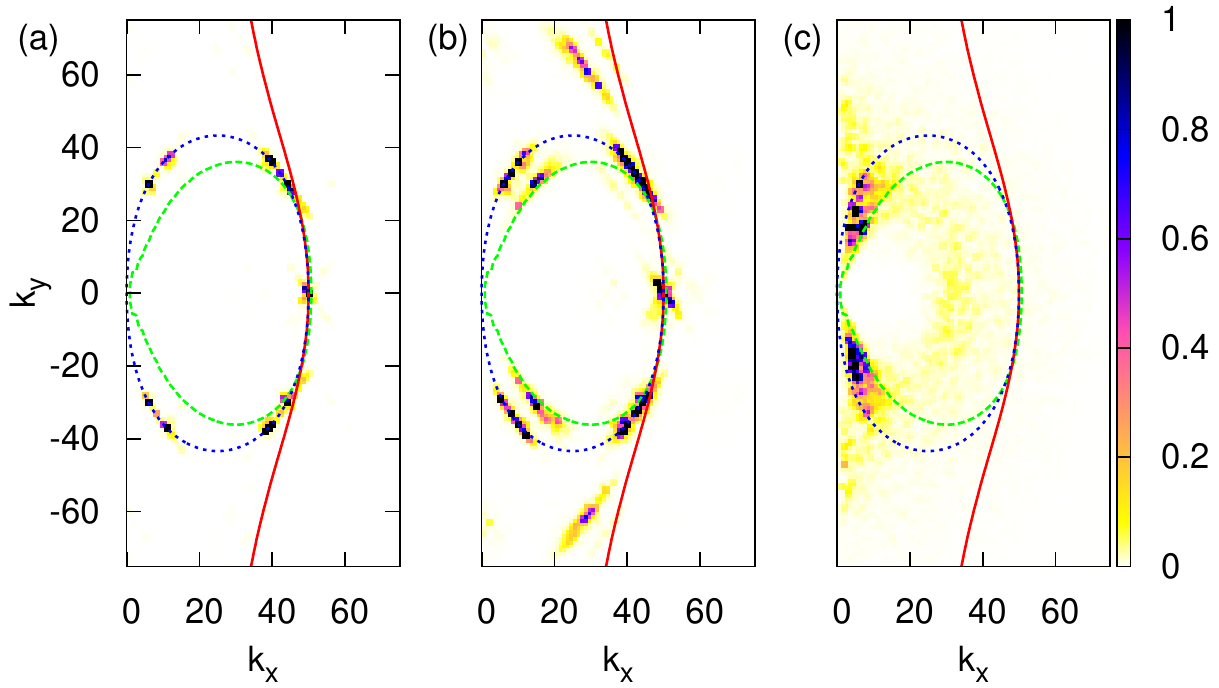}
\end{center}
\caption{ Evolution of the energy spectrum out of a random seed in the 2D $\bf k$-space for narrowband forcing.}
\label{2dspectrum_idealised}
\end{figure}

\subsubsection{Broadband forcing}
An expression for $\gamma_{\bf k}$ which is more physically relevant can be obtained by considering the linear dynamics of a higher level model which does contain an intrinsic instability such as the HW model in the case of plasmas or the two-layer model in GFD.  In this case, analysis of a mean zonal baroclinic flow in its simplest configuration as a two-layer QG model~\cite{McWilliams2006} derives an expression for the baroclinic instability.  Two layers of equal height and sliding relatively to each other with a horizontal slip velocity $V$.
%with a vertical, but no horizontal, shear between them is assumed.  $U(y)$ is the constant mean flow.  
By analysing the baroclinic mode, an exponentially growing modal solution is found to exist, which represents the baroclinic instability.  The growth rate of this unstable mode is  ~\cite{James1987,McWilliams2006}:
\begin{equation} 
\label{gamma_GFD}
\gamma^+_{\bf k}= V k_x \sqrt{\frac{F-k^2}{F+k^2}}.
\end{equation}
A  derivation of this instability can be found in~\cite{McWilliams2006}.  The expression in Eq.~(\ref{gamma_GFD}) is used in the numerical simulations and is shown graphically in figure~\ref{gamma}.  Points to note are that the growth rate ($\gamma^+_{\bf k} > 0$) is in a region adjacent to the $k_x$ axis and it reaches a maximum on this axis. The $\gamma^+_{\bf k}=0$ contour line passes through ${\bf k}={\bf 0}$ and 
${\bf k} = (\sqrt F, 0)$.
%$\gamma^+ \rightarrow -\infty$ as $|k| \rightarrow \infty$.

\begin{figure}
\begin{center}
\includegraphics[width=0.5\textwidth]{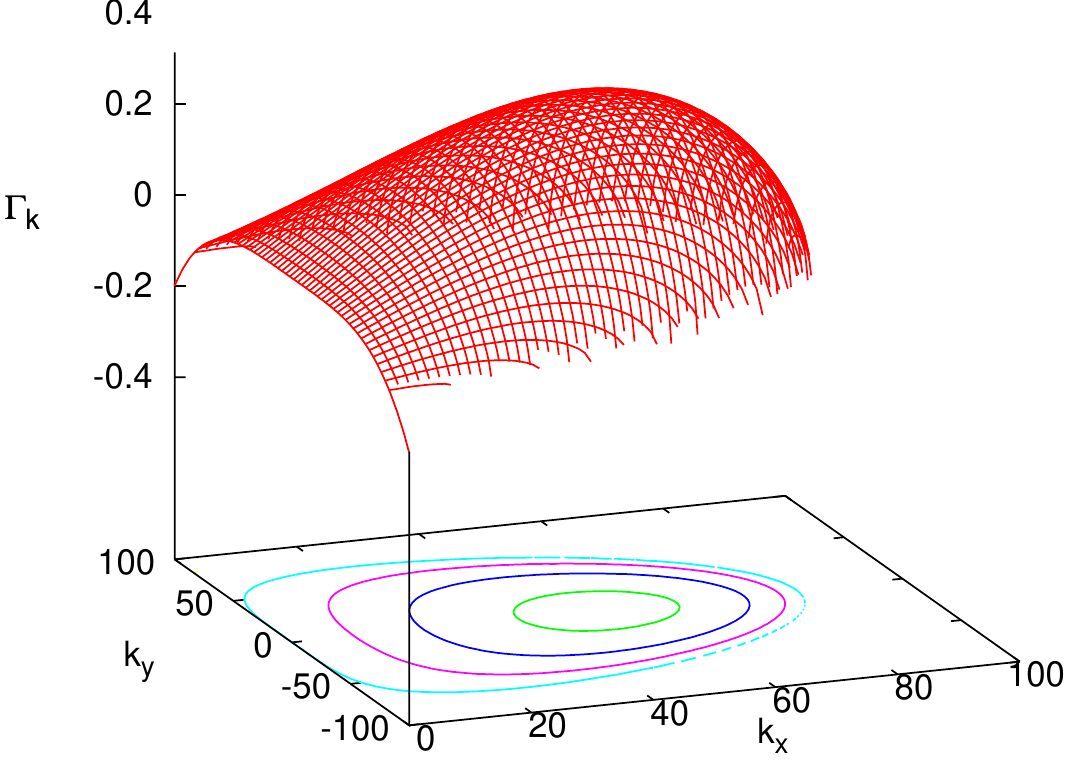}
\caption{\label{gamma} Graphical representation for $\gamma_{\bf k}=\gamma^+_{\bf k}-\nu_2 k^4$ with $\gamma^+_{\bt k}$ defined by Eq.~\eqref{gamma_GFD} with $V=0.01$, $F=7000$ and $\nu_2 = 2.\times 10^{-9}$.  The outermost contour is $\gamma=0$.}
\end{center}
\end{figure}
Analogous to the baroclinic instability in GFD, is the ITG instability in plasmas, which is the most important instability for fusion turbulence.  The centre of a magnetically confined plasma could reach a temperature of $10-20 keV$ with colder edge temperatures of the order $1 eV$.   This large thermodynamic gradient in a tokamak plasma is a perfect source of free energy for instabilities.  It arises in magnetically confined plasmas due to the particles in the hotter side of the plasma having higher kinetic energy which eventually causes perturbations along the temperature gradient.  The higher drift velocity of these particles across the magnetic field adds to the perturbations which then can become unstable.  The typical expression for the ITG instability~\cite{hammett2007} is 
\begin{equation} 
\label{gamma_ITG}
\gamma^+_{\bf k}= {\rm const} \frac{v_{t_i}}{\sqrt{R L_T}},
\end{equation}
where $v_{t_i}$ is the ion thermal velocity, $R$ is the tokamak major radius and $L_T=\frac{\nabla T}{T}$ and $T$ is temperature.  It can be derived from the electrostatic gyrokinetic equation~\cite{hammett2007}.

For our simulations we have chosen the instability shape as in~\eqref{gamma_GFD} with
 $V=0.01$ and $F=7000$.
 %for various $\beta$. 
 For dissipation at high wavenumbers we chose the usual viscosity with $m=2$, $\nu_m=2.0 \times 10^{-9}$.   The initial condition is gaussian, centered at the maximum of the forcing mode. Figure~\ref{energy_beta_10E6} shows the time evolution of the energy contained within each sector where the forcing sector now encompasses the modes within the variance of the Gaussian distribution.
\begin{figure}
\begin{center}
\includegraphics[width=0.65\textwidth]{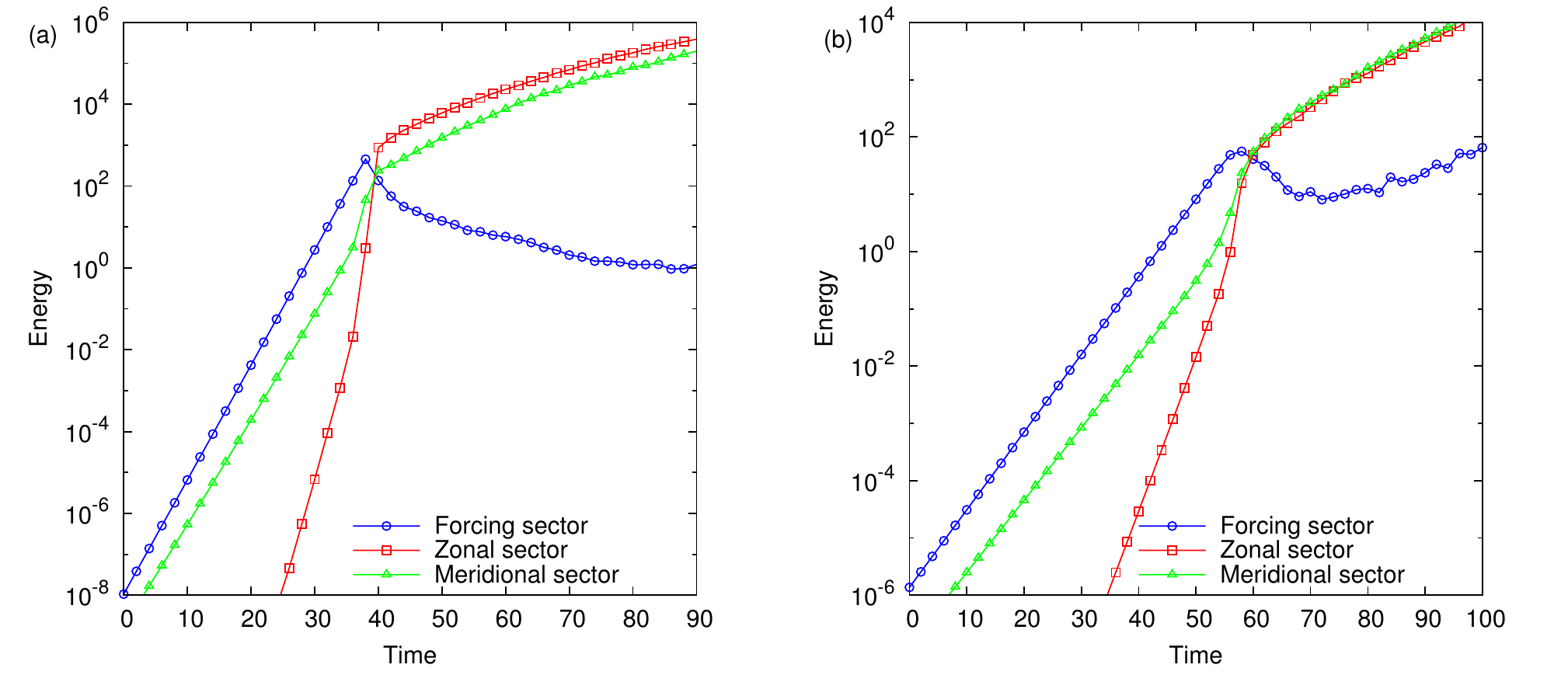}
\end{center}
\caption{The evolution of the energy contained in the forcing mode, zonal and meridional sectors for broadband forcing for (a) $\chi=2.5 \times 10^{-5}$ and (b) $\chi=25 $.}
\label{energy_beta_10E6}
\end{figure}

\begin{figure}
\begin{center}
\includegraphics[width=0.65\textwidth]{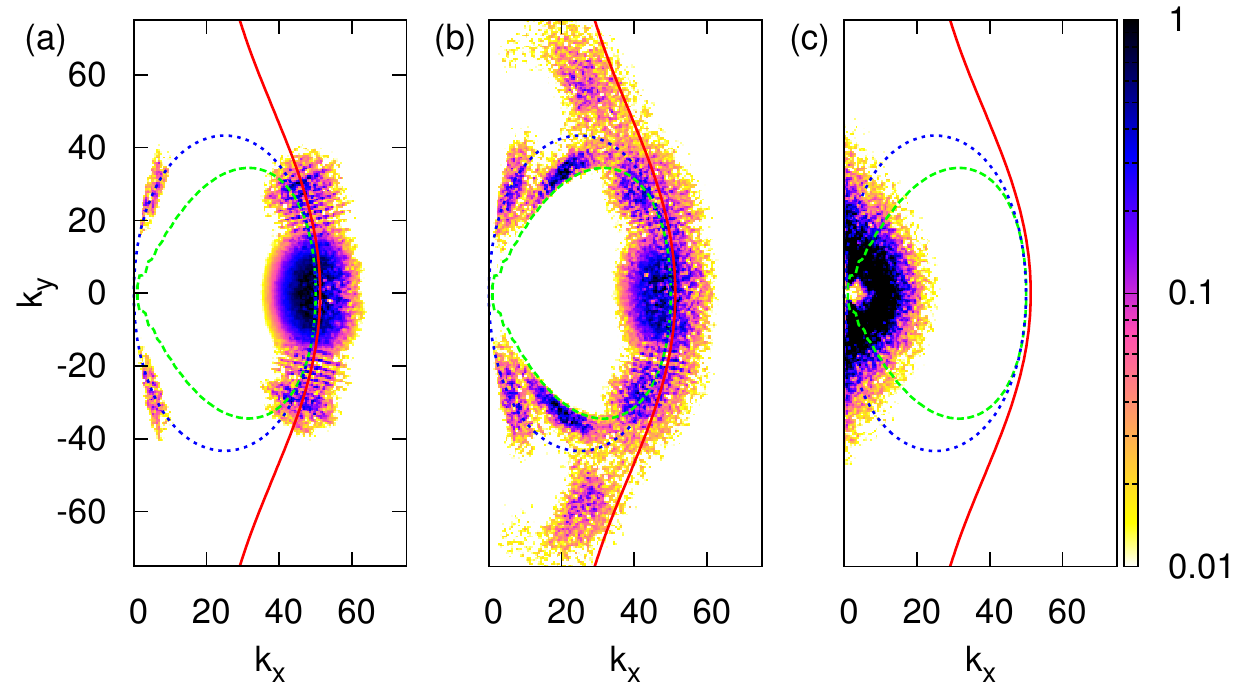}
\end{center}
\caption{The 2D energy spectrum for $\chi=2.5 \times 10^{-5}$ at times (a) 38.6, (b) 39.2 and (c) 70 normalised by the maximum energy 0.25, 0.25 and 5 respectively.}
\label{2dspectrum_beta10E6} 
\end{figure}
As before, the energy of the forcing sector grows exponentially at the beginning.  The energies in the zonal and meridional sectors also grow from the outset because both of these sectors now contain some linearly unstable modes.  The forcing energy continues to grow until the resonant nonlinear interactions occur and the forcing is subsequently suppressed.  The zonal energy is the dominant sector.  At large time, we see saturation of the zonal and meridional energies for the run with small nonlinearity; see the results for the case 
$\chi=2.5 \times 10^{-5}$ in figure~\ref{energy_beta_10E6}(a).  However, there is no saturation for high levels
of nonlinearity; see the results for the case 
$\chi =25$ in Figure~\ref{energy_beta_10E6}(b).
%is not observed here but a definite reduction in its growth rate is. Panels (a), (b) and (c) of figure~\ref{energy_beta_10E6} show that as the nonlinearity becomes stronger, the energies in the zonal and meridional sectors become more-or-less equal.  The horizontal dashed line in the plots represents an approximate saturated value as estimated from Eq.~(\ref{Uzf_sat}).  It is clearly a very good estimate of the ideal level where saturation would be deemed to occur, according to the point where the growth rate slows down at $t \approx 40$.  This estimate decreases with $\beta$.
Note that to compare the nonlinearity levels $\chi$ in the single mode forcing and the broadband forcing cases, one has to take into account that there are many more modes excited in the latter case (by the factor $ \sim F$ in our case) and, therefore, there is much more energy pumped into the system. Roughly, at the same value of $\chi$ the nonlinearity level is $\sim F$ times greater in the broadband forcing case than in the single mode case.

\begin{figure}
\begin{center}
\includegraphics[width=0.9\textwidth]{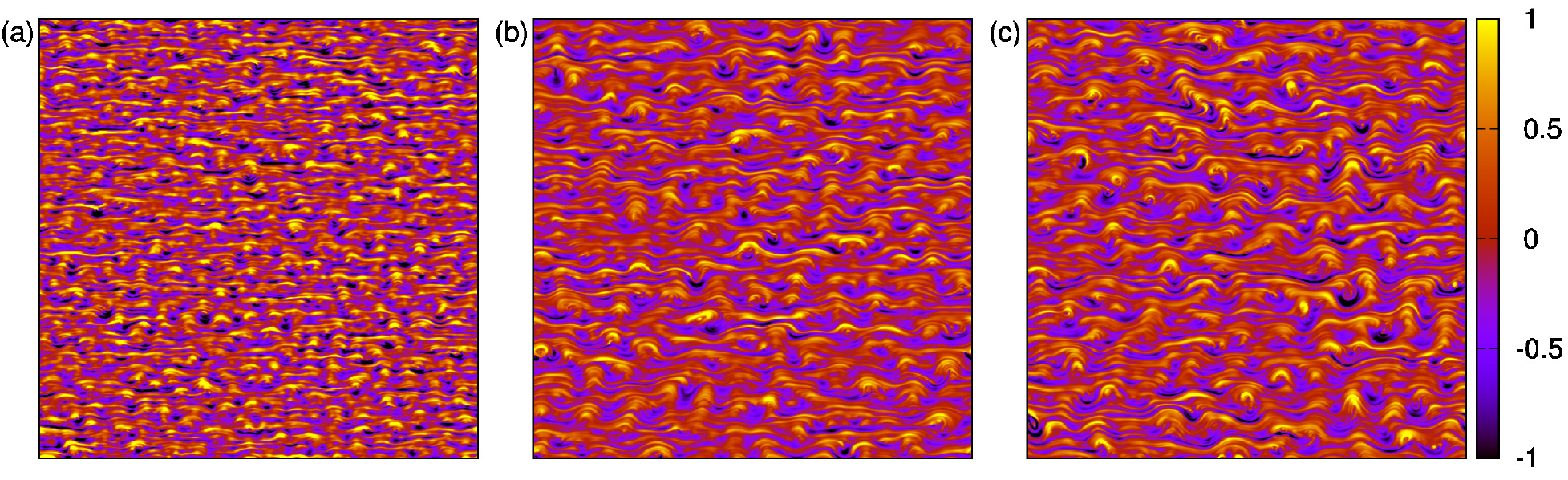}
\end{center}
\caption{A sequence of vorticity snapshots for a weakly nonlinear case $\chi=2.5 \times 10^{-5}$.}
%$\chi=25$ at times (a) 30, (b) 60 and (c) 80 normalised by the maximum vorticity 50, 500 and 750 respectively. Horizontal axis is~$x$ and the vertical~$y$.}
\label{w_beta10E6a} 
\end{figure}

\begin{figure}
\begin{center}
\includegraphics[width=0.9\textwidth]{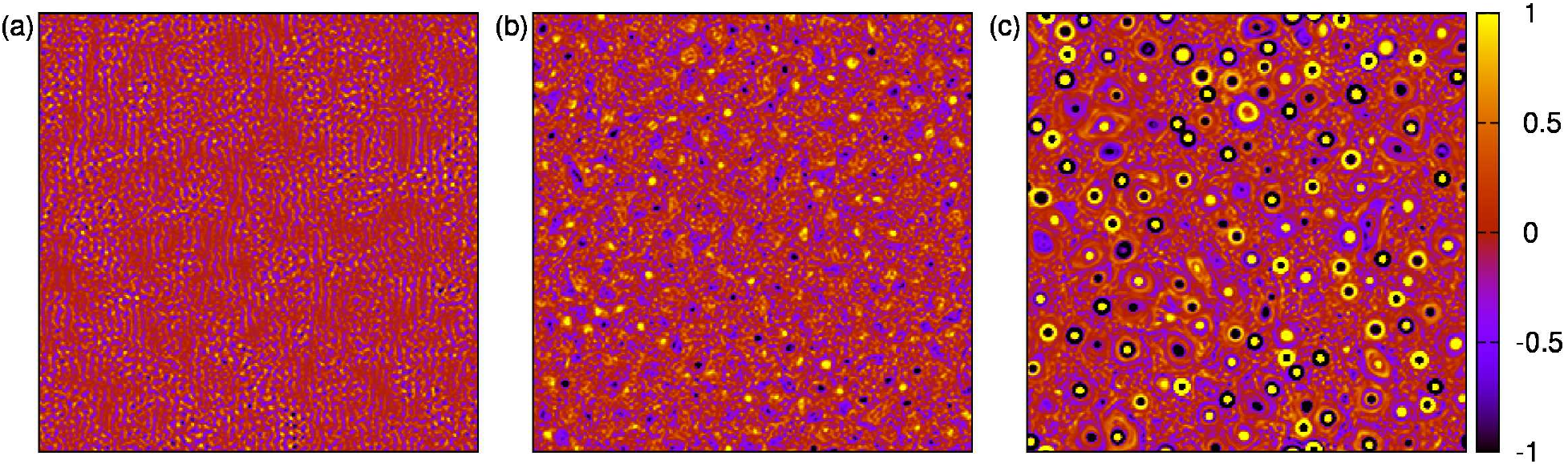}
\end{center}
\caption{A sequence of vorticity plots for a strongly nonlinear case $\chi=25$.}
%$\chi=25$ at times (a) 30, (b) 60 and (c) 80 normalised by the maximum vorticity 50, 500 and 750 respectively. Horizontal axis is~$x$ and the vertical~$y$.}
\label{w_beta10E0} 
\end{figure}

\begin{figure}
\begin{center}
\includegraphics[width=0.65\textwidth]{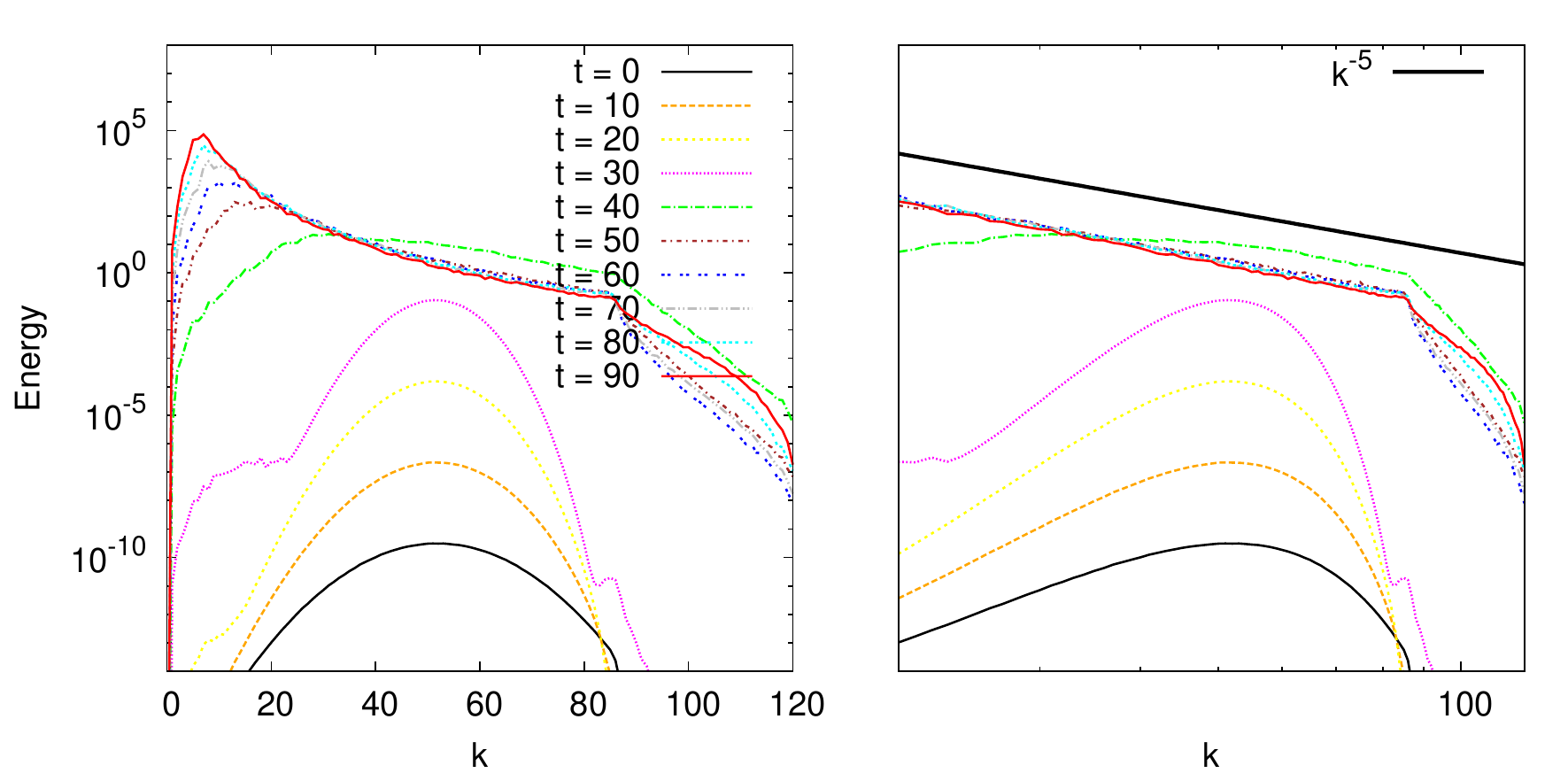}
\end{center}
\caption{The 1D averaged spectra for the weakly nonlinear case.  Both plots are for the same data but RHS is a log-log plot.}
\label{1dspectra_beta10E6}
\end{figure}

The initial resonant interactions in the weakly nonlinear case ($\chi=2.5 \times 10^{-5}$) are again evident in the 2D energy spectrum plots in figure~\ref{2dspectrum_beta10E6}(a).  Subsequently in a later frame shown in figure~\ref{2dspectrum_beta10E6}(b) the small scale spectrum begins to diffuse along the open curves of constant $\Omega_{\bt k}$ and also diffusion along the shell-like curve of constant ${Z}_{\bt k}$ to large scales, showing the two unconnected components of the turbulence spectrum. Finally in figure~\ref{2dspectrum_beta10E6}(c), the energy has been suppressed at the forcing modes and is now concentrated around the zonal modes. The corresponding vorticity plots in figure~\ref{w_beta10E6a} show these zonal-type structures at late times.

The angle-averaged spectra in the weakly nonlinear case ($\chi=2.5 \times 10^{-5}$) are shown in figure~\ref{1dspectra_beta10E6}. We see a tendency to a $k^{-5}$ law.  The isotropic power law with this exponent was originally proposed by Rhines~\cite{Rhines1975} who then dismissed it as being physically impossible. However, there are arguments, based on data from the Voyager spacecraft, that this spectrum exists for the zonal flows on Jupiter and Saturn~\cite{Galperin2001}, albeit for the zonal component of the wavenumber only.
Of course, our system is strongly anisotropic and the angle-averaged spectra are of a limited value. However, the zonal component dominates in the angle-averaged spectrum at the late stages and, therefore, the $-5$ spectrum we observe appears to be consistent with the previous observations on Jupiter and Saturn.

For higher levels of the nonlinearity measure, as expected, the vorticity plots are characterised by vortices rather than zonal structures, evident in figure~\ref{w_beta10E0} for the case $\chi=25$.  The $\beta$ term can be neglected so that we have the system equivalent of 2D Navier-Stokes turbulence forced by an instability. In this case, condensation occurs into strong round vortices with radial profiles of alternating vorticity sign rather than zonal flows, similar to the structures which have been identified in simulations of the Navier--Stokes--Kuramoto--Sivashinsky equation \cite{gama_two-dimensional_1991} which also contains an instability term.  In the unforced system such vortices would be highly unstable. The instability stabilises them and makes them grow exponentially without a visible sign of saturation.

A study with the idealised single-mode forcing in the EHM model reveals a qualitatively similar behaviour for the spectrum diffusion in $k$-space and for the suppression of the forcing mode.  It does however, display somewhat better zonation properties in that the zonal sector saturates at a slightly higher level. 

\subsection{Feedback loop in HW \& EHW models}
The drift turbulence - ZF feedback loop in HW and EHW was first studied by~\cite{Numata2007}. 
ZF generation followed by suppression of the primary instability scales was demonstrated for EHW,
but not for HW. They concluded that there is no ZF generation and feedback mechanism in HW and therefore,
the extension of the model is crucial. However, this assertion is true only for relatively small coupling 
paramenter values, $\alpha \lesssim 1$.
Recently~\cite{Pushkarev2013} computed the HW system exploring a wide range of $\alpha$.
They showed that for large $\alpha$'s the HW system is perfectly capable of generating pronounced ZF's
which suppress the primary instability and saturate. For example, for $\alpha$ the the HW system was shown to saturate with the zonal-to-meridional energy ratio of about 10. This result seems natural considering that for    
large $\alpha$ the HW system becomes identical to CHM. However, to observe the WT-ZF feedback 
mechanism in CHM one had to introduce an extra instability
forcing term, whereas in the HW system such an instability is present naturally, even though it is weak in the large $\alpha$ case.

%%%%%%%%%%%%%%%%%%%%%%%%%%%%%%%%%%%%%%%%%%%%%%%%%%%%%%%%%%%%%%%%%%%%%%%%%%%%%%%%%%%%%%%%%%%%%%%%%%%%
\section{Conclusion}
\label{conclusion}

 In this review we have summarised the ideas and findings of Rossby and drift wave turbulence
together with zonal flows within the simplest PDE model which describes such systems, the Charney-Hasegawa-Mima (CHM) equation and its extensions -- extended Hasegawa-Mima (EHM), forced-dissipated CHM and EHM, 
Hasegawa-Wakatani (HW) and extended Hasegawa-Wakatani (EHW) equations.
Our focus has been on the most important processes and stages in the life cycle  for such basic systems: generation of small-scale, predominantly meridional, waves by a primary instability, nonlinear interaction of the small-scale waves leading to generation of zonal flows, and feedback of the zonal flows on the small-scale waves. Understanding these processes in such basic nonlinear models helps gain a qualitative (and sometimes even quantitative) understanding of the real nonlinear phenomena in  plasmas and geophysical fluids.

The first important stage in the nonlinear evolution of the Rossby and drift wave system (after the initial generation of meridional waves) is the zonal flow (ZF) generation.
There are two classical scenarios of this process, the modulational instability (MI) and the anisotropic inverse cascade. These two mechanisms are not completely unrelated -- MI is expected when the primary meridional waves have a relatively narrow spectrum whereas a cascade is more likely for a broadband initial spectrum.  In real, non-idealised situations, one should expect a mixture of the two.

We have reviewed the theory and the numerical simulation results on the MI of the Rossby and drift waves. We
have described the linear theory based  on the four-mode truncation (4MT) and emphasised the role of the non-dimensional carrier
wave amplitude $M$ (effective Rossby number), the role of the deformation
radius $\rho$ and the role of resonant wave interactions in the weakly nonlinear case. 
We described a bifurcation
of the most unstable modulation from zonal to off-zonal when
$M$
falls
below a critical value: 0.53 for CHM with $\rho=\infty$ and a lower ($\rho$-dependent) value for EHM.
 This effect has its signature in the late nonlinear evolution stage too where numerically we observe generation of off-zonal random jets
in the $M=0.1$ case, and it is likely to play a role in the formation of slightly off-zonal jets in both GFD and plasmas.
%By testing the 4MT system for both CHM and EHM against DNS, we have shown that the  4MT  works very well for small nonlinearities $M$ including the case of the purely zonal modulations. Also, the 4MT  works well for the initial evolution in the strongly nonlinear cases, $M \gtrsim 1$, including the linear growth stage and the prediction of the critical value of $M$  for which a transition from the oscillatory to  the saturated  nonlinear regimes is observed.

The initial nonlinear evolution of MI 
%for small $M$ is is characterized by quasi-periodic oscillations, whereas 
for large
$M$  leads to a rolling-up of the carrier wave
vorticity into K\'arm\'an-like vortex streets. Such hydrodynamic vortices behave very differently from waves, and
it it precisely at the moment of roll-up that the full system's evolution strongly departs from the prediction of
the 4MT. After the roll-up the full system enters into a saturated quasi-stable state which persists for a relatively long time but eventually decays due
to presence of hyper-viscosity. 
For the zonal velocity component,  we observe the formation of stable, narrow
 jets. These jets are more stable
than one would expect based on the Rayleigh-Kuo criterion  because their 2D structure consists of stable
vortex streets. Such narrow jets represent very effective
transport barriers in both plasmas and GFD.
If $M$ 
is small and the
roll-ups do not occur (or are delayed), the full system may
follow its 4MT counterpart for much longer whereby the initial modulation growth reverses and may exhibit the nonlinear
oscillations as in the 4MT.

The EHM model differs from the CHM only
for finite $\rho$ and only for the purely zonal modulations.
In general,   ZFs form easier in EHM than in CHM.

We have also reviewed the cascade scenario of the ZF generation within the wave turbulence (WT) approach. We have briefly reviewed the general WT theory and its application to the Rossby and drift waves. We have discussed Kolmogorov-Zakharov scaling solutions and established their non-realisability due to their nonlocality and instability. We have described additional quadratic invariants, in particular the zonostrophy invariant, the existence of which makes the energy cascade strongly anisotropic and directed towards the zonal scales via the triple cascade behaviour. We have described a WT theory based on the hypothesis of nonlocality of interactions with ZFs.

The second important stage in the nonlinear evolution of the Rossby and drift wave system
is the feedback of the ZFs onto the small-scale WT leading to suppression of the energy input to the system (via the primary instability) via the suppression of the motions at the instability scales.
This process closes the evolution loop in the Rossby and drift turbulence life cycle.
The feedback loop has been described based on the the nonlocal WT theory of drift and Rossby waves and has been theoretically investigated and numerically validated.  This is one candidate for the explanation of the LH transition previously mentioned.  The instability, modelled here as an extra forcing term in the CHM equation at the scale of the ion Larmor or deformation radius, generates the (predominantly meridional) small-scale drift or Rossby waves, whose energy initially grows exponentially.  The nonlinear evolution of the energy spectrum progresses as diffusion along 1D curves in $\bf k$-space, increasing the diffusion coefficient as it does so, resulting in faster diffusion to the large scales.  The increasing diffusion battles with the growth of the instability until a balance is reached whereby the instability is effectively suppressed and the large scale zonal flow saturates at a level of the order defined by Eq.~(\ref{Uzf_sat}). We have also extended the analysis on a qualitative level beyond the weak WT approach and derived another saturation estimate, Eq.~(\ref{Uzf_satrd}), which is valid for strong WT produced in the cases when the instability is strong.

Using an idealised narrowband instability forcing, we have presented a numerical validation of the theoretical predictions for the instability suppression and the ZF saturation both in the case of the weak and the strong forcings given by Eqs.~(\ref{Uzf_sat}) and~(\ref{Uzf_satrd})
respectively. 
The idealised forcing produces well the saturation mechanism.  We have then made numerical simulations with a broadband instability forcing mimicking the baroclinic or ITG instability. We have observed qualitatively similar behavior as in the case of the narrowband forcing in the weak instability case. Namely, the instability get suppressed by the emerging ZFs and the latter saturate. However, for strong instabilities the ZFs fail to suppress the primary intability
and an unbounded growth of the small-scale turblence is observed. This behavior is a prototype of a well-discussed phenomena in the nuclear fusion community, the Dimits shift -- a numerically observed effect of the instability threshold being higher in terms of the temperature gradient than that predicted by the linear ITG theory \cite{dimits_comparisons_2000}. Indeed, our results suggest that in a certain range above the linear instability threshold the emerging ZFs act to suppress the instability growth, whereas for strong enougth instability (strong temperature gradient) the ZFs are not able to suppress the growth at the small scales. 

We have also discussed the results for drift wave and ZF feedback loop in the extended models, EHM, HW and EHW, and have shown that the qualitative picture in these models remains similar to the basic forced-dissipated CHM model. 

Thus we see that the CHM, being the most basic PDE model, is capable of capturing, at least at a qualitative level, the most nontrivial and profoud effects observed in the real Rossby and drift wave systems.
Whether or not these results could be of use on a quantitative level for real tokamak experiments depends on whether the weak turbulence regime is observed.  The strong turbulence scaling in Eq.~(\ref{Uzf_satrd}), appropriate for $U \gtrsim \frac{\beta}{F}$ has been widely used in tokamak theory~\cite{Kingsbury1994} for some time now (usually without the finite gyroradius correction present in Eq.~(\ref{Uzf_satrd})). On the other hand, the weak turbulence estimate in Eq.~~(\ref{Uzf_sat}) is more recent~\cite{Nazarenko2011} and should be used when $U \lesssim\frac{\beta}{F}$.  Zonal flows in the DIIID and JET tokamaks have been reported with $U \sim 5-35\,{\rm km\,sec}^{-1}$~\cite{Burrell2004,Crombe2006_proc}, whereas $\frac{\beta}{F}$ for both systems could range from $5\,{\rm km\,sec}^{-1}$ in the core to $100\,{\rm km\,sec}^{-1}$ in the pedestal regions.  Thus, both the weak and the strong turbulent regimes should be realistically realisable.

\subsection*{Acknowledgments}
This research was supported in part by the National Science Foundation under Grant No. NSF PHY11-25915. CC is grateful for the hospitality of the Okinawa Institute of Science and Technology Graduate University and the Kavli Institute for Theoretical Physics during the preparation of this manuscript. Part of the manuscript preparation time SN spent at 
the SPEC laboratory at CEA Saclay in France being supported by the grant
Chaire du Labex PALM, which is gratefully acknowledged.

%%%%%%%%%%%%%%%%%%%%%%%%%%%%%%%%%%%%%%%%%%%%%%%%%%%%%%%%%%%%%%%%%%%%%%%%%%%%%%%%%%%%%%%%%%%%%%%%%%%%
\appendix 
\renewcommand*{\thesection}{\Alph{section}}

%\section{The Charney-Hasegawa-Mima Equation}
%\label{apxCHM}
%\input app-CHM.tex

%%%%%%%%%%%%%%%%%%%%%%%%%%%%%%%%%%%%%%%%%%%%%%%%%%%%%%%%%%%%%%%%%%%%%%%%%%%%%%%%%%%%%%%%%%%%%%%%%%%%
\section{Nonlocal interaction with  zonal flows}
\label{apx_nonlocal}
In this appendix we explain the details of the reduction of the kinetic equation to a diffusion equation in k-space in the case of nonlocal turbulence and derive the curves along which spectral transport of small scale turbulence occurs due to nonlocal interactions. The starting point for the analysis is the Doppler-shifted kinetic equation, Eq.~(\ref{kinetic_eqn_doppler}), in the frame moving with the drift velocity. The doppler shifted frequency is
\begin{equation}
\label{eq-Omegak}
\Omega_\kv = \frac{\beta}{F}\,k_x + \w_\kv = \frac{\beta}{F}\frac{k_x\,k^2}{k^2+F}.
\end{equation}
The resonant manifolds are not altered by this shift due to the presence of the momentum delta function. If we consider small scale turbulence having typical wave vector $\kv$, the corresponding resonant manifold intersects the $k_{1x}=0$ axis at two points:
\begin{eqnarray}
\label{eq-P1}{\mathrm P}_1 &=& (0,0)\, ,\\
\label{eq-P2}{\mathrm P}_2 &=& (0,-2\,k_y).
\end{eqnarray}
The point $P1$ corresponds to an interaction with a large scale zonal flow. The second corresponds to interaction with a relatively small scale zonal flow. In the latter case, "small scale" is not meant in any asymptotic sense since the wavenumber of the zonal flow is only a factor of two different from the wavenumber of the small scale turbulence. The zonal flow described by the point $P_2$ is certainly small scale with respect to that described by the point $P_1$ although it may only be moderately small scale with respect to the wave number $\kv$, of the small scale turbulence itself. Nonlocal interaction with each of the points $P_1$ and $P_2$ produce spectral transport which is diffusive in $\kv$-space although the details differ considerably. We will now consider each case in turn beginning with the large scale case and then moving on to the small scale case.

\subsection{Nonlocal interaction with large scale zonal flows}
\label{sec-app-largeScaleZF}
Let us introduce a large scale reference wavenumber, $K$, such that $K\ll k$ and $K \ll \sqrt{F}$. We can split the right hand side of Eq.~(\ref{kinetic_eqn_doppler}) as 
\begin{equation}
\label{eq-KE3}\pd{n_\kv}{t} =  \int_{k_1<K}\mathrm{Coll}\left[n_\kv, n_{\kv}, \kv,\kv_1\right]\,d\kv_1 + \int_{k_1\geq K}\mathrm{Coll}\left[n_\kv, n_{\kv_1}, \kv,\kv_1\right]\,d\kv_1,
\end{equation}
where the collision integral is as written in Eq.~(\ref{kinetic_eqn_doppler}). If the interaction is nonlocal then contributions to the collision integral for $n_\kv$ come mainly from scales in the neighbourhood of the point $P_1$. Therefore we neglect the second term compared with the first. The collision integral can be further simplified since we assume that the amplitude of the small scale turbulence, $n_\kv$, is small compared to the amplitude of the large scales, $n_{\kv_1}$. Since $k_1 \ll k$, $n_{\kv-kv_1}$ is also small compared to $n_{\kv_1}$. Therefore we can neglect the term $n_\kv n_{\kv-\kv_1}$. Furthermore, $\mathrm{sign}(\omega_{\bt k} \omega_{{\bt k} - {\bt k}_1})=1$ as ${\bt k}_1 \to 0$.  The term in the square brackets of Eq.~(\ref{kinetic_eqn_doppler}) then becomes
\begin{displaymath}
(n_{{\bt k}-{\bt k}_1} - n_{\bt k})n_{{\bt k}_1}\, .
\end{displaymath}
Let us denote the simplified integrand by
\begin{equation}
\label{collisionF1}
F({\bt k},{\bt k}_1) = 4\pi \left|  V^{\bt k}_{{\bt k}_1\, {\bt k}-{\bt k}_1} \right|^2  \delta( \Omega_{\bt k} - \Omega_{{\bt k}_1} - \Omega_{{\bt k}-{\bt k}_1})(n_{{\bt k}-{\bt k}_1} - n_{\bt k})n_{{\bt k}_1},
\end{equation}
Using the symmetries of $ V^{\bt k}_{{\bt k}_1\, {\bt k}-{\bt k}_1}$ and $\Omega_{\bt k}$, it can be shown that 
\begin{equation}
F({\bt k},{\bt k}_1) = - F({\bt k}-{\bt k}_1,-{\bt k}_1)
\end{equation} 
Eq.~(\ref{kinetic_eqn_doppler}) can then be written as
\begin{eqnarray}
\nonumber
\frac{\partial n_{\bt k}}{\partial t} & = & \frac{1}{2} \int_{k_1<K} F({\bt k},{\bt k}_1)\, d{\bt k}_1 
  -\frac{1}{2} \int_{k_1<K} F({\bt k}-{\bt k}_1,-{\bt k}_1)\, d{\bt k}_1 \\
\label{F_kin_eq2}
& = & \frac{1}{2}\int_{k_1<K}  F({\bt k},{\bt k}_1)\, d{\bt k}_1 
  -F({\bt k}+{\bt k}_1,{\bt k}_1)\, d{\bt k}_1\,.
\end{eqnarray} 
Since $k_1$ is small, we can Taylor expand $F({\bt k}+{\bt k}_1,{\bt k}_1)$ with respect to ${\bt k}_1$ in the first argument:
\begin{equation}
\label{taylorFkp}
F({\bt k}+{\bt k}_1,{\bt k}_1) = F({\bt k},{\bt k}_1) + {\bt k}_1 \cdot \nabla_{\bt k} F({\bt k},{\bt k}_1) + \mathrm{O}(k_1^2)\,,
\end{equation}
where $\nabla_\kv = (\partial_{k_x},\partial_{k_y})$ is
the gradient operator in $\kv$-space. Substituting back into Eq.~\ref{F_kin_eq2} and neglecting second-order terms gives
\begin{equation}
\label{F_kin_eq3}
\frac{\partial n_{\bt k}}{\partial t} = -\frac{1}{2} \int_{{\bt k}_1<{\bt k}} {\bt k}_1 \cdot \nabla_{\bt k} F({\bt k},{\bt k}_1)\,.
\end{equation} 
We now Taylor expand $F({\bt k},{\bt k}_1)$ with respect to ${\bt k}_1$ 
To do so, we first Taylor expand the argument of the $\delta$-function. Taylor
expanding $\Omega_{\kv-\kv_1}$ with respect to $\kv_1$ gives:
\begin{displaymath}
\Omega_{\kv-\kv_1} = \Omega_\kv - \kv_1\cdot\nabla_\kv\Omega_\kv + O(k_1^2)
\end{displaymath} 
as $\kv_1\to 0$. Thus the difference 
$\Omega_{\vv{k}} - \Omega_{\kv-\kv_1}$ behaves as $\kv_1\cdot\nabla_\kv\Omega_\kv$ as $\kv_1\to 0$.
On the other hand, the remaining term $\Omega_{\kv_1}$ behaves as $k_1^2k_{1x}$ as
$\kv_1 \to 0$ as can be seen directly from Eq.~(\ref{eq-Omegak}). Thus the latter
can be neglected compared with the former in the limit $\kv_1\to 0$ and
we can replace
\begin{eqnarray}
\nonumber
\delta(\Omega_{\bt k} - \Omega_{{\bt k}-{\bt k}_1}) &=& \delta(\Omega_{\bt k} - (\Omega_{\bt k} - {\bt k}_1 \cdot \nabla_{\bt k} \Omega_{\bt k} + \mathrm{O}(k_1^2)) \\
\label{taylorOmega}
&\approx& \delta({\bt k}_1 \cdot \nabla_{\bt k} \Omega_{\bt k}).
\end{eqnarray}
Performing a similar expansion for the $n_{{\bt k}-{\bt k}_1}$ term gives
\begin{equation}
\label{collisionF2}
F({\bt k},{\bt k}_1) \approx 4\pi \left|  V^{\bt k}_{{\bt k}_1\, {\bt k}-{\bt k}_1} \right|^2  \delta({\bt k}_1 \cdot \nabla_{\bt k} \Omega_{\bt k}){\bt k}_1 \cdot \nabla_{\bt k} n_{\bt k}\,,
\end{equation}
and substituting Eq.~(\ref{collisionF2}) back into Eq.~(\ref{F_kin_eq3}), the kinetic equation for the small scales can then be written as an anisotropic diffusion equation in ${\bt k}$-space:
\begin{equation}
\label{diffusion_kxky}
\frac{\partial n_{\bt k}}{\partial t} = \frac{\partial}{\partial k_i} D_{ij}(k_x,k_y)\frac{\partial n_{\bt k}}{\partial k_j}
\end{equation} 
with the components of the diffusion tensor given by
\begin{equation}
\label{diffTensorD1}
D_{ij}(k_x,k_y) = 2\pi \int_{{k}_1<{K}} \left| V^{\bt k}_{{\bt k}_1\, {\bt k}-{\bt k}_1} \right|^2 \delta({\bt k}_1 \cdot \nabla_{\bt k} \Omega_{\bt k}), k_{1i}k_{1j} n_{{\bt k}_1} d{\bt k}_1\,.
\end{equation}
If we assume that the large scale turbulence is principally supported at scales $k_1 \ll K$, then the reference wavenumber, $K$, can  be extended to infinity in Eq.~\ref{diffTensorD1}. The components of the diffusion tensor are then
\begin{eqnarray}
\nonumber
\label{diffTensorD2}
D_{ij}(k_x,k_y) & = & 2\pi \int^{\infty}_{-\infty} \left| V^{\bt k}_{{\bt k}_1\, {\bt k}-{\bt k}_1} \right|^2 \delta \left(k_{1x}\frac{\partial \Omega_{\bt k}}{\partial k_x} + k_{1y}\frac{\partial \Omega_{\bt k}}{\partial k_y}\right) k_{1i}k_{1j} n_{{\bt k}_1}d{\bt k}_1 \\
\nonumber
& = & 2\pi \int^{\infty}_{-\infty} \left| V^{\bt k}_{{\bt k}_1\, {\bt k}-{\bt k}_1} \right|^2 \left|\frac{\partial \Omega_{\bt k}}{\partial k_x}\right|^{-1} \delta \left(k_{1x} + k_{1y} \sigma_\kv \right) k_{1i}k_{1j} n_{{\bt k}_1} d{\bt k}_1\\
\nonumber & = & 2\pi \left|\frac{\partial \Omega_{\bt k}}{\partial k_x}\right|^{-1} \int^{\infty}_{-\infty}  \left[ \left| V^{\bt k}_{{\bt k}_1\, {\bt k}-{\bt k}_1} \right|^2  \begin{pmatrix}
k_{1x}^2 & k_{1x}k_{1y}\\ k_{1x}k_{1y} & k_{1y}^2
\end{pmatrix} \,\delta \left(k_{1x} + k_{1y} \sigma_\kv \right)\, n_{{\bt k}_1}\right]\,dk_{1x}dk_{1y} 
\end{eqnarray}
where we have introduced the parameter $\sigma_{\bt k}$ for brevity, defined as,
\begin{equation}
\label{eq-sigmak}
\sigma_{\bt k} =\frac{\partial \Omega_{\bt k}}{\partial k_y} / \frac{\partial \Omega_{\bt k}}{\partial k_x} =  \frac{2 F k_x k_y}{k^4 +F(3k_x^2+k_y^2)}.
\end{equation}
We can now perform the integration with respect to $k_{1x}$ to obtain the final expression for the diffusion tensor:
\begin{equation}
\label{diffTensorD3}
D(\kv) =  \mathcal{D}_\kv\,\pd{\Omega_\kv}{k_{1x}}\, \begin{pmatrix}
\sigma_{\bt k}^2 & -\sigma_{\bt k}\\ -\sigma_{\bt k} & 1
\end{pmatrix}
\end{equation}
where the scalar function, $\mathcal{D}(\kv)$, contains all the integrations over the large scale zonal flow:
\begin{displaymath}
\mathcal{D}_\kv = 2\pi \left|\frac{\partial \Omega_{\bt k}}{\partial k_x}\right|^{-2} \int^{\infty}_{-\infty}  \left[ \left| V^{\bt k}_{{\bt k}_1\, {\bt k}-{\bt k}_1} \right|^2  n_{{\bt k}_1} \right]_{k_{1_x}=-\sigma_{\bt k} k_{1_y}}\, k_{1_y}^2 \,dk_{1_y}.
\end{displaymath}

A crucial point is that the matrix in Eq.~\ref{diffTensorD3} is singular.  The eigenvalues are $0$ and $1+\sigma_{\bt k}^2$ with corresponding eigenvectors $(\frac{1}{\sigma_{\bt k}},1)$ and $(-\sigma_{\bt k},1)$.  This suggests that diffusion is along one-dimensional curves, i.e. in the $(-\sigma_{\bt k},1)$ direction.  We now aim to find a change of coordinates which will make the one-dimensional nature of the diffusion explicit so that we can identify the curves along which diffusive transport occurs.  
The general structure of Eq.~\ref{diffusion_kxky} is
\begin{equation}
\label{genform_k}
\left(\frac{\partial}{\partial k_x}, \frac{\partial}{\partial k_y}\right)\Lambda(k_x,k_y)
\begin{pmatrix} \frac{\partial}{\partial k_x} \\ \frac{\partial}{\partial k_y} \end{pmatrix} n(k_x,k_y)\,.
\end{equation}
Under a general change of coordinates,
\begin{equation}
\label{eq-coordinateChange}
F : (k_x,k_y) \mapsto (\kappa_1(k_x,k_y), \kappa_2(k_x,k_y)),
\end{equation}
this equation transforms to  \cite{smith_change_1934} 
\begin{equation}
\label{genform_q} \pd{n_\kv}{t} = \left|\det J\right| \left(\pd{}{\kappa_1},\pd{}{\kappa_2}\right) \left|\det J\right|^{-1} J\, D(\kv)\, J^T \left( \begin{array}{c}\pd{}{\kappa_1}\\\pd{}{\kappa_2}\end{array}\right) \ n_\kv,
\end{equation}
where $J$ is the Jacobian matrix for the change of variables,
\begin{equation}
\label{jacobianq}
J=\begin{pmatrix}
   \frac{\partial q_1}{\partial k_x} & \frac{\partial q_1}{\partial k_y}\\ 
   \frac{\partial q_2}{\partial k_x} & \frac{\partial q_2}{\partial k_y}
\end{pmatrix}\,.
\end{equation}
and our notation tacitly assumes that in Eq.~(\ref{genform_q}) we have used the inverse of the coordinate transformation, Eq.~(\ref{eq-coordinateChange}), to express all $\kv$-dependence of $n_\kv$ and $D(\kv)$ in terms of $\kappa_1$ and $\kappa_2$. The task is to find the correct change of variables, $F$, so that Eq.~(\ref{genform_q}) becomes a one-dimensional diffusion. If we denote the Jacobian of $F$ schematically by
\begin{equation}
% \label{jacobian_abcd}
\nonumber
J=\begin{pmatrix}
   a & b\\ 
   c & d
\end{pmatrix}\,
\end{equation}
then we seek a transformation which satisfies
\begin{equation}
% \label{diagonalJLambdaJT}
\nonumber
J\Lambda J^T=\begin{pmatrix}
   (b-a\sigma_{\bt k})^2 & (b-a\sigma_{\bt k})(d-c\sigma_{\bt k})\\ 
   (b-a\sigma_{\bt k})(d-c\sigma_{\bt k}) & (d-c\sigma_{\bt k})^2
\end{pmatrix} = 
\begin{pmatrix} 1 & 0 \\ 0 & 0
\end{pmatrix}\,.
\end{equation}
This is true if $b-a\sigma_{\bt k} = 1$ and $d-c\sigma_{\bt k} = 0$.
The new coordinates must therefore satisfy the equations
\begin{eqnarray}
\nonumber
\frac{\partial \kappa_1}{\partial k_y} - \sigma_{\bt k} \frac{\partial \kappa_1}{\partial k_x} & = & 1\\ 
\nonumber
\frac{\partial \kappa_2}{\partial k_y} - \sigma_{\bt k} \frac{\partial \kappa_2}{\partial k_x} & = & 0\,.
\end{eqnarray}

Keeping in mind Eq.~(\ref{eq-sigmak}), the solution of these equations is 
\begin{eqnarray}
\nonumber
\kappa_1(k_x,k_y) & = & k_y\\
\label{changevar_q}
\kappa_2(k_x,k_y) & = & \Omega(k_x,k_y).
\end{eqnarray}
The Jacobian matrix is
\begin{equation}
% \label{jacobian_kyOmega}
\nonumber
J=\begin{pmatrix}
   0 & 1 \\ 
   \frac{\partial \Omega_{\bt k}}{\partial k_x} & \frac{\partial \Omega_{\bt k}}{\partial k_y} 
\end{pmatrix}\hspace{1.0cm}\mbox{and}\hspace{1.0cm} \left|\det J\right| = \left|\pd{\Omega_\kv}{k_x}\right|\,.
\end{equation}
In the new coordinates,  the diffusion Eq.~(\ref{diffusion_kxky}) is now one-dimensional
\begin{equation}
\label{eq-diffusion3}
\pd{n_\kv}{t} = \pd{\Omega_\kv}{k_x} \pd{}{\kappa_1} \left(D(\kv) \pd{n_\kv}{\kappa_1}\right).
\end{equation}
In the $(\kappa_1,\kappa_2)$ plane Eq.~(\ref{eq-diffusion3}) describes one dimensional 
diffusion in the $\kappa_1$ direction with $\kappa_2$ being constant. Translating this back into the original $\kv$-plane, we have
\begin{equation}
\label{eq-diffusion_kyOmega}
\frac{\partial n_{\bt k}}{\partial t} =\frac{\partial \Omega_{\bt k}}{\partial k_x} \frac{\partial}{\partial k_y}\left( D(\kv)\left(\frac{\partial n_{\bt k}}{\partial k_y}\right)_\Omega\right)_\Omega \, ,
\end{equation} 
where $\left(\pd{\cdot }{k_y}\vphantom{\frac{1}{2}}\right)_\Omega$ means differentiation with respect to $k_y$ with $\Omega_\kv$ being held constant and the diffusion coefficient is defined by Eq.(\ref{diffTensorDtilde}). Illustrative examples of the curves $\Omega_\kv=$constant are plotted in Fig.~\ref{levelsets_Omega}(a).

\subsection{Nonlocal interaction with small scale zonal flows}
\label{sec-app-smallScaleZF}

We now ouline the analogous calculation assuming that the evolution
is dominated by interaction with the point $P_2=(0,2 k_y)$ corresponding to
small scale zonal flows. These results were first
presented in \cite{Nazarenko1991}.

Following the same reasoning as before, we introduce a reference scale, $K$,
satisfying $K \ll k$ and $K\ll 1/\rho$, and neglect the contributions to
the collision integral coming from outside of a ball of radius $K$
of the point $\pv=(0, 2\,k_y)$. We thus approximate Eq.~(\ref{kinetic_eqn_doppler}) by
\begin{eqnarray}
\nonumber \pd{n_\kv}{t}&=& \int_{\left|\pv-\kv_1\right|<K} d\kv_1\,F(\kv,\kv_1)\\
\label{eq-ssKE1} &=&-\int_{\left|\pv-\kv_1\right|<K} d\kv_1\,F(\kv-\kv_1,-\kv_1).
\end{eqnarray}
Taylor expanding the first argument about $\kv_1=\pv$ gives
\begin{eqnarray}
\nonumber \pd{n_\kv}{t}&=& -\int_{\left|\pv-\kv_1\right|<K} d\kv_1\left[F(\kv-\pv,-\kv_1) + (\kv_1-\pv)\cdot\nabla_{\kv_{1}} \left. F(\kv-\kv_1,\kv_1)\right|_{\kv_1=\pv} + O(\left|\pv-\kv_1\right|^2)\right]\\
\label{eq-ssKE2} &\approx& = -\int_{\left|\pv-\kv_1\right|<K} d\kv_1\left[ F(\kv^*,-\kv_1) - (k_{1x}, k_{1y}-2\,k_y) \cdot
\left(\partial_{k_x}F(\kv^*,-\kv_1), \partial_{k_y}F(\kv^*,-\kv_1)\right)\right].
\end{eqnarray}
Here for brevity we used the shorthand notation
\begin{equation}
\kv^* = \kv-\pv = (k_x,-k_y).
\end{equation}
We now expand $F(\kv^*,-\kv_1)$ about $\kv_1=\pv$. Since $V_{\kv_1\,\kv\!-\!\kv_1\,\kv}$
and $n_{\kv_1}$ both vary rapidly near $\kv_1=\pv$, we should only expand the
$n_{\kv^*+\kv_1}-n_{\kv^*}$ and $\delta(\Omega_{\kv*}-\Omega_{-\kv_1}-\Omega_{\kv^*+\kv_1})$ terms. Let us look at these two terms in turn.
\begin{eqnarray}
\nonumber n_{\kv^*+\kv_1}-n_{\kv^*} &=& n_{\kv^*+\pv}-n_{\kv^*}
+ (\kv_1-\pv)\cdot \nabla_{\kv_1^\prime}\left. n(\kv^*+\kv_1^\prime)\right|_{\kv_1^\prime=\pv} + O(\left|\pv-\kv_1\right|^2)\\
\label{eq-TaylorF}&\approx& n_{\kv}-n_{\kv^*} + (k_{1x}, k_{1y}-2\,k_y) \cdot \left(\partial_{k_x}n_\kv,\partial_{k_y}n_\kv\right).
\end{eqnarray}
Unlike the expansion of the collision integral about the point
$(0,0)$ in Sec.~\ref{sec-app-largeScaleZF} above, the leading order term in the Taylor expansion of
$F(\kv^*,-\kv_1)$ about $\kv_1=\pv$ is not necessarily zero. It vanishes only if the
spectrum is symmetric about the $k_x$ axis (ie $n(k_x,k_y)=n(k_x,-k_y)$). We shall
return to this point below. Next, let us look at the argument of the 
$\delta$-function near $\kv_1=\pv$.
\begin{eqnarray}
\nonumber \Omega_{\kv^*}-\Omega_{-\kv_1}-\Omega_{\kv^*+\kv_1}  &=& \Omega_{\kv^*}-\Omega_{-\pv}-\Omega_{\kv} - k_{1x}\left[\pd{\Omega}{q^\prime_1}(-q^\prime_{1},-q^\prime_{2}) + \pd{\Omega}{q^\prime_{1}}(k_x+q^\prime_{1},-k_y+q^\prime_{2}) \right]_{(q^\prime_{1},q^\prime_{2}) = (0, 2\,k_y)}\\
\nonumber & & - (k_{1y}-2\,k_y)\,\left[\pd{\Omega}{q^\prime_{2}}(-q^\prime_{1},-q^\prime_{2}) + \pd{\Omega}{q^\prime_{2}}(k_x+q^\prime_{1},-k_y+q^\prime_{2}) \right]_{(q^\prime_{1},q^\prime_{2}) = (0, 2\,k_y)} + O(\left|\pv-\kv_1\right|^2)\\
&\approx& -\pd{\Omega_\kv}{k_y}\left( k_{1y} - 2\,k_y+\xi_\kv\,k_{1x}\right),
\end{eqnarray}
where
\begin{equation}
\label{eq-xi}
\xi_\kv = \frac{3F(k_x^2 -k_y^2) + k_x^4+6k_x^2k_y^2-3k_y^4}{2 k_x k_y (F+4 k_y^2)}.
\end{equation}
The various derivatives have been computed from Eq.~(\ref{eq-Omegak}). For example: 
\begin{displaymath}
\left[ \pd{\Omega}{q^\prime_{2}}(k_x+q^\prime_{1},-k_y+q^\prime_{2})\right]_{(q^\prime_{1},q^\prime_{2}) = (0, 2\,k_y)} =\frac{2 \beta k_xk_y}{( k^2+F)^2} = \pd{\Omega_\kv}{k_y},
\end{displaymath}
and so forth.  The leading order term is zero since $\kv,\pv$ and 
$\kv^*$ are resonant. Putting this together, the Taylor expansion of the 
$\delta$-function is
\begin{equation}
\label{eq-TaylorDeltaFn}
\delta(\Omega_{\kv^*}-\Omega_{-\kv_1}-\Omega_{\kv^*+\kv_1}) \approx \left| \pd{\Omega_\kv}{k_y} \right|^{-1}\,\delta \left( k_{1y} - 2\,k_y+\xi_\kv\,k_{1x}\right).
\end{equation}
From Eqs.~(\ref{eq-ssKE2}), (\ref{eq-TaylorF}) and (\ref{eq-TaylorDeltaFn}) we
see that the leading order term in the kinetic equation coming from nonlocal
interaction with small scale zonal flows is:
\begin{equation}
\label{eq-ssKE3}
\pd{n_\kv}{t} = Y_\kv\, \left[n(k_x,-k_y)-n(k_x,k_y)\right],
\end{equation}
where
\begin{eqnarray}
\nonumber Y_\kv &=& 4\pi \left| \pd{\Omega_\kv}{k_y} \right|^{-1} \int_{\left|\pv-\kv_1\right|<K} dk_{1x} dk_{1y}  \left|V_{-\kv_1\,\kv\!+\!\kv_1\,\kv^*}\right|^2 \,\delta(k_{1y} - 2\,k_y+\xi_\kv\,k_{1x})\, n(k_{1x},k_{1y})\\
\label{eq-Yk} &=& 4\pi \left| \pd{\Omega_\kv}{k_y} \right|^{-1} \int_{\left|k_{1x}\right|<K} \left[\left|V_{\kv_1\,\kv\!-\!\kv_1\,\kv}\right|^2 n_{\kv_1}\right]_{k_{1y}=2\,k_y} dk_{1x}.
\end{eqnarray}
We have used the symmetries, Eqs.~(\ref{eq-Vpqr}) and (\ref{eq-Vsymmetries1}), 
of $V^{\kv_1}_{\kv\!-\!\kv_1\,\kv}$ and relabelled the integration variable $q_1 \to -q_1$
to bring the interaction coefficient to a neater form. At the final step the
delta function has been used to integrate out $q_2$ to leading order. Eq.~(\ref{eq-ssKE3}) tells us that the leading order 
effect of interactions with small scale zonal flows is to cause the spectrum
to relax to a state which is symmetric about the $k_x$ axis, a point which
was first made in \cite{Nazarenko051990,Nazarenko081990}.

If we wish to observe any redistribution of spectral energy density due to
the interactions with small scale zonal flows, we need to consider the higher
order terms in Eq.~(\ref{eq-ssKE2}). The first order term vanishes after 
integration over $\kv_1$ since the integrand is an odd function of $\pv-\kv_1$. The
next contribution is therefore the second order one. From Eqs.~(\ref{eq-ssKE2})
and (\ref{eq-TaylorF}) the second order contribution can again be
presented as an anisotropic diffusion equation in $\kv$-space:
\begin{equation}
\label{eq-ssDiffusion1}
\pd{n_\kv}{t} = \pd{}{k_i}\left(\,B_{ij}(\kv) \pd{n_\kv}{k_j}\right)
\end{equation}
where the diffusion tensor is given by the matrix
\small{
\begin{eqnarray}
\nonumber B(\kv) &=& 4 \pi \left| \pd{\Omega_\kv}{k_y} \right|^{-1} \int_{\left|\pv-\kv_1\right|<K} dk_{1x} dk_{1y}  \left|V_{-\kv_1\,\kv\!+\!\kv_1\,\tilde{\kv}}\right|^2 \,\delta(k_{1y} - 2\,k_y+\xi_\kv\,k_{1x})\, n(k_{1x},k_{1y}) \left(\begin{array}{cc}k_{1x}^2&k_{1x}(k_{1y}-2\,k_y) \\ k_{1x}(k_{1y}-2\,k_y)&(k_{1y}-2\,k_y)^2\end{array} \right)\\
\label{eq-B}&=& \left| \pd{\Omega_\kv}{k_y} \right|^{-1}\, \mathcal{B}(\kv)\, \left(\begin{array}{cc}1&-\xi_\kv\\-\xi_\kv&\xi_\kv^2\end{array} \right)  
\end{eqnarray}
}
where we have performed the same manipulations as those used to arrive at
Eq.~(\ref{eq-Yk}) above and, in common with $\mathcal{D}(\kv)$ in Eq.~(\ref{diffTensorD3}), defined the scalar quantity
\begin{equation}
\label{eq-Btilde}
\mathcal{B}(\kv) = 4\pi\,\int_{\left|k_{1x}\right|<K} \left[ \left|V_{\kv_1\,\kv\!-\!\kv_1\,\kv}\right|^2 n(-k_{1x},k_{1y})\, k_{1x}^2\right]_{k_{1y}=2\,k_y} dk_{1x}.
\end{equation}
The diffusion tensor, $B(\kv)$, is again singular
indicating that Eq.~(\ref{eq-ssDiffusion1}) should be reducible to
a one-dimensional diffusion equation by an appropriate change of
variables. Following the same proceedure as in Sec.~\ref{sec-app-largeScaleZF}, the change of coordinates (\ref{eq-coordinateChange}) will bring this equation to one-dimensional form provided that the new coordinates, $(\kappa_1,\kappa_2)$, satisfy the equations 
\begin{eqnarray*}
\label{eq-kappa1Eqn}\pd{\kappa_1}{k_x} - \xi_\kv\,\pd{\kappa_1}{k_y} &=& 1\\
\label{eq-kappa2Eqn} \pd{\kappa_2}{k_x} - \xi_\kv\, \pd{\kappa_2}{k_y} &=& 0.
\end{eqnarray*}
The first equation can be easily solved by inspection while the second requires a 
little more effort (using the method of characteristics). The result is
\begin{eqnarray}
\label{eq-kappa1}\kappa_1(\kv) &=& k_x\\
\nonumber \kappa_2(\kv) = Z_\kv &=& \arctan \left( \frac{(k_y + \sqrt{3}\, k_x)\sqrt{F}}{k^2}\right)
-\arctan \left( \frac{(k_y - \sqrt{3}\, k_x)\sqrt{F}}{k^2}\right)\\
\label{eq-Z} &&-\frac{2\sqrt{3\,F}\,\,k_x}{k^2+F}.
\end{eqnarray}
With this done, $\det J = \pd{Z_\kv}{k_y}$, so that, 
according to Eq.~(\ref{genform_q}), the diffusion equation, 
Eq.~(\ref{eq-ssDiffusion1}), transforms into
\begin{equation}
\label{eq-ssDiffusion2}
\pd{n_\vv{\kappa}}{t} =   \pd{Z_\kv}{k_y} \pd{}{\kappa_1} \left[ \left( \pd{Z_\kv}{k_y} \right)^{-1} \left( \pd{\Omega_\kv}{k_y} \right)^{-1}\, \tilde{B}(\kv)\, \pd{n_\vv{\kappa}}{\kappa_1}\right],
\end{equation}
where $\tilde{B}(\kv)$ is given by Eq.~(\ref{eq-Btilde}).
In the $(\kappa_1,\kappa_2)$ plane Eq.~(\ref{eq-ssDiffusion2}) describes one 
dimensional diffusion in the $\kappa_1$ direction with $\kappa_2$ constant. 

Similarly to what we have done before in Eq.~(\ref{eq-diffusion_kyOmega}), we translate Eq.~(\ref{eq-ssDiffusion2}) back into the $(k_x,k_y)$ plane
%, using Eqs.~(\ref{eq-kappa1}) and (\ref{eq-Z}), Eq.~(\ref{eq-ssDiffusion2}) therefore describes one-dimensional 
making explicit the process of 
diffusion  along lines of constant $Z_\kv$:
\begin{equation}
\label{eq-ssDiffusion3}
\pd{n_\vv{\kappa}}{t} =   \pd{Z_\kv}{k_y} \pd{}{k_x} \left( \left( \pd{Z_\kv}{k_y} \right)^{-1} \left( \pd{\Omega_\kv}{k_y} \right)^{-1}\, \tilde{B}(\kv)\, \left(\pd{n_\vv{\kappa}}{k_x}\right)_Z\right)_Z,
\end{equation}
where $\left(\pd{\cdot }{k_x}\vphantom{\frac{1}{2}}\right)_Z$ means differentiation with respect to $k_x$ with $Z_\kv$ being held constant. 
Illustrative 
examples of the curves $Z_\kv=$constant are plotted in Fig.~\ref{levelsets_Omega}(b).

%%%%%%%%%%%%%%%%%%%%%%%%%%%%%%%%%%%%%%%%%%%%%%%%%%%%%%%%%%%%%%%%%%%%%%%%%%%%%%%%%%%%%%%%%%%%%%%%%%%%

%% References


%\section*{References}
\bibliographystyle{ieeetr}  % want neuron
\bibliography{bibliography,additional}






% \end{thebibliography}

\end{document}